\documentclass[
11pt,
openany,
twoside,
english,
b5paper
]{memoir}

\epigraphfontsize{\small\itshape}
\makeatletter
\let\ifprint\if@twoside
\makeatother

\makeatletter
\def\ignorecitefornumbering#1{%
     \begingroup
         \@fileswfalse
         #1
    \endgroup
}
\makeatother

\newcommand{\thesistitlefront}{%
  { \centering\fontsize{18pt}{21pt}\selectfont
    \spread{Transport and information in~open~quantum~systems}\par
  }
}

\usepackage[utf8]{inputenc}
\usepackage[T1]{fontenc}
\usepackage[english]{babel}

\usepackage{graphicx}
\usepackage{sidecap}
\newsubfloat{figure}

\usepackage{lmodern} 

\usepackage{microtype}

\usepackage[hyperref]{xcolor}

\usepackage{amsmath}
\usepackage{amssymb}
\usepackage{amsfonts}

\usepackage{bm}
\usepackage{esvect}
\usepackage{upgreek}

\usepackage{mathtools}
\usepackage{physics}

\usepackage[amsmath,thmmarks,hyperref]{ntheorem}

\usepackage{xspace}

\usepackage{varioref}

\usepackage{url}

\usepackage{soul}

\usepackage{bibentry}

\usepackage{enumitem}

\usepackage{lipsum}
\usepackage[square,numbers,sort&compress]{natbib}

\usepackage{mflogo}

\usepackage{refcount}

\ifpdf
  \usepackage[
    pdftitle={KasperPoulsenPhDThesis},
    pdfauthor={Kasper Poulsen},
    pdfkeywords={Quantum mechanics, quantum transport, open quantum system, rectification, Maxwell's demon, quantum simulation},
    pdfdisplaydoctitle=true,
    bookmarksopen=true,
    pdftex]{hyperref}[2024/01/11]
\fi

\setlxvchars[\normalfont]

\settypeblocksize{*}{1.08\lxvchars}{1.61803}
\settypeblocksize{19cm}{1.1\lxvchars}{*}

\ifprint
  \setlrmargins{*}{*}{1.33}
\else
  \setlrmargins{*}{*}{1}
\fi
\setulmargins{*}{*}{1.1}

\checkandfixthelayout

\newcommand{\foliofont}{\sffamily}
\newcommand{\headerfont}{\sffamily}

\makepagestyle{rvthesis}
\makeevenhead{rvthesis}{\foliofont\thepage}    {} {\headerfont\leftmark}
\makeoddhead {rvthesis}{\headerfont\rightmark} {} {\foliofont\thepage}
\makeevenfoot{rvthesis}{}{}{}
\makeoddfoot {rvthesis}{}{}{}

\makeatletter
\makepsmarks {rvthesis}{
  \createmark{chapter}    {both}  {shownumber} {\@chapapp\ } {\ $\cdot$\ }
  \createmark{section}    {right} {shownumber} {}            { \ }
  \createplainmark{toc}   {both}  {\contentsname}
  \createplainmark{lof}   {both}  {\listfigurename}
  \createplainmark{lot}   {both}  {\listtablename}
  \createplainmark{bib}   {both}  {\bibname}
  \createplainmark{index} {both}  {\indexname}
}
\makeatother
\nouppercaseheads
\copypagestyle{chapter}{empty}
\makeoddfoot{chapter}{}{\foliofont\thepage}{}
\makeevenfoot{chapter}{}{\foliofont\thepage}{}

\maxsecnumdepth{subsection}
\settocdepth{subsection}

\setsecheadstyle{\Large\bfseries\sffamily\raggedright}
\setsubsecheadstyle{\large\bfseries\sffamily\raggedright}

\definecolor{CHAPblue}{HTML}{5B6B9A} 

\definecolor{CHAPgray}{HTML}{111F33}
\makechapterstyle{thesischap}{%
  \chapterstyle{default}
   \setlength{\afterchapskip}{40pt}

  \setlength{\beforechapskip}{5mm}
  \setlength{\midchapskip}{\paperwidth}
  \addtolength{\midchapskip}{-\textwidth}
  \addtolength{\midchapskip}{-\spinemargin}
   \makeoddfoot{plain}{}{}{\thepage}}

\chapterstyle{thesischap}

\subcaptionstyle{\raggedright}

\captionwidth{.9\linewidth}
\changecaptionwidth

\captionnamefont{\small}
\captiontitlefont{\small}

\makeatletter
\@ifpackageloaded{hyperref}{
    \hypersetup{colorlinks=false,
      pdfborder=0 0 0}
  \addto\extrasenglish{
  }
}{}
\makeatother

\sodef\spread{}{.2em}{.9em plus.4em}{1em plus.1em minus.1em}

\usepackage{braket}
\newcommand{\us}{\uparrow}
\newcommand{\ds}{\downarrow}

\newcommand\identity{1\kern-0.25em\text{l}}

\makeatletter
\newcommand*\bigcdot{\mathpalette\cdot@{.5}}
\newcommand*\bigcdot@[2]{\mathbin{\vcenter{\hbox{\scalebox{#2}{$\m@th#1\bullet$}}}}}
\makeatother

\strictpagechecktrue

\usepackage{bibentry}
\usepackage{bbm}
\nobibliography*

\begin{document}

\frontmatter
\pagenumbering{Roman}
\pagenumbering{gobble}

\begin{titlingpage}
  \newlength{\frontpagecorrection}
  \calccentering{\frontpagecorrection}
  \begin{adjustwidth*}{\frontpagecorrection-2cm}{-\frontpagecorrection-2cm}
    
    \centering
    \scshape
    
    {\color{white} blank}
	\vspace{0.5cm}    
    
    \thesistitlefront
    
    \vspace{1.7cm}
    \begin{figure}[h!]
	\centering
	\hspace{0.5cm}\scalebox{+1}[1]{\includegraphics[width=0.85\columnwidth]{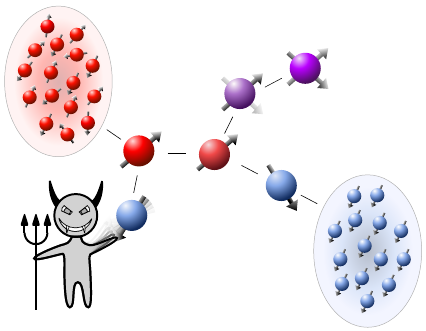}}
	\end{figure}

    \vspace{1cm}

    \fontsize{14pt}{18pt}\selectfont
    \spread{Kasper Poulsen}\par

    \bigskip
    
    \spread{PhD thesis}\par
    2nd edition\par
    
    \vfill
    
    \fontsize{12pt}{14.5pt}\selectfont
    Department of Physics and Astronomy\par
    Aarhus University 
  \end{adjustwidth*}
  
\clearpage  
 
{\color{white} blank}

\vspace{10.5cm} 

\noindent \copyright \, Copyright, Kasper Poulsen \\
1st edition, October 2023 \\
2nd edition, January 2024 \\

\noindent Department of Physics and Astronomy\\
Aarhus University\\
Ny Munkegade 120\\
8000 Aarhus C\\
Denmark\\
E-mail: phys@au.dk\\
Tel.: +45 8715 0000\\

\noindent Coverimage: Visualization of an open \\
quantum system. Made with Inkscape.\\

\noindent Typeset in \LaTeX\\
Figures in Python and Inkscape

\end{titlingpage}



\clearpage
\thispagestyle{plain}
\par\vspace*{.45\textheight}{\centering \textit{\large{
Dedicated to Kim, Anne-Kathrine, Sophie, and Lia,\\who made this journey possible.
}}\par}

\clearpage\mbox{}\clearpage

\pagenumbering{roman}

\chapter*{English summary}

With the approaching second quantum revolution, the study of quantum thermodynamics, particularly heat flow, has become even more relevant for two main reasons. First, understanding heat and other types of noise is essential for protecting quantum information and preventing decoherence. Second, the ability to manufacture and control quantum systems developed for the quantum computer allows for experimental study of quantum thermodynamics in entirely new settings. 

In this thesis, several systems involving quantum systems in contact with baths are studied theoretically in experimentally available settings. First, two rectification or diode setups for heat currents are proposed using a dark-state mechanism. In one system, the dark-state mechanism is imperfect but very robust. In the other system, the dark-state mechanism relies on quantum entanglement and is much better but more fragile towards decoherence. Next, a quantum version of the Wheatstone bridge is built using the same entanglement-powered dark state mechanism. The mechanism is broken at the balance point, resulting in a dramatic change in the entangled state, which makes it possible to measure a system coupling strength. After having studied several boundary-driven quantum systems, the lessons learned are generalized into resonance conditions using a general linear chain of weakly interacting chains of strongly interacting spins. 

The final two chapters focus on the ability to study statistical physics in realizable quantum systems. First, a Maxwell's demon setup is proposed. A demon-controlled qutrit is coupled to two non-Markovian baths. The information back-flow from the non-Markovian baths allows the demon to more effectively transfer heat from the cold bath to the hot bath. Second, the Mott insulator to superfluid phase transition in a lattice of transmons is examined. The ground state has a variable particle number and is prepared using adiabatic state preparation. This allows for the exploration of the entire phase diagram.

\chapter*{Dansk resumé}
\hyphenation{sig-nal-for-sin-kel-ser}
\hyphenation{sy-stem-dy-na-mik}

Studiet af kvantetermodynamik, og specielt varmes bevægelse, er kun blevet mere relevant med tilnærmelsen af den anden kvanterevolution af to hovedårsager. For det første er det essentielt at forstå varme, og andre former for støj, for at kunne beskytte kvanteinformation of forhindre dekohærens. For det andet muliggør den nyfundne evne til at bygge og kontrollere kvantesystemer til kvantecomputere eksperimentelle studier af kvantetermodynamik i helt nye situationer. 

Denne afhandling undersøger flere systemer teoretisk som involverer kvantesystemer i kontakt med bade og er indenfor eksperimental rækkevidde. Først bliver der foreslået, to forskellige dioder for varmestrømme som fungerer ved hjælp af en dark-state mekanisme. I det første system er mekanismen uperfekt men meget robust. I det andet system bygger mekanismen på en kvantesammenfiltringsmekanisme, og den er derfor langt bedre men også mere skrøbelig overfor dekohærens. Derefter bliver en kvantemekanisk version af Wheatstone broen bygget ved hjælp af den samme samfiltringsdrevne dark-state mekanisme. Denne mekanisme bliver ødelagt på balancepunktet hvilket resulterer i en dramatisk ændring i tilstanden og gør det muligt at måle en kvantemekanisk kobling. Efter at have studeret flere systemer koblet til bade i begge ender bliver de lærte principper generaliseret til resonans mekanismer. Dette gøres ved at kigge på et system a svagt interagerende kæder af kraftigt interagerende spin. 

De sidste to kapitler fokuserer på evnen til at studere statistisk fysik i realiserbare kvantesystemer. Et Maxwells dæmon system foreslås først. En dæmonstyret qutrit kobles til to ikke Markoviske bade. Information kan passere tilbage fra de ikke Markoviske bade hvilket tillader dæmonen at overføre varme fra det kolde bad til det varme bad mere effektivt. Derefter undersøges faseovergangen fra Mott isolator til supervæske i et gitter af transmon qubits. Grundtilstanden har et variabelt partikelnummer og bliver forberedt ved hjælp af adiabatisk tilstandsforberedelse. Dette tillader udforskning af hele fasediagrammet.

\begin{KeepFromToc}
\chapter{Preface}

This thesis was submitted to the Graduate School of Natural Sciences at Aarhus University, Denmark, in partial fulfillment of the requirement for a PhD degree in physics.
The research was carried out during the period between September 2018 and October 2023 under the supervision of Nikolaj T. Zinner.

A full list of my publications in chronological order follows the acknowledgments section. All sections in chapters \ref{chapter:QutritDiode}-\ref{chapter:Maxwell} have been reproduced from Refs.~\cite{PhysRevLett.126.077203, PhysRevA.105.052605, PhysRevE.105.044141, PhysRevLett.128.240401, PhysRevE.106.034116, poulsen2022heat}, with only minor changes in the following way: chapter \ref{chapter:QutritDiode} is based on Refs.~\cite{PhysRevE.106.034116, poulsen2022heat}, chapter \ref{chapter:Interference} is based on Ref.~\cite{PhysRevA.105.052605}, chapter \ref{chapter:WB} is based on Ref.~\cite{PhysRevLett.128.240401}, chapter \ref{chapter:GMR} is based on Ref.~\cite{PhysRevLett.126.077203}, and chapter \ref{chapter:Maxwell} is based on Ref.~\cite{PhysRevE.105.044141}.

In accordance with GSNS rules, parts of this thesis were also used in the progress
report for the qualifying examination.

\subsection*{Second edition}

After successfully defending this thesis on January 11, 2024, I have decided to put this thesis on arXiv so that anyone interested can have access to it.
The second edition includes some minor corrections and a few additional references. An error was corrected in Eqs.~\eqref{eq:MasterEquationsRevisitingTwoLevelRateEquation1}-\eqref{eq:MasterEquationsRevisitingTwoLevelRateEquation2}, and the absolute value was added to the current in several figures.

\newpage
\section*{Acknowledgments}

First and foremost, I would like to thank my supervisor, Nikolaj T. Zinner, for his guidance during my study and for introducing me to the wonders of quantum engineering. Thank you for allowing me to carry out a PhD under your supervision. Nikolaj has allowed me to follow my own ideas and the natural evolution of my research while always being ready with advice and guidance when things went wrong. Nikolaj's curiosity and attitude to physics have pushed me to always look deeper and do better.

I am grateful for all the people I have worked with and my fellow PhD students, in particular, Alan C. Santos, Kasper S. Christensen, Lasse B. Kristensen, Marco Majland, Morten Kjaergaard, Seth Lloyd, Stig E. Rasmussen, and Thomas Bækkegaard for the many discussions, insightful inputs, and papers. 

I would also like to thank William D. Oliver for hosting me in his group at Massachusetts Institute of Technology. I had the pleasure of spending four months with all the brilliant people in the Engineering Quantum Systems group in Boston. Thank you all for making the stay so memorable, for your passion for quantum engineering, and for our many discussions.
A special thanks goes to Sarah Muschinske and Ilan Rosen for bestowing me with some of your knowledge on superconducting circuits and for trusting me to do the theory for your experiment.

I am also grateful to all my friends from physics, Claus, Frederik, Joakim, Line, Niklas, Peter, and Simon, for making my studies a breeze. 

Finally, I would like to thank my family. To my parents and Sophie, thank you for the eternal support for all my weird projects, hobbies, and life in general. Without you, I would never have made it this far. To my fiancée, Lia, thank you for believing in me when I did not. Thank you for caring for me and supporting me when I could not. I can not wait to spend the rest of my life with you.

\smallskip
\begin {flushright}
  \textit{Kasper Poulsen}
  \\
  Aarhus, October 2023
\end {flushright}

\newpage

\vspace{0.8cm}
\section*{List of Publications}
\bibliographystyle{../bibstyles/apsrev_thesis.bst}

\begin{itemize}[align=left,labelsep = 0.25em]
\setlength\itemsep{0.5em}
\item[\cite{PhysRevLett.126.077203}] \bibentry{PhysRevLett.126.077203}.
\item[\cite{PhysRevA.105.052605}] \bibentry{PhysRevA.105.052605}.
\item[\cite{PhysRevE.105.044141}] \bibentry{PhysRevE.105.044141}.
\item[\cite{PhysRevLett.128.240401}] \bibentry{PhysRevLett.128.240401}.
\item[\cite{PhysRevE.106.034116}] \bibentry{PhysRevE.106.034116}.
\item[\cite{poulsen2022heat}] \bibentry{poulsen2022heat}.
\end{itemize}

\section*{Thesis notation}
Throughout the thesis, $\hbar = k_B = 1$ to simplify the notation. Operators are shown with a hat, and the Dirac notation is used for quantum states. Additionally, $[\bullet, \bullet]$, $\{\bullet, \bullet\}$, and $\tr\{\bullet\}$ correspond to the commutator, anti-commutator, and trace, respectively. If the trace is only partial over some subsystem $\alpha$, it is denoted by $\tr_{\alpha}\{\bullet\}$. The Pauli matrices are denoted $\hat{\sigma}^\alpha$ for $\alpha \in \{x, y, z\}$ and $\hat{\sigma}^\pm = (\hat{\sigma}^x \pm i \hat{\sigma}^y)/2$, and the harmonic oscillator annihilation operator is $\hat{a}$ throughout. For a qubit (spin 1/2 particle), the convention used is $\hat{\sigma}^- \ket{1(\uparrow)} = \ket{0(\downarrow)}$.

\end{KeepFromToc}

\newpage
\begin{KeepFromToc}
\tableofcontents 
\end{KeepFromToc}

\mainmatter

\cleardoublepage
\chapter{Introduction and outline}
\label{chapter:Introduction}


Quantum mechanics is the famously counter-intuitive, probabilistic theory developed to describe the properties of atoms and molecules. 
Although the main pillars of the theory were developed almost a century ago, quantum mechanics is becoming ever more important, with consequences reaching far into our everyday lives. Technological advances have put us on the brink of exploiting this theory for performing calculations and simulations of systems that would otherwise be impossible. Yet, the quantum world seems diametrically opposite to ours.

At the center of our difficulty in understanding quantum mechanics, and to physics as a science, lies the act of measurement. While probabilistic observations might be reasonable to anyone who has ever rolled a die, the path toward these observations is still clouded in mystery.  
Arguably, the best exemplification of the counter-intuitive nature of quantum mechanics is the thought experiment of Schrödinger's cat. 
Most people, physicists or not, have heard of Schrödinger's cat. However, stated briefly, a cat is placed in a box with poison. 
The poison is coupled to an atom in an equal superposition of the ground and excited states. The box is closed, and the poison is released if the atom is excited. 

When the box is opened, and the total system of cat, poison, and atom is measured, there is a 50 percent chance the cat will be dead, the poison will be released, and the atom will be found in the excited state. 
As a consequence, Schrödinger concluded that, prior to the measurement, the cat inherited the superposition of the atom and was both dead and alive. 
This final statement, most likely, seems entirely wrong for anyone unfamiliar with quantum mechanics. 

As humans living in the same macroscopic world as cats, we, of course, know that cats cannot be both alive and dead at the same time, something that Schrödinger also points out in his original formulation of the thought experiment \cite{Schrodinger1935}. 
Instead, the encounter of a quantum superposition with the complex collection of infinitely many degrees of freedom of our macroscopic world has resulted in something that quantum mechanics cannot describe. 

We can understand the actual fate of Schrödinger's cat by modifying the thought experiment slightly. In the ideal setting summarized above, there are two possible outcomes upon opening the box. In outcome one, the cat is dead, the poison is released, and the atom is in the excited state, while in outcome two, the cat is alive, the poison is intact, and the atom is in the ground state. Let us imagine that a single air molecule is added to the box. The air molecule is allowed to bounce around the box several times during the entangling phase, ending up in either the left or right half of the box. Consequently, the outcome must now include the two possible final positions of the air molecule. Outcome one has the molecule in the left half of the box, and outcome two has the molecule in the right half of the box. The air molecule is entirely out of our control, and we cannot recapture and measure it. In order to observe the cat as both dead and alive, we would have to ask the question: Is the cat dead, the poison released, the atom excited, and the air molecule in the left half of the box? The answer to this question is unattainable to us, and the cat is, therefore, effectively either dead or alive and never both. 

Schrödinger's cat is a thought-provoking introduction to quantum mechanics, which allows for interesting discussions on the meaning of quantum mechanics as well as the emergence of classical physics on the macroscopic scale. However, for our purpose, it also sets the tone and theme of challenges that will be explored in this thesis. The ability for quantum systems to be in multiple states at once can be exploited to try many possible solutions to a problem in parallel \cite{Bennett2000, PhysRevLett.79.325, nielsen2010quantum}. A device capable of such calculations is called a quantum computer. 

However, if a single air molecule can kill Schrodinger's cat, how can we ever hope to build such a computer? The answer is clear; only if we can nearly perfectly control every interaction of the computer with any air molecule, photon, and imperfection in the material will this be possible \cite{Devitt_2013, PhysRevA.51.992}. 

Throughout my five years of PhD studies, I have explored several aspects of statistical physics and heat control in quantum systems. In this thesis, the majority of these results and my contribution to the advancement of the field will be presented. In this chapter, the different research subjects will be introduced, and the outline for the rest of the thesis will be laid out.

\section{Statistical ensembles of quantum systems}

The language of statistical physics is spoken through probabilities, i.e., $P(X)$ for some outcome $X$. If a die is rolled in a box, the chance of observing a six upon opening the box is $P(X=6)=1/6$. Contrary to the discussion above, the die will only take one path, and prior to the observation, the number on the die is fixed. The observation changes nothing in the real world; It is only a description of our knowledge of the outcome, not the state of the die itself. In the limit of rolling infinitely many dice, the probabilities describe the total population of six-eyed dice. This is called an ensemble, and the probabilities describe the population of each microstate, in this case each possible face on the die. 

The language of quantum mechanics is spoken through wave functions, $\Psi(x)$, or quantum states $\ket{\Psi}$. Let $\ket{g}$ and $\ket{e}$ symbolize the ground and first excited state of an atom. This atom is put into an equal superposition of the ground and first excited state $\ket{+}$, where $\ket{\pm} = (\ket{g} \pm \ket{e})/\sqrt{2}$. As a consequence, a measurement has a 50 percent chance of finding the atom in the ground state. This might seem similar to the dice roll above; However, in quantum mechanics, the question asked by the measurement is crucial. If we instead ask if the atom is in the state $\ket{+}$, the observation would yield the same result every single time. Suddenly, there is no randomness anymore. 

Finally, imagine the ensemble of dice from before. For every die with the one-eyed face pointing up, an atom is put into the quantum state $\ket{+}$; for any other outcome, an atom is put into the quantum state $\ket{-}$. What is the probability of measuring the excited state on any random atom? In this example, we have a mixture of classical and quantum randomness. This is exactly where the mathematical framework of quantum statistical physics is needed. Such an ensemble of quantum states can be described using the density matrix, which is given by
\begin{align}
\hat{\rho} = \frac{1}{6} |+\rangle \langle + | + \frac{5}{6} |-\rangle \langle - |.
\end{align}
The probability of any random atom being in the excited state is now given by the expectation value of the projection operator for the appropriate state, $\hat{P} = \op{e}$,
\begin{align*}
\langle \hat{P} \rangle = \tr \{ \hat{P} \hat{\rho} \} = \frac{1}{2}.
\end{align*}
For a general ensemble of quantum systems with a fraction of $p_n$ systems in the state $\ket{\psi_n}$, the density matrix is
\begin{align}
\hat{\rho} = \sum_n p_n |\psi_n \rangle \langle \psi_n |.
\end{align}
The time evolution of a quantum state is governed by the Schrödinger equation. However, an equivalent equation can be found for the density matrix, which is called the Von-Neumann equation
\begin{align}
\frac{d\hat{\rho}}{dt} = -i[\hat{H}, \hat{\rho}] .
\label{eq:IntroductionLiouvilleEquation}
\end{align}
This framework is mostly useful in the presence of one or more baths, which can cause random transitions in the quantum system. The evolution of the density matrix, in this case, is the theme of chapter \ref{chapter:MasterEquations}. However, one simpler example will be examined later in the introduction as well.

\begin{figure}[t]
\centering
\includegraphics[width=0.9\linewidth, angle=0]{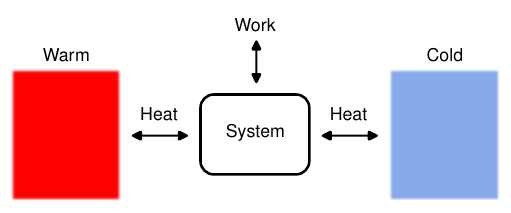}
\caption{General setup for a boundary-driven (quantum) system, where a (quantum) system is connected to two baths at either end. The baths impose an imbalance, driving heat to travel from one bath through the system to the other bath. Additionally, energy can be added to the system through external work.}\label{figure:ChapterIntroductionSketch}
\end{figure}

\section{Boundary-driven quantum systems}

The most common setup for studying heat flow is a system connected to two baths at different temperatures; see Fig~\ref{figure:ChapterIntroductionSketch}. The temperature gradient will induce energy transport through the system in the form of heat. Additionally, energy can be added from the outside through work as sketched in Fig.~\ref{figure:ChapterIntroductionSketch}. The subject of transport in these systems has been studied for more than 200 years. The simplest version is a one-dimensional homogeneous metal rod. Here, the temperature at any point $x$ along the rod and any time $t$, $T(x,t)$, is determined by the heat equation
\begin{align}
\frac{\partial T}{\partial t} = a \frac{\partial^2 T}{\partial x^2},
\end{align}
where $a$ is a constant. This equation can be solved using separation of variables and a Fourier series ansatz \cite{arfken2011mathematical}. In fact, this problem of heat transport was one of the first problems solved using the Fourier series by Jean Baptiste Joseph Fourier himself in 1822 \cite{baron2003analytical}. With the increased ability to do numerical experimentation in the 20th century, it became possible to look at larger and larger systems of coupled harmonic oscillators as toy models for solids. Under certain conditions, these models were shown to obey Fourier's heat law, which is a precursor to the heat equation \cite{Casati1986, Prosen_1992, PhysRevE.57.2992, doi:10.1080/00018730802538522}. After further examination of these classical systems, they were shown to exhibit rectification of heat flow \cite{PhysRevLett.88.094302, PhysRevLett.93.184301}. In other words, the heat current through the system is not symmetric under a permutation of the two baths. Recently, boundary-driven quantum systems have received significant attention \cite{RevModPhys.94.045006}. One of the most studied models is a linear XXZ spin chain model in the presence of a magnetic field given by
\begin{align}
\hat{H} = \sum_i^{N-1} J \left( \hat{\sigma}_i^x \hat{\sigma}_{i+1}^x + \hat{\sigma}_i^y \hat{\sigma}_{i+1}^y + \Delta \hat{\sigma}_i^z \hat{\sigma}_{i+1}^z \right) + \sum_i^{N} h_i \hat{\sigma}_{i}^z,
\end{align}
where $J$ is the spin coupling, $\Delta$ is the anisotropy, and $h_i$ is the strength of the magnetic field on the $i$th spin. Both heat and spin transport can be defined for spin chains \cite{PhysRevE.97.022115, PhysRevE.90.042142}, which we will discuss in much more detail in chapter \ref{chapter:MasterEquations}. This system's long-time or steady-state solution has been found analytically for certain bath interactions \cite{PhysRevLett.106.217206, PhysRevLett.107.137201, PhysRevLett.110.047201}. The first and most important criterion for rectification is a left-right asymmetry. Rectification has been found in $XXZ$ spin chains where this symmetry is broken through the magnetic field \cite{PhysRevE.90.042142, PhysRevB.79.014207, PhysRevB.80.172301} or the anisotropy \cite{PhysRevLett.120.200603, PhysRevE.99.032136}. Going beyond the one-dimensional spin chain, rectification effects have been found in quantum systems ranging from only one anharmonic system \cite{PhysRevB.73.205415, PhysRevB.103.155434} and a system of two spins \cite{PhysRevE.89.062109, PhysRevApplied.15.054050} to larger two-dimensional geometries \cite{PhysRevE.103.032108}.

\begin{figure}[t]
\centering
\includegraphics[width=0.9\linewidth, angle=0]{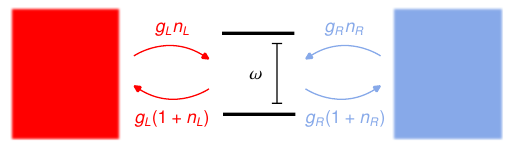}
\caption{The smallest example of a boundary-driven quantum system, i.e., a simple two-level system. The left and right excitation and decay rates are shown with arrows.}\label{figure:ChapterIntroductionQubit}
\end{figure}

\section{Two-level system as a rectifier}
\label{sec:IntroductionTwoLevelRectifier}

The simplest boundary-driven quantum system to exhibit rectification is a single two-level system. As an introduction to the analysis of these types of systems, we will solve the two-level system coupled to two thermal baths as seen in Fig.~\ref{figure:ChapterIntroductionQubit}. One bath is labeled L for left and the other R for right. Each bath can either add or subtract energy from the two-level system. Since we only include two levels, energy can only be added or subtracted in one way, simplifying the analysis significantly. In fact, the system can be described by the classical master equation
\begin{align}
\frac{d \vec{P}}{dt} = W \vec{P}. 
\label{eq:IntroductionMasterEquation}
\end{align}
The state of the two-level system is described by the vector $\vec{P} = \{P(\ket{0}), P(\ket{1})\}^T$ with each entry describing the probability of finding the two-level system in each level. Thus, $P(\ket{0})$ is the probability of finding the two-level system in the ground or lowest state, and $P(\ket{1})$ is the probability of finding the two-level system in the first excited or highest state. The time evolution is governed by the rate matrix $W$. The meaning of each entry can be gleamed by writing out the two resulting equations
\begin{align}
\frac{d }{dt} P(\ket{0}) &= W_{0,0}P(\ket{0}) + W_{0,1}P(\ket{1}),\\
\frac{d }{dt} P(\ket{1}) &= W_{1,0}P(\ket{0}) + W_{1,1}P(\ket{1}).
\end{align}
The entry $W_{i,j}$ is the rate of population transfer from state $j$ to state $i$ for $i\neq j$. The entry $W_{i,i}$ is the rate of population transfer into the state $i$. By adding the two rate equations above, we get
\begin{align}
\frac{d }{dt} (P(\ket{0}) + P(\ket{1})) &= (W_{0,0} + W_{1,0})P(\ket{0}) + (W_{0,1} + W_{1,1}) P(\ket{1}) .
\end{align} 
Since the total population is conserved, or the probabilities always add to one, we see that $W_{0,0} = -W_{1,0}$ and $W_{1,1} = -W_{0,1}$. More generally it is true that $W_{i, i} = -\sum_{j\neq i} W_{j, i}$. For the left bath, we let the excitation rate be $g_L n_L$ and the decay rate be $g_L (1+n_L)$. 
This way, $g_L$ is the coupling strength between the bath and the two-level system, while $n_L = [\exp (\omega/T_L) -1]^{-1}$ is the Bose-Einstein distribution. Here, $\omega$ is the energy difference between the two levels, and $T_L$ is the temperature of the left bath. Likewise, the rates are $g_L n_L$ and $g_L (1+n_L)$ for the right bath. This nature of this interaction will be motivated later, but for now, just note that $n_L > n_R$ if and only if $T_L > T_R$, in which case, heat will flow from left to right. The entries of the rate matrix are, therefore,
\begin{align}
W_{1,0} &= \Gamma_+ = g_L n_L + g_R n_R,\\
W_{0,1} &= \Gamma_- = g_L (1+n_L) + g_R (1+n_R).
\end{align}
We have introduced $\Gamma_+$ as the excitation rate and $\Gamma_-$ as the decay rate for convenience. Now that the problem has been set up, it is straightforward to solve the master equation \eqref{eq:IntroductionMasterEquation} and obtain a solution
\begin{align}
\vec{P} (t) = \vec{P}_{\mathrm{ss}}
+  e^{-[\Gamma_+ + \Gamma_-] t} \vec{G},
\end{align}
where
\begin{align}
\vec{P}_\mathrm{ss} &= \begin{pmatrix}
\frac{\Gamma_-}{\Gamma_+ + \Gamma_- } \\
\frac{\Gamma_+}{\Gamma_+ + \Gamma_- }
\end{pmatrix}, \quad
\vec{G} = 
\begin{pmatrix}
 \frac{\Gamma_+}{\Gamma_+ + \Gamma_- } - P(\ket{1}, t=0)  \\
-\left[\frac{\Gamma_+}{\Gamma_+ + \Gamma_- } - P(\ket{1}, t=0) \right]
\end{pmatrix}.
\end{align}
$P(\ket{1}, t=0)$ is the initial population of the excited state. The solution consists of two parts: a time-dependent part containing $\vec{G}$ and a time-independent part consisting of $\vec{P}_{\mathrm{ss}}$. Evidently, every initial state will decay exponentially toward the same state, namely $\vec{P}_{\mathrm{ss}}$, which is named the steady state. Additionally, the steady state is unique unless both bath couplings are zero. This is usually the case, but as we will see later, there are situations where the steady state becomes dependent on the initial state. 

\begin{figure}[t]
\centering
\includegraphics[width=1.0\linewidth, angle=0]{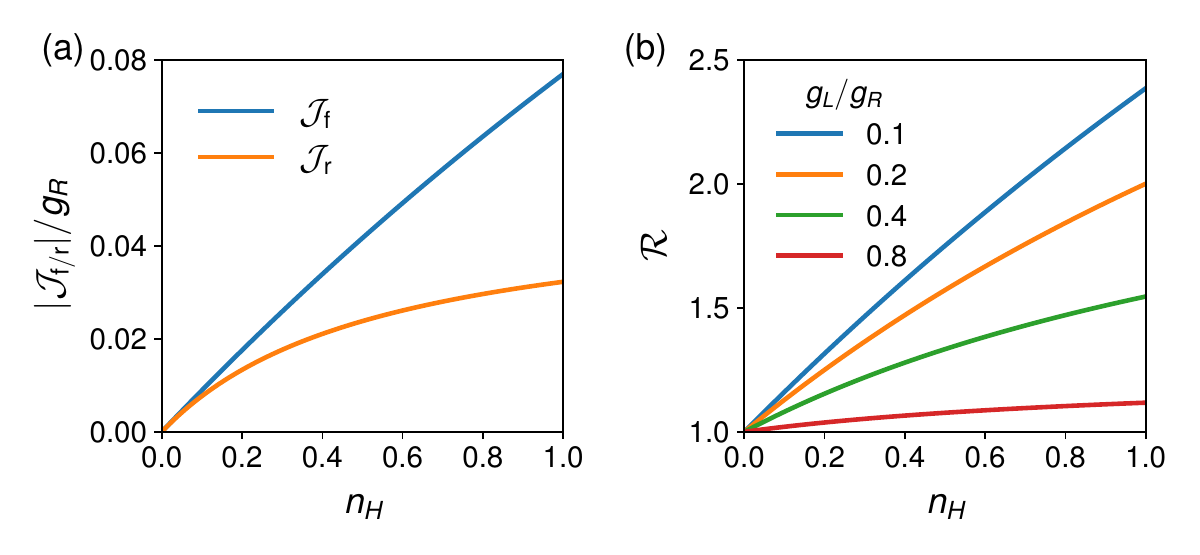}
\caption{(a) Forward and reverse bias excitation current for a two-level system coupled to two baths at different temperatures. The bath couplings are chosen to be $g_L/g_R = 0.1$, and the cold bath temperature is picked to be $n_C = 0$. (b) Rectification for the same two-level system.}\label{figure:ChapterIntroductionQubitR}
\end{figure}

We are now able to calculate the heat transport through the two-level system. Since energy can only be transported in quanta of $\omega$, we can first calculate the number of quanta or excitations being transported per unit of time $\mathcal{J}$. From this, the energy transported per unit of time can be calculated as $\mathcal{K} = \mathcal{J} \omega$; $\mathcal{J}$ is called the excitation current, and $\mathcal{K}$ is called the heat current. Unless an explicit time-dependence is stated, the two currents are calculated in steady state. The excitation current between the right bath and the two-level is
\begin{align}
\mathcal{J} &= g_L n_L P_{\mathrm{ss}}(\ket{0}) - g_L (1+n_L ) P_{\mathrm{ss}}(\ket{1})\\
&= \frac{ (n_L - n_R) g_L g_R }{g_L (1+2n_L) + g_R (1+2n_R) } .
\end{align}
We have chosen the positive direction to be from left to right. This current is the difference between the excitations going into the two-level system from the left bath, first part, and the excitations going out of the two-level, second part. Each term is the product of the transfer rate, e.g., $g_L n_L$, and the fraction of ensemble systems in the initial state of the transition, e.g., $P(\ket{0})$. Likewise, we can write out the excitation current between the two-level system and the right bath. However, in steady-state, the energy into the two-level system from the left is equal to the energy out of the two-level system to the right, so the two currents are equal. As a sanity check, we can look at the three temperature cases
\begin{align}
T_L > T_R \quad &\Leftrightarrow \quad \mathcal{J} > 0, \\
T_L = T_R \quad &\Leftrightarrow \quad \mathcal{J} = 0, \\
T_L < T_R \quad &\Leftrightarrow \quad \mathcal{J} < 0. 
\end{align}
As expected, heat flows from hot to cold. Finally, we will explore the rectification of this minimal system, i.e., does heat flow more easily one way compared to the other? For this, we pick two sets of temperatures $T_C$ and $T_H$ with corresponding $n_C$ and $n_H$ with $n_H > n_C$. We denote the two cases
\begin{align}
\mathrm{Forward\, bias} : \quad n_L = n_H \quad \mathrm{and} \quad n_R = n_C \\
\mathrm{Reverse\, bias} : \quad n_L = n_C \quad \mathrm{and} \quad n_R = n_H
\end{align}
We can denote the bias using either the subscript f for forward bias or r for reverse bias. In forward bias, heat flows to the right ($\mathcal{J}_{\mathrm{f}}>0$) and, in reverse bias, heat flow to the left ($\mathcal{J}_{\mathrm{r}}<0$). A useful way of measuring the asymmetry in current between the two situations is called the rectification, $\mathcal{R} = - \mathcal{J}_\mathrm{f} / \mathcal{J}_\mathrm{r}$. For a perfect diode, the rectification tends to infinity, while for a symmetric system, the rectification is unity. For the two-level system, the rectification becomes
\begin{align}
\mathcal{R} = \frac{g_L (1+2n_C) + g_R (1+2n_H)}{g_L (1+2n_H) + g_R (1+2n_C)}.
\end{align}
For $g_L=g_R$, the two-level setup is completely symmetric, and the rectification is indeed one. The forward and reverse bias currents are plotted in Fig.~\ref{figure:ChapterIntroductionQubitR}(a), and the rectification is plotted in Fig.~\ref{figure:ChapterIntroductionQubitR}(b). Surprisingly, this small system does exhibit rectification, albeit with very small rectification values. 

\section{Maxwell's demon and information as a resource}

Another old discussion is the role of information in statistical physics. This is best exemplified by yet another thought experiment, Maxwell's demon, which was first thought up by Maxwell more
than 150 years ago \cite{leff2002maxwell, knott1911life}. There are slight variations in the thought experiment, but here, the most relevant is briefly introduced \cite{bennett1987demons}. 

Imagine two gases separated by a wall with a demon-controlled door. The two gasses initially have the same temperature, and the demon can open and close the door without performing work. The demon opens the door for fast air molecules from the right side while closing the door for slow air molecules. Likewise, the demon opens the door for slow air molecules from the left side while closing the door for fast air molecules. As the demon keeps operating, the air molecules are slowly sorted, putting all the fast molecules in the left half and all the slow molecules in the right half. Equivalently, the temperature of the left half is increased, and the temperature of the right half is decreased. Therefore, the demon has managed to lower the entropy of the two gasses without performing any work. 

This is in apparent contradiction with the second law of thermodynamics. However, the contradiction can be solved by accounting for the entropy of the demon itself \cite{Szilard1929}. While the classical solution is convoluted, the quantum solution is more straightforward and will be presented in chapter \ref{chapter:Maxwell}.

Due to the counterintuitive nature of Maxwell's demon, it has motivated the development of many variations, including the Szilard engine \cite{Szilard1929}. The Szilard engine relies on information as a resource to extract work instead of lower entropy. Similar to the case of transport, Maxwell's demon has been brought into the quantum world with theoretical proposals using a quantum system \cite{PhysRevA.56.3374, PhysRevLett.106.070401}.

James Clerk Maxwell probably never imagined that his thought experiment would become more than that. However, with our newfound ability to measure and control quantum degrees of freedom, it is now possible to build Maxwell's demon using several different quantum technology platforms \cite{doi:10.1073/pnas.1704827114, PhysRevLett.113.030601, PhysRevLett.121.030604, PhysRevResearch.2.032025, Masuyama2018, PhysRevLett.115.260602}. We will not get further into how this was done; however, some of them are very similar to the discussion in chapter~\ref{chapter:Maxwell}.

\section{Experimental techniques}

The experimental side of quantum computing has seen immense progress in recent years, and a diverse set of platforms has emerged. These include but are not limited to trapped ions \cite{10.1063/1.5088164}, neutral atoms in an optical lattice \cite{Barnes2022, Evered2023}, silicon and germanium quantum dots \cite{Hendrickx2021, Madzik2021}, photonics \cite{Larsen2021}, and superconducting circuits \cite{Kim2023}. Most quantum computing platforms only function at very low temperatures. For a qubit of frequency $\omega$, thermodynamics demand that excitation of the qubit can be prevented with a temperature, $T$, for which $T \ll \omega$. Therefore, studying heat transport using high-temperature baths $T \sim \omega$ with these platforms might seem counterintuitive. However, with controlled interactions, this is not a problem. Effective bath dynamics can be simulated using randomized interactions. Specifically, a zero-temperature bath forces the adjacent qubit to decay, equivalent to a bad qubit. An infinite temperature bath randomly flips the qubit, which can be simulated by applying randomized NOT gates. Therefore, all the systems studied in this thesis could probably be built in most of the above platforms. 
\begin{figure}[t]
\centering
\includegraphics[width=0.5\linewidth, angle=0]{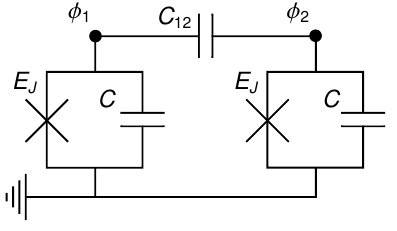}
\caption{Circuit diagram for two coupled transmons. $E_J$ is the Josephson energy for the Josephson junctions, $C$ is the capacitance for the two transmons, and $C_{12}$ is the coupling capacitance between the two transmons. $\phi_1$ and $\phi_2$ are the fluxes at the two flux nodes.}\label{figure:ChapterIntroductionCircuits}
\end{figure}
However, they were designed for superconducting circuits, which will be the main focus of this section. Superconducting circuits consist of a layer of aluminum on top of a substrate. The aluminum is then edged away in certain places, forming islands. These islands form natural capacitive couplings with one another. The exact architecture varies significantly, but for the very common transmon qubit, these islands are coupled to a common ground through a Josephson junction and an additional stronger shunting capacitance. The circuit diagram for two coupled transmon qubits can be seen in Fig.~\ref{figure:ChapterIntroductionCircuits}. A Josephson junction is a circuit element that is quantum in nature. It consists of two superconducting wires connected by a thin insulator, allowing electrons to tunnel from one wire to the other \cite{PRXQuantum.2.040204}. A classical Hamiltonian can be constructed for these circuits. For the circuit in Fig.~\ref{figure:ChapterIntroductionCircuits}, this Hamiltonian becomes
\begin{align}
\mathcal{H} = \sum_{i=1}^2 \left[ \frac{1}{2} C \dot{\phi}_i^2  -E_J \cos \left(2\pi \frac{\phi_i}{\Phi_0} \right) \right] + \frac{1}{2} C_{12} (\dot{\phi}_2 - \dot{\phi}_1)^2,
\end{align}
where $\dot{\phi}_i = \partial \phi_i /\partial t$ and $\Phi_0$ is the magnetic flux quantum. $\phi_i$ is the node flux for the $i$th node, defined through
\begin{align*}
\frac{d}{dt} \phi_i(t) = V_i(t),
\end{align*}
where $V_i$ is the voltage over the capacitance $C_i$. Next, the fluxes are quantized, substituting $\phi_i \rightarrow \hat{\phi}_i$, and the cosine is Taylor expanded to fourth order, assuming $\langle \hat{\phi}_i/\Phi_0 \rangle \ll 1$. The resulting quantum mechanical Hamiltonian becomes
\begin{align}
\hat{H} = \sum_{i=1}^2 \left[ \omega \hat{a}_i^\dag \hat{a}_i - \frac{U}{2} \hat{a}_i^\dag \hat{a}_i (\hat{a}_i^\dag \hat{a}_i - 1) \right] - J \left( \hat{a}_1 - \hat{a}_1^\dag \right) \left( \hat{a}_2 - \hat{a}_2^\dag \right). \label{eq:IntrodutionTransmonHamiltonian}
\end{align}
$\omega$ is the transmon frequency, $U$ is the anharmonicity, and $J$ is the coupling between the two transmons. Due to the anharmonicity, the Hilbert space for each transmon is truncated to the lowest two states
\begin{align}
\hat{H} = \sum_{i=1}^2 \frac{\omega}{2} \hat{\sigma}_i^z + J \left( \hat{\sigma}_1^- \hat{\sigma}_2^+ + \hat{\sigma}_1^+ \hat{\sigma}_2^- \right). \label{eq:IntrodutionTransmonQubitHamiltonian}
\end{align}
For transmons, the system parameters usually take values around
\begin{align}
&\omega/2\pi \sim 10\,\mathrm{GHZ}, \quad U/2\pi \sim 100\,\mathrm{MHZ}, \quad J/2\pi \sim 10\,\mathrm{MHZ}.
\end{align}
The parameters can be manufactured to take values different from this, and some architectures allow for the parameters to be varied during operation. For transmons, a SQUID, effectively a tunable Josephson junction, can make the qubit frequency tunable, while a tunable coupler can achieve a controllable $J$ \cite{PhysRevApplied.10.054062}. The coherence times for decay and dephasing are usually around $T_1 \sim 10\, \mathrm{\mu s}, $ and $T_2 \sim 10\, \mathrm{\mu s}$, respectively. Although, much longer coherence times have been achieved \cite{doi:10.1146/annurev-conmatphys-031119-050605}. Single qubit gates can be performed using a drive line capacitively coupled to the transmon. 

This is just the basics. Many additional elements make superconducting circuits a very versatile platform. Both readout and control of higher levels in Eq.~\eqref{eq:IntrodutionTransmonHamiltonian} are possible \cite{PhysRevX.11.021010}, which we will use in chapter \ref{chapter:PhaseTransition}. Additionally, a bath at zero temperature can be implemented using a leaky resonator \cite{Ma2019, PRXQuantum.3.020305}. A bath at some finite temperature can be implemented using a resonator coupled to a small piece of copper, forcing the correlation functions of the resonator to decay \cite{Gubaydullin2022, Senior2020}.

\section{Outline}

Despite all the work that has gone into studying the intersection of quantum mechanics and statistical physics, there are still many open questions. Some of the questions are: To what extent can quantum effects such as entanglement be used to control heat flow? How well do heat components such as diodes generalize to larger circuits? Does the Wheatstone bridge generalize to quantum transport? Is there a general framework for understanding the transport properties of boundary-driven quantum systems? Can Maxwell's demon benefit from the information flow of non-Markovian baths? And finally, how is quantum statistical physics simulated on a quantum computer? These are precisely the diverse set of questions that this thesis aims to answer one at a time. In the following, the content of each chapter is briefly outlined.

\textbf{Chapter 2} goes through the master equation used for most of the remaining chapters. No new results are presented here. Instead, both the local and global master equations are introduced, and the various assumptions that are needed for the derivation are clearly stated. The two-level rectifier is briefly revisited using the complete quantum treatment, and finally, the different master equations are numerically compared.

\textbf{Chapter 3} explores the first instance of a novel boundary-driven quantum system in this thesis, consisting of a single qutrit with engineered interaction to the two baths, achieving a perfect rectifier. An implementation using a qutrit and two harmonic oscillators is then introduced and solved analytically in the strong bath system interaction limit. Finally, a larger circuit is built, forming a full wave bridge rectifier, and it is shown that the output bias is independent of the input bias.

\textbf{Chapter 4} expands on the methods of chapter 3 to utilize entanglement for a better rectifier. This model consists of two linear chains of spins connected by a two-way interface. This allows for interference between the two channels through an entangled state, completely blocking transport in reverse bias while allowing a current to run in forward bias. It is explained how the interface is driven into the entangled state. Finally, the same system is modeled using the global master equation, and it is shown that the rectification persists.

\textbf{Chapter 5} explores the same two-way setup as chapter 4 but with the aim of measuring an unknown coupling strength. This makes for a quantum version of the classical Wheatstone bridge for spin-spin couplings and spin or heat currents. The sensitivity towards the unknown parameter is studied using the quantum Fisher information, and it is found that the sensitivity can be controlled via an external magnetic field. This sensitivity is due to an abrupt change in the entangled state, which can be measured using the spin current.

\textbf{Chapter 6} looks at a more general linear chain consisting of weakly coupled chains of strongly coupled spins. The current is then studied as a function of an external magnetic field, and it is shown that the current changes dramatically. This is done for different chains, and the locations of the large current resonances are found analytically. Finally, a Z-coupling is added, and the current profile becomes much more complicated.

\textbf{Chapter 7} studies a Maxwell's demon setup with two non-Markovian baths. First, a single operation of the demon acquiring, using, and discarding information on the two baths is studied. Second, a double operation of the demon is tried, and the effectiveness of the demon is studied as a function of the demon's timing. Finally, the demon operates periodically, and the long time limit is examined. The non-Markovian baths allows for information to flow back from the bath into the system, which the demon can exploit.

\textbf{Chapter 8} explores the ability to simulate statistical physics and phase transitions on a modern superconducting circuits chip. A lattice of transmons has a very similar Hamiltonian to the Bose-Hubbard model, which exhibits a quantum phase transition. This phase transition is first studied in the canonical ensemble, i.e., for a constant number of particles. Next, the phase transition is studied in the grand canonical ensemble, i.e., for a variable number of particles, and a method for performing the simulation is proposed. Finally, adiabatic state preparation is examined as a method for preparing the ground state, and the phase diagram is simulated both with and without decoherence.

\textbf{Chapter 9} briefly outlines the main results of the previous chapters, and it gives an outlook on the future of the field.

\cleardoublepage
\chapter{Quantum master equations}
\label{chapter:MasterEquations}

\noindent The most important tool for studying open quantum systems is arguably the master equation. In the introduction, we briefly introduced the classical master equation to solve a two-level system coupled to two baths, not allowing any quantum effects to be present in the system. To properly describe such a system, we need a quantum version of the master equation. As we will see in later chapters, it can take many forms, and even more variations exist. For our purposes, there are two general categories of master equations: local and global. These names refer to how they behave and act on the quantum system.

The local master equation couples the baths locally to the quantum system. For this terminology to make sense, the quantum system has to be comprised of many smaller interacting quantum systems. The local master equation can then couple to as little as a single subsystem at a time. This makes transport more intuitive since heat can only be added to or subtracted from the subsystem coupled to the bath. Furthermore, the local master equation usually has fewer terms, and as a consequence, it is much faster to simulate.

The global master equation couples any bath globally to the entire quantum system, even if the bath interacts through a single subsystem operator. More specifically, the global master equation drives transitions between every pair of eigenstates of the entire quantum system \cite{breuer2002theory}. Therefore, for a large system of $N$ eigenstates, there are $N(N-1)/2$ pairs, making the master equation large and time-consuming to build and simulate. For a system undergoing periodic driving, Floquet theory can be used to derive a global master equation in much the same way. Since the eigenstates can be delocalized, energy can be added or subtracted globally, making transport unintuitive. The main advantage of the global master equation is that it can be derived from a general setup with minimal assumptions. In fact, it can be used as an intermediary to derive the local master equation \cite{Cattaneo_2019, Hofer_2017}.

Understanding the validity of these master equations is essential for understanding the validity and generality of most of the results in other chapters. Therefore, this chapter will introduce both master equations in a setting useful for later chapters. The local master equation is first introduced, and transport is briefly discussed. This is exemplified by revisiting the two-level rectifier introduced in the introduction using a fully quantum approach. Next, the global master equation is derived from a general setup using the Born-Markov approximation. It is shown that the global master equation reduces to the local master equation for rapidly decaying bath coherence functions. The derivation of the master equation is modified slightly to allow for time-dependent system Hamiltonians. Finally, we examine the validity of the two master equations and when the two overlap.

\section{Local Master equation}
\label{sec:MasterEquationsLocal}

A commonly used master equation for modeling thermal baths in everything from out-of-equilibrium studies to decoherence of qubits is the local Master equation. The main advantages are the ease of use and the relatively easy intuition when used in boundary-driven quantum systems. For a local Master equation to make sense, the Hamiltonian has to be comprised of $N$ smaller quantum systems, e.g., two-level systems or harmonic oscillators. Generally, we can write the Hamiltonian of such a network of smaller systems as
\begin{align}
\hat{H}_S = \sum_{i_0, i_1,...,i_N} \varepsilon_{i_0,i_1,...,i_N} \hat{A}_0^{i_0} \otimes \hat{A}_1^{i_1} \otimes ... \otimes \hat{A}_N^{i_N}. \label{eq:MasterEquationsLocalHamiltonian}
\end{align}
The operator $A_n^{i_n}$ is the $i_n$th operator in the Hilbert space for the $n$th subsystem. Generally, $\{A_n^{i_n}\}_{i_n}$ must form a complete set of operators for the $n$th subsystem. However, in practice, we rarely need this to be the case. Likewise, we will usually only include nearest-neighbor interactions, which will dramatically reduce the number of terms in the sum. For two-level systems, these operators could be Pauli matrices and the identity, $A_n^{i_n} \in \{\hat{\identity}, \hat{\sigma}^x, \hat{\sigma}^y, \hat{\sigma}^z\}$. The local master equation in Lindblad form is defined as
\begin{equation}
\frac{d \hat{\rho}}{dt} = \mathcal{L}[\hat{\rho}] = -i[ \hat{H}_S, \hat{\rho}] +  \sum_n \mathcal{D}_n[\hat{\rho} ]. \label{eq:MasterEquationsLocal}
\end{equation}
$\mathcal{L}[\hat{\rho}]$ is called the Lindbladian and governs the evolution of the system, similar to the role usually played by the Hamiltonian. The first part is the usual evolution due to the Schrödinger equation, and the second part is a non-unitary part describing the interaction with the environment. It is called the local master equation because the non-unitary parts only involve operators local to each subsystem, e.g., the environment acts locally. The dissipative evolution for each subsystem is determined by
\begin{equation}
\mathcal{D}_n[\hat{\rho} ] = \sum_{i_n} \Gamma_{n,i_n} \mathcal{M}[\hat{A}_n^{i_n}, \hat{\rho}],
\label{eq:MasterEquationsLocalDissipator}
\end{equation}
where we have introduced the shorthand, $\mathcal{M}[\hat{A}, \hat{\rho}] = \hat{A} \hat{\rho} \hat{A}^{\dag} - \{\hat{A}^{\dag} \hat{A}, \hat{\rho}\}/2 $. The first operator argument of $\mathcal{M}$ is called a jump operator. While, for a network of qubits, the Hamiltonian contains $4^N$ terms, the dissipative part of the local master equation contains only $4N$ terms. This clearly shows the power of the local master equation. In practice, the sum will be much smaller, and only two terms will be included per bath. To quantify the energy transport into or out of the system, we can examine the change in internal energy
\begin{equation}
\frac{d}{dt} \langle \hat{H}\rangle = \text{tr}\left\{ \frac{d \hat{H}}{dt} \hat{\rho} \right\} + \sum_n  \text{tr} \{ \hat{H} \mathcal{D}_n[\hat{\rho}] \}.   
\end{equation}
The usual interpretation of this result is that the first term is energy added due to work, while the second term is energy added due to heat. These are important quantities that we denote
\begin{align}
\mathcal{W}(t) &= \text{tr}\left\{ \frac{d \hat{H}}{dt} \hat{\rho} \right\},\ \\
\mathcal{K}(t) &= \sum_n \text{tr} \{ \hat{H} \mathcal{D}_n[\hat{\rho}] \}.
\end{align}
Both quantities are in units of energy per time. $\mathcal{W}$ is work done per unit of time, and $\mathcal{K}$ is heat exchanged per unit of time. Both are defined to be positive when the system energy is increased, leading to a quantum version of the first law of thermodynamics
\begin{align}
\frac{d}{dt} \langle \hat{H}\rangle = \mathcal{W} + \mathcal{K}.
\end{align}
Since work comes from external manipulation of the Hamiltonian, it is easy to control and can be very useful. On the other hand, heat comes from interactions with unknown degrees of freedom, which is more complicated to control. Further analyzing the different contributions to the heat, we can easily single out the heat contributions from each bath or the heat into each subsystem, e.g.,
\begin{equation}
\mathcal{K}_n = \text{tr} \{ \hat{H} \mathcal{D}_n[\hat{\rho}] \}.
\end{equation}
For a time-independent Hamiltonian, the system will reach a steady state denoted $\hat{\rho}_{\text{ss}}$ obeying the relation
\begin{align}
\mathcal{L}[\rho_{\text{ss}}] = 0.
\end{align}
Usually, the steady state is unique and does not depend on initial conditions. However, in certain situations, the null space of $\mathcal{L}$ has a higher dimension, and the steady state is not unique. In steady state, the energy is constant, and since the work is zero, we have $\mathcal{K}=0$.

\subsection{Revisiting the two-level rectifier}
\label{subsec:ChapterMasterEquationRevisitingTwoLevel}

Before turning to the global master equation, we will revisit the two-level rectifier to see how a fully quantum description of the system can look. The Hamiltonian for a two-level system is $\hat{H} = \omega \op{1}$, and the master equation is
\begin{align}
\frac{d \hat{\rho}}{dt} &= -i[ \hat{H}, \hat{\rho}] + \mathcal{D}[\hat{\rho}],\\ 
\mathcal{D}[\hat{\rho}] &= \Gamma_- \mathcal{M}[\hat{\sigma}^-, \hat{\rho}] + \Gamma_+ \mathcal{M}[\hat{\sigma}^+, \hat{\rho}].
\end{align}
The first term in $\mathcal{D}$ describes decay, and the second term describes excitation due to the baths. The rates are, like before,
\begin{align}
\Gamma_- &= g_L (1+n_L) + g_R (1+n_R), \\
\Gamma_+ &= g_L n_L + g_R n_R.
\end{align}
Letting $\ket{0}$ denote the ground state and $\ket{1}$ be the excited state, we can determine the differential equations for the matrix elements of $\hat{\rho}$,
\begin{align}
\frac{d}{dt} \langle 0|\hat{\rho}|0 \rangle &= \Gamma_- \langle 1|\hat{\rho}|1 \rangle - \Gamma_+ \langle 0|\hat{\rho}|0 \rangle, \\
\frac{d}{dt} \langle 0|\hat{\rho}|1 \rangle &= i\omega \langle 0|\hat{\rho}|1\rangle -\Gamma_-/2 \langle 0|\hat{\rho}|1 \rangle - \Gamma_+/2 \langle 0|\hat{\rho}|1 \rangle, \label{eq:MasterEquationsRevisitingTwoLevelRateEquation1} \\
\frac{d}{dt} \langle 1|\hat{\rho}|0 \rangle &= -i\omega \langle 1|\hat{\rho}|0\rangle -\Gamma_-/2 \langle 1|\hat{\rho}|0 \rangle - \Gamma_+/2 \langle 1|\hat{\rho}|0 \rangle, \label{eq:MasterEquationsRevisitingTwoLevelRateEquation2} \\
\frac{d}{dt} \langle 1|\hat{\rho}|1 \rangle &= -\Gamma_- \langle 1|\hat{\rho}|1 \rangle + \Gamma_+ \langle 0|\hat{\rho}|0 \rangle .
\end{align}
The diagonals $\langle 0|\hat{\rho}|0 \rangle$ and $\langle 1|\hat{\rho}|1 \rangle$ are the populations $P_0$ and $P_1$ from section \ref{sec:IntroductionTwoLevelRectifier}, respectively. The off-diagonals $\langle 0|\hat{\rho}|1 \rangle$ and $\langle 1|\hat{\rho}|0 \rangle$ contain the quantum coherence of the state and are, therefore, called coherences. The first and last equations above are the same as derived previously in section \ref{sec:IntroductionTwoLevelRectifier}. The equations for the coherences can easily be solved
\begin{align}
\langle 0|\hat{\rho}(t)|1 \rangle &= e^{-[\Gamma_- + \Gamma_+]t/2} e^{i\omega t} \langle 0|\hat{\rho}(0)|1 \rangle, \\
\langle 1|\hat{\rho}(t)|0 \rangle &= e^{-[\Gamma_- + \Gamma_+]t/2} e^{-i\omega t} \langle 1|\hat{\rho}(0)|0 \rangle.
\end{align}
The only difference between the descriptions using the classical master equation and the quantum master equation is the exponential decay of the coherences. Likewise, it is possible to derive the quantum version of the heat current, which we will do for a more general case later.

\section{Global master equation}
\label{sec:MasterEquationsGlobal}
The local master equation discussed in the previous section is easy to use and fairly intuitive in its effects. However, it is only applicable in a limited number of situations and is difficult to derive. The global master equation is the opposite. It can be derived directly from a general setup of a small quantum system and a big environment consisting of many unknown degrees of freedom. The basic Hamiltonian at the beginning of most master equation derivations is
\begin{align}
\hat{H} = \hat{H}_S \otimes \identity + \identity \otimes \hat{H}_B + \hat{H}_I. \label{eq:MasterEquationsHamiltonian}
\end{align}
The first part is the quantum system Hamiltonian, which is controllable and measurable to some degree. The second part is the bath Hamiltonian, which is usually inaccessible and much bigger than the system. The bath could technically be anything, but later, we will make assumptions restricting what the bath can be. Since the bath Hamiltonian can contain several different baths, we can stick to a single bath Hamiltonian without loss of generality. The final part is the interaction Hamiltonian connecting the system to one or more baths. 
The global master equation comes in different forms, but here, we will derive the basic version as derived in Ref.~\cite{breuer2002theory}. 
This derivation can later be changed slightly to fit different situations. Throughout, we will keep track of all assumptions made and discuss what they mean. 
For this, we define $\tau_B$ as the relaxation timescale of the baths, $\tau_S$ as the timescale of the system evolution, and $\tau_R$ as the relaxation timescale of the system. 

\subsection{Time-independent Hamiltonian}

First, we assume the Hamiltonian to be time-independent.
The derivation initially follows Ref.~\cite{breuer2002theory} but deviates in a few minor places. First, the Hamiltonian is changed into the interaction picture with respect to the system and bath Hamiltonian
\begin{equation}
\hat{H}_I(t) = e^{i(\hat{H}_S \otimes \identity + \identity \otimes \hat{H}_B)t} \hat{H}_I e^{-i(\hat{H}_S \otimes \identity + \identity \otimes \hat{H}_B)t}.
\end{equation}
The interaction picture is shown through the explicit time dependence of operators. The equation of motion for the density matrix only involves this time-dependent interaction Hamiltonian 
\begin{equation}
\frac{d}{dt} \hat{\rho}(t) = -i [\hat{H}_I(t), \hat{\rho}(t)].
\end{equation}
This differential equation is integrated, $\hat{\rho}(t) = \hat{\rho}(0) -i \int_0^t ds [\hat{H}_I(s), \hat{\rho}(s)].$, and plugged into the right-hand side of the equation. Additionally, a trace is performed over the bath Hilbert space, resulting in an equation for the system density matrix
\begin{equation}
\frac{d}{dt} \hat{\rho}_S(t) =-i \mathrm{tr}_B\{ [\hat{H}_I(t), \hat{\rho} (0)]\} - \int_{0}^{t} ds\, \mathrm{tr}_B\{ [\hat{H}_I(t), [\hat{H}_I(t-s),\hat{\rho}(t-s)]]\}.
\label{eq:MasterEquationsvonNeumann2}
\end{equation}
A substitution of $s \rightarrow t-s$ was also performed. Since the von Neumann equation was used twice, it might seem like a second-order approximation, but it is important to remember that the above equation is still exact. However, to proceed, the Born approximation is required, which consists of writing the density matrix as a product state 
\begin{equation}
\hat{\rho}(t) = \hat{\rho}_S(t) \otimes \hat{\rho}_B.
\end{equation}
Ironically, coherence between the system and bath is necessary to mediate any population changes. The von Neumann equation coupled populations with coherences and coherences with populations, so the Born approximation would completely fail with the von Neumann equation applied only once. This is why the integral form of the von Neumann equation was plugged into the original von Neumann equation, allowing for two steps of evolution seen through the two applications of the Hamiltonian in Eq.~\eqref{eq:MasterEquationsvonNeumann2}. While Eq.~\eqref{eq:MasterEquationsvonNeumann2} is exact, the addition of the Born approximation can be seen as second order in the allowed system-bath coherence. The Born approximation is valid when the bath-system correlations decay quickly on the timescale of the system dynamics, i.e., $\tau_S \gg \tau_B$. Additionally, the bath state is assumed time-independent, and $[\hat{H}_B, \hat{\rho}_B]=0$, which is trivially true for a thermal state. Furthermore, it is assumed that $\mathrm{tr}_B\{ [\hat{H}_I(t), \hat{\rho} (0)]\}=0$. We will examine this assumption later when we have the tools to understand its meaning properly. 
Without loss of generality, the interaction Hamiltonian is written as a sum of product operators
\begin{equation}
\hat{H}_I = \sum_\alpha \hat{A}_\alpha \otimes \hat{B}_\alpha.
\end{equation}
Unlike in Ref.~\cite{breuer2002theory}, we do not assume that $\hat{A}_\alpha$ and $\hat{B}_\alpha$ are all Hermitian. However, since $\hat{H}_I$ is Hermitian, the Hermitian conjugate must be present in the sum. First, we decompose $\hat{H}_I$ into eigenoperators. This is done by letting $\epsilon$ be eigenvalues of $\hat{H}_S$, with $\hat{\Pi}(\epsilon)$ being the projection operator into the space of eigenvectors with eigenvalue $\epsilon$. We define the operators
\begin{equation}
\hat{A}_\alpha(\omega) = \sum_{\epsilon'-\epsilon = \omega} \hat{\Pi} (\epsilon) \hat{A}_{\alpha} \hat{\Pi} (\epsilon').
\end{equation}
Where the sum is over all pairs $\epsilon$ and $\epsilon'$ for which $\epsilon'-\epsilon = \omega$. This operator includes all possible transitions of $\hat{A}_\alpha$ with transition frequency $\omega$. Since the system Hamiltonian can be written as $\hat{H}_S = \sum_\epsilon \epsilon\, \hat{\Pi} (\epsilon)$, we easily see that
\begin{equation}
[\hat{H}_S, \hat{A}_\alpha(\omega)] = -\omega \hat{A}_\alpha(\omega).
\end{equation}
These are called eigenoperators of $\hat{H}_S$, which in the interaction picture evolve as 
\begin{align}
e^{i \hat{H}_S t} \hat{A}_\alpha(\omega) e^{-i \hat{H}_S t} = e^{-i\omega t} \hat{A}_\alpha(\omega),\\
e^{i \hat{H}_S t} \hat{A}_\alpha^\dagger(\omega) e^{-i \hat{H}_S t} = e^{+i\omega t} \hat{A}_\alpha^\dag(\omega).
\end{align}
The interaction Hamiltonian is now written 
\begin{equation}
\hat{H}_I = \sum_{\alpha, \omega} e^{-i\omega t } \hat{A}_\alpha (\omega) \otimes \hat{B}_\alpha (t) = \sum_{\alpha, \omega} e^{i\omega t } \hat{A}_\alpha^\dagger (\omega) \otimes \hat{B}_\alpha^\dagger (t),
\end{equation}
where the bath operators are also in the interaction picture, $\hat{B}_\alpha (t) = e^{i \hat{H}_B t} \hat{B}_\alpha e^{-i \hat{H}_B t}$. With this little detour, Eq.~\eqref{eq:MasterEquationsvonNeumann2} can be written
\begin{align}
\frac{d}{dt} \hat{\rho}_S(t) &= \sum_{\omega, \omega'} \sum_{\alpha, \beta} e^{i(\omega' -\omega) t } \int_{0}^{t} ds \left[ \hat{A}_\beta (\omega) \hat{\rho}_S(t-s) \hat{A}_\alpha^\dagger (\omega') \right. \\
& \hspace{2cm} \left. - \hat{A}_\alpha^\dagger (\omega') \hat{A}_\beta (\omega) \hat{\rho}_S(t-s) \right] e^{i\omega s } \langle \hat{B}_\alpha^\dagger (s)  \hat{B}_\beta (0) \rangle_B + \textrm{h.c.}, \nonumber
\end{align}
where h.c. is the Hermitian conjugate. Next, we are ready to invoke the Markov approximation, which is a combination of two changes to the above equation. First, we make the substitution. 
\begin{align}
\hat{\rho}_S(t-s) \langle \hat{B}_\alpha^\dagger (s)  \hat{B}_\beta (0) \rangle_B = \hat{\rho}_S(t) \langle \hat{B}_\alpha^\dagger (s)  \hat{B}_\beta (0) \rangle_B.
\end{align}
This is valid when the system density matrix is constant during the time $\tau_B$ in which the bath correlation functions are non-zero. So by the time the density matrix starts changing, the correlation functions should be zero. Second, following the same logic, the integral is changed to be from zero to infinity. For $t<\tau_B$, the system has, per the assumption, not changed, and the integration is small and does not contribute significantly. For $t>\tau_B$, the correlation functions are already zero, and the rest of the integral from $t$ to infinity is zero. Collectively, the two main assumptions used are called the Born-Markov approximation. The master equation now takes the form
\begin{align}
\frac{d}{dt} \hat{\rho}_S(t) &=\sum_{\omega, \omega'} \sum_{\alpha, \beta} e^{i(\omega' -\omega) t } \Gamma_{\alpha \beta} (\omega) \left( \hat{A}_\beta (\omega) \hat{\rho}_S(t) \hat{A}_\alpha^\dagger (\omega') \right. \label{eq:MasterEquationsGlobalPre} \\ 
& \hspace{5cm}\left. - \hat{A}_\alpha^\dagger (\omega') \hat{A}_\beta (\omega) \hat{\rho}_S(t) \right) + \textrm{h.c.}, \nonumber
\end{align}
where we have defined the one-sided Fourier transform of the bath correlation functions
\begin{align}
\Gamma_{\alpha \beta} (\omega) = \int_{0}^{\infty} ds\,e^{i\omega s } \langle \hat{B}_\alpha^\dagger (s)  \hat{B}_\beta (0) \rangle.
\end{align}
All of the unknown structure of the bath has now been moved into this function of $\omega$. The one-sided Fourier transform is usually used to define the two new functions
\begin{align}
S_{\alpha \beta}(\omega) &= \frac{1}{2i}\left( \Gamma_{\alpha \beta} (\omega) - \Gamma_{\beta \alpha}^* (\omega) \right),\\
\gamma_{\alpha \beta}(\omega) &= \Gamma_{\alpha \beta} (\omega) + \Gamma_{\beta \alpha}^* (\omega) = \int_{-\infty}^{\infty} ds\,e^{i\omega s } \langle \hat{B}_\alpha^\dagger (s)  \hat{B}_\beta (0) \rangle.
\end{align}
The second function is the normal Fourier transform of the bath correlation functions. Normally, $S_{\alpha \beta}$ is neglected essentially substituting $\Gamma_{\alpha \beta}$ for $\gamma_{\alpha \beta}/2$. Likewise, cross terms between different operators are neglected. This is trivially true when only one operator is used per bath, but it can also be true in other situations. The final version of the global master equation is
\begin{align}
\frac{d}{dt} \hat{\rho}_S(t) &= -i[\hat{H}_S, \hat{\rho}_S] + \sum_{\alpha} \mathcal{N} [\hat{A}_\alpha, \hat{\rho}_S, \gamma_{\alpha \alpha}] , \label{eq:MasterEquationsGlobal}\\
\mathcal{N} [\hat{A}, \hat{\rho}, \gamma] &= \sum_{\omega, \omega'} \frac{\gamma (\omega)}{2} \left( \hat{A} (\omega) \hat{\rho}(t) \hat{A}^\dagger (\omega') + \hat{A} (\omega') \hat{\rho}(t) \hat{A}^\dagger (\omega) \right. \label{eq:MasterEquationsGlobalN} \\ 
& \hspace{2.7cm}\left. - \hat{A}^\dagger (\omega') \hat{A} (\omega) \hat{\rho}(t) - \hat{\rho}(t) \hat{A}^\dagger (\omega) \hat{A} (\omega') \right) , \nonumber
\end{align}
where $\mathcal{N}$ is a shorthand taking two operators and a function as input. 
Going back to Eq.~\eqref{eq:MasterEquationsGlobalPre}, one last approximation can be performed called the secular approximation. This approximation will not be valid for the majority of the systems in later chapters; However, the result is still quite instructive. The secular approximation is a rotating wave approximation in which all terms containing $\omega \neq \omega'$ are neglected. This works if all pairs of frequencies $\omega$ and $\omega'$ are sufficiently different, such that $e^{i(\omega' -\omega) t }$ rotates fast enough to not contribute on average. Sometimes, only a partial secular approximation is performed. Here, only terms where $|\omega - \omega'|$ is larger than some threshold are thrown away. This makes numerical simulations easier. With the full secular approximation, the master equation becomes
\begin{align}
\frac{d}{dt} \hat{\rho}_S(t) &= -i [\hat{H}_{LS}, \hat{\rho}_S] + \sum_\omega \sum_{\alpha} \gamma_{\alpha \alpha} (\omega) \mathcal{M}[\hat{A}_\alpha (\omega), \hat{\rho}_S], \label{eq:MasterEquationsGlobalSecular}
\end{align}
where we have ignored cross terms between different operators setting $\alpha = \beta$, we have used $\mathcal{M}[\hat{A}, \hat{\rho}] = \hat{A} \hat{\rho} \hat{A}^{\dag} - \{\hat{A}^{\dag} \hat{A}, \hat{\rho}\}/2 $ as defined in section~\ref{sec:MasterEquationsLocal}, and we have defined 
\begin{align}
\hat{H}_{LS} &= \sum_{\omega} \sum_{\alpha, \beta} S_{\alpha \beta}(\omega) \hat{A}_\alpha^\dagger (\omega) \hat{A}_\beta (\omega).
\end{align}
Going back to the Schrödinger picture, the Hamiltonian is $H = H_S + H_{LS}$ such that $H_{LS}$ gives a Lamb shift due to the interaction. Since {$[H_S, H_{LS}]=0$}, the additional Hamiltonian only perturbs the energies of the eigenstates of $\hat{H}_S$. Therefore, in many situations, it does not affect the dynamics or steady state of the system, so it is usually neglected. Note that the above master equation is on Lindblad form similar to the local master equation from Sec~\ref{sec:MasterEquationsLocal}.
Finally, we shall address the one assumption that we have not discussed yet
\begin{align}
0&= \mathrm{tr}_B\{ [\hat{H}_I(t), \hat{\rho} (0)]\} \label{eq:MasterEquationsAssumption} \\
&= \sum_\alpha \langle \hat{B}_\alpha \rangle_B \sum_\omega e^{-i\omega t} \left[ \hat{A}_\alpha (\omega) \hat{\rho}_S(0) - \hat{\rho}_S(0) \hat{A}_\alpha (\omega) \right]. \nonumber
\end{align}
Above, we have rewritten the assumption using the eigen operators, $\hat{A}_\alpha (\omega)$. Without the above being true, the time derivative would depend on the initial state at all later times. Intuitively, we would expect most systems to lose all memory of the initial state as time evolves. To actually prove this, the bath correlation functions can be used
\begin{equation}
\langle \hat{B}_\alpha^{\dag}(t) \hat{B}_\beta (0) \rangle_B = \mathrm{tr}_B\{  \hat{B}_\alpha^{\dag} e^{-i\hat{H}_B t} \left(\hat{B}_\beta  \hat{\rho}_B \right) e^{i\hat{H}_B t} \}.
\end{equation}
Letting $t\rightarrow \infty$, the bath state has to return to the steady state $\hat{\rho}_B$ after the application of the bath operator $\hat{B}_\beta$, i.e., $\hat{B}_\beta \hat{\rho}_B \rightarrow \mathrm{tr}_B\{ \hat{B}_\beta \hat{\rho}_B \} \hat{\rho}_B$. Since the trace of the density matrix is preserved under time evolution, the density matrix goes to $\mathrm{tr}_B\{ \hat{B}_\beta \hat{\rho}_B \} \hat{\rho}_B$ instead of $ \hat{\rho}_B$. Finally, we use $\mathrm{tr}_B\{ \hat{B}_\beta \hat{\rho}_B \} = \langle \hat{B}_\beta \rangle_B$ to get
\begin{equation}
\lim_{t \rightarrow \infty} \langle \hat{B}_\alpha^{\dag}(t) \hat{B}_\beta (0) \rangle_B = \langle \hat{B}_\alpha^{\dag} \rangle_B \langle \hat{B}_\beta \rangle_B.
\end{equation}
For the Markov approximation to be valid, the correlation functions have to approach zero for longer times. As a consequence, the expectation value of all the bath interaction operators has to be zero
\begin{equation}
\langle \hat{B}_\alpha \rangle_B = 0.
\end{equation}
Therefore, Eq.~\eqref{eq:MasterEquationsAssumption} is fulfilled as a direct consequence of the Markov approximation. As a summary, the assumptions left are $[\hat{H}_B, \hat{\rho}_B]=0$, which is always true for a thermal state, and the Born-Markov approximation.

\subsection{Reduction to the local master equation}
\label{subsec:MasterEquationsReductionToLocal}

In certain situations, the global master equation reduces to the local master equation on Lindblad form discussed in section \ref{sec:MasterEquationsLocal}. To examine when this can happen, we will look at a single bath coupling to the system through a single operator, $\hat{A}$. The bath correlation functions are picked to be exponentially decaying
\begin{equation}
\langle \hat{B}^\dag (t) \hat{B} \rangle = \frac{\Gamma \kappa}{2} e^{-\kappa |t|} .
\end{equation}
Here $\Gamma$ is not the one-sided Fourier transform but rather a constant rate. The Lamb shift and decay rates can now easily be found
\begin{align*}
S(\omega) &= 0,\\
\gamma (\omega) &= \frac{\Gamma \kappa^2/4}{\kappa^2/4 + \omega^2}.
\end{align*}
Next, we let $\kappa \gg \epsilon_{\mathrm{max}} - \epsilon_{\mathrm{min}}$, where $\epsilon_{\mathrm{min}}$ is the minimum eigenenergy and $\epsilon_{\mathrm{max}}$ is the maximum eigenenergy of the system Hamiltonian. Therefore, $\gamma(\omega)$ is constant for all possible inputs $\omega$ in the sum, in Eq~\eqref{eq:MasterEquationsGlobalPre}, such that it can be substituted for $\gamma (0)$
\begin{align}
\frac{d}{dt} \rho_S(t) &= \frac{\Gamma}{2} \sum_{\omega, \omega'}  e^{i(\omega' -\omega) t } \left( \hat{A} (\omega) \rho_S(t) \hat{A}^\dagger (\omega')  - \hat{A}^\dagger (\omega') \hat{A} (\omega) \rho_S(t) \right) + \textrm{h.c.}.
\end{align}
Moving back into the Schrödinger picture, we get
\begin{align}
\frac{d}{dt} \rho_S(t) &= -i[\hat{H}_S, \hat{\rho}_S] + \Gamma  \left( \hat{A} \hat{\rho}_S(t) \hat{A}^\dagger - \frac{1}{2}\{\hat{A}^\dagger \hat{A}, \hat{\rho}_S(t) \} \right),
\end{align}
where we have used the identity $\sum_\omega \hat{A}(\omega) = \hat{A}$. This is now on the same form as the local master equation, namely on Lindblad form. However, for this to be local, the system must be comprised of many smaller systems, and the operator $\hat{A}$ must act on only one subsystem. More generally, the same result can be achieved by a combination of $\hat{A}$ only allowing certain transitions and $\gamma(\omega)$ being approximately constant around the frequencies of these allowed transitions. Fortunately, this will be the case in most of the systems in other chapters. In section \ref{sec:MasterEquationsComparison}, we will compare the local and global master equations and examine when they are each valid.

\subsection{Time-dependent Hamiltonian}
\label{subsec:MasterEquationsTimeDependentH}

The global master equation derived in Eq.~\eqref{eq:MasterEquationsGlobal} can be expanded to include time-dependent Hamiltonians. The time dependence can be included through either the system or interaction Hamiltonians. First, the interaction Hamiltonian is generalized to be time dependent through
\begin{align*}
\hat{H}_I = \sum_\alpha f_\alpha (t) \hat{A}_\alpha \otimes \hat{B}_\alpha.
\end{align*}
$f_\alpha(t)$ is a general function of time. We let the function $f_\alpha(t)$ follow $\hat{B}_\alpha$ through the derivation while decomposing $\hat{A}_\alpha$ into eigenoperators $\hat{A}_\alpha(\omega)$. The result only deviates through the one-sided Fourier transform
\begin{align}
\Gamma_{\alpha \beta} (\omega) = \int_{0}^{\infty} ds\,e^{i\omega s } f_\alpha (t)^* f_\beta (t-s) \langle \hat{B}_\alpha^\dagger (s)  \hat{B}_\beta (0) \rangle, \label{eq:MasterEquationsTimeDependentInteraction}
\end{align}
where $f_\alpha (t)^*$ is the complex conjugate of the function. This one-sided Fourier transform can then be used in conjunction with Eq.~\eqref{eq:MasterEquationsGlobalPre}, which will be done in chapter \ref{chapter:QutritDiode}.

Finally, the system Hamiltonian can be time dependent. In this case, the interaction operator $\hat{A}_\alpha$ can no longer be decomposed into eigenoperators of $\hat{H}_S$, and the derivation no longer applies. The derivation can be adapted in general \cite{PhysRevA.98.052129}; However, we will assume that the time dependence is periodic, $\hat{H}_S(t + \Theta) = \hat{H}_S(t)$, with period $\Theta$.
In this case, the bath operators are decomposed using Floquet theory \cite{PhysRevA.98.052129, PhysRevE.85.061126, PhysRevA.73.052311}
\begin{align}
\hat{A}_\alpha(t) = \sum_q \sum_\omega e^{i(\omega + q\Omega)t} \hat{A}_\alpha(\omega, q), \label{eq:MasterEquationsFloquet}
\end{align}
where $\Omega = \frac{2\pi}{\Theta}$ and $q$ is a whole number. $\omega$ is a difference of Floquet energies, and $\hat{A}_\alpha(\omega, q)$ drive transitions between Floquet states of frequency $\omega$. The result is a master equation on the form of Eq.~\eqref{eq:MasterEquationsGlobalPre} with the following replacements
\begin{align}
\sum_{\omega,\omega'} \sum_{\alpha \beta} e^{i(\omega'-\omega)t} \Gamma_{\alpha \beta} (\omega) &\rightarrow \sum_{q,q'} \sum_{\omega,\omega'} \sum_{\alpha \beta} e^{i(\omega' +q'\Omega -\omega -q\Omega)t} \Gamma_{\alpha \beta} (\omega+q \Omega), \\
\hat{A}_\alpha (\omega) &\rightarrow \hat{A}_\alpha(\omega, q).
\end{align}
This new master equation is more difficult to use; However, for our purposes, it is only important to know that it exists and reduces to the local master equation under certain conditions. For example, if $\Gamma_{\alpha \beta}$ is constant, Eq.~\eqref{eq:MasterEquationsFloquet} can be put back in, resulting in the local master equation.

\section{Comparison of master equation validity}
\label{sec:MasterEquationsComparison}

\begin{figure}[h]
\centering
\includegraphics[width=1.0\linewidth, angle=0]{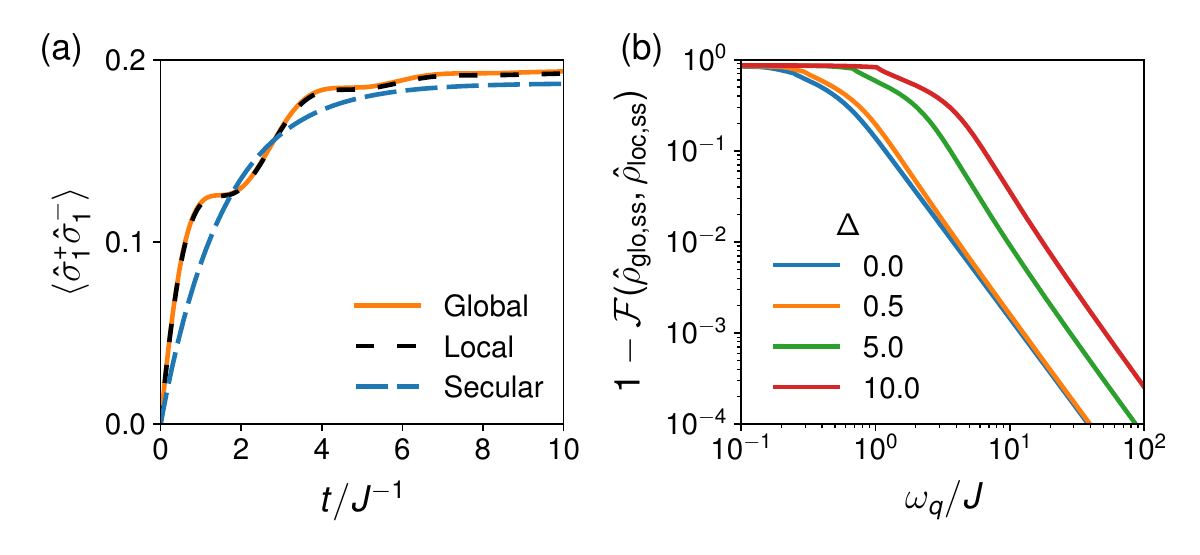}
\caption{(a) Population of the excited state for qubit 1 as a function of time using the global master equation, global master equation with full secular approximation, and the local master equation. For this $\gamma = 0.4J$, $\omega_q = 10J$ and $\Delta =0.3$. (b) The fidelity $\mathcal{F}$ between the steady states using the global and local master equations as a function of two-level-system frequency with $\gamma = J$. }\label{figure:ChapterMasterEquationsComparison}
\end{figure}

As mentioned in section \ref{subsec:MasterEquationsReductionToLocal}, the global master equation reduces to the local master equation in certain situations. In fact, this frequently happens for chains of spins or qubits when the longitudinal magnetic field or qubit frequency is large enough. To exemplify this, we define the following toy model comprised of two coupled qubits
\begin{equation}
\hat{H} = \frac{\omega_1}{2} \hat{\sigma}_1^z + \frac{\omega_2}{2} \hat{\sigma}_2^z + J \left( \hat{\sigma}_1^+ \hat{\sigma}_2^- + \hat{\sigma}_1^- \hat{\sigma}_2^+ + \frac{\Delta}{4} \hat{\sigma}_1^z \hat{\sigma}_2^z \right).
\end{equation}
The two qubit frequencies are $\omega_1 = \omega_q + \epsilon$ and $\omega_2=\omega_q$, where $\omega_q$ is a constant frequency and $\epsilon = 0.2J$ is a small frequency breaking the degeneracy between the two qubits. $J$ is the hopping between the two sites, and $\Delta$ sets the relative strength of the Z-coupling. When $J\neq 0$, an excitation can oscillate between the two sites. Next, the two master equations are set up. The global master equation is defined in Eq.~\eqref{eq:MasterEquationsGlobal} and takes the form
\begin{align} 
\frac{d \hat{\rho}_{\mathrm{glo}}}{dt} &= -i[ \hat{H}, \hat{\rho}_{\mathrm{glo}}] + \mathcal{D}_{\mathrm{glo}}[\hat{\rho}_{\mathrm{glo}}], \\
\mathcal{D}_{\mathrm{glo}}[\hat{\rho}] &= \sum_{i=1}^2 \mathcal{N}[\hat{\sigma}^x_i, \hat{\rho}, \gamma_{i}] ,
\end{align}
where the frequency-dependent decay rate is
\begin{align}
\gamma_{i}(\omega) &= \Gamma \frac{\omega}{\omega_i} \big(1+n_i(\omega)\big).
\end{align}
When it takes this form, proportional to $\omega$, the baths are called Ohmic. The other part is given by $n_i(\omega) = [\exp(\omega/T_i)-1]^{-1}$, where $T_1=\omega_1$ and $T_2 = 0.2\omega_2$ are the temperatures of the baths. 
The local master equation is defined in Eq.~\eqref{eq:MasterEquationsLocal} and takes the form
\begin{align}
\frac{d \hat{\rho}_{\mathrm{loc}}}{dt} &= -i[ \hat{H}, \hat{\rho}_{\mathrm{loc}}] + \mathcal{D}_{\mathrm{loc}}[\hat{\rho}_{\mathrm{loc}}], \\
\mathcal{D}_{\mathrm{loc}}[\hat{\rho}] &= \sum_{i=1}^2 \Gamma_{i, -} \mathcal{M}[\hat{\sigma}^-_i, \hat{\rho}] + \Gamma_{i, +} \mathcal{M}[\hat{\sigma}^+_i, \hat{\rho}].
\end{align}
When $\omega_q$ is much larger than all other parameters, any excitation added or subtracted by a bath will add or subtract the energy $\omega_1$ or $\omega_2$. The equivalent rates for the local master equation can be found by evaluating $\gamma_i$ at $\pm \omega_i$ 
\begin{align}
\Gamma_{1, -} &= \gamma_1(\omega_1), \quad \quad 
\Gamma_{1, +} = \gamma_1(-\omega_1) , \\
\Gamma_{2, -} &= \gamma_2(\omega_2), \quad \quad
\Gamma_{2, +} = \gamma_2(-\omega_2).
\end{align}
To see the difference between the master equations, we plot $\langle \hat{\sigma}^+_1 \hat{\sigma}^-_1 \rangle$ as a function of time for an initial state of $\hat{\rho}_\mathrm{loc}(0) = \hat{\rho}_\mathrm{glo}(0) = |0_1 0_2\rangle$ in Fig.~\ref{figure:ChapterMasterEquationsComparison}(a). In addition to the standard local and global master equations, the global master equation with the full secular approximation, defined in Eq.~\eqref{eq:MasterEquationsGlobalSecular}, is also plotted. Even though $\omega_q$ is only $10J$, the global and local master equations yield very similar results. The global master equation with secular approximation reaches a similar steady state. However, it does not capture any oscillations for smaller times. This is as expected since $\Gamma$ is too large for the approximation to be valid. In the context of the global master equation, the secular approximation is quite unreliable, especially for larger systems, where the Hilbert space is more likely to be close to degenerate. Finally, the fidelity between the steady states obtained using the global and local master equations is plotted in Fig.~\ref{figure:ChapterMasterEquationsComparison}(b). The fidelity used is defined as
\begin{equation}
\mathcal{F} (\hat{\rho}, \hat{\sigma}) = \left( \mathrm{tr} \left\{ \sqrt{ \sqrt{\hat{\rho}} \hat{\sigma} \sqrt{\hat{\rho}} } \right\} \right)^2,
\end{equation}
which satisfies $0 \leq \mathcal{F} \leq 1$ with $\mathcal{F}=1$ if and only if the two states are identical. To better resolve the fidelity, one minus the fidelity is plotted instead. The fidelity is lower for larger $\Delta$ as expected. However, a fidelity of $0.9999$ is reached already for $\omega_q \sim 10^2J$. These results are consistent with more involved studies comparing the local and global master equations \cite{Cattaneo_2019, Hofer_2017}. To conclude, the local master equation is valid whenever a well-defined excitation frequency, here $\omega_q$, exists, and all other parameters are much smaller than this frequency, here $\omega_q \gg \epsilon, J, \Delta$. Note that to a large extend, only Hamiltonian parameters affect the validity of the local master equation, while bath parameters can be as large as the Born-Markov approximation permits.

\cleardoublepage
\chapter{Dark-state-induced rectification}
\label{chapter:QutritDiode}
\epigraph{\footnotesize This chapter is based on Refs.~\cite{PhysRevE.106.034116} and \cite{poulsen2022heat}. In particular, sections \ref{sec:QutritDiodeIdealized}-\ref{sec:QutritDiodeImplementation} and all figures herein have been reproduced with permission from Ref.~\cite{PhysRevE.106.034116} with only minor changes. Likewise, section \ref{sec:QutritDiodeFWBR}, and all figures herein, have been reproduced with permission from Ref.~\cite{poulsen2022heat} with only minor changes. Copyright 2022 by the American Physical Society.
}{}

\noindent After having set up the mathematical framework for studying open quantum systems in chapter \ref{chapter:MasterEquations}, we can start looking at some boundary-driven quantum systems. The first transport phenomenon that we are going to study is called rectification. Rectification was first introduced in the introduction or chapter \ref{chapter:Introduction} and is characterized by a non-symmetric transport. In the introduction, we saw rectification in a simple two-level system caused by asymmetric bath coupling strengths. However, the rectification values are fairly small for this system. 

Since rectification occurs in such a minimal classical system, one might overestimate the general presence of unidirectional transport in quantum systems. In fact, starting from quantum mechanics, rectification might instead seem impossible. To see why, let us consider a simple routing problem. Let $\hat{H}$ be the Hamiltonian for a multipartite system, consisting of many nearest neighbor connected two-level systems or harmonic oscillators. The Hamiltonian is assumed to be real, which is true for most systems studied in this thesis. Now, let $\ket{L} = \ket{1,...,0}$ be the state with one excitation at the leftmost system and none at the rightmost system and similarly for $\ket{R} = \ket{0,...,1}$. The transition probability between these two states is 
\begin{align}
P(L \rightarrow R) &= |\langle R | \hat{U} (t, 0) | L \rangle|^2 \\
&= \sum_{n,m} \langle m | R \rangle \langle R | n \rangle \langle n | L \rangle \langle L | m \rangle e^{-i(\epsilon_n - \epsilon_m) t}, \nonumber
\end{align}
where $\{|\hat{n}\rangle\}$ form a complete basis of energy eigenstates. Since $\hat{H}$ is real, the coefficients $\langle n |L \rangle$ and $\langle n |R \rangle$ can all be picked real. Calculating the reverse transition probability, it is easy to see that
\begin{align}
P(L \rightarrow R) = P(R \rightarrow L).
\end{align}
This can be generalized to allow the middle subsystems to contain excitations as well, but we will keep it simple. So even for very asymmetric systems, transport through this constrained yet large class of quantum systems is completely symmetric. Coming from this perspective, rectification in quantum systems might seem impossible. 

This chapter will introduce one strategy to achieve large rectification without relying on engineering the system-bath interaction. Instead, the rectification is achieved through a dark state being populated in reverse bias but not in forward bias. First, the general idea is introduced for a simplified yet unrealistic setting. Second, a realistic implementation inspired by superconducting circuits as a platform is introduced. This is solved approximately in the limit of strong system-bath coupling compared to inter-subsystem couplings, and the robustness of the rectification is studied. Finally, four diodes are put into a bridge to study how they behave in a larger setup. The ideas of this chapter will be expanded in chapter \ref{chapter:Interference} using interference and entanglement to achieve even larger rectification.

\section{Idealized setup}
\label{sec:QutritDiodeIdealized}

\begin{figure}[h]
\centering
\includegraphics[width=1.0\linewidth, angle=0]{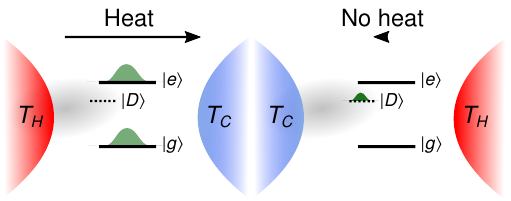}
\caption{ Idealized setup for dark-state-induced rectification. (a) In forward bias, excitations are transported through the channel $|g\rangle \leftrightarrow |e\rangle$. (b) In reverse bias, the diode is closed by the dark state $|D\rangle$. Reproduced with permission from Phys. Rev. E \textbf{106}, 034116 (2022) \cite{PhysRevE.106.034116}. Copyright 2022 American Physical Society.}
\label{figure:ChapterQutritDiodeSetupIdealized}
\end{figure}

The main strategy for rectification explored in this and the next chapter can be summed up by introducing three states coupled to two baths as seen in Fig.~\ref{figure:ChapterQutritDiodeSetupIdealized}. In forward bias, the left bath is hot and the right bath is cold as seen in Fig.~\ref{figure:ChapterQutritDiodeSetupIdealized}(a), while reverse bias is seen in Fig.~\ref{figure:ChapterQutritDiodeSetupIdealized}(b). The system is engineered to exploit the unidirectionality of a cold bath to drive the system into the state $\ket{D}$ in reverse bias. $\ket{D}$ is a dark state of the right bath, and if it is populated, it will completely block any transport between the two baths. In forward bias, the two remaining states $\ket{g}$ and $\ket{e}$ facilitate transport as usual, thus implementing a perfect heat diode. The mechanism can easily be understood through the master equation from section \ref{sec:IntroductionTwoLevelRectifier}
\begin{align}
\frac{d}{dt} \vec{P}(t) = W \vec{P}(t),
\end{align}
where the probability vector is now $\vec{P} = \{P(\ket{D}), P(\ket{g}), P(\ket{e})\}^T$ and the matrix of rates $W$ takes the form
\begin{align}
W= \begin{pmatrix}
-\Gamma_{D\rightarrow g}-\Gamma_{D\rightarrow e} & \Gamma_{g \rightarrow D} & \Gamma_{e \rightarrow D}\\
\Gamma_{D\rightarrow g} & -\Gamma_{g \rightarrow D}- \Gamma_{g \rightarrow e} & \Gamma_{e \rightarrow g}\\
\Gamma_{D\rightarrow e} & \Gamma_{g \rightarrow e} & -\Gamma_{e \rightarrow D}-\Gamma_{e \rightarrow g}
\end{pmatrix} .
\end{align}
Here $\Gamma_{a\rightarrow b}$ is the transition rate from $\ket{a}$ to $\ket{b}$. The left bath interaction is engineered such that, in reverse bias, the cold bath allows for transitions into the dark state but not out of it i.e. $\Gamma_{D\rightarrow g}, \Gamma_{D\rightarrow e} =0$ and $\Gamma_{g\rightarrow D}, \Gamma_{e\rightarrow D} > 0$. The other rates are kept general for now, but an example is given in section \ref{sec:QutritDiodeImplementation}. In reverse bias, the rate matrix becomes
\begin{align}
W= 
\begin{pmatrix}
0 & \Gamma_{g \rightarrow D} & \Gamma_{e \rightarrow D}\\
0 & -\Gamma_{g \rightarrow D}- \Gamma_{g \rightarrow e} & \Gamma_{e \rightarrow g}\\
0 & \Gamma_{g \rightarrow e} & -\Gamma_{e \rightarrow D}-\Gamma_{e \rightarrow g}
\end{pmatrix} .
\end{align}
In the long time limit, the system will go towards the steady-state solution $d \vec{P}_\text{ss}/dt = 0$, which is easily found to be
\begin{align}
P_\text{ss} (\ket{D}) = 1.
\end{align}
With one state fully populated no heat is transferred, and the system implements a perfect diode.

\section{Minimal implementation of dark-state-induced rectification}
\label{sec:QutritDiodeImplementation}

\begin{figure}[t]
\centering
\includegraphics[width=.9\linewidth, angle=0]{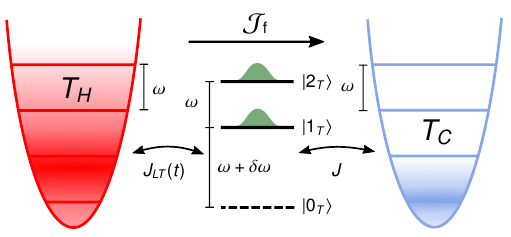}
\caption{ Possible setting for dark-state-induced rectification consisting of a qutrit and two harmonic oscillators. The ground state acts as the dark state, which is isolated from the right bath due to energy conservation. Reproduced with permission from Phys. Rev. E \textbf{106}, 034116 (2022) \cite{PhysRevE.106.034116}. Copyright 2022 American Physical Society.}\label{figure:ChapterQutritDiodeSetup}
\end{figure}

To discuss the heat transferred in forward bias and to discuss one possible implementation of this idea, the example seen in Fig.~\ref{figure:ChapterQutritDiodeSetup} is studied. 

\subsection{Hamiltonian and master equation}

The Hamiltonian for this implementation is inspired by superconducting circuits as a platform
\begin{align}
\hat{H} &= \omega \Big(\hat{a}_L^\dag \hat{a}_L + \hat{a}_R^\dag \hat{a}_R\Big) + (\omega +\delta \omega) \hat{a}_T^\dag \hat{a}_T  - \frac{\delta \omega}{2} \hat{a}_T^\dag \hat{a}_T \Big(\hat{a}_T^\dag \hat{a}_T -1\Big)  \\ &\hspace{0.8cm} + J_{LT}(t) \Big(\hat{a}_L + \hat{a}_L^\dag \Big) \Big(\hat{a}_T + \hat{a}_T^\dag \Big) + J \Big(\hat{a}_T + \hat{a}_T^\dag \Big) \Big(\hat{a}_R + \hat{a}_R^\dag \Big) , \nonumber
\end{align}
where the subscripts L, T, and R are used for the left harmonic oscillator, the qutrit, and the right harmonic oscillator, respectively. $\hat{a}_\alpha$ and $\hat{a}^\dag_\alpha$ for $\alpha \in \{L, T, R\}$ are ladder operators. $\omega$ is the frequency of the oscillators, and $\delta \omega$ is the anharmonicity of the qutrit. The left hopping is time dependent $J_{LT} (t) = J + J' \cos(\delta \omega t)$ and $J$ sets the energy scale of the system. Transforming into the interaction picture with respect to $\hat{H}_0 = \omega \sum_{\alpha \in \{L, T, R\}} \hat{a}_\alpha^\dag \hat{a}_\alpha$, performing a rotating wave approximation on terms rotating with frequency $2 \omega$, and truncating the qutrit to the three lowest levels the Hamiltonian becomes
\begin{align}
\hat{H}_I &= -\delta \omega \op{0_T} + J_{LT}(t) \Big(\hat{a}_L \hat{a}_T^\dag + \hat{a}_L^\dag \hat{a}_T \Big) + J \Big(\hat{a}_T \hat{a}_R^\dag + \hat{a}_T^\dag \hat{a}_R \Big), 
\end{align}
where we have introduced the notation $\ket{0_T}$, $\ket{1_T}$, and $\ket{2_T}$ for the three qutrit states. The two harmonic oscillators act as the two baths in Fig.~\ref{figure:ChapterQutritDiodeSetupIdealized}. The decay of the harmonic oscillator correlation functions is modeled through the local master equation on Lindblad form
\begin{align}
\frac{d \hat{\rho}}{d t} = \mathcal{L}[\hat{\rho}] &= -i [\hat{H}, \hat{\rho}] + \mathcal{D}_L[\hat{\rho}] + \mathcal{D}_R[\hat{\rho}]. \label{eq:ChapterQutritDiodeLiouvillian}
\end{align}
The dissipative terms, $\mathcal{D}_{L}[\hat{\rho}]$ and $\mathcal{D}_{R}[\hat{\rho}]$, describe the action of the left and right baths, respectively,
\begin{align}
\mathcal{D}_{L}[\hat{\rho}] &= \Gamma (n_{L} + 1) \mathcal{M}[ \hat{a}_{L}, \hat{\rho}]
+ \Gamma n_{L} \mathcal{M}[\hat{a}_{L}^\dag, \hat{\rho}] , \\
\mathcal{D}_{R}[\hat{\rho}] &= \Gamma (n_{R} + 1) \mathcal{M}[ \hat{a}_{R}, \hat{\rho}]
+ \Gamma n_{R} \mathcal{M}[\hat{a}_{R}^\dag, \hat{\rho}] ,
\end{align}
where $\mathcal{M}[\hat{A}, \hat{\rho}] = \hat{A} \hat{\rho} \hat{A}^\dag - \{\hat{A}^\dag \hat{A}, \hat{\rho}\}/2$ as defined in Eqs.~\ref{eq:MasterEquationsLocal}-\ref{eq:MasterEquationsLocalDissipator}. As discussed in chapter \ref{chapter:MasterEquations}, the local master equation can describe thermal baths for $\omega$ much larger than any other parameter in the Hamiltonian. $\Gamma$ is the coupling strength between the baths and harmonic oscillators and $n_{L(R)}~=~\big(e^{\omega/T_{L(R)}}-1\big)^{-1}$ is the mean number of excitation in the left (right) harmonic oscillator in the absence of the qutrit. By forward bias, we denote the case where $n_L = n_H$ and $n_R = n_C$, and heat flows from left to right. By reverse bias, we denote the case where $n_L = n_C$ and $n_R = n_H$, and heat flows from right to left. We assume that $n_H > n_C$. Once again, we are interested in the the long time limit, where the steadystate $\hat{\rho}_{\mathrm{ss}}$ is reached. Unless otherwise stated throughout this chapter, we will use the set of parameters 
\begin{align}
\delta \omega = 300J, \quad J'=0.5J, \quad \Gamma = 10J, \quad n_H = 0.5, \quad n_C = 0. \label{eq:QutritDiodeDefaultValues}
\end{align} 
To study transport, we write the change in total energy of the system
\begin{align}
0=\frac{d\ev{H}_\text{ss}}{dt}  = \ev{\frac{d\hat{H}}{dt} }_\text{ss}  + \tr \left\{\hat{H} \mathcal{D}_L [\hat{\rho}_\text{ss}] \right\} + \tr \left\{\hat{H} \mathcal{D}_R [\hat{\rho}_\text{ss}] \right\}, \label{eq:heat}
\end{align}
where $\ev{\bullet}_\text{ss}= \tr \left\{\bullet \hat{\rho}_\text{ss} \right\}$ is the steady state expectation value. Similar to section \ref{sec:MasterEquationsLocal}, the first part is identified as the work, the second term is the heat increase due to the left bath, and the third term is the heat increase due to the right bath. From this, we define the two transport measures:
\begin{align}
\mathcal{J}_{L(R)} &= \Gamma n_{L(R)} \ev{\hat{a}_{L(R)} \hat{a}_{L(R)}^\dag}_\text{ss} - \Gamma (n_{L(R)}+1) \ev{\hat{a}_{L(R)}^\dag \hat{a}_{L(R)}}_\text{ss}, \label{eq:ChapterQutritDiodeCurrentDef} \\
\mathcal{W} &= \frac{J' \delta \omega}{2i} \ev{ \hat{a}_L \hat{a}_T^\dag e^{-i \delta \omega t} - \hat{a}_L^\dag \hat{a}_T e^{i \delta \omega t} }_\text{ss} .
\end{align}
Since the steady-state density matrix is independent of $\omega$, we use the $\omega$-independent excitation current $\mathcal{J}_{L(R)}$ instead of the heat current $\omega \mathcal{J}_{L(R)}$. Likewise, we will focus on the number of excitations added through work $\mathcal{W}/\delta \omega$. We have used that $\omega, \delta \omega \gg J, J'$ such that the substitution $\hat{H} \rightarrow \hat{H}_0$ can be made in the second and third term in Eq.~\eqref{eq:heat}. We define the forward bias excitation current to be $\mathcal{J}_\textrm{f} = -\mathcal{J}_R$, while the reverse bias excitation current is $\mathcal{J}_\textrm{r} = \mathcal{J}_L$. The quality of the diode is quantified using the rectification
\begin{align}
\mathcal{R} = -\frac{\mathcal{J}_\textrm{f}}{\mathcal{J}_\textrm{r}},
\end{align}
which tends to infinity for a perfect diode. 

In summary, the system is designed using the methodology seen in Fig.~\ref{figure:ChapterQutritDiodeSetupIdealized}. This is seen through the equivalences
\begin{align}
\ket{D} \equiv \ket{0_T}, \quad \ket{g} \equiv \ket{1_T}, \quad \ket{e} \equiv \ket{2_T}.
\end{align}
In forward bias, excitations can propagate through the system through, e.g., transitions like
\begin{align}
\text{Forward:}& \quad \ket{1_L 1_T 0_R} \leftrightarrow \ket{0_L 2_T 0_R} \leftrightarrow \ket{0_L 1_T 1_R}.
\end{align}
In reverse bias, the qutrit is trapped in the dark state $\ket{0_T}$ through the transitions
\begin{align}
\text{Reverse:}& \quad \ket{0_L 1_T 0_R} \leftrightarrow \ket{1_L 0_T 0_R} \rightarrow \ket{0_L 0_T 0_R}. \label{eq:QutritDiodeReverseTransition}
\end{align}
The first part is allowed when $J' > 0$, and the second part is due to the cold bath. When the qutrit is in the dark state, excitations are not allowed to propagate from the hot bath to the qutrit due to energy conservation.

\subsection{Rectification values} 

\begin{figure}[t]
\centering
\includegraphics[width=1.0\linewidth, angle=0]{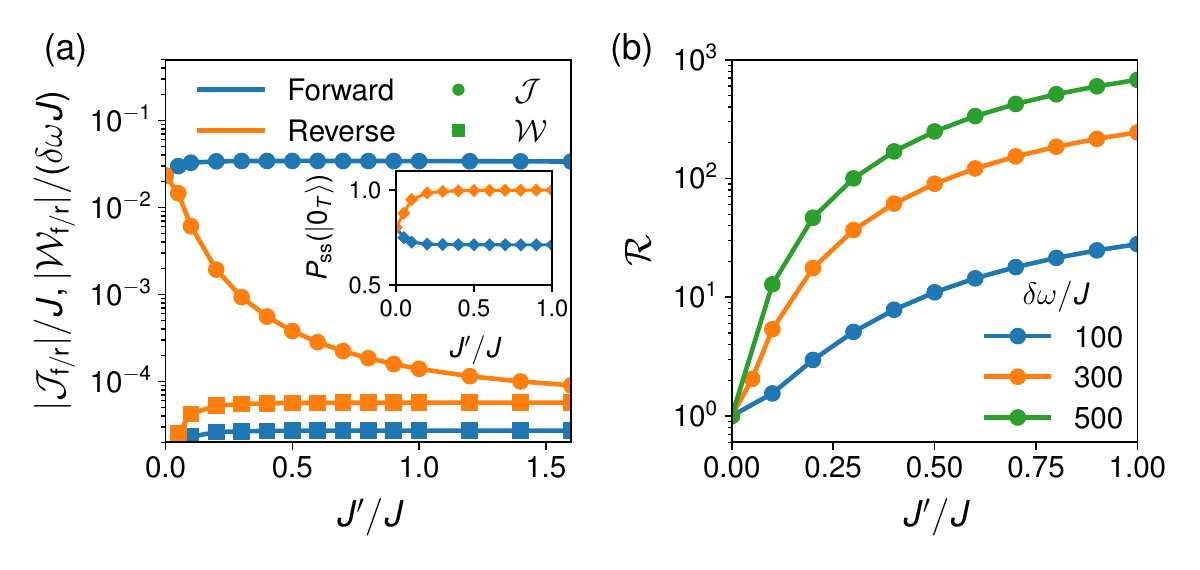}
\caption{ (a) Excitation current and work as the diode is turned on through $J'$. The inset to the left panel has the steady-state population of the dark state as a function of $J'$ for both forward and reverse bias. (b) Rectification $\mathcal{R}$ as a function of $J'$ for different anharmonisities $\delta \omega$. See Eq.~\eqref{eq:QutritDiodeDefaultValues} for the values of the remaining system parameters. Reproduced with permission from Phys. Rev. E \textbf{106}, 034116 (2022) \cite{PhysRevE.106.034116}. Copyright 2022 American Physical Society.}\label{figure:ChapterQutritDiodeJAndR}
\end{figure}

With the context set up, we are almost in a position to calculate the currents, work, and rectification factors. However, in order to achieve numerical results, the harmonic oscillators are truncated such that the highest excited state $\ket{m_\text{max}}$ has $P_{\text{th}}(\ket{m_\text{max}}) < 10^{-3}$ where $m_\text{max} \in \{0, 1, 2, 3, ...\}$. Furthermore, $P_{\text{th}}(\ket{m})$ is the population of $\ket{m}$ in the harmonic-oscillator thermal state
\begin{align}
P_\text{th}(\ket{m}) &= \frac{\tr \{ \op{m} e^{-\omega/T \hat{a}^\dag \hat{a}}\}}{\tr \{e^{-\omega/T \hat{a}^\dag \hat{a}}\}}  = \frac{n^m}{(1+n)^{m+1}},
\end{align} 
where $n$ is $n_L$ for the left bath and $n_R$ for the right bath. The highest kept excited state can then be found to be 
\begin{align}
m_\text{max} = \left\lceil \frac{\ln\big\{ (n+1) 10^{-3} \big\} }{\ln n - \ln (n+1)} \right\rceil ,
\end{align}
where $\lceil \bullet \rceil$ is the function that returns the smallest integer greater than or equal to the input. A minimum of three levels are always kept per harmonic oscillator. The initial state is picked to be diagonal with the approximate populations of the Markovian solution from the next section. The density matrix is then evolved for a time $t_{\text{final}} = 5000J^{-1}$ and the quantities are averaged over times $t = 4900J^{-1} - 5000J^{-1}$.

Now the currents, work, rectification can be found. The excitation current and work results are plotted in Fig.~\ref{figure:ChapterQutritDiodeJAndR}(a) as the diode is turned on via an increase in $J'$. The current, in reverse bias, becomes suppressed, and the number of excitations added through work per unit of time is orders of magnitude smaller than the excitation current in either bias. This is further verified by Fig.~\ref{figure:ChapterQutritDiodeJAndR}(b) where the rectification is plotted. To verify that the current in reverse bias is indeed blocked due to the dark state being populated, we plot the dark state population, $P_{\text{ss}}(\ket{0_T}) = \ev{\op{0_T}}_{\text{ss}}$ in the inset of Fig.~\ref{figure:ChapterQutritDiodeJAndR}(a). The dark state population in reverse bias does indeed approach unity as the diode is turned on.

\subsection{Markovian solution} 
\label{subsec:QutritDiodeMarkovianSolution}

\begin{figure}[t]
\centering
\includegraphics[width=0.625\linewidth, angle=0]{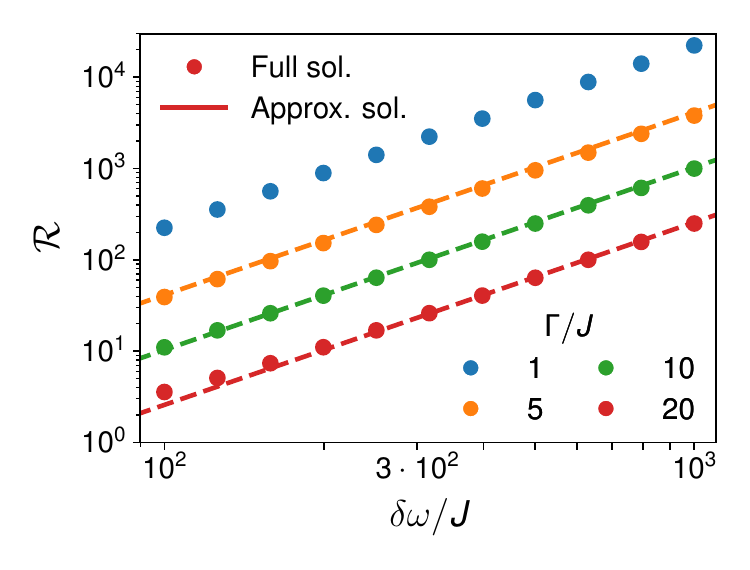}
\caption{ Rectification as a function of the anharmonisity $\delta \omega$ for different values of $\Gamma$. The full solution is plotted using points, and the approximate solution, Eq.~\eqref{eq:ChapterQutritDiodeRMarkov}, is plotted using a dashed line. See Eq.~\eqref{eq:QutritDiodeDefaultValues} for the values of the remaining system parameters. Reproduced with permission from Phys. Rev. E \textbf{106}, 034116 (2022) \cite{PhysRevE.106.034116}. Copyright 2022 American Physical Society.}\label{figure:ChapterQutritDiodeR}
\end{figure}

Knowing that the diode mechanism functions as intended, we can turn to one regime where an approximate solution is achievable. This solution can be found for $\omega, \delta \omega \gg \Gamma \gg J,J'$, $J'/J\gg \Gamma/\delta \omega$, $n_C=0$, and $n_H \leq 1$. In this regime, all coherences in $\hat{\rho}$ will decay rapidly, and the harmonic oscillators can be seen as baths with Lorenzian spectral densities through the Born-Markov approximation. To do this, we have to take a step back to chapter \ref{chapter:MasterEquations}, redefine the bath and system Hamiltonians, and rederive a master equation. Fortunately, since the general derivation has been done, it is not necessary to redo. The bath, interaction, and system Hamiltonians, respectivelly, become
\begin{align}
\hat{H}'_B &= \hat{H}_B + \hat{H}_I + \omega (\hat{a}_L^\dag \hat{a}_L + \hat{a}_R^\dag \hat{a}_R), \\
\hat{H}'_I &= J_{LT}(t) \left(\hat{a}_L + \hat{a}_L^\dag \right) \left(\hat{a}_T + \hat{a}_T^\dag \right) + J \left(\hat{a}_T + \hat{a}_T^\dag \right) \left(\hat{a}_R + \hat{a}_R^\dag \right), \\
\hat{H}'_S &= (\omega +\delta \omega) \hat{a}_T^\dag \hat{a}_T  - \frac{\delta \omega}{2} \hat{a}_T^\dag \hat{a}_T \Big(\hat{a}_T^\dag \hat{a}_T -1\Big) ,
\end{align}
where the unprimed Hamiltonians are the old bath, interaction, and system Hamiltonians. In order to write out the new master equation, the correlation functions for the two bath operators $\hat{B}_L = \hat{a}_L + \hat{a}_L^\dag$ and $\hat{B}_R = \hat{a}_R + \hat{a}_R^\dag$ have to be found. These can easily be found using the old master equation in the Heisenberg picture with no interaction to the qutrit
\begin{align}
\frac{d }{dt} \hat{B}_L (t) &= i[ \omega \hat{a}_L^\dag \hat{a}_L, \hat{B}_L (t)] + \Gamma (1+n_L) \left( \hat{a}_L^\dag \hat{B}_L (t) \hat{a}_L - \frac{1}{2} \{ \hat{a}_L^\dag \hat{a}_L , \hat{B}_L(t)\} \right) \nonumber \\
&\hspace{1cm} + \Gamma n_L \left( \hat{a}_L \hat{B}_L (t) \hat{a}_L^\dag - \frac{1}{2} \{ \hat{a}_L \hat{a}_L^\dag , \hat{B}_L (t)\} \right), \label{eq:ChapterQutritDiodeHeisenberg}
\end{align}
and similarly for $\hat{B}_R$. These can be solved 
\begin{align}
\hat{B}_L(t) &= e^{-\frac{\Gamma t}{2}} \left( e^{-i\omega t} \hat{a}_L + e^{i\omega t} \hat{a}_L^\dag \right), \\
\hat{B}_R(t) &= e^{-\frac{\Gamma t}{2}} \left( e^{-i\omega t} \hat{a}_R + e^{i\omega t} \hat{a}_R^\dag \right). \label{eq:ChapterQutritDiodeBathOpB}
\end{align}
From this, the correlations functions can be found
\begin{align}
\langle \hat{B}_L(t) \hat{B}_L \rangle &= e^{-\frac{\Gamma t}{2}} \left( (1+n_L) e^{-i\omega t} + n_L e^{i\omega t} \right),\\
\langle \hat{B}_R(t) \hat{B}_R \rangle &= e^{-\frac{\Gamma t}{2}} \left( (1+n_R) e^{-i\omega t} + n_R e^{i\omega t} \right).
\end{align}
Since the harmonic oscillators are now incorporated into the Born-Markov approximation, their states are assumed to be thermal, i.e., $\langle \hat{a}_L^\dag \hat{a}_L \rangle = n_L$ and $\langle \hat{a}_R^\dag \hat{a}_R \rangle = n_R$. 

Knowing the correlation functions, we can return to setting up the new master equation. Since the coupling strength for the left bath depends on time, the one-sided Fourier transforms from Eq.~\eqref{eq:MasterEquationsTimeDependentInteraction} is used
\begin{align}
\Gamma_L(\omega') &= \int_0^\infty ds e^{i\omega' s} J_{LT} (t) J_{LT} (t-s) \langle \hat{B}_L(s) \hat{B}_L \rangle, \\
\Gamma_R(\omega') &= \int_0^\infty ds e^{i\omega' s} J^2 \langle \hat{B}_R(s) \hat{B}_R \rangle.
\end{align}
Before moving on, the time-dependence of $J_{LT}(t)$ has to be resolved. To do this, we compose the product into its frequency components
\begin{align}
& J_{LT} (t) J_{LT} (t-s) = J^2 + JJ' \cos(\delta \omega t ) + JJ' \cos[\delta \omega (t-s) ]\\
& \hspace{5cm} + \frac{J'^2}{2} \cos[ \delta \omega (2t-s)] + \frac{J'^2}{2} \cos( \delta \omega s). \nonumber
\end{align}
The second to last term rotates with frequency $2 \delta \omega$ in $t$ which is fast enough that it can safely be neglected through a secular approximation. The second and third term rotate with frequency $\delta \omega$ in $t$ which is also fast enough that it can be neglected. However, the rest of Eq.~\eqref{eq:MasterEquationsGlobalPre} contains the terms $e^{i(\omega'-\omega'')t}$ where $\omega'$ and $\omega''$ are summed over all possible transition frequencies, i.e., $\omega$ and $\omega+\delta \omega$. Since the difference between these two frequencies are $\pm \delta \omega$, the second and third term does contribute to terms in the master equation where $\omega' = \omega$ and $\omega'' = \omega + \delta \omega$. These terms drive transitions between, e.g., $\langle 1_T | \hat{\rho}_T | 0_T \rangle$ and $\langle 2_T | \hat{\rho}_T | 1_T \rangle$. These are both coherences, which decay exponentially, and therefore, they are irrelevant. As a conclusion, the second and third term can also be neglected. This leaves the first and last term only. Putting all of this into the one-sided Fourier transforms, we get
\begin{align}
\Gamma_L(\omega') &= J^2  \left( \frac{(1+n_R)}{i(\omega-\omega') + \Gamma/2} + \frac{n_R}{-i(\omega+\omega') + \Gamma/2} \right) \\
& \hspace{1cm} + \frac{J'^2}{4} \left( \frac{(1+n_R)}{i(\omega-\omega'-\delta \omega) + \Gamma/2} + \frac{n_R}{-i(\omega+\omega'-\delta \omega) + \Gamma/2} \right) \nonumber \\
& \hspace{1cm} + \frac{J'^2}{4} \left( \frac{(1+n_R)}{i(\omega-\omega'+\delta \omega) + \Gamma/2} + \frac{n_R}{-i(\omega+\omega'+\delta \omega) + \Gamma/2} \right), \nonumber \\
\Gamma_R(\omega') &= J^2 \left( \frac{(1+n_R)}{i(\omega-\omega') + \Gamma/2} + \frac{n_R}{-i(\omega+\omega') + \Gamma/2} \right).
\end{align}
From this the $\omega$-dependent rates can be found
\begin{align}
\gamma_L(\omega') &= J^2  \left( \frac{(1+n_R) \Gamma}{(\omega'-\omega)^2 + \Gamma^2/4} + \frac{n_R \Gamma}{(\omega'+\omega)^2 + \Gamma^2/4} \right) \\
& \hspace{1cm} + \frac{J'^2}{4} \left( \frac{(1+n_R) \Gamma}{(\omega'-\omega - \delta \omega)^2 + \Gamma^2/4} + \frac{n_R \Gamma}{(\omega'+\omega - \delta \omega)^2 + \Gamma^2/4} \right) \nonumber \\
& \hspace{1cm} + \frac{J'^2}{4} \left( \frac{(1+n_R) \Gamma}{(\omega'-\omega + \delta \omega)^2 + \Gamma^2/4} + \frac{n_R \Gamma}{(\omega'+\omega + \delta \omega)^2 + \Gamma^2/4} \right), \nonumber \\
\gamma_R(\omega') &= J^2  \left( \frac{(1+n_R) \Gamma}{(\omega'-\omega)^2 + \Gamma^2/4} + \frac{n_R \Gamma}{(\omega'+\omega)^2 + \Gamma^2/4} \right), \\
S_R(\omega') &= J^2  \left( \frac{(1+n_R) (\omega'-\omega)}{(\omega'-\omega)^2 + \Gamma^2/4} + \frac{n_R (\omega'+\omega)}{(\omega'+\omega)^2 + \Gamma^2/4} \right).
\end{align}
The Lamb shift for the right bath was also calculated. However, as mentioned in chapter \ref{sec:MasterEquationsComparison}, it does not affect the evolution or steady state, so we will ignore it. Finally, we use these with the global master equation after the secular approximation, i.e., Eq.~\eqref{eq:MasterEquationsGlobalSecular}. This will give terms describing the decay of all coherences for the qutrit similar to section \ref{subsec:ChapterMasterEquationRevisitingTwoLevel}. Therefore, the populations for the qutrit, $P(\ket{\alpha_T}) = \tr \{\op{\alpha_T} \hat{\rho}\}$, can be written in the form of the classical master equation, $d \vec{P}(t)/dt = W \vec{P}(t)$. Adding $\gamma_L(\omega')$ and $\gamma_R(\omega')$ together, the transition rates become
\begin{subequations}
\label{rates}
\begin{alignat}{1}
\Gamma_{2_T \rightarrow 0_T} = 0, \quad \Gamma_{1_T \rightarrow 0_T} = \frac{(1+n_L) J'^2 }{\Gamma} + \frac{(1+n_R ) J^2 \Gamma}{\delta \omega^2 + \Gamma^2/2}, \hspace{0.55cm}\\
\Gamma_{2_T \rightarrow 1_T} = \frac{8(1+n_L) J^2}{\Gamma} + \frac{8(1+n_R) J^2}{\Gamma},\quad \Gamma_{0_T \rightarrow 2_T} = 0, \hspace{0.5cm}\\
\Gamma_{0_T \rightarrow 1_T} = \frac{n_L J'^2 }{\Gamma} + \frac{n_R J^2 \Gamma}{\delta \omega^2 + \Gamma^2/2}, \,\, \Gamma_{1_T \rightarrow 2_T} = \frac{8n_L J^2}{\Gamma} + \frac{8n_R J^2}{\Gamma}. 
\end{alignat}
\end{subequations}
The first term in each rate is due to the left bath, while the second term is due to the right term. The solution to $W \vec{P}_{\text{ss}}=0$ can be found to be
\begin{equation}
\vec{P}_{\text{ss}} = \mathcal{N} \begin{pmatrix}
(2+n_R+n_L) \Big((1+n_L) J'^2 + (1+n_R) J^2 \frac{\Gamma^2}{\delta \omega^2}\Big) \\
(2+n_R+n_L) \Big( n_L J'^2 + n_R J^2 \frac{\Gamma^2}{\delta \omega^2}\Big) \\
(n_R + n_L)\Big(n_L J'^2 + n_R J^2 \frac{\Gamma^2}{\delta \omega^2}\Big)
\end{pmatrix}, \label{eq:steadystate_mark}
\end{equation}
where $K$ is a constant ensuring $P_{\text{ss}}(\ket{0_T}) + P_{\text{ss}}(\ket{1_T}) + P_{\text{ss}}(\ket{2_T})=1$, and we have used the assumption $\delta \omega \gg \Gamma$. The current is now found as the number of excitations decaying due to the cold bath, e.g., $\mathcal{J}_{\textrm{f}} \simeq P(\ket{2_T}) \frac{8J^2}{\Gamma}$. Since one quanta of energy $\delta \omega$ is added through work every time the left bath causes a transition between $\ket{0_T}$ and $\ket{1_T}$, the excitation work is found as the number of excitations exchanged with the left bath through the $J'$-interaction, e.g., $\mathcal{W}_\textrm{r}/\delta \omega =- P(\ket{1_T}) \frac{J'^2}{\Gamma}$. Under the stated assumptions the currents and work become
%
\begin{align}
\mathcal{J}_\textrm{f} &= \frac{8 n_H^2}{2+5n_H+3n_H^2} \frac{J^2}{\Gamma},\\
\mathcal{J}_\textrm{r} &= -\frac{n_H}{2+n_H} \frac{8n_H J^2 + (2+n_H)J'^2}{J'^2} \frac{J^2 \Gamma}{\delta \omega^2},\\
\mathcal{W}_\textrm{f}/\delta \omega &= \frac{n_H (2+n_H)}{2 + 5n_H+3n_H^2 } \frac{J^2\Gamma}{\delta \omega^2},  \\
\mathcal{W}_\textrm{r}/\delta \omega &= - n_H \frac{J^2 \Gamma}{\delta \omega^2} .
\end{align}
The rectification can be found to be
\begin{equation}
\mathcal{R} = \frac{8 n_H (2+n_H)}{2+5n_H+3n_H^2} \frac{J'^2}{8n_H J^2 + (2+n_H)J'^2} \frac{\delta \omega^2 }{ \Gamma^2}. \label{eq:ChapterQutritDiodeRMarkov}
\end{equation}
This approximate expression for the rectification and the full solution are plotted in Fig.~\ref{figure:ChapterQutritDiodeR} for different values of $\Gamma$. There is a clear overlap between the two solutions, and we see that $\mathcal{R} \propto \delta \omega^2$. Furthermore, in the limit $\delta \omega \rightarrow \infty$, this model approaches the idealized model in Fig.~\ref{figure:ChapterQutritDiodeSetupIdealized}, and we achieve an ideal diode. The work both in forward and reverse bias is suppressed as $1/\delta \omega^{2}$, and therefore, the work done on the system is small. This is also verified from Fig.~\ref{figure:ChapterQutritDiodeJAndR}(a). Thus the work done acts as a catalyst, and it does not contribute excitations so as to keep $\mathcal{J}_R \approx - \mathcal{J}_L$ both in forward and revers bias, see Eq.~\eqref{eq:ChapterQutritDiodeCurrentDef}.

\subsection{Robustness} 

\begin{figure}[ht]
\centering
\includegraphics[width=1\linewidth, angle=0]{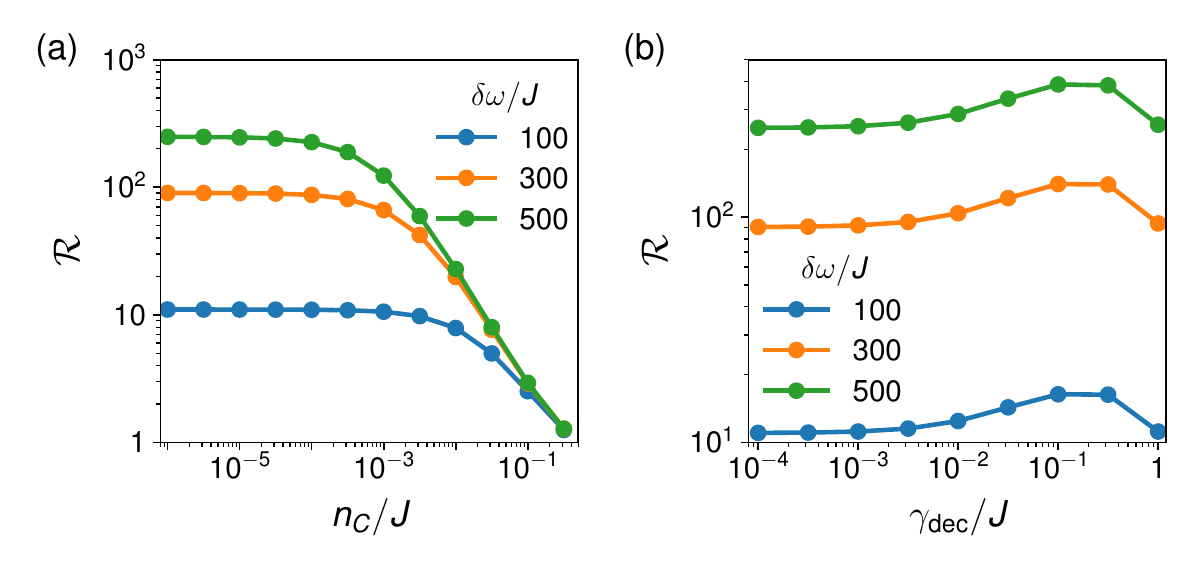}
\caption{ (a) Rectification as a function of $n_C$ for different values of the anharmonicity, $\delta \omega$. (b) Rectification as a function of the decoherence rate $\gamma_{\text{dec}}$ for different values of $\delta \omega$. See Eq.~\eqref{eq:QutritDiodeDefaultValues} for the values of the remaining system parameters. Reproduced with permission from Phys. Rev. E \textbf{106}, 034116 (2022) \cite{PhysRevE.106.034116}. Copyright 2022 American Physical Society.}\label{figure:ChapterQutritDiodencAndDec}
\end{figure}

Finally, we study the robustness towards excitations from the cold bath and decoherence. First, we let the cold bath introduce excitations by letting $n_C > 0$. The rectification is plotted in Fig.~\ref{figure:ChapterQutritDiodencAndDec}(a) as a function of $n_C$. The diode functionality is clearly diminished for larger $n_C$. This can be explained by looking at the reverse-bias rate out of the dark state
\begin{align}
\Gamma_{0_T \rightarrow 1_T} = \frac{J'^2 n_C}{\Gamma} + \frac{n_H J^2 \Gamma}{\delta \omega^2}.
\end{align}
Since this process decreases the population of the dark state, it results in a decrease in rectification. Therefore, in addition to large $\delta \omega$, we need a small $n_C$. The assumption $n_C=0$ is valid when $n_C \ll n_H \frac{J^2}{J'^2} \frac{\Gamma^2}{ \delta \omega^2}$. For the default values, this corresponds to $T_C \ll 0.16\omega$. Second, we add decoherence in the form of decay and dephasing to the qutrit. This is done through an updated Lindbladian
\begin{align}
\mathcal{L}_{\text{dec}}[\hat{\rho}] &= \mathcal{L}[\hat{\rho}] + \gamma_\text{dec} \left( \hat{a}_T \hat{\rho} \hat{a}_T^\dag - \frac{1}{2} \{ \hat{a}_T^\dag \hat{a}_T, \hat{\rho} \} \right) \\
& \hspace{2cm} + \gamma_\text{dec} \left( \hat{a}_T^\dag \hat{a}_T \hat{\rho} \hat{a}_T^\dag \hat{a}_T - \frac{1}{2} \{ \hat{a}_T^\dag \hat{a}_T \hat{a}_T^\dag \hat{a}_T, \hat{\rho} \} \right), \nonumber
\end{align}
where $\mathcal{L}[\hat{\rho}]$ is the Lindbladian from Eq.~\eqref{eq:ChapterQutritDiodeLiouvillian}. This results in decay and dephasing coherence times of $T_1=T_2=\gamma^{-1}_\text{dec}$ for the lowest two states. On the other hand, the second excited state of the qutrit has decay coherence time $T_1 = \gamma^{-1}_\text{dec}/2$. It is important to note that the use of the local master equation of Lindblad form here is not derived from a general model of Markovian thermal baths. In fact, noise is complicated and not always Markovian or due to interactions with a reservoir \cite{Bylander2011, 10.1063/1.5089550}. 

In Fig.~\ref{figure:ChapterQutritDiodencAndDec}(b), the rectification is plotted as a function of the decoherence, $\gamma_{\text{dec}}$. State-of-the-art quantum platforms can achieve $\gamma_\text{dec}/J < 10^{-3}$ \cite{doi:10.1146/annurev-conmatphys-031119-050605}. However, the dark-state-induced rectification is clearly not sensitive to decoherence, and other parameters can be focused on, e.g., a larger anharmonicity can be picked even if it results in larger decoherence. The stability of the rectification towards decoherence is clearly a result of the dark state being the qutrit ground state.

\section{Full wave bridge rectifier}
\label{sec:QutritDiodeFWBR}

\begin{figure}[ht]
\centering
\includegraphics[width=1\linewidth, angle=0]{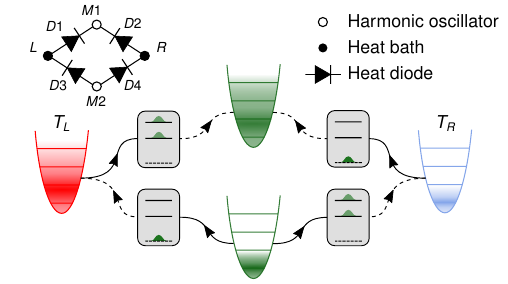}
\caption{ Schematic of the full wave bridge rectifier using the diode introduced in section \ref{sec:QutritDiodeImplementation}. In the top left corner, the circuit is shown using pictograms defined in the top right corner. In the center, the schematic is drawn with four qutrit diodes connected through four harmonic oscillators. The left and right harmonic oscillator is connected to thermal baths while the top and bottom ones are not. The arrows show the direction of allowed heat flow through the diode. The figure is adopted from \cite{poulsen2022heat}.}\label{figure:ChapterQutritDiodeBridgeRectifier}
\end{figure}

This chapter is finalized with an example circuit containing the diode introduced in section \ref{sec:QutritDiodeImplementation}. The circuit is inspired by the full wave bridge rectifier from classical electronics, which creates a positive vertical voltage bias independent of the sign of the horizontal voltage bias \cite{patent}. The circuit for the bridge rectifier consists of four diodes and two additional harmonic oscillators as seen in Fig.~\ref{figure:ChapterQutritDiodeBridgeRectifier}. If the qutrit diodes function as intended, the bias of the top and bottom harmonic oscillators will remain the same independent of the bias of the left and right harmonic oscillators. Since the circuit is horizontally symmetric, we can assume $T_L \geq T_R$. 

\subsection{Hamiltonian and master equation}

The Hamiltonian becomes
\begin{align}
\hat{H} &=  \hat{H}^{D1}_{L\rightarrow M1} + \hat{H}^{D2}_{R\rightarrow M1} + \hat{H}^{D3}_{M2 \rightarrow L} + \hat{H}^{D4}_{M2\rightarrow R} \\
&\hspace{2.5cm} + \omega \Big( \hat{a}_{L}^\dag \hat{a}_{L} + \hat{a}_{M1}^\dag \hat{a}_{M1} + \hat{a}_{M2}^\dag \hat{a}_{M2} + \hat{a}_{R}^\dag \hat{a}_{R} \Big), \nonumber\\
\hat{H}^{\alpha}_{A\rightarrow B} &= (\omega +\delta \omega) \hat{a}_\alpha^\dag \hat{a}_\alpha  - \frac{\delta \omega}{2} \hat{a}_\alpha^\dag \hat{a}_\alpha \Big(\hat{a}_\alpha^\dag \hat{a}_\alpha -1\Big)  \\ &\hspace{0.8cm} + J_{LT}(t) \Big(\hat{a}_A + \hat{a}_A^\dag \Big) \Big(\hat{a}_\alpha + \hat{a}_\alpha^\dag \Big) + J \Big(\hat{a}_\alpha + \hat{a}_\alpha^\dag \Big) \Big(\hat{a}_B + \hat{a}_B^\dag \Big) ,\nonumber
\end{align}
where $\hat{a}_{L}$, $\hat{a}_{M1}$, $\hat{a}_{M2}$, and $\hat{a}_{R}$ are annihilation operators for the left, upper, lower, and right harmonic oscillator, respectively. Likewise, $\hat{a}_{D1}$, $\hat{a}_{D2}$, $\hat{a}_{D3}$, and $\hat{a}_{D4}$ are the four qutrits; see the upper left corner of Fig.~\ref{figure:ChapterQutritDiodeBridgeRectifier} for more information on the labels. The parameters for all diodes have been chosen the same, i.e., 
\begin{align}
\delta \omega = 300J, \quad J'=0.5J, \quad \Gamma = 10J, \quad T_L = \omega, \quad T_R = 0.1\omega. \label{eq:QutritDiodeDefaultValues2}
\end{align} 
The temperatures enter through $n_{L(R)}~=~\big(e^{\omega/T_{L(R)}}-1\big)^{-1}$. The master equation is
\begin{align}
\frac{d\hat{\rho}}{dt} &= -i[\hat{H}, \hat{\rho}] + \mathcal{D}_{L}[\hat{\rho}]  + \mathcal{D}_{R}[\hat{\rho}] + \mathcal{D}_{\mathrm{dec}}[\hat{\rho}] , 
\label{eq:}
\end{align}
where the dissipators are the same as in section \ref{sec:QutritDiodeImplementation}
\begin{align}
\mathcal{D}_{L(R)}[\hat{\rho}] &= \Gamma (n_{L(R)} + 1) \mathcal{M}[ \hat{a}_{L(R)}, \hat{\rho}]
+ \Gamma n_{L(R)} \mathcal{M}[\hat{a}_{L(R)}^\dag, \hat{\rho}] ,\\
\mathcal{D}_{\mathrm{dec}}[\hat{\rho}] &= \gamma_{\mathrm{dec}} \sum_\alpha \left[ \mathcal{M}[ \hat{a}_{\alpha}, \hat{\rho}] + \mathcal{M}[ \hat{a}_{\alpha}^\dag \hat{a}_{\alpha}, \hat{\rho}] \right].
\label{eq:QutritDiodeFWBRMasterEquation}
\end{align}
The constant $\gamma_\text{dec}$ is the decoherence rate, which unless otherwise stated, is set to $\gamma_\text{dec}=10^{-3}J$. The sum over $\alpha$ in Eq.~\eqref{eq:QutritDiodeFWBRMasterEquation} is carried out for $\alpha\in \{D1, M1, D2, D3, M2, D4\}$. This system is too large to solve on a normal computer, so instead, the left and right harmonic oscillators are traced away using the assumption $\Gamma \gg J$ the same way it was done in section \ref{sec:QutritDiodeImplementation}. This dicouples to upper and lower part of the circuit. The upper part becomes time-independent, and the steady state can be solved directly, i.e., solving $\mathcal{L}[\hat{\rho}_\mathrm{ss}]=0$. The lower part is still time-dependent, and the steady state is found by evolving the system in time until the current converges. The eight lowest states are kept for $M1$, and the seven lowest states are kept for $M2$. 
To unsure that the current is allowed to converge, we use the average current
\begin{equation*}
\mathcal{J}_{\text{f/r}, m} = \frac{1}{1000J^{-1}} \int_{(m+1)5000J^{-1} -1000J^{-1}}^{(m+1)5000J^{-1}} \mathcal{J}_{\text{f/r}}(t) dt,
\end{equation*}
where $m$ determines for how long the density matrix is evolved. We say that the current is converged when 
\begin{equation*}
\frac{\mathcal{J}_{\text{f/r}, m} - \mathcal{J}_{\text{f/r}, m-1}}{\mathcal{J}_{\text{f/r}, m-1}} < 10^{-4},
\end{equation*}
and we set $\mathcal{J}_{\text{f/r}}=\mathcal{J}_{\text{f/r}, m}$. For more information on the effective master equation after the assumption $\Gamma \gg J$ and the convergence of the current, see Ref.~\cite{poulsen2022heat}.

\subsection{Functionality}

\begin{figure}[t]
\centering
\includegraphics[width=1\linewidth, angle=0]{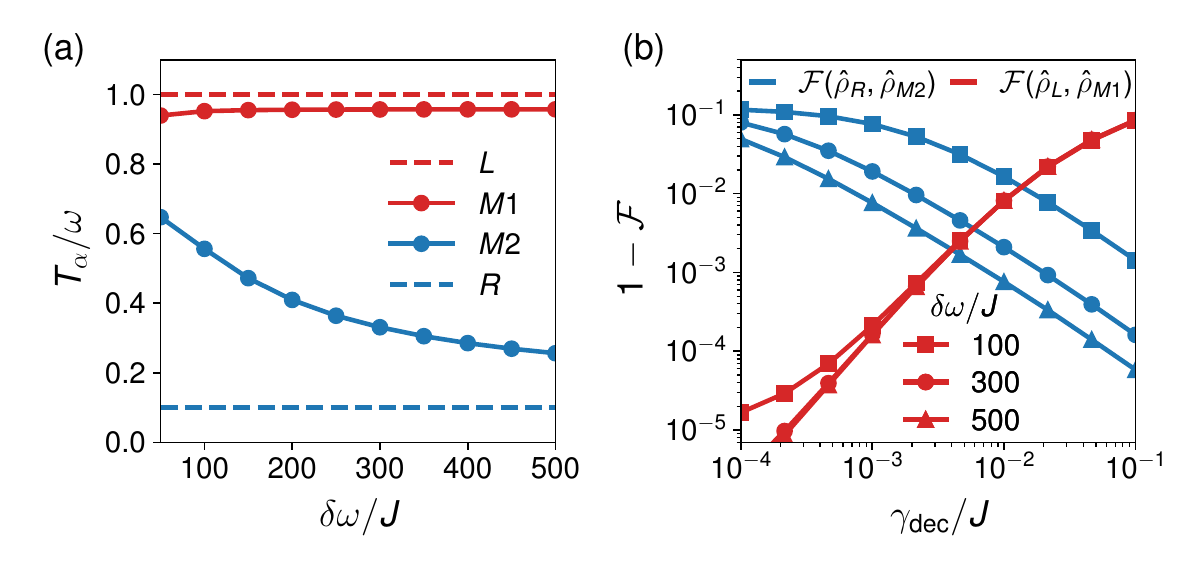}
\caption{ (a) Effective temperature of the harmonic oscillators as a function of the qutrit anharmonicity for the bridge rectifier. (b) Fidelities between the states of the harmonic oscillators pairwise as a function of the decoherence rate $\gamma_\text{dec}$. See Eq.~\eqref{eq:QutritDiodeDefaultValues2} for the values of the remaining system parameters. The figure is adopted from \cite{poulsen2022heat}. }\label{figure:ChapterQutritDiodeBridgeRectifier1}
\end{figure}

To quantify the functionality of the bridge rectifier, we use the temperature of the two harmonic oscillators $M1$ and $M2$. Since the states of $M1$ and $M2$ are not necessarily a thermal state, we use the effective temperature
\begin{equation}
T_\alpha/\omega = \Big[ \ln\left(\langle \hat{n}_\alpha \rangle_{\text{ss}} + 1 \right) - \ln \langle \hat{n}_\alpha \rangle_{\text{ss}}\Big]^{-1},
\end{equation}
which is the temperature of a thermal state with the same occupation number $\langle \hat{n}_\alpha \rangle_{\text{ss}}$ as the harmonic oscillators. In Fig.~\ref{figure:ChapterQutritDiodeBridgeRectifier1}(a), the effective temperature of the four harmonic oscillators is plotted as a function of the diode quality parametrized through $\delta \omega$. This is compared to the temperatures of the left and right baths, $T_L = \omega$ and $T_R = 0.1\omega$, plotted with dashed lines. A perfect rectifier bridge would result in:
\[T_L = T_{M1}, \quad \text{and} \quad T_R = T_{M2}.\]
From Fig.~\ref{figure:ChapterQutritDiodeBridgeRectifier1}(a), we see that $T_{M1}$ quickly approaches $T_L$. The discrepancy between $T_{M1}$ and $T_{L}$ for large $\delta \omega$ is mainly due to decay through $\gamma_\text{dec}$. For M1 to get cold in the long time limit, diode D3 needs to close i.e. be driven into the ground state $|0_{D3}\rangle$. This can either be done by the cold right bath through M2 or through the decoherence rate $\gamma_{\text{dec}}$. To test the importance of $\gamma_{\text{dec}}$, we look at the fidelity between the relevant pairs of states, i.e., between L and M1 and between M2 and R. The fidelity between two density matrices is
\begin{equation}
F(\hat{\rho}, \hat{\sigma}) = \left[ \text{tr} \sqrt{\sqrt{\hat{\rho}} \hat{\sigma} \sqrt{\hat{\rho}}} \right]^2.
\end{equation}
The density matrix for $\alpha \in \{L, M1, M2, R\}$ is $\hat{\rho}_\alpha = \text{tr}_{S \backslash \alpha}\{\hat{\rho}\}$ where the trace is over the entire system except $\alpha$. The fidelities between two pairs of states $F(\hat{\rho}_L, \hat{\rho}_{M1})$ and $F(\hat{\rho}_R, \hat{\rho}_{M2})$ are plotted in Fig.~\ref{figure:ChapterQutritDiodeBridgeRectifier1}(b). The fidelity between the state of L and M1 becomes smaller for larger $\gamma_\text{dec}$, while the state of M2 becomes closer to the state of R. So picking $\gamma_{\text{dec}}$ is clearly a balance between getting $T_L \simeq T_{M1}$ for small $\gamma_{\text{dec}}$ and $T_R \simeq T_{M2}$ for larger $\gamma_{\text{dec}}$.

\cleardoublepage
\addtocontents{toc}{\protect\newpage}
\chapter{Rectification due to interference}
\label{chapter:Interference}
\epigraph{\footnotesize This chapter is based on Ref.~\cite{PhysRevA.105.052605}.
Sections \ref{sec:InterferenceLocal}-\ref{sec:InterferenceGlobal} and all figures herein have been reproduced with permission from Ref.~\cite{PhysRevA.105.052605}.}{}

\noindent In chapter \ref{chapter:QutritDiode}, we used steady-state engineering to achieve rectification through a dark state. The rectification was achieved by designing a system that, in reverse bias, is driven into a dark state of the warm bath, preventing transport. The dark-state mechanism was energy conservation, which is limited by the anharmonicity of the qutrit and does not rely on any quantum effects. A more effective dark state can be found using quantum entanglement, usually in a $\Lambda$-type three-level system \cite{PhysRevA.65.022314, RevModPhys.77.633}. Therefore, the task is to design a system where the cold bath drives the system into an entangled dark state $\ket{D}$ in reverse bias. This dark state should be designed such that any excitation coming from the warm bath is reflected due to destructive interference. An attempt at such a system can be seen in Fig.~\ref{figure:ChapterInterferenceSetup}. 

Usually, entanglement in open systems decays exponentially, as we saw in subsection \ref{subsec:ChapterMasterEquationRevisitingTwoLevel} and later used in section \ref{sec:QutritDiodeImplementation}. However, this does not have to be the case. As a simple example, we define the Hamiltonian for two two-level systems coupled through a hopping term
\begin{align}
\hat{H} = J (\hat{\sigma}_1^x \hat{\sigma}_2^x + \hat{\sigma}_1^y \hat{\sigma}_2^y + \Delta \hat{\sigma}_1^z \hat{\sigma}_2^z).
\end{align}
To preempt the next section, we let the two-level systems be spin-1/2 particles with states $\ket{\uparrow}$ and $\ket{\downarrow}$, and the above Hamiltonian is called an XXZ coupling. The Hamiltonian couples $\ket{\uparrow \downarrow}$ and $\ket{\downarrow \uparrow}$ while leaving $\ket{\downarrow \downarrow}$ and $\ket{\uparrow \uparrow}$ alone. The anharmonicity $\Delta$ decreases the energy of $\ket{\uparrow \downarrow}$ and $\ket{\downarrow \uparrow}$ while increasing the energy of $\ket{\uparrow \uparrow}$ and $\ket{\downarrow \downarrow}$ for $\Delta > 0$. The opposite is true for $\Delta <0$. Coupling this simple system to a thermal bath with temperature $T$, the steady state is
\begin{align}
\hat{\rho}_\mathrm{th} = \frac{e^{-\hat{H}/T}}{\tr \{e^{-\hat{H}/T} \}}.
\end{align}
There are many ways to quantify entanglement, some of which will be used later in this chapter. However, for simplicity, the population is instead computed
\begin{align}
\bra{\Psi_-} \hat{\rho}_\mathrm{th} \ket{\Psi_-} = \frac{e^{2(\Delta +1)J/T}}{2 + e^{2 (\Delta+1)J /T} + e^{2 (\Delta -1)J /T} },
\end{align}
where $\ket{\Psi_-} = (\ket{\uparrow \downarrow} - \ket{\downarrow \uparrow})/\sqrt{2}$ is one of the four Bell states. For $T \rightarrow 0$, this tends to a step function
\begin{align}
\lim_{T \rightarrow 0} \bra{\Psi_-} \hat{\rho}_\mathrm{th} \ket{\Psi_-} = \left\{ \begin{matrix}
1 & \Delta > -1 \\
1/3 & \Delta = -1 \\
0 & \Delta < -1
\end{matrix} \right. .
\end{align}
So for this simple system, entanglement develops naturally because the ground state is entangled. Only entanglement between eigenstates will decay exponentially. When more baths are added, it gets more complicated. It has been shown that entanglement can be present in a non-equilibrium steady state \cite{Bohr_Brask_2015, Khandelwal_2020}. However, a much larger degree of entanglement is needed for an entangled dark state to induce rectification.

This chapter will expand on the ideas of dark-state-induced rectification in chapter \ref{chapter:QutritDiode} to achieve a better dark-state mechanism using entanglement and interference. First, the model is set up as a chain of spins coupled to two spin baths at either end using the local master equation. Therefore, the setting is quite different from the one studied in chapter \ref{chapter:QutritDiode} even though the idea is the same. Next, the mechanism is examined and finally understood as a dark-state mechanism. The system is explored for a large set of system and bath parameter values, as well as including decoherence. Finally, the same system is modeled using the global master equation, and even larger rectification factors are found. However, due to the non-locality of the global master equation, it was not possible to connect this to a dark-state mechanism in the same way. The dark-state mechanism for a two-way spin chain is also explored in chapter \ref{chapter:WB} for sensing coupling strengths.

\section{Spin rectification using the local master equation}
\label{sec:InterferenceLocal}

\begin{figure}[t]
\centering
\includegraphics[width=1.0\linewidth, angle=0]{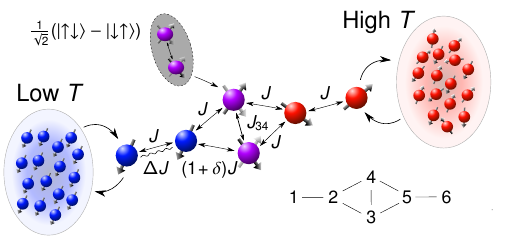}
\caption{ Illustration of a few-spin model of a quantum rectification device consisting of two segments, an $XXZ$ chain on the left and an $XX$ chain on the right, connected by a two-way interface. The device is connected to thermal baths at each end, one at low and one at high temperature. The solid arrows signify a hopping term, while a zigzag line signify a Z-coupling. During operation in reverse bias as shown here, the central interface spins are in the maximally-entangled Bell state illustrated in the top left-hand corner. In the bottom right-hand corner, the numbering of the spins is shown. Reproduced with permission from Phys. Rev. A \textbf{105}, 052605 (2022) \cite{PhysRevA.105.052605}. Copyright 2022 American Physical Society.}\label{figure:ChapterInterferenceSetup}
\end{figure}

To illustrate the mechanism of entanglement enhanced rectification via a dark state, we concentrate on the few-spin example shown in Fig.~\ref{figure:ChapterInterferenceSetup}. 
It consists of six spin-1/2 particles in a two-segment chain connected by a two-way interface
and described by an $XXZ$ Heisenberg Hamiltonian of the form
\begin{equation}
\label{eq:InterferenceHamiltonianMain}
\hat{H}/J = \hat{X}_{12} + (1+\delta) \hat{X}_{23} + \hat{X}_{24} + J_{34}/J \hat{X}_{34} + \hat{X}_{35} + \hat{X}_{45} + \hat{X}_{56} + \Delta \hat{Z}_{12},
\end{equation}
where $\hat{X}_{ij} = \hat{\sigma}^x_{i} \hat{\sigma}^x_{j} + \hat{\sigma}^y_{i} \hat{\sigma}^y_{j}$ 
is the $XX$ spin exchange operator, while 
$\hat{Z}_{ij} = \hat{\sigma}^z_{i} \hat{\sigma}^z_{j}$ is the $Z$ coupling that induces relative
energy shifts. The Pauli matrices for the $i$th spin are denoted $\hat{\sigma}^\alpha_{i}$ for $\alpha = x,y,z$.  The exchange coupling $J$ gives the overall scale of the problem, while the 
exchange between the interface spins is $J_{34}$. A prerequisite of rectification 
is a breaking of left-right symmetry which we implement by a non-zero $Z$ coupling parametrized
by $\Delta$, although this may as well have been provided by local magnetic fields 
applied to spins 1 and 2. 
Due to the interface, we also have to consider up-down symmetry, i.e. the symmetry between the upper and lower part, and we parametrize its breaking by adding $\delta$ to the exchange between spins 2 and 3 in Fig. \ref{figure:ChapterInterferenceSetup}. 

Once again the system is coupled locally to thermal baths on the left and right, see Fig. \ref{figure:ChapterInterferenceSetup}. One bath is cold and forces the adjacent spin to point down, while the other is hot and forces the adjacent spin into a statistical mixture of up and down. 
The evolution of the density matrix $\hat{\rho}$ of the system is determined by the local master equation on Lindblad form
\begin{align}
\frac{d \hat{\rho}}{d t} = \mathcal{L}[\hat{\rho}] = -i [\hat{H}, \hat{\rho}] + \mathcal{D}_1 [\hat{\rho}] + \mathcal{D}_6 [\hat{\rho}].
\label{eq:InterferenceMasterEquationMain}
\end{align}
The dissipative terms describing the action of the baths are
\begin{equation}
\begin{aligned}
\mathcal{D}_n [\hat{\rho}] &= \gamma \left[ \lambda_n \mathcal{M}[ \hat{\sigma}^+_{n}, \hat{\rho}] + (1- \lambda_n) \mathcal{M}[ \hat{\sigma}^-_{n}, \hat{\rho}] \right].
\end{aligned}
\end{equation}
$\gamma$ is the strength of the interaction with the baths. Unlike in previous sections, the nature of the system-bath interaction is determined by $\lambda_n$. If $\lambda_n = 0$, the bath will force the corresponding spin to tend down ($\ket{\downarrow}_n \! \bra{\downarrow}$) corresponding to a low entropy or low temperature bath, and if $\lambda_n = 0.5$, the bath will force the spin into a statistical mixture of up and down ($(\ket{\downarrow}_n \! \bra{\downarrow} + \ket{\uparrow}_n \! \bra{\uparrow}) /2$) corresponding to a high entropy or high temperature bath. 

Unless otherwise stated throughout this chapter, we will use the set of parameters 
\begin{align}
&\Delta =5, \quad \delta = 0.01, \quad J_{34}=-(\Delta + 1.3)J, \label{eq:InterferenceDefaultValues} \\
& \hspace{1.7cm} \gamma=J, \quad \lambda_n \in \{0, 0.5\}. \nonumber
\end{align}
This choice for $J_{34}$ will be explained in subsection \ref{subsection:InterferenceJAndRev}.

The baths induce currents, and since the Hamiltonian is time-independent, the system will eventually reach a steady state, $\mathcal{L}[\hat{\rho}_{\mathrm{ss}}] = 0$. 
It is this steady state that determines the rectification properties. 
For $\delta\neq 0$, the steady state will be unique and independent of the initial state \cite{PhysRevA.105.052605}.

Instead of heat transport, we will use something called the spin current. The spin current can be motivated by looking at the rate of change of $\hat{\sigma}_n^z$. Due to the complex geometry of the spin chain these are all different, but a few examples are
\begin{align}
\frac{d\ev{\hat{\sigma}_1^z}}{dt} &= 2J  \ev*{\hat{s}_{21}} + 2\gamma \lambda_1 \ev*{\hat{\sigma}_1^- \hat{\sigma}_1^+} - 2\gamma (1-\lambda_1) \ev*{\hat{\sigma}_1^+ \hat{\sigma}_1^-},\\
\frac{d\ev{\hat{\sigma}_2^z}}{dt} &= 2J  \ev*{\hat{s}_{12}} + 2(1+\delta)J \ev*{\hat{s}_{32}} + 2J  \ev*{\hat{s}_{42}}, \\
\frac{d\ev{\hat{\sigma}_3^z}}{dt} &= 2(1+\delta)J \ev*{\hat{s}_{23}} + 2J_{34} \ev*{\hat{s}_{43}} + 2J  \ev*{\hat{s}_{53}},
\end{align}
where $\hat{s}_{ij} = \hat{\sigma}^{x}_{i}\hat{\sigma}^{y}_{j} - \hat{\sigma}^{y}_{i} \hat{\sigma}^{x}_{j}$. In steady state, all three equations are equal to zero. Therefore, each term can be interpreted as a source of spin excitations from either a bath or other sites. Accordingly, the second and third terms in the first line is the contribution from the left bath, and the first term, $2J\ev*{\hat{s}_{21}}$, is the spin current from spin 2 and into spin 1. Spin 2 is connected to three other sites, and therefore, the three spin currents into site 2 has to add to zero. 

Since we are interested in the spin current from the left bath to the right bath, we define the spin current as the steady-state expectation value
\begin{align}
\mathcal{J} = 2J \ev*{\hat{s}_{12} }_\mathrm{ss}.
\end{align}
However, due to spin conservation, the current can be calculated in several ways e.g. $\mathcal{J} = 2J \ev*{\hat{s}_{56}}_\mathrm{ss}$, $\mathcal{J} = 2(1+\delta)J \ev*{ \hat{s}_{23}}_\mathrm{ss} + 2J \ev*{\hat{s}_{24} }_\mathrm{ss}$, or through a bath term. In chapter \ref{chapter:QutritDiode}, the bath term was used to calculate the heat current whereas the inter-site term is used here. This is only a matter of preference, and both approaches are equal. 

Similar to previous chapters, we denote the situation $\lambda_1 > \lambda_6$ as forward bias, and we denote the situation $\lambda_1 < \lambda_6$ as reverse bias; see Fig. \ref{figure:ChapterInterferenceSetup}. To obtain a well-functioning diode, we must demand that
\begin{enumerate}
\item no spin current is allowed to flow in reverse bias $\mathcal{J}_\mathrm{r} \sim 0$,
\item an appreciable spin current can flow in forward bias $\mathcal{J}_\mathrm{f} \gg -\mathcal{J}_\mathrm{r}$.
\end{enumerate}
A measure of quality that contains both requirements is the rectification from previous chapters
\begin{equation}
\mathcal{R} = - \frac{\mathcal{J}_\mathrm{f}}{\mathcal{J}_\mathrm{r}}.
\end{equation}
An alternative quality measure
is the contrast defined as
\begin{equation}
\mathcal{C} = \left| 
\frac{\mathcal{J}_\mathrm{f} + \mathcal{J}_\mathrm{r}}{\mathcal{J}_\mathrm{f} - \mathcal{J}_\mathrm{r}} \right|,
\end{equation}
such that $\mathcal{C} = 0$ is equivalent to $\mathcal{R} = 1$, while 
$\mathcal{C} = 1$ for $\mathcal{J}_\mathrm{r}\to 0$, i.e. for the perfect diode.

\subsection{Current and rectification values}
\label{subsection:InterferenceJAndRev}

\begin{figure}[t]
\centering
\includegraphics[width=1.\linewidth, angle=0]{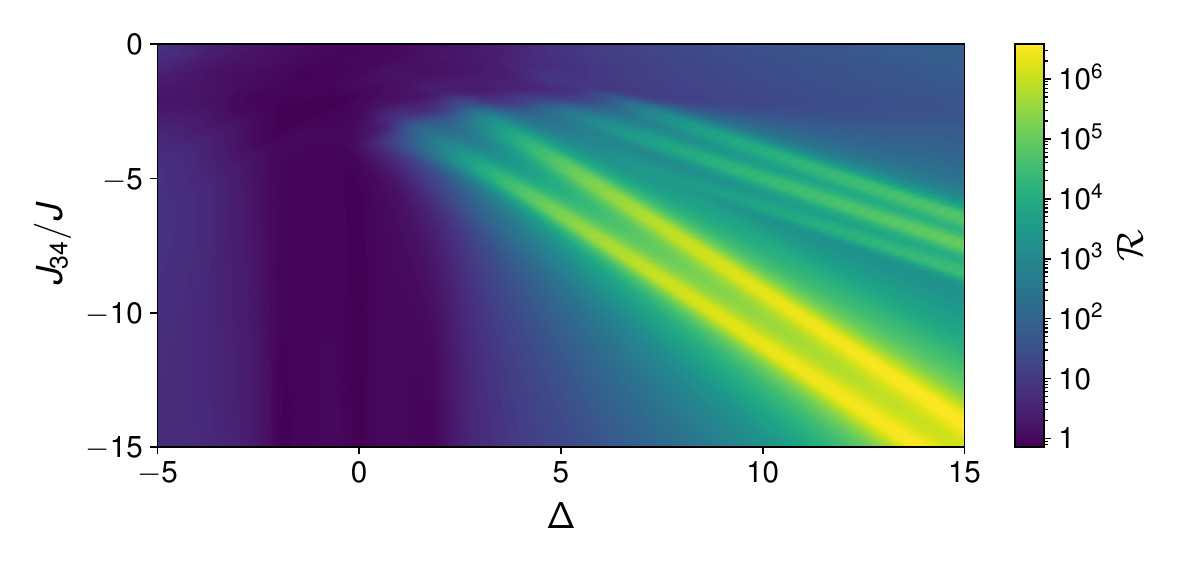}
\caption{ Rectification $\mathcal{R}$ as a function of $\Delta$ and $J_{34}$. See Eq.~\eqref{eq:InterferenceDefaultValues} for the values of the remaining system parameters. Reproduced with permission from Phys. Rev. A \textbf{105}, 052605 (2022) \cite{PhysRevA.105.052605}. Copyright 2022 American Physical Society.}\label{figure:ChapterInterferenceRecD12vsJ34}
\end{figure}

\begin{figure}[t]
\centering
\includegraphics[width=1.\linewidth, angle=0]{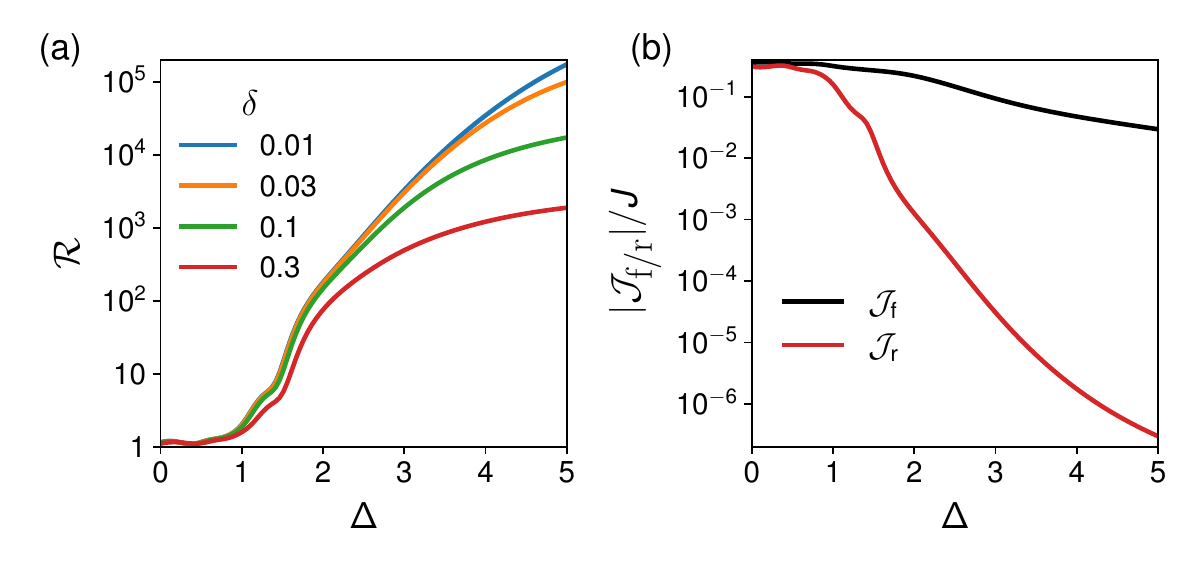}
\caption{ (a) $\mathcal{R}$ as a function of $\Delta$ for different values of $\delta$. (b) Steady-state currents $\mathcal{J}_\mathrm{f}$ and $\mathcal{J}_\mathrm{r}$. See Eq.~\eqref{eq:InterferenceDefaultValues} for the values of the remaining system parameters. Reproduced with permission from Phys. Rev. A \textbf{105}, 052605 (2022) \cite{PhysRevA.105.052605}. Copyright 2022 American Physical Society.}\label{figure:ChapterInterferenceRecD12}
\end{figure}

The rectification results for the six-spin implementation of Fig.~\ref{figure:ChapterInterferenceSetup} 
are shown in 
Fig.~\ref{figure:ChapterInterferenceRecD12vsJ34} as a function of the relevant parameters of 
the model. The contour plot in Fig.~\ref{figure:ChapterInterferenceRecD12vsJ34} shows 
$\mathcal{R}$ as a function of $J_{34}$ and $\Delta$. There is a clear region in the bottom right corner where values of $\mathcal{R} > 10^6$ are reached. Further inspection of the two lines of large $\mathcal{R}$ shows that they occur for $J_{34} = -(\Delta \pm 1)J$ for large $\Delta$. This precise number will be justified later.  However, very large anisotropi values can be experimentally challenging. Therefore, to keep the model general, we mostly stick to $\Delta \leq 5$ for which the lower line of large rectification is better parametrizated by 
\begin{equation}
J_{34} = -(\Delta + 1.3)J.
\end{equation}
Fig.~\ref{figure:ChapterInterferenceRecD12}(a) demonstrates the dependence on $\Delta$ using $J_{34} = -(\Delta +1.3)J$ showing that $\delta\ll 1$ gives higher $\mathcal{R}$ as a function of $\Delta$. We also confirm that the large rectification values are mainly due to suppression of $\mathcal{J}_\mathrm{r}$; see Fig.~\ref{figure:ChapterInterferenceRecD12vsJ34}(b).

\begin{figure}[t]
\centering
\includegraphics[width=1.\linewidth, angle=0]{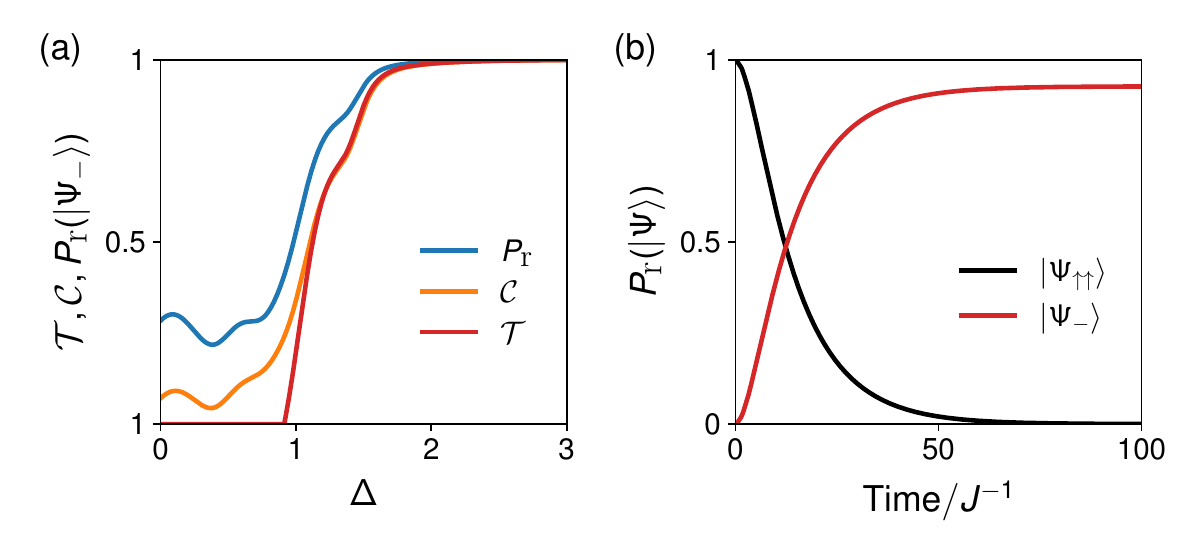}
\caption{ (a) Population $P_r( |\Psi_- \rangle )$, Contrast $\mathcal{C}$, and concurrence $\mathcal{T}$ as a function of $\Delta$. (b) Interface population as a function of time in reverse bias for an initial state of $\ket{\uparrow \uparrow}$, $\delta = 0.1$, $\Delta = 50$, and $J_{34} = -(\Delta + 1)J$. See Eq.~\eqref{eq:InterferenceDefaultValues} for the values of the remaining system parameters. Reproduced with permission from Phys. Rev. A \textbf{105}, 052605 (2022) \cite{PhysRevA.105.052605}. Copyright 2022 American Physical Society.}\label{figure:ChapterInterferencePopulations}
\end{figure} 

\subsection{Understanding the mechanism}
\label{subsec:InterferenceUnderstanding}
To explain the above observations and connect it to a dark-state mechanism, we first note that the biggest change in current occurs in reverse bias. Therefore, this is the situation we will focus on. 

To motivate entanglement as a cause of the large rectification, we plot the entanglement measure $\mathcal{T}$ for the interface alongside the contrast $\mathcal{C}$ in Fig.~\ref{figure:ChapterInterferencePopulations}(a). The entanglement measure used is called the concurrence \cite{PhysRevLett.78.5022, PhysRevLett.80.2245}
\begin{equation} 
\mathcal{T}(\hat{\rho}_{\textrm{ss},\textrm{r}}^{34}) = \max (0, \lambda_1 - \lambda_2 - \lambda_3 - \lambda_4), 
\end{equation}
where $\lambda_1, ..., \lambda_4$ are eigenvalues, in decreasing order, of the non-hermitian matrix 
\begin{equation*} 
\hat{\rho}_{\textrm{ss},\textrm{r}}^{34} \left(\sigma_{3}^y \sigma_{4}^y\right) \hat{\rho}_{\textrm{ss},\textrm{r}}^{34*} \left(\sigma_{3}^y \sigma_{4}^y\right).
\end{equation*}
The concurrence is a widely used measure of entanglement which is $1$ only for a maximally entangled state. The state of the interface $\hat{\rho}_{\text{ss,r}}^{34} = \tr_{1,2,5,6} \{\hat{\rho}_{\text{ss,r}}\}$ is found by tracing over the Hilbert space of spins 1, 2, 5, and 6. In Fig.~\ref{figure:ChapterInterferencePopulations}(a), we observe a strong correlation between the amount of entanglement and the diode being in a working regime. An inspection of the density matrix shows that the entanglement is in the form of the entangled bell state $\ket{\Psi_-}~=~\left(\ket{\uparrow \downarrow}-\ket{\downarrow \uparrow}\right)/\sqrt{2}$. This is further backed by the steady-state population for the interface $P_{\text{r}}(\ket{\Psi_-})~=~\langle \Psi_- | \hat{\rho}_{\text{ss,r}}^{34}| \Psi_-\rangle$ plotted in Fig.~\ref{figure:ChapterInterferencePopulations}(a). 

To see that the rectification is indeed due to entanglement through $\ket{\Psi_-}$, the explanation can be broken into two parts:
\begin{enumerate}
\item The entangled state $\ket{\Psi_-}$ prevents transport between the baths, i.e., $\ket{\Psi_-}$ is a dark state of the right bath.
\item In reverse bias, the interface is driven into the dark state $\ket{\Psi_-}$.
\end{enumerate}
For the first part, the Hamiltonian is used on a compound state where spins 1 and 2 are down due to the cold bath, the interface is in the entangled state, and spins 5 and 6 are in a general state
\begin{equation}
\hat{H} \ket{\downarrow \downarrow \Psi_-\, S} = E \ket{\downarrow \downarrow \Psi_-\, S} + \sqrt{2} \delta J \ket{\downarrow \uparrow \downarrow \downarrow S},\label{eq:InterferenceInterference}
\end{equation}
where $S \in \{\ket{\downarrow \downarrow}, \ket{\Psi_-}, \ket{\Psi_+}, \ket{\uparrow \uparrow}\}$. This state is unaffected by the left bath, and the right bath can only couple these four states to each other. Remarkably, the state $\ket{\downarrow \downarrow \Psi_- \, S}$ is close to being a stationary state of $\hat{H}$ for $\delta \ll 1$ with energy $E$. Therefore, the entangled state can not propagate to spin 2 due to destructive interference, and the transition, $\ket{\downarrow \downarrow \Psi_-\, S} \leftrightarrow \ket{\downarrow \uparrow \downarrow \downarrow S}$, is further forbidden by energy conservation. Furthermore, any spin excitation at spin 5 cannot propagate to the interface due to perfect destructive interference. This destructive interference can be summed up by the relations
\begin{align} 
\left(\hat{X}_{23} + \hat{X}_{24} \right) \ket{\downarrow \downarrow \! \Psi_- \, S} &= 0, \label{eq:InterferenceInterferenceCondition1} \\
\left(\hat{X}_{45} + \hat{X}_{35} \right) \ket{\downarrow \downarrow \! \Psi_- \, S} &= 0, \label{eq:InterferenceInterferenceCondition2}
\end{align}
which show up directly when deriving Eq.~\eqref{eq:InterferenceInterference}. The second equation proves that $\ket{\Psi_-}$ is indeed a dark state of the hot side of the system. Therefore, a spin excitation is prevented from traveling between the baths resulting in a suppressed spin current. Due to the second part of Eq.~\eqref{eq:InterferenceInterference}, the entangled state will decay weakly, and therefore, we expect $\delta \ll 1$ to be preferable. 

For the second part of the explanation, we already verified that the interface is indeed driven into the entangled dark state $\ket{\Psi_-}$. The reason for the system being driven into the entangled state can be found by looking at the transition
\begin{equation}
\label{eq:InterferenceReverseTransition}
\ket{\downarrow \downarrow \uparrow \uparrow S} \leftrightarrow \ket{\Psi_\pm \Psi_- \,S} \rightarrow \ket{\downarrow \downarrow \Psi_- \,S}.
\end{equation}
This transition is completely analogous to Eq.~\eqref{eq:QutritDiodeReverseTransition} from chapter \ref{chapter:QutritDiode}. In Fig.~\ref{figure:ChapterInterferencePopulations}(b), the population of the interface spins is plotted as a function of time for an initial state of $\ket{\downarrow \downarrow \uparrow \uparrow \downarrow \downarrow}$. A population of $P_r(\ket{\Psi_-}) \simeq 0.9$ is quickly reached, $t<50J^{-1}$, at timescales which are much shorter than the usual relaxation time $t \sim 10^3J^{-1}$. 

In the system studied in chapter \ref{chapter:QutritDiode}, it is more obvious which transitions are allowed and which obey energy conservation. 
The current system is more opaque, and we have to write out the relevant matrix elements of the Hamiltonian
\begin{align}
\langle \downarrow \downarrow \uparrow \uparrow S | \hat{H} | \downarrow \downarrow \uparrow \uparrow S\rangle &= \Delta J + E_S, \\ 
\langle \downarrow \downarrow \uparrow \uparrow S|\hat{H}|\Psi_\pm \Psi_- \,S\rangle &= \mp \delta J, \\ \langle \Psi_\pm \Psi_- \,S|\hat{H}|\Psi_\pm \Psi_- \,S\rangle &= -\Delta J - 2J_{34} \pm 2J + E_S,
\end{align}
where $E_S$ is the energy of $\ket{S}$. For the first part of the transition, $\ket{\downarrow \downarrow \uparrow \uparrow S} \leftrightarrow \ket{\Psi_\pm \Psi_- \, S}$, to be favorable, it needs to obey energy conservation, and the transition matrix element needs to be large. The two states are at resonance when
\begin{align}
\langle \downarrow \downarrow \uparrow \uparrow S|\hat{H}|\downarrow \downarrow \uparrow \uparrow S\rangle &= \langle \Psi_\pm \Psi_- \,S|\hat{H}|\Psi_\pm \Psi_- \,S\rangle, \\ 
 \text{or} \quad J_{34} &= -(\Delta \pm 1)J.
\end{align}
These two solutions correspond to the two bands of large rectification in Fig.~\ref{figure:ChapterInterferenceRecD12vsJ34} mentioned previously. The positive solution is almost the same as the default value $J_{34}=-(\Delta+1.3)$ with a discrepancy of $0.3$. From the matrix element $\langle \downarrow \downarrow \uparrow \uparrow S|\hat{H}|\Psi_\pm \Psi_- \,S\rangle$, we would expect a larger $\delta$ to yield a larger rectification. However, picking $\delta$ is clearly a balance. While a larger $\delta$ results in the state $\ket{\Psi_-}$ recovering faster, it also results in a decay of $\ket{\Psi_-}$, as can be seen in Eq.~\eqref{eq:InterferenceInterference}. Apparently, the present setup requires the smallest non-zero value of $\delta$ achievable for this balance to be optimal i.e. $\delta \rightarrow 0$. 

If decoherence is included, this balance is changed and a larger $\delta$ is required to compensate. The same is true for other imperfections e.g. a magnetic field on spin 3 or 4. Now that the mechanism is understood, many alternative versions and an expansions of the setup can be found using the same logic. Some of these are studied in the appendices of Ref.~\citep{PhysRevA.105.052605}.

\subsection{Exploring different system parameters and decoherence}
\label{subsec:InterferenceDec}

\begin{figure}[t]
\centering
\includegraphics[width=1.\linewidth, angle=0]{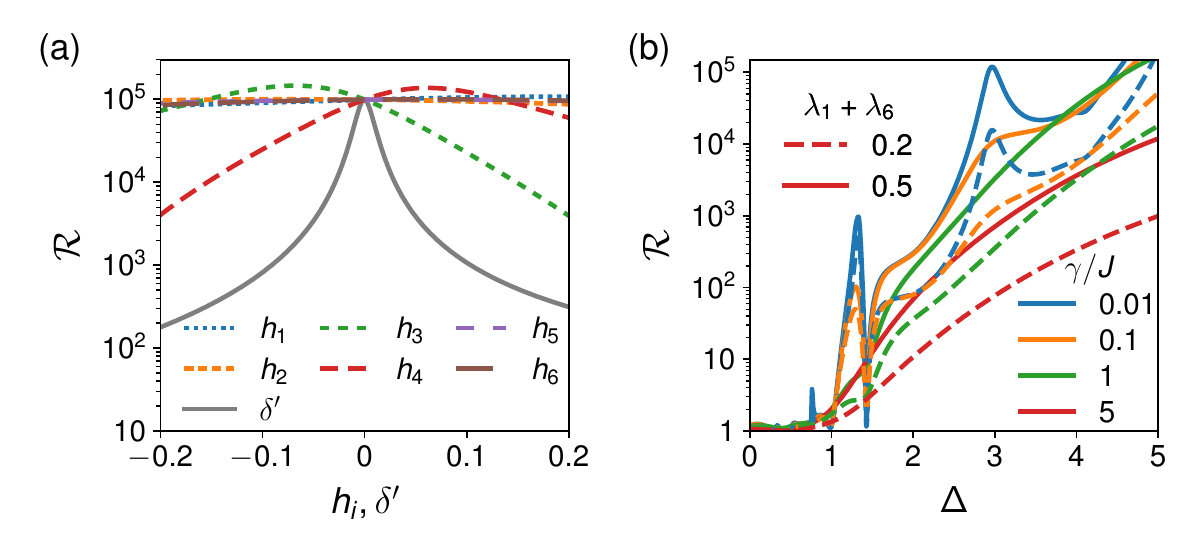}
\caption{ (a) $\mathcal{R}$ as a function of $h_n$ and $\delta'$ with 
$\delta = 0.03$. (b) Rectification $\mathcal{R}$ as a function of $\Delta$ for different interaction strengths $\gamma$ between the system and the bath. The dashed lines denotes the case $\lambda_n \in \{0, 0.2\}$, and the solid lines denotes the case $\lambda_n \in \{0, 0.5\}$. See Eq.~\eqref{eq:InterferenceDefaultValues} for the values of the remaining system parameters. Reproduced with permission from Phys. Rev. A \textbf{105}, 052605 (2022) \cite{PhysRevA.105.052605}. Copyright 2022 American Physical Society.}\label{figure:ChapterInterferenceMagnetic}
\end{figure}

\begin{figure}[t]
\centering
\includegraphics[width=0.625\linewidth, angle=0]{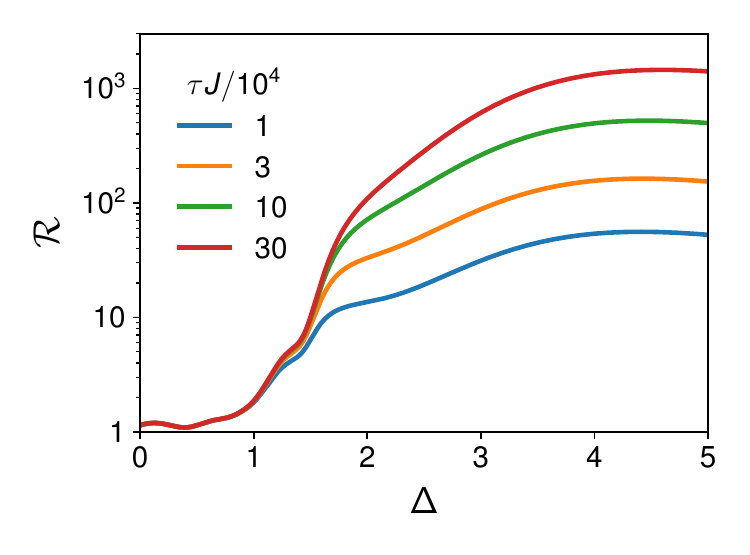}
\caption{ Rectification $\mathcal{R}$ as a function of $\Delta$ for different values of $\tau$ where $\delta = 0.1$. See Eq.~\eqref{eq:InterferenceDefaultValues} for the values of the remaining system parameters. Reproduced with permission from Phys. Rev. A \textbf{105}, 052605 (2022) \cite{PhysRevA.105.052605}. Copyright 2022 American Physical Society.}
\label{figure:ChapterInterferenceDecoherence}
\end{figure}

\noindent Next, we study the sensitivity of the rectification to local magnetic fields, coupling strength perturbations, and finite coherence times. 
Since the rectification mechanism relies on entanglement and interference, we can expect the rectification to be sensitive towards decoherence and pertubations that break the interference conditions in Eqs.~\eqref{eq:InterferenceInterferenceCondition1}-\eqref{eq:InterferenceInterferenceCondition2} and, therefore, the dark-state mechanism. Therefore, we can expect spins 3 and 4 to be most sensitive to magnetic fields, 
while the coupling of spins 4 and 5 should be the more sensitive coupling
parameter.
Hence, we add to Eq.~\eqref{eq:InterferenceHamiltonianMain} perturbations of the form
\begin{align}
\hat{H}' = \sum_{n = 1}^6 h_n \hat{\sigma}^z_{n} + \delta' J \hat{X}_{45}.
\end{align}
Fig.~\ref{figure:ChapterInterferenceMagnetic}(a) shows $\mathcal{R}$ as a function of $h_n$ and $\delta'$, where for each line, the other perturbations are kept zero.
As expected, the rectification is stable towards changes in $h_1$, $h_2$, $h_5$, and $h_6$. The largest $\mathcal{R}$ requires magnetic fields $h_3$ and $h_4$ of 
less than $20\%$ of $J$. 
Fig.~\ref{figure:ChapterInterferenceMagnetic}(a) also shows $\mathcal{R}$ as a function of $\delta'$ and indicates that $\delta' < \delta$ is the region of 
large rectification. The rapid decrease in $\mathcal{R}$ could 
be used to detect variations in couplings in the system. In fact, this sensitivity is explored in chapter \ref{chapter:WB} as a method for measuring coupling strengths. 

Next, we study the effect of changing the interaction strength $\gamma$ between the baths and the system, as defined in Eq.~\eqref{eq:InterferenceMasterEquationMain}, as well as the nature of the baths $\lambda_n$. This can be seen in Fig.~\ref{figure:ChapterInterferenceMagnetic}(b), where $\mathcal{R}$ is plotted for different interaction strengths and $\lambda_n$. We see that the general behavior of the rectification is still achieved. However, for small $\gamma$, the rectification becomes more sensitive to the inner structure of the system. Generally, the rectification is increased slightly for weaker interaction strengths or larger $\lambda_1 + \lambda_6$. Since a larger $\gamma$ makes energy conservation less important, the entangled state $\ket{\Psi_-}$ becomes more fragile for larger $\gamma$. This is discussed in much more detail in a similar setup in chapter \ref{chapter:WB}. 

Finally, decoherence is included by adding to Eq.~\eqref{eq:InterferenceMasterEquationMain} the perturbation 
\begin{align}
\mathcal{L}'[\hat{\rho}] &= \frac{1}{\tau} \sum_{n = 1}^6 \mathcal{M} [\hat{\sigma}^-_{n}, \hat{\rho}] + \frac{1}{4 \tau} \sum_{n = 1}^6 \mathcal{M} [\hat{\sigma}^z_{n}, \hat{\rho}],
\end{align}
which insures that, if $\mathcal{L}[\hat{\rho}] = 0$, then the lifetime for all spins for decay ($T_1$) and dephasing ($T_2$) is $\tau = T_1 = T_2$.
In Fig.~\ref{figure:ChapterInterferenceDecoherence}, the rectification for the diode is plotted as a function of $\Delta$ for different values of $\tau$. The rectification is clearly much smaller with decoherence included, and rectification values similar to the ones found in chapter \ref{chapter:QutritDiode} are observed. With these coherence times, chapter \ref{chapter:QutritDiode} and \ref{chapter:Interference} achieve similar rectification values. However, both approaches has their pros and cons. For better coherence time, the entanglement approach will improve. For larger anharmonicities, the qutrit approach will perform better.

\section{Heat rectification using the global master equation}
\label{sec:InterferenceGlobal}

\begin{figure}[t]
\centering
\includegraphics[width=1.\linewidth, angle=0]{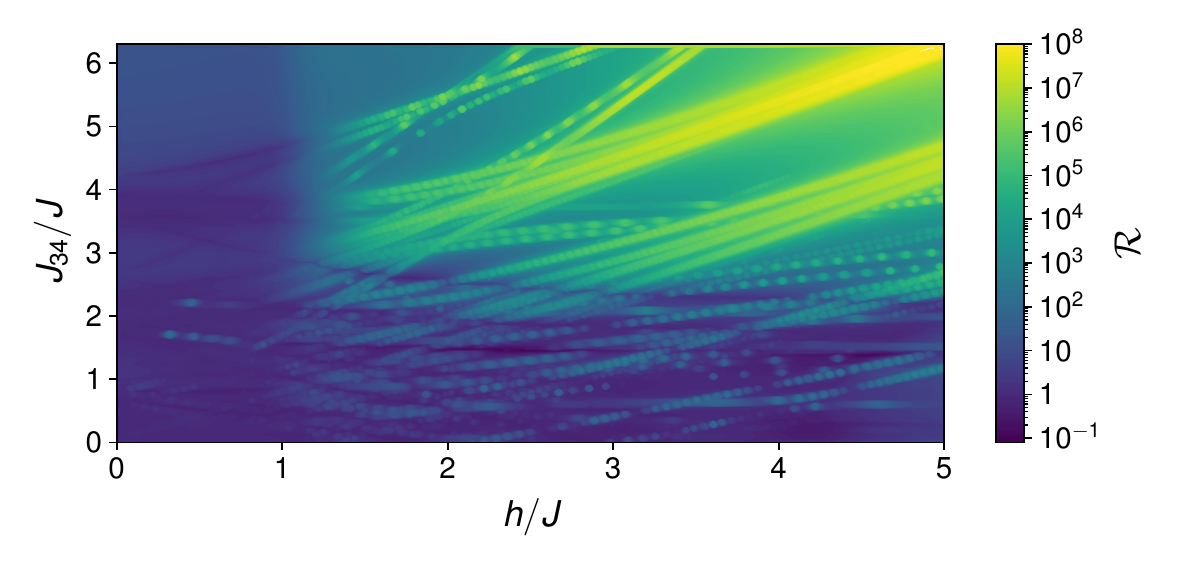}
\caption{ $\mathcal{R}_Q$ as a function of $h$ and $J_{34}$. See Eq.~\eqref{eq:InterferenceGlobalDefaultValues} for the values of the remaining system parameters. Reproduced with permission from Phys. Rev. A \textbf{105}, 052605 (2022) \cite{PhysRevA.105.052605}. Copyright 2022 American Physical Society.}\label{figure:ChapterInterferenceGlobalRechvsJ34}
\end{figure}

\begin{figure}[t]
\centering
\includegraphics[width=1.\linewidth, angle=0]{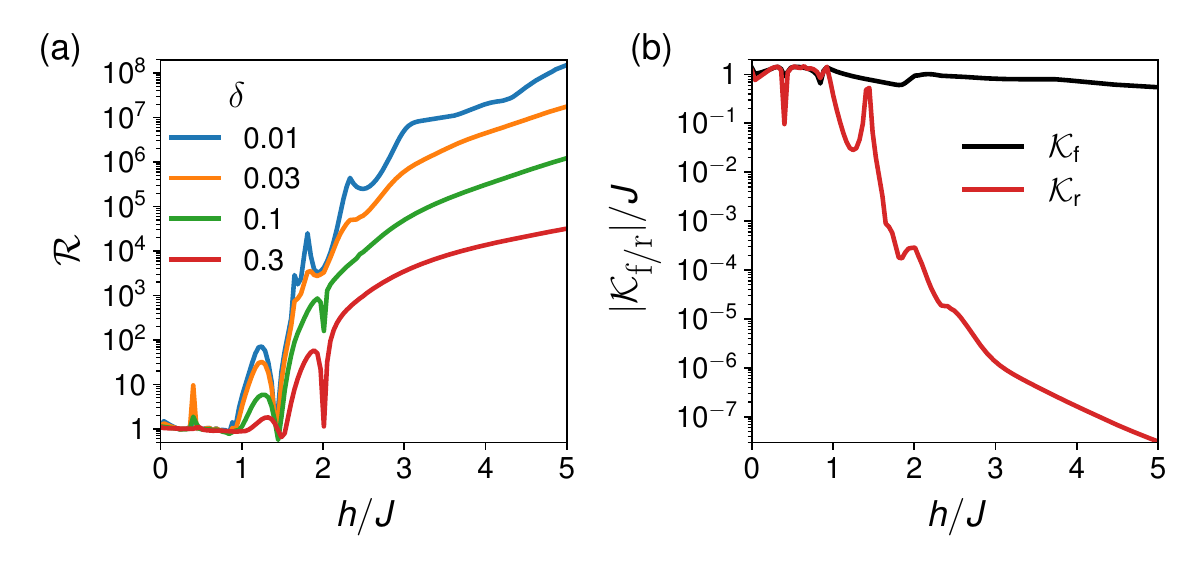}
\caption{ (a) Heat current rectification as a function of $h$ for different values of $\delta$. (b) Steady-state heat currents as a function of $h$. See Eq.~\eqref{eq:InterferenceGlobalDefaultValues} for the values of the remaining system parameters. Reproduced with permission from Phys. Rev. A \textbf{105}, 052605 (2022) \cite{PhysRevA.105.052605}. Copyright 2022 American Physical Society.}\label{figure:ChapterInterferenceGlobalRech2}
\end{figure}

\begin{figure}[t]
\centering
\includegraphics[width=1.\linewidth, angle=0]{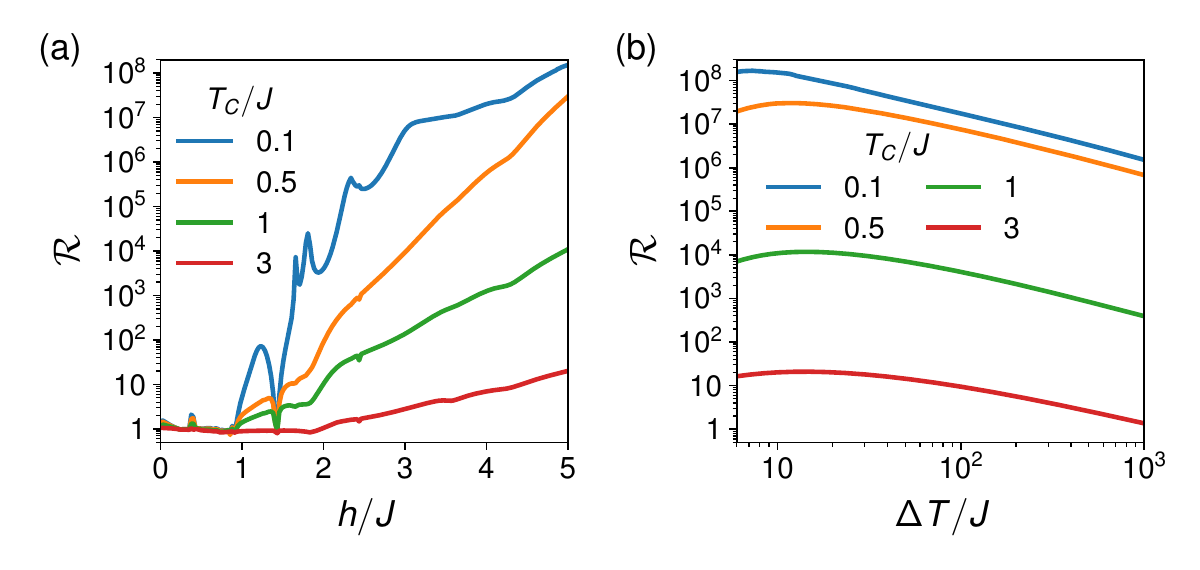}
\caption{ The rectification $\mathcal{R}_Q$ plotted for different cold bath temperatures $T_C$ with hot bath temperature $T_H = T_C + \Delta T$. First, $h$ is varied keeping $\Delta T = 10J$, (a), and next $\Delta T$ is varied keeping $h=5J$, (b). See Eq.~\eqref{eq:InterferenceGlobalDefaultValues} for the values of the remaining system parameters. Reproduced with permission from Phys. Rev. A \textbf{105}, 052605 (2022) \cite{PhysRevA.105.052605}. Copyright 2022 American Physical Society.}\label{figure:ChapterInterferenceGlobalRecTemp}
\end{figure}

\noindent In this section, the setup in Fig.~\ref{figure:ChapterInterferenceSetup} is studied in the context of heat currents using the global master equation. For this, the Hamiltonian is modified such that the energy gab created by the Z-coupling is still present, but the spin flip symmetry is broken
\begin{equation}
\label{eq:InterferenceHamiltonianGlobalMain}
\hat{H}_Q = \hat{H}(\Delta = 0) + h \left( \hat{\sigma}^z_{1} + \hat{\sigma}^z_{2} \right) + h_\mathrm{offset} \sum_{i = 1}^6 \hat{\sigma}^z_{i}.
\end{equation}
The system is coupled to the two baths at finite temperature using the global master equation
\begin{align}
\frac{d \hat{\rho}}{d t} = \mathcal{L}_Q[\hat{\rho}] = -i [\hat{H}_Q, \hat{\rho}] + \mathcal{D}_1 [\hat{\rho}] + \mathcal{D}_6 [\hat{\rho}],
\label{eq:InterferenceMasterEquationGlobalMain}
\end{align}
where the dissipative terms are now on the form defined in Eq.~\eqref{eq:MasterEquationsGlobalN}
\begin{align}
\mathcal{D}_1[\hat{\rho}] &= \mathcal{N} [\hat{\sigma}_1^x, \hat{\rho}, \gamma_1] ,\\
\mathcal{D}_6[\hat{\rho}] &= \mathcal{N} [\hat{\sigma}_6^x, \hat{\rho}, \gamma_6].
\end{align}
The coupling strength for transitions of frequency $\omega$ is
\begin{equation} 
\gamma_i(\omega) = \gamma_Q \omega \big(1 + n_i(\omega) \big),
\end{equation}
where $n_i(\omega) = \big( e^{\omega/T_i} -1 \big)^{-1}$ is the usual Bose-Einstein distribution. A partial secular approximation is perform keeping only terms where $\omega-\omega' < 0.1J$; see  Eq.~\eqref{eq:MasterEquationsGlobalN}. 

Unless otherwise stated, the values are set to the default
\begin{align}
& h = 5J, \quad \delta = 0.01, \quad J_{34}=h + 1.3J, \label{eq:InterferenceGlobalDefaultValues} \\
& \hspace{0.8cm} \gamma_Q = J, \quad T_i \in \{0.1J, 10.1J\}.  \nonumber
\end{align}
Once again, we are interested in the steady state $\hat{\rho}_\mathrm{ss}$, defined through $\mathcal{L}_Q[\hat{\rho}_\mathrm{ss}]=0$. The heat current is defined as the heat exchanged between the system and the corresponding bath, 
\begin{equation}
\mathcal{K} = \mathrm{tr}\left\{\hat{H}_{Q} \mathcal{D}_1[\hat{\rho}_{ss}]\right\} =- \mathrm{tr}\left\{ \hat{H}_{Q} \mathcal{D}_6[ \hat{\rho}_{ss}]\right\}.
\end{equation}
Like before the heat current rectification is defined as $\mathcal{R}_Q = - \mathcal{K}_\mathrm{f}/\mathcal{K}_\mathrm{r}$. For $h_\mathrm{offset} \gg h, J_{34}, J, \delta J$, the baths are approximately local similar to the original model, as proven in subsection \ref{subsec:MasterEquationsReductionToLocal}. As a consequence, the rectification values are similar to those seen in Fig.~\ref{figure:ChapterInterferenceRecD12}(a). Instead, we shall focus on $h_\mathrm{offset} = 0$, where the global master equation becomes important.

The contour plot in Fig.~\ref{figure:ChapterInterferenceGlobalRechvsJ34} shows $\mathcal{R}_Q$ as a function of $J_{34}$ and $h$. Unlike for the spin current case, we clearly see multiple resonances that makes the plot chaotic for small $J_{34}$ and $h$. However, in the upper right corner many of the resonance merge and create thicker more stable lines of large rectification of $> 10^8$. We note that the region of largest $\mathcal{R}_Q$ follows a similar parametrization to before, namely $J_{34} = h + 1.3J$. 
If one goes through the methods of subsection \ref{subsec:InterferenceUnderstanding}, using the new Hamiltonian from Eq.~\eqref{eq:InterferenceHamiltonianGlobalMain}, this is the parametrization that would be found. 

In Fig.~\ref{figure:ChapterInterferenceGlobalRech2}(a), the rectification is plotted using this parametrization for different values of $\delta$. In Fig.~\ref{figure:ChapterInterferenceGlobalRech2}(b), it is verified that large rectification is due to a suppression of $\mathcal{K}_r$. 
The proposed diode thus generalizes very well to heat currents where rectification values of $>10^8$ can be reached. 

Finally, we study the effects of the finite temperatures by letting $T_i \in \{T_C, T_C+\Delta T\}$, where $T_C$ is the cold bath temperature, and $\Delta T$ is the temperature gradient. In Fig.~\ref{figure:ChapterInterferenceGlobalRecTemp}(a), rectification is plotted for different values of the cold bath temperature. 
It can be seen that the largest rectification is achieved for $T_C < J$. Since $J$ sets the energy scale of the diode, for $T_C < J$ the cold bath will predominantly induce decay while the hot bath will induce both decay and excitation in the energy levels. Therefore, we expect a better diode for smaller $T_C$. In Fig.~\ref{figure:ChapterInterferenceGlobalRecTemp}(b), we plot the rectification as a function of temperature bias.  The rectification is stable over the first order of magnitude in $\Delta T$ but decreases slightly for very large $\Delta T$.

\cleardoublepage
\chapter{Quantum Wheatstone bridge}
\label{chapter:WB}
\epigraph{\footnotesize This chapter is based on Ref.~\cite{PhysRevLett.128.240401}.
Sections \ref{sec:WBHamiltonian}-\ref{sec:WBCurrent} and all figures herein have been reproduced with permission from Ref.~\cite{PhysRevLett.128.240401}.}{}

\noindent Through exploring rectification in chapter \ref{chapter:Interference}, we found a system that, in reverse bias, was driven into an interference-based dark state. However, interference mechanisms are often very fragile since small perturbations can break the interference condition. On the other hand, this makes interference mechanisms ideal for metrology in experiments such as the Michelson-Morley experiment \cite{ohanian2001special}. This sensitivity of the interference mechanism is also present for the entanglement-enhanced diode, as discovered in subsection \ref{subsec:InterferenceDec}. 
This can be exploited and expanded into a quantum version of the classical Wheatstone bridge. 

In order to set the scene, some context is needed in two main areas. First, since the classical counterpart inspired some of the choices for the quantum Wheatstone bridge, a basic introduction to the Wheatstone bridge from electronics is needed. Second, since the quantum Wheatstone bridge is used to measure coupling strengths, we will go through the most straightforward way of measuring coupling strengths for comparison. 

\begin{figure}[t]
\centering
\includegraphics[width=0.92\linewidth, angle=0]{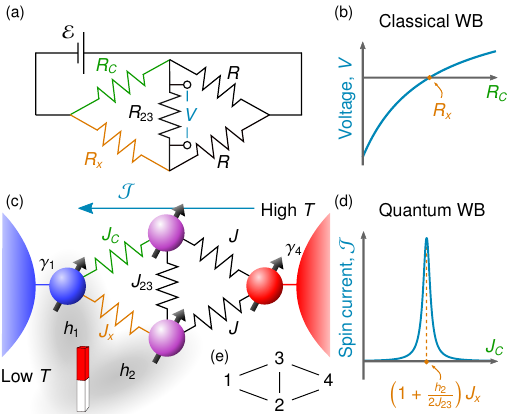}
\caption{ (a) Classical Wheatstone bridge consisting of three resistors with known resistances $R$ and $R_C$ and one resistor with unknown resistance $R_x$. (b) Sketch of the voltage dependence used in the classical Wheatstone bridge. (c) Quantum Wheatstone bridge consisting of four spins interacting with three known coupling strengths $J_C, J$, and $J_{23}$ and one unknown coupling strength $J_x$. Spins 1 and 4 are interacting with two thermal baths of different temperature, and two magnetic fields of strength $h_1$ and $h_2$ are applied to spins 1 and 2, respectively. (d) Sketch of the spin current dependence used in the Quantum Wheatstone bridge. (e) Numbering for the 4 spins used throughout the text. Reproduced with permission from Phys. Rev. Lett. \textbf{128}, 240401 (2022) \cite{PhysRevLett.128.240401}. Copyright 2022 American Physical Society.}\label{figure:ChapterWBSetup}
\end{figure}

The bridge, as seen in Fig.~\ref{figure:ChapterWBSetup}(a), is a device used to precisely determine an unknown resistance, $R_x$. In a circuit of four resistances, there are two known resistances $R$, a tunable resistance $R_C$, and the resistance $R_x$. A fifth resistance $R_{23}$ is inside a voltmeter and, therefore, much larger than other resistances, resulting in a negligible current through $R_{23}$. To determine the values of $R_x$, the voltage over $R_{23}$ is measured as a function of $R_C$. This voltage can be found to be
\begin{align*}
V = \mathcal{E}R \frac{R_C - R_x}{(R_C + R)(R_x + R)}.
\end{align*}
This voltage is sketched in Fig.~\ref{figure:ChapterWBSetup}(b). Therefore, when the voltage $V$ is zero, the resistance is known to be $R_x = R_C$. This is called the balance point.

In quantum systems, transport is governed by coupling strength dependent matrix elements, which can be seen as analogous to the conductance, or the reciprocal of resistance, in electronics. For two spins connected through an XX coupling, the coupling strength $J$ enters the Hamiltonian through
\begin{align}
\hat{H} = J \left( \hat{\sigma}_1^x \hat{\sigma}_2^x + \hat{\sigma}_1^y \hat{\sigma}_2^y \right).
\end{align}
For such a system, one way of determining $J$ is to measure the transition energies. This system is diagonalized by the four states $\ket{\ds \ds}, \ket{\Psi_-}, \ket{\Psi_+}$, and $\ket{\us \us}$ with energies $0, -2J, 2J$, and $0$, respectively. Therefore, by measuring the transition spectrum, $J$ can be extrapolated directly. This is common practice \cite{Gubaydullin2022, PhysRevLett.129.220501} and a good initial step for determining $J$ to greater precision. Additionally, the state $\ket{\us \ds}$ can be prepared, and the resulting oscillations can be measured. The state at later times will then be
\begin{align}
\ket{\us \ds, t} = \cos(2Jt) \ket{\us \ds} + i \sin(2Jt) \ket{\ds \us}.
\end{align}
If the first spin is measured, the probability of measuring up is $P(\ket{\us_1}) = \cos(2Jt)^2$, which can be used to determine $J$. A useful way to quantify the sensitivity of a given measurement towards slight variations in $J$, and therefore, our ability to determine $J$, is the Fisher information $\mathcal{I}$.
The Fisher information for the above setup and measurement of $\ket{\ds_1}$ is
\begin{align}
\mathcal{I}(\ket{\ds_1}) = 16t^2.
\end{align}
No other measurement would result in a larger Fisher information, and the measurement is, therefore, optimal. More information on how the Fisher information is calculated and what it means will be given in section \ref{sec:WBQuantumFisherInformation}.

This chapter will set up a quantum version of the Wheatstone bridge described above for measuring coupling strengths instead of resistances. Expanding on the interference mechanism from chapter \ref{chapter:Interference}, a system of 4 spin-1/2 particles connected to two thermal baths is set up. The two baths have a temperature gradient driving a spin current through the system. First, the state of the interface is shown to be very sensitive towards small changes in an unknown coupling. Next, the state is solved analytically in the Markovian regime, resulting in an equation for the sensitivity. The sensitivity is then quantified using the quantum Fisher information, and the optimal measurement for determining the unknown coupling is found. Finally, the state sensitivity is connected to the spin current, allowing for indirect measurement of the state using the spin current. Therefore, the spin current plays the role of the voltage $V$ for the quantum Wheatstone bridge.

\section{Hamiltonian and master equation}
\label{sec:WBHamiltonian}

The model studied is two spins connected through a double-spin interface in a diamond configuration similar to the classical Wheatstone bridge as seen in Fig~\ref{figure:ChapterWBSetup}(c). The Hamiltonian for the four spins is
\begin{equation}
\begin{aligned}
\hat{H} &= \frac{\omega + 2h_1}{2} \hat{\sigma }^z_{1} +  \frac{\omega + 2h_2}{2} \hat{\sigma }^z_{2} +  \frac{\omega}{2} \hat{\sigma }^z_{3} +  \frac{\omega}{2} \hat{\sigma }^z_{4} \\ &\hspace{1.5cm} +  J_x \hat{X}_{12} +  J_C \hat{X}_{13} + J_{23} \hat{X}_{23} + J\hat{X}_{24} + J\hat{X}_{34}, 
\end{aligned}
\end{equation}
where $\hat{X}_{ij} = \hat{\sigma}^x_{i} \hat{\sigma}^x_{j} + \hat{\sigma}^y_{i} \hat{\sigma}^y_{j}$ 
is the $XX$ spin exchange operator. The Pauli matrices for the $i$th spin are denoted $\hat{\sigma}^\alpha_{i}$ for $\alpha = x,y,z$. The parameter $J_x$ is unknown, and $J_C$ is a controllable coupling. 
The angular frequency $\omega$ corresponds to a homogeneous magnetic field, and $h_n$ is an offset in the magnetic field acting on the $n$th spin. The exchange coupling $J$ gives the overall scale of the problem while the exchange between the interface spins is $J_{23}$. 
Here, we focus on the regime $\omega \gg J_{23}, h_1 \gg h_2,J$ and $J_{23} \simeq h_1$. 

Spin 1 is coupled to a Markovian thermal bath at very low temperature $T_1 \ll \omega + 2h_1$ forcing spin 1 to decay. Similarly, spin 4 is coupled to a Markovian thermal bath of higher temperature $T_4 \sim \omega$ inducing both decay and excitation of spin 4.
The evolution of the system is described by the local master equation on Lindblad form
\begin{equation}
\frac{d \hat{\rho}}{d t} = -i [\hat{H}, \hat{\rho}] + \mathcal{D}_1[\hat{\rho}] + \mathcal{D}_4[\hat{\rho}]. \label{me:1}
\end{equation}
The baths are modeled using the non-unitary parts
\begin{equation}
\begin{aligned}
\mathcal{D}_{1}[\hat{\rho}] &= \gamma_1 \mathcal{M}[\hat{\sigma}^-_{1},\hat{\rho}],\\
\mathcal{D}_{4}[\hat{\rho}] &= \gamma_4 (n + 1)\mathcal{M}[\hat{\sigma}^-_{4},\hat{\rho}] + \gamma_4 n \mathcal{M}[\hat{\sigma}^+_{4}, \hat{\rho}], \nonumber
\end{aligned}
\end{equation}
where $\mathcal{M}[\hat{A}, \hat{\rho}] = \hat{A} \hat{\rho} \hat{A}^\dag - \{ \hat{A}^\dag \hat{A}, \hat{\rho} \}/2$. The coupling strength between the cold bath and spin 1 is $\gamma_1$; Likewise, the coupling strength between the hot bath and spin 4 is $\gamma_4$. The mean number of excitations in the hot bath mode of energy $\omega$ is $n = \big( e^{\omega/T_4} -1 \big)^{-1}$, where $T_4$ is the temperature of the hot bath. The two baths will induce heat flow and generally drive the system into a non-equilibrium steady state $\hat{\rho}_\mathrm{ss}$. It is the steady state that we will use to probe the value of $J_x$. 

Unless otherwise stated throughout this chapter, we will use the set of parameters 
\begin{align}
&J_{23} = 20J, \quad h_1 = 20J, \quad h_2 = 0.5J, \quad J_x = J, \label{eq:WBDefaultValues} \\
&\hspace{0.5cm} J_C = J, \quad n = 0.5, \quad \gamma_1 = J, \quad \gamma_4 = 10J. \nonumber
\end{align}
The excitation energy $\omega$ has no impact on the steady state and is not set. As we have seen before, this can be seen by transforming the Lindblad master equation into the interaction picture with respect to $\hat{H}_0 = \frac{\omega}{2} \sum_{i=1}^4 \hat{\sigma}^z_{i}$. Since the Hamiltonian is spin preserving, $\hat{\rho}_{\text{ss}}$ is unchanged.

\begin{figure}[t]
\centering
\includegraphics[width=1.\linewidth, angle=0]{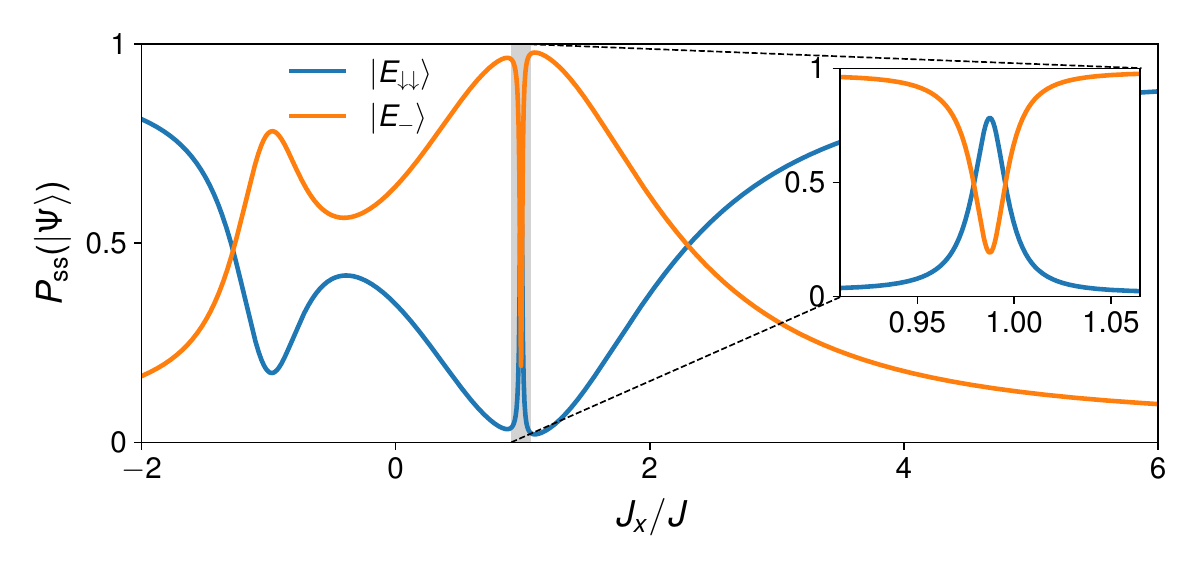}
\caption{Population of the two states $\ket{E_{\downarrow \downarrow}}$ and $\ket{E_-}$ as a function of the unknown parameter $J_x$. See Eq.~\eqref{eq:WBDefaultValues} for the values of the remaining system parameters. Reproduced with permission from Phys. Rev. Lett. \textbf{128}, 240401 (2022) \cite{PhysRevLett.128.240401}. Copyright 2022 American Physical Society.}\label{figure:ChapterWBPopulations1}
\end{figure}

\begin{figure}[t]
\centering
\includegraphics[width=0.92\linewidth, angle=0]{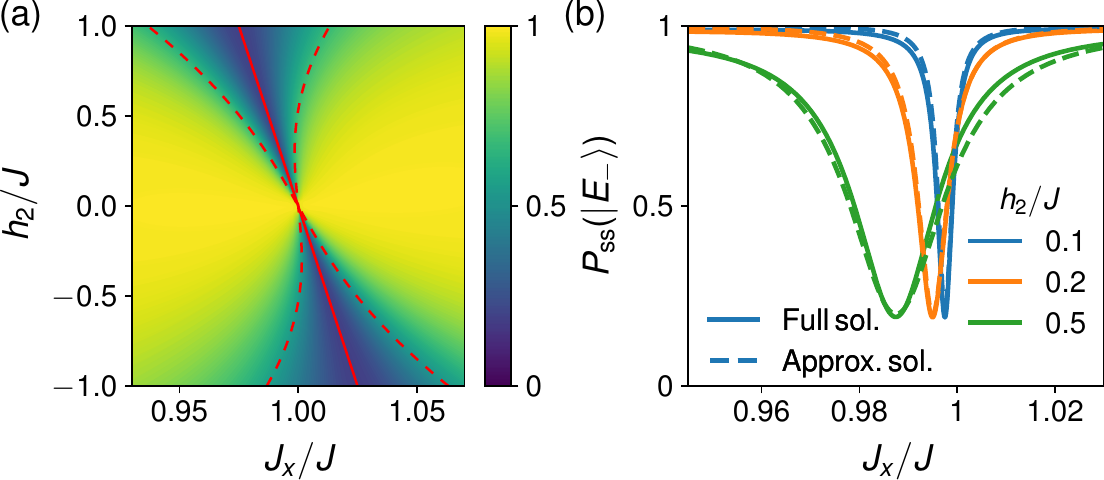}
\caption{ (a) Population of $\ket{E_-}$ as a function of both $J_x$ and $h_2$. The solid red line shows the points of destructive interference, Eq.~\eqref{eq:WBResonance}, while the dashed lines denote the width of the population drop, Eq.~\eqref{eq:WBWidth}. (b) Population of $\ket{E_-}$ as a function of $J_x$ for both the full numerical solution (solid line) and the approximate solution (dashed line) given by Eq.~\eqref{eq:WBPopulation1}. See Eq.~\eqref{eq:WBDefaultValues} for the values of the remaining system parameters. Reproduced with permission from Phys. Rev. Lett. \textbf{128}, 240401 (2022) \cite{PhysRevLett.128.240401}. Copyright 2022 American Physical Society.}\label{figure:ChapterWBPopulations2}
\end{figure}

\section{State sensitivity towards variations in $J_x$} 

The steady state of the two interface spins is sensitive to changes in the unknown parameter $J_x$. This sensitivity can be used to determine the value of $J_x$. Because of the two baths, the coherences between the two interface spins and the outer spins are negligible, and the density matrix is well described by the populations of the interface. Therefore, the Hamiltonian for spins 2 and 3 is diagonalized and the eigenstates, to linear order in $h_2/J_{23}$, are found to be
\begin{subequations}
\label{states}
\begin{alignat}{1}
\ket{E_{\uparrow \uparrow}} = \ket{\uparrow \uparrow},\quad
&\ket{E_+} = \ket{\Psi_+} + \frac{h_2}{4J_{23}} \ket{\Psi_-},\\
\ket{E_{\downarrow \downarrow}} = \ket{\downarrow \downarrow},\quad &\ket{E_-} = \ket{\Psi_-} - \frac{h_2}{4J_{23}} \ket{\Psi_+} ,
\end{alignat}
\end{subequations}
where $\ket{\Psi_\pm} = (\ket{\uparrow \downarrow} \pm \ket{\downarrow \uparrow})/\sqrt{2}$. 
For the reduced density matrix $\hat{\rho}_{\text{ss}}^{23} = \tr_{14} \{\hat{\rho}_{\text{ss}}\}$, where the Hilbert spaces of spins 1 and 4 have been traced away. The populations for the two states $\ket{E_{\downarrow \downarrow}}$ and $\ket{E_-}$ are shown in Fig.~\ref{figure:ChapterWBPopulations1} as a function of $J_x$. 
The populations are given by $P_{\text{ss}}(\ket{E_{\downarrow \downarrow}}) = \tr_{23} \{\hat{\rho}_{\text{ss}}^{23} \op{E_{\downarrow \downarrow}}\}$ and $P_{\text{ss}}(\ket{E_{-}}) = \tr_{23} \{\hat{\rho}_{\text{ss}}^{23} \op{E_{-}}\}$. This shows that the two interface spins are predominantly in these two states.

The two spins are driven into the entangled state $\ket{E_-}$ as $J_x$ approaches $J_C$ as in chapter \ref{chapter:Interference}. This can be explained through the two series of transitions
\begin{align}
\ket{\downarrow \uparrow \uparrow \ds} \leftrightarrow \ket{\uparrow E_- \ds} \rightarrow \ket{\downarrow E_- \ds}, \label{eq:WBTransition1} \\
\ket{\downarrow \uparrow \uparrow \us} \leftrightarrow \ket{\uparrow E_- \us} \rightarrow \ket{\downarrow E_- \us}. \label{eq:WBTransition2} 
\end{align}
Learning from chapter \ref{chapter:Interference}, the parameters of the model are chosen such that the transition $\ket{\downarrow \uparrow \uparrow} \leftrightarrow \ket{\uparrow E_-}$ is energetically allowed, and the transition $\ket{\uparrow E_-} \rightarrow \ket{\downarrow E_-}$ makes the process irreversible. However, the present scheme is very different from chapter \ref{chapter:Interference}, when $J_x$ gets even closer to $J_C$. The population of $\ket{E_-}$ drops suddenly, which can be seen in the inset of Fig.~\ref{figure:ChapterWBPopulations1}. 
This sudden change in the populations is what can be exploited for precise determination of $J_x$.

In Fig. \ref{figure:ChapterWBPopulations2}(a), the population of $\ket{E_-}$ is plotted as a function of both the magnetic field $h_2$ and the unknown parameter $J_x$. It is seen that the value of $J_x$ for which the population is minimal is linear in $h_2$. This can be explained by looking at the matrix element for the first transition in Eqs.~\eqref{eq:WBTransition1}-\eqref{eq:WBTransition2} to linear order in $h_2/J_{23}$
\begin{equation}
\begin{aligned}
&\mel{\uparrow E_- \uparrow}{\hat{H}}{\downarrow \uparrow \uparrow \uparrow} = \mel{\uparrow E_- \downarrow}{\hat{H}}{\downarrow \uparrow \uparrow \downarrow} \\
&\hspace{2cm}= -\sqrt{2} \left[ J_x \left( 1+\frac{h_2}{4J_{23}} \right) - J_C \left( 1-\frac{h_2}{4J_{23}} \right) \right]. 
\end{aligned}
\end{equation}
The transition is forbidden to first order when the above matrix element is zero. This is equivalent to destructive interference occurring between the spin excitation located at spin 2 and the spin excitation located at spin 3. The destructive interference condition is $J_x = J_{x,0}$, where
\begin{equation}
J_{x,0} = J_C \left( 1- \frac{h_2}{2J_{23}} \right). \label{eq:WBResonance} 
\end{equation}
This parametrization is plotted as a solid red line in Fig. \ref{figure:ChapterWBPopulations2}(a). 

\subsection{Markovian solution}

To analytically approximate the width of the dip in population shown in Fig.~\ref{figure:ChapterWBPopulations1}(a), we will follow the same approach as in subsection \ref{subsec:QutritDiodeMarkovianSolution} to obtain the density matrix of the interface in steady state. First, the Hamiltonian for spins 2 and 3 is diagonalized to linear order in $h_2/J_{23}$
\begin{align}
E_{\us \us} &= \omega + h_2,\\
E_+ &= 2J_{23},\\
E_- &= -2J_{23},\\
E_{\ds \ds} &= -\omega - h_2,
\end{align}
with corresponding states $\ket{E_{\us \us}}, \ket{E_+}, \ket{E_-}$ and $\ket{E_{\ds \ds}}$.
The two baths drive transitions between these four states with rates that are approximately
\begin{align}
\Gamma_{\ket{E} \rightarrow \ket{E'}}^C &= \frac{\big|M_{\ket{E} \rightarrow \ket{E'}}^C\big|^2 \gamma_1}{\big|M_{\ket{E} \rightarrow \ket{E'}}^C\big|^2 + (\omega + 2h_1 + E' - E)^2 + \gamma_1^2/4}, \\\ 
\Gamma_{\ket{E'} \rightarrow \ket{E}}^C &= 0, \\
\Gamma_{\ket{E} \rightarrow \ket{E'}}^H &= \frac{\big|M_{\ket{E} \rightarrow \ket{E'}}^H\big|^2 \gamma_4 (n+1)}{(\omega + E' - E)^2 + \gamma_4^2 (2n+1)^2 /4}, \label{eq:WBRateFullHot1} \\
\Gamma_{\ket{E'} \rightarrow \ket{E}}^H &= \frac{\big|M_{\ket{E'} \rightarrow \ket{E}}^H\big|^2 \gamma_4 n}{(\omega + E - E')^2 + \gamma_4^2 (2n+1)^2 /4}, \label{eq:WBRateFullHot2}
\end{align}
for $(E, E') \in \{(E_{\us \us}, E_-), (E_{\us \us}, E_+), (E_{-}, E_{\ds \ds}), (E_+, E_{\ds \ds})\}$.  The two matrix elements are 
\begin{align}
M_{\ket{E} \rightarrow \ket{E'}}^C &= 2\mel*{E'}{J_x \hat{\sigma}^-_{2} + J_C \hat{\sigma}^-_{3}}{E}, \\
M_{\ket{E} \rightarrow \ket{E'}}^H &= 2J\mel*{E'}{\hat{\sigma}^-_{2} + \hat{\sigma}^-_{3}}{E}.
\end{align}
The first two rates are for the cold bath, while the last two rates are for the hot bath. Therefore, the total rates are defined by 
\begin{align}
\Gamma_{\ket{E} \rightarrow \ket{E'}} = \Gamma_{\ket{E} \rightarrow \ket{E'}}^C + \Gamma_{\ket{E} \rightarrow \ket{E'}}^H.
\end{align}
The cold bath rates are found by adding the rate of interaction between the interface and the cold qubit from Fermi's golden rule with the rate of decay for the cold qubit \cite{Kapit_2017}. The hot bath rates are found by using the Markov approximation for the hot qubit, thus assuming that the hot qubit correlation functions decay faster than the coupling to the interface \cite{breuer2002theory}. This is exactly the same as the Markovian solution in chapter \ref{chapter:QutritDiode} with a qubit instead of a harmonic oscillator. The full derivation of Eqs.~\eqref{eq:WBRateFullHot1}-\eqref{eq:WBRateFullHot2} is essential to the topic of chapter \ref{chapter:Maxwell}, and it will, therefore, be derived in subsection \ref{subsec:MaxwellMarkovianLimit}. 

Writing the populations of the four states in a vector \linebreak$\vec{P} = \{P(\ket{E_{\ds \ds}}), P(\ket{E_{-}}), P(\ket{E_{+}}), P(\ket{E_{\us \us}})\}^T$, the time evolution is again governed by the master equation
\begin{equation}
\frac{d}{dt} \vec{P} = \mathbf{W} \vec{P},
\end{equation}
where $\mathbf{W}$ is a matrix of rates. The system is small enough that the steady state could be found through $\mathbf{W} \vec{P}_\mathrm{ss}=0$ already, however, the solution is too involved to be useful.
Instead, we wish to obtain an approximate solution to the populations close to the point of interest 
\begin{equation}
J_{x,0} = J_C \left( 1-\frac{h_2}{2J_{23}} \right).
\end{equation}
In the previous section, we found that a dramatic change in populations occur for $J_x \sim J_{x,0}$. In order to resolve and explain this, we write the distance from this value as
\begin{equation}
\delta J_x = J_x - J_{x,0},
\end{equation}
and assume $|\delta J_x| \ll J$. Remembering the assumptions $J,h_2 \ll J_{23}$ and $J_x \sim J_C \sim J \sim \gamma_1$, we can find an approximate form for the rates. 

As an example, we use $\Gamma_{\ket{E_{\us \us}} \rightarrow \ket{E_-}}$. First, the matrix elements are calculated
\begin{align}
M^C_{\ket{E_{\uparrow \uparrow }} \rightarrow \ket{E_-}} &= 2 \left( \bra{\Psi_-} - \frac{h_2}{4J_{23}} \bra{\Psi_+} \right)\left[ J_x \hat{\sigma}_-^{(2)} + J_C \hat{\sigma}_-^{(3)}\right] \ket{\uparrow \uparrow} \\
&\simeq -\sqrt{2} \left( 1+\frac{h_2}{4J_{23}} \right) \delta J_x, \nonumber \\ 
M_{\ket{E_{\us \us}} \rightarrow \ket{E_-}}^H &= 2J \left( \bra{\Psi_-} - \frac{h_2}{4J_{23}} \bra{\Psi_+} \right)\left[ \hat{\sigma}_-^{(2)} + \hat{\sigma}_-^{(3)}\right] \ket{\uparrow \uparrow} \\
&\simeq -\frac{h_2 J}{\sqrt{2} J_{23}}. \nonumber
\end{align}
The rates become
\begin{align}
\Gamma^C_{\ket{E_{\uparrow \uparrow }} \rightarrow \ket{E_-}} &= \frac{2 \delta J_x^2 \gamma_1}{2\delta J_x^2 + (2h_1-2J_{23}-h_2)^2 + \gamma_1^2/4} \\
&\simeq \frac{2\delta J_x^2 \gamma_1}{(2h_1-2J_{23}-h_2)^2 + \gamma_1^2/4}, \nonumber\\
\Gamma^H_{\ket{E_{\uparrow \uparrow }} \rightarrow \ket{E_-}} &= \frac{\frac{h_2^2 J^2}{2 J_{23}^2} (n+1)\gamma_4}{(2J_{23}+h_2)^2 + \gamma_4^2 (2n+1)^2/4} \\
&\simeq \frac{ h_2^2 J^2 (n+1) \gamma_4}{2J_{23}^2 \left(4J_{23}^2 + (2n+1)^2\gamma_4^2 /4\right)} . \nonumber
\end{align}
Here, the expression for $E_- - E_{\uparrow \uparrow}$ is only taken to linear order in $h_2$, and we have assumed that $\delta J_x^2 \ll (2h_1 -2J_{23} - h_2)^2/2 +\frac{\gamma_1^2}{8}$. We will see later that $|\delta J_x|/J$ is of order $h_2/J_{23}$, and therefore, this is a valid assumption. We have assumed $h_1 \simeq J_{23}$, but we keep $h_1$ in the expressions such that small deviations from this value are allowed. The contribution from the hot bath is small compared to the contribution from the cold bath, and therefore, it is neglected. This is seen by noting that $|\delta J_x|/J$ is of the order $h_2/J_{23}$ and $J/J_{23}$. The term could also be kept for now and thrown away by direct comparison later. 
The full list of rates becomes:
\begin{align}
\Gamma_{\ket{E_{\us \us}} \rightarrow \ket{E_-}} &\simeq \frac{2 \gamma_1 \delta J_x^2}{ \eta_1^2 }, \quad \Gamma_{\ket{E_{\ds \ds}} \rightarrow \ket{E_{-}}} \simeq \Gamma_{\ket{E_-} \rightarrow \ket{E_{\us \us}}} \simeq  \frac{h_2^2 J^2 n \gamma_4 }{2 J_{23}^2 \eta_4^2},  \\
\Gamma_{\ket{E_-} \rightarrow \ket{E_{\ds \ds}}} &\simeq \frac{ \left(\sqrt{2} \delta J_x - \frac{\sqrt{2} h_2 J_C}{J_{23}}\right)^2 \gamma_1 }{(2h_1+2J_{23})^2} + \frac{ h_2^2 J^2 (n+1) \gamma_4}{2J_{23}^2 \eta_4^2}, \\
\Gamma_{\ket{E_{\us \us}} \rightarrow \ket{E_+}} &\simeq \frac{8 J_C^2 \gamma_1}{(2h_1 + 2J_{23})^2} + \frac{ 8J^2 (n+1) \gamma_4}{\eta_4^2}, \\
\Gamma_{\ket{E_{+}} \rightarrow \ket{E_{\ds \ds}}} &\simeq \frac{8 J_C^2 \gamma_1}{8 J_C^2 + \eta_1^2},   \quad 
\Gamma_{\ket{E_{\ds \ds}} \rightarrow \ket{E_{+}}} \simeq \Gamma_{\ket{E_+} \rightarrow \ket{E_{\us \us}}} \simeq \frac{ 8J^2 n \gamma_4}{\eta_4^2},
\end{align}
where $\eta_1^2 = (2h_1 - 2J_{23} - h_2)^2 + \gamma_1^2/4$ and $\eta_4^2 = 4J_{23}^2 + (2n+1)^2\gamma_4^2 /4$. 

Since the quantum Wheatstone bridge would ideally function the same for all values of $J_x$, the final result should depend only on $\delta J_x$ and not $J_C$ or $J_x$. This is achieved by picking $\gamma_4$ such that the second term in $\Gamma_{\ket{E_-} \rightarrow \ket{E_{\downarrow \downarrow}}}$ and $\Gamma_{\ket{E_{\uparrow \uparrow}} \rightarrow \ket{E_{+}}}$ is larger than the first term. 
For this, we pick $\gamma_1 \ll \gamma_4 \leq 4 J_{23}/(2n+1)$, although $\gamma_4$ could be much larger than the last inequality as long as the assumption on $\Gamma_{\ket{E_-} \rightarrow \ket{E_{\downarrow \downarrow}}}$ and $\Gamma_{\ket{E_{\uparrow \uparrow}} \rightarrow \ket{E_{+}}}$ is fulfilled. 
Finally, we assume that $8J_C^2 \gg (2h_1 - 2J_{23} - h_2)^2 + \gamma_1^2/4$. The new rates are
\begin{align}
\Gamma_{\ket{E_{\us \us}} \rightarrow \ket{E_-}} &\simeq \frac{2 \gamma_1 \delta J_x^2}{ \eta_1^2 }, \quad \Gamma_{\ket{E_{\ds \ds}} \rightarrow \ket{E_{-}}} \simeq \Gamma_{\ket{E_-} \rightarrow \ket{E_{\us \us}}} \simeq  \frac{h_2^2 J^2 n \gamma_4 }{2 J_{23}^2 \eta_4^2} , \label{eq:WBRatesFirst} \\ 
\Gamma_{\ket{E_-} \rightarrow \ket{E_{\ds \ds}}} &\simeq \frac{ h_2^2 J^2 (n+1) \gamma_4}{2J_{23}^2 \eta_4^2}, \quad \Gamma_{\ket{E_{\us \us}} \rightarrow \ket{E_+}} \simeq \frac{ 8J^2 (n+1) \gamma_4}{\eta_4^2}, \\
\Gamma_{\ket{E_{+}} \rightarrow \ket{E_{\ds \ds}}} &\simeq \gamma_1 , \quad \Gamma_{\ket{E_{\ds \ds}} \rightarrow \ket{E_{+}}} \simeq \Gamma_{\ket{E_+} \rightarrow \ket{E_{\us \us}}} \simeq \frac{ 8J^2 n \gamma_4}{\eta_4^2}. \label{eq:WBRatesLast}
\end{align}

With these more manageable approximate rates, the steady state, $\mathbf{W} \vec{P}_{\text{ss}} = 0$, can be obtained
\begin{equation}
\vec{P}_{\text{ss}} \simeq \frac{1}{(K_1+K_2-1)\delta J_x^2 + \frac{\Lambda^2}{4}}\begin{pmatrix}
(K_1 - 1) \delta J_x^2 + \frac{2n+1}{3n+1} \frac{\Lambda^2}{4} \\
K_2 \delta J_x^2 + \frac{n}{3n+1} \frac{\Lambda^2}{4}  \\
\frac{(n+1) h_2^2   }{16 nJ_{23}^2 } \delta J_x^2 + \frac{8 n(2n+1) J^2  \gamma_4}{(3n+1) \gamma_1 \eta_4^2} \frac{\Lambda^2}{4} \\
K_3 \frac{n^2 (2n+1)}{(n+1)(3n+1)}\frac{8  J^2 \gamma_4}{ \gamma_1 \eta_4^2} \frac{\Lambda^2}{4}
\end{pmatrix}.
\end{equation}
We defined the three constants $K_1$, $K_2$ and $K_3$ to be
\begin{align}
K_1 &= 1 + \frac{n+1}{n^2} \frac{  h_2^2 \gamma_1 }{32 J^2 \gamma_4} \frac{\eta_2^2 }{ 4 J_{23}^2 } \simeq 1 ,\\
K_2 &=  1 + \frac{1}{n} \frac{h_2^2 \gamma_1}{32 J^2 \gamma_4}\frac{\eta_2^2}{4J_{23}^2} \simeq 1 ,\\
K_3 &= 1 + \frac{1}{2n+1} \frac{h_2^2 \gamma_1 }{32J^2 \gamma_4} \frac{\eta_2^2 }{4J_{23}^2} \simeq 1,
\end{align}
which are approximately unity within the approximations. The new steady state is
\begin{align}
\vec{P}_{\text{ss}} \simeq \frac{1}{\delta J_x^2 + \frac{\Lambda^2}{4}}\begin{pmatrix}
 \frac{2n+1}{3n+1} \frac{\Lambda^2}{4} \\
\delta J_x^2 + \frac{n}{3n+1} \frac{\Lambda^2}{4}  \\
\frac{(n+1) h_2^2   }{16 nJ_{23}^2 } \delta J_x^2 + \frac{8 n(2n+1) J^2  \gamma_4}{(3n+1) \gamma_1 \eta_4^2} \frac{\Lambda^2}{4} \\
 \frac{n^2 (2n+1)}{(n+1)(3n+1)}\frac{8  J^2 \gamma_4}{ \gamma_1 \eta_4^2} \frac{\Lambda^2}{4} \label{eq:WBSteadyStatePopulations}
\end{pmatrix}.
\end{align}
All the populations are now of Lorentzian form, with full width at half maximum
\begin{equation}
\Lambda = \sqrt{\frac{(n+1)(3n+1)}{2n^2}} \frac{ h_2  \sqrt{\left[2h_1 - 2J_{23} - h_2\right]^2 + \gamma_1^2/4 }  }{2 J_{23} }, \label{eq:WBWidth}
\end{equation}
Since the solution was found to lowest order in $h/J_{23}, J/J_{23}$, and $|\delta J_x|$, other parameters have to be dominant i.e. $n \gg J/J_{23}$. To summarize, the approximations are $J\sim J_C \sim J_x$; $h_2,J \gg J_{23}$; $h_1 \simeq J_{23}$; $\delta J_x^2 \ll \eta_1^2 \ll 8J_C^2$; $\gamma_1 \ll \gamma_4 \leq 4 J_{23}/(2n+1)$; $n\gg J/J_{23}$; and $|\delta J_x| \ll J$. 
Even though we have the full population vector, it is still worth writing out the population for $\ket{E_-}$
\begin{equation}
P_{\text{ss}}(\ket{E_-}) \simeq \frac{\left(J_x - J_{x,0} \right)^2 + P_{\text{ss},-}^0 \frac{\Lambda^2}{4}}{\left(J_x - J_{x,0} \right)^2 + \frac{\Lambda^2}{4}}. \label{eq:WBPopulation1}
\end{equation}
$P_{\text{ss},-}^0 = \frac{n}{3n+1}$ is the population at $J_x = J_{x,0}$, and $\Lambda$ is again the width of the Lorenzian.
The dashed red lines in Fig.~\ref{figure:ChapterWBPopulations2}(a) correspond to the two lines $J_x = J_{x,0} \pm \Lambda /2$.

To further explore the validity of the approximate expression for $P_{\text{ss}}(\ket{E_-})$, we plot both the exact numerical solution to the master equation and the approximate solution in Fig~\ref{figure:ChapterWBPopulations2}(b). The two solutions have small deviations which become greater for larger $h_2$ as expected from the assumptions. Overall, there is good agreement between the two solutions further justifying the assumptions.
Furthermore, the width, and thus the sensitivity to $J_x$, can be tuned through the ratio $h_2/J_{23}$ in agreement with Eq.~\eqref{eq:WBWidth}. This is particularly useful for calibration of $h_2$, since $h_2\simeq 0$ can be found by minimizing the width of the population with respect to $h_2$.

\begin{figure}[t]
\centering
\includegraphics[width=1.\linewidth, angle=0]{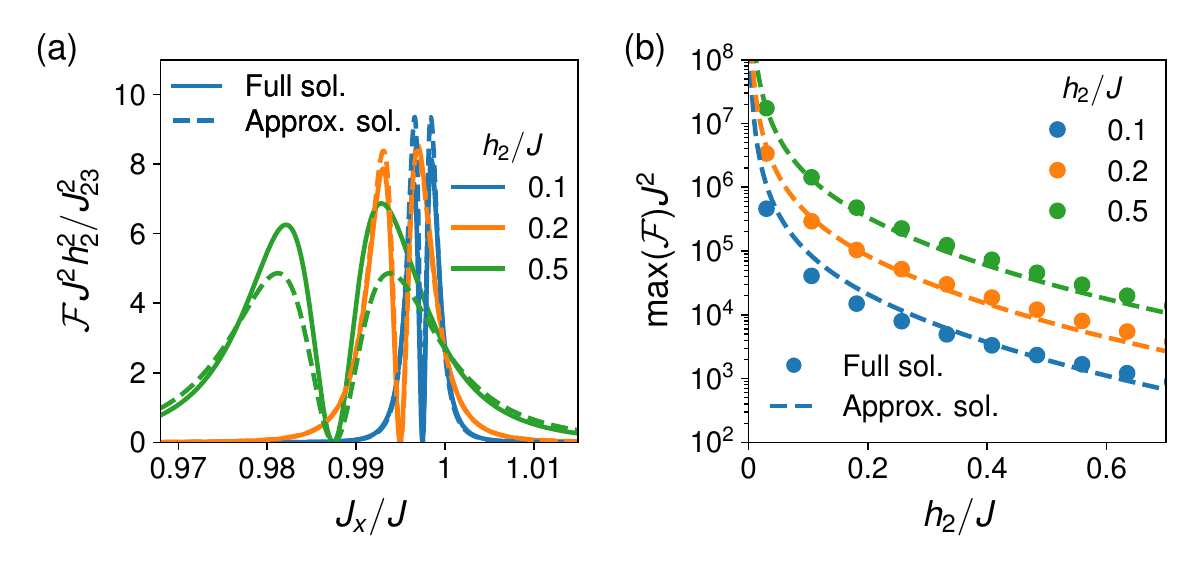}
\caption{ (a) Quantum Fisher information, $\mathcal{F}$, as a function of the unknown parameter $J_x$ for different values of the magnetic field $h_2$. The dashed lines denote the approximate solution in Eq.~\eqref{eq:WBQFIEntangledState}. (b) Quantum Fisher information maximized over $J_x$ for both the full numerical solution and the approximate solution given in Eq.~\eqref{eq:WBMaxFI}. See Eq.~\eqref{eq:WBDefaultValues} for the values of the remaining system parameters. Reproduced with permission from Phys. Rev. Lett. \textbf{128}, 240401 (2022) \cite{PhysRevLett.128.240401}. Copyright 2022 American Physical Society.}\label{figure:ChapterWBFisherInformation}
\end{figure}

\begin{figure}[t]
\centering
\includegraphics[width=0.625\linewidth, angle=0]{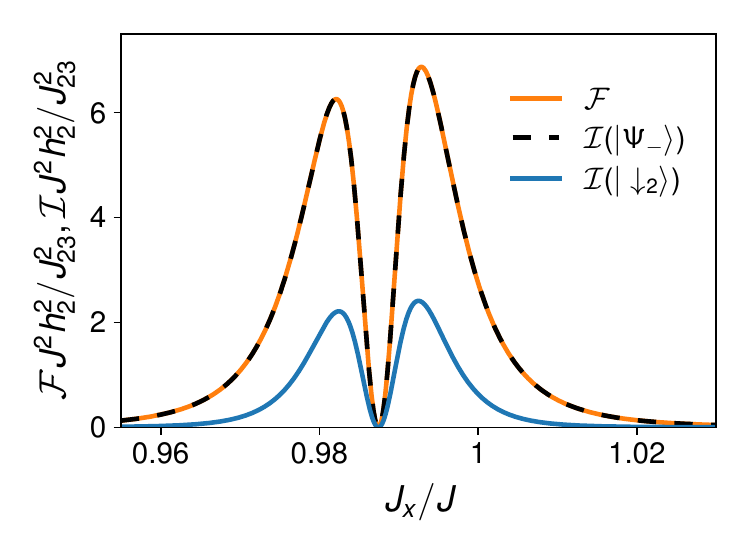}
\caption{Quantum Fisher information and classical Fisher information as a function of $J_x$. The classical Fisher information is calculated for a measurement of either $\ket{E_-}$ or $\ket{\ds_2}$. See Eq.~\eqref{eq:WBDefaultValues} for the values of the remaining system parameters. Reproduced with permission from Phys. Rev. Lett. \textbf{128}, 240401 (2022) \cite{PhysRevLett.128.240401}. Copyright 2022 American Physical Society.}\label{figure:ChapterWBClassicalFisherInformation}
\end{figure}

\section{Quantum Fisher information.} 
\label{sec:WBQuantumFisherInformation}

A measure of the sensitivity of the density matrix to small variations in the unknown parameter $J_x$ is the quantum Fisher information \cite{PhysRevA.97.042322, CHAPEAUBLONDEAU20171369}, which for a diagonal density matrix, $\hat{\rho}_{\text{ss}} = \sum_k p_{k} \op{k}$, is
\begin{equation}
\mathcal{F} = 2\sum_{p_k+p_l > 0} \frac{\mel{k}{\partial_{J_x} \hat{\rho}_{\text{ss}}}{l} \mel{l}{\partial_{J_x} \hat{\rho}_{\text{ss}}}{k}}{p_k + p_l}.
\end{equation}
The ultimate limit in precision is then given by the Cramér-Rao bound $\text{Var}(J_x) \geq 1/\mathcal{F}$ for a single shot measurement. The quantum Fisher information is plotted in Fig.~\ref{figure:ChapterWBFisherInformation}(a) for different values of $h_2$. Comparing Fig.~\ref{figure:ChapterWBPopulations2}(b) and Fig.~\ref{figure:ChapterWBFisherInformation}(a), we see that the largest quantum Fisher information overlaps with the largest change in the populations. 
Furthermore, we see that the maximum quantum Fisher information is $\text{max}(\mathcal{F}) \propto J_{23}^2/h_2^2$. 

To explain this, we note that the populations $P_{\text{ss}}(\ket{E_+})$ and $P_{\text{ss}}(\ket{E_{\us \us}})$ are of order $J^2/J_{23}^2$. 
Therefore, the quantum Fisher information can be found using the population in Eq.~\eqref{eq:WBPopulation1} and $P_{\text{ss}}(\ket{E_{\ds \ds}}) \simeq 1-P_{\text{ss}}(\ket{E_{-}})$,
\begin{align}
\mathcal{F} &\simeq \frac{1}{P_{\text{ss}}(\ket{E_{\ds \ds}})} \left( \frac{\partial}{\partial J_x} P_{\text{ss}}(\ket{E_{\ds \ds}}) \right)^2 + \frac{1}{P_{\text{ss}}(\ket{E_{-}})} \left( \frac{\partial}{\partial J_x} P_{\text{ss}}(\ket{E_{-}}) \right)^2 \label{eq:WBQFIEntangledState} \\
&\simeq \frac{4 \left(J_x - J_{x,0}\right)^2 \Lambda^2 \left(2 n + 1 \right)}{\left(\left(J_x - J_{x,0}\right)^2 + \frac{\Lambda^2}{4}\right)^2 \left(n\Lambda^2+4(3n + 1)\left(J_x - J_{x,0}\right)^2 \right)}. \nonumber
\end{align}
The maximum quantum Fisher information is found by maximizing the above expression over $J_x$
\begin{equation}
\text{max}(\mathcal{F}) \simeq \frac{4 N(n)}{\Lambda^2}, \label{eq:WBMaxFI}
\end{equation}
where $N(n)$ is a function of $n$ only
\begin{align}
N(n) = \frac{64 \left(2\,n+1\right)\left(3\,n+1\right) \left(\sqrt{n(25n+8)} - n\right)}{\left(3\,n+\sqrt{n(25n+8)}\right)\,{\left(11\,n+\sqrt{n(25n+8)}+4\right)}^2}. \label{eq:WBN}
\end{align}
This function is bounded, $\frac{3}{4} < N(n) \leq 4$, and for the default value we have $N(n=0.5) \simeq 1.14$ \cite{PhysRevLett.128.240401}. In Fig.~\ref{figure:ChapterWBFisherInformation}(b), the maximum quantum Fisher information is plotted for both the full numerical solution and the approximate expression above. The full numerical solution is found by optimizing the quantum Fisher information with respect to $J_x$ around $J_x \sim J_C$. We observe that the agreement is generally good, and that it is better in the expected limit $h_2, J \ll J_{23}$.

\subsection{Optimal measurement basis}
\label{subsec:WBOptimalMeasurement}

The quantum Fisher information does not tell us which measurement is optimal for achieving the values found above. To find the optimal measurement basis, we need to use the classical fisher information \cite{doi:10.1142/S0219749909004839, PhysRevLett.106.153603}
\begin{align}
\mathcal{I} = \sum_a P(a|J_x)\left( \frac{\partial}{\partial J_x} \ln \big(P(a|J_x) \big) \right)^2,
\end{align}
where $P(a|J_x)$ is the probability of the outcome $a$ of the measurement given $J_x$. Therefore, the classical Fisher information depends on the measurement performed. It is related to the quantum fisher information through the inequality $\mathcal{F} \geq \mathcal{I}$. The optimal measurement is the one for which this is an equality. 

For a measurement that yields $a=1$ for the state $\ket{\psi}$ and $a=0$ otherwise, the conditional probabilities become
\begin{align}
P(1|J_x) = P_{\text{ss}}(\ket{\psi}) \quad \text{and} \quad P(0|J_x) = 1-P_{\text{ss}}(\ket{\psi}),
\end{align}
where $P_{\text{ss}}(\ket{\psi}) = \tr_{23} \{\hat{\rho}_{\text{ss}}^{23} \op{\psi} \}$. The classical Fisher information for this measurement becomes
\begin{align}
\mathcal{I}(\ket{\psi}) = \frac{1}{P_\text{ss}(\ket{\psi})} \left( \frac{\partial }{\partial J_x} P_\text{ss}(\ket{\psi}) \right)^2 + \frac{1}{1-P_\text{ss}(\ket{\psi})} \left( \frac{\partial }{\partial J_x} \left[1-P_\text{ss}(\ket{\psi})\right] \right)^2.
\end{align}
Under the assumption $P_{\text{ss}}(\ket{\ds \ds}) + P_{\text{ss}}(\ket{E_-}) \simeq 1$, a diagonal density matrix, and comparing with Eq.~\eqref{eq:WBQFIEntangledState}, we see that $\mathcal{F} \simeq \mathcal{I}(\ket{E_-}) \simeq \mathcal{I}(\ket{\ds \ds})$. Therefore, the measurement of either $\ket{\ds \ds}$ or $\ket{E_-}$ constitutes an optimal measurement. Both the quantum Fisher information and the classical Fisher information for the state $\ket{E_-}$ are plotted in Fig.~\ref{figure:ChapterWBClassicalFisherInformation}. 

$\ket{E_-}$ is an entangled state, and therefore, it can be challenging to measure. Alternatively, the states $\ket{E_{\ds \ds}}$ and $\ket{E_-}$ can be distinguished by measuring $\hat{\sigma}_z$ for either spin 2 or spin 3. If the measurement yields $-1$ every time, the state is $\ket{E_{\ds \ds}}$, and otherwise, it is $\ket{E_-}$. This corresponds to measuring for spin up on spin 2 or $\ket{\us_2}$. In Fig.~\ref{figure:ChapterWBClassicalFisherInformation}, we see that the classical Fisher information for this measurement is smaller, $\mathcal{I}(\ket{\ds_2}) < \mathcal{I}(\ket{E_-})$. However, measuring locally on spin 2 is still effective and experimentally easier.

\section{Measuring the interface state.} 
\label{sec:WBCurrent}

\begin{figure}[t]
\centering
\includegraphics[width=1\linewidth, angle=0]{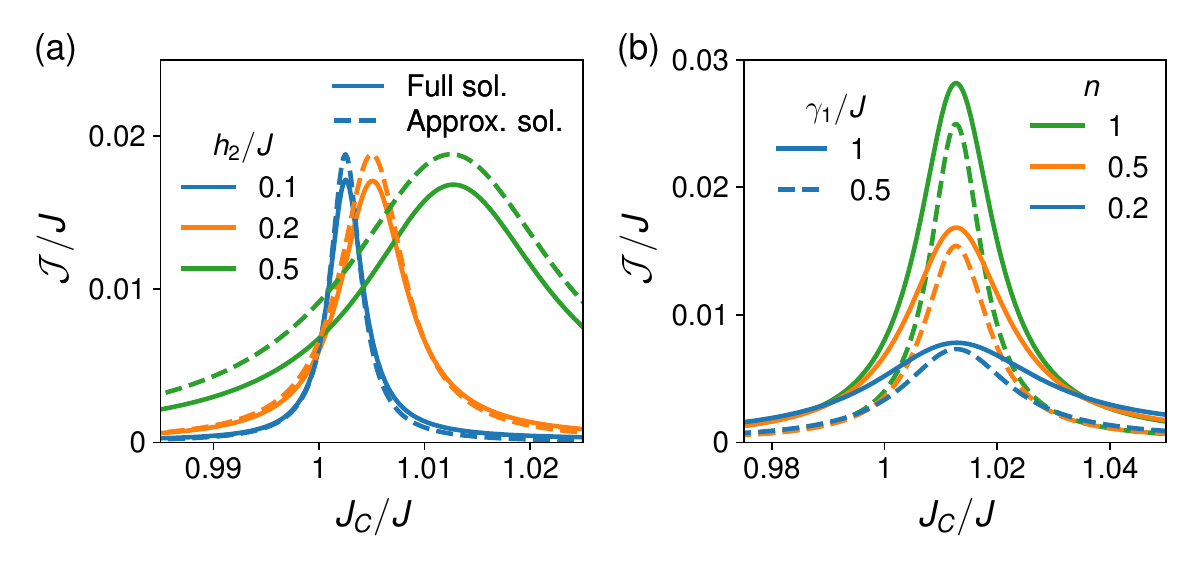}
\caption{(a) Spin current as a function of $J_C$ for different values of $h_2$. The solid line denotes the full theory while the dashed line shows the approximate solutions \eqref{eq:WBApproximateCurrent}. (b) $\mathcal{J}$ as a function of $J_C$ for different values of the cold bath coupling strength $\gamma_1$ and hot bath parameter $n$. See Eq.~\eqref{eq:WBDefaultValues} for the values of the remaining system parameters. Reproduced with permission from Phys. Rev. Lett. \textbf{128}, 240401 (2022) \cite{PhysRevLett.128.240401}. Copyright 2022 American Physical Society.}\label{figure:ChapterWBCurrents1}
\end{figure}

\begin{figure}[t]
\centering
\includegraphics[width=1\linewidth, angle=0]{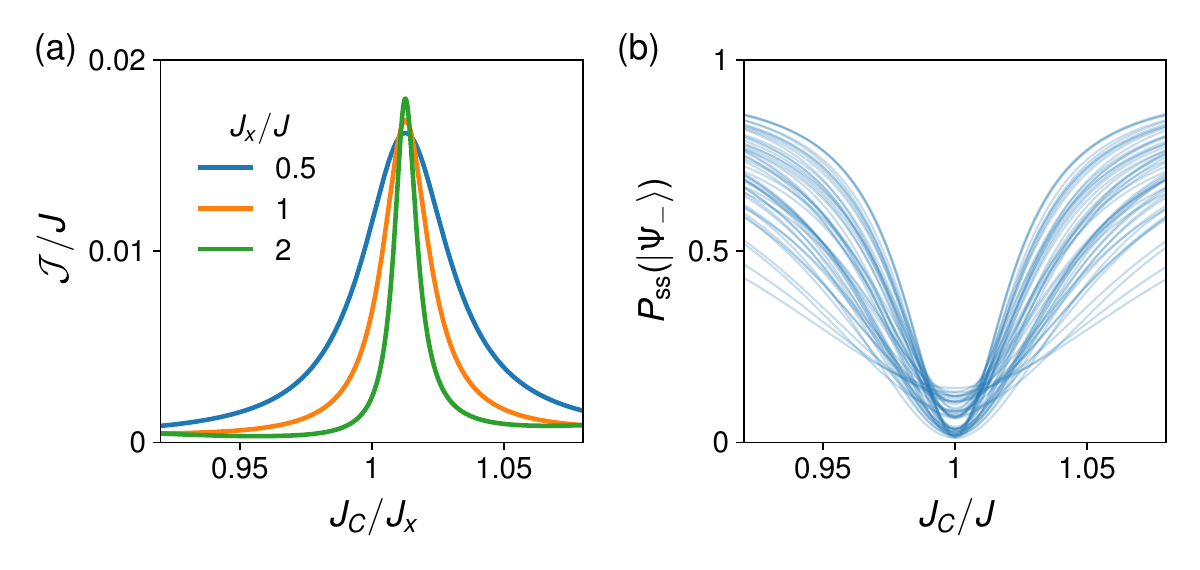}
\caption{ (a) Spin current, $\mathcal{J}$, as a function of $J_C$ for different values of the unknown parameter $J_x$. (b) $P_\text{ss}(\ket{\Psi_-})$ as a function of $J_C$ for 50 sets of random errors on all parameters except $h_2$ and $J_x$ (see text) where $h_2 = 0$. See Eq.~\eqref{eq:WBDefaultValues} for the values of the remaining system parameters. Reproduced with permission from Phys. Rev. Lett. \textbf{128}, 240401 (2022) \cite{PhysRevLett.128.240401}. Copyright 2022 American Physical Society.}\label{figure:ChapterWBCurrents2}
\end{figure}

The general operation of the quantum Wheatstone bridge is to slowly vary $J_C$ until the population of $\ket{E_-}$ drops.
The unknown parameter $J_x$ is then either determined by finding the minimum population at the balance point and subsequently using the relation \eqref{eq:WBResonance}, or it is determined to greater precision by using the slope of the population thus exploiting the full potential found through the quantum Fisher information. In subsection \ref{subsec:WBOptimalMeasurement}, we showed that a measurement of the state $\ket{E_-}$ saturates the quantum Fisher information. 
While this is a difficult measurement, it is possible to perform such a measurement through a shadow spin \cite{PhysRevLett.128.240401}. Alternatively, the states $\ket{E_{\ds \ds}}$ and $\ket{E_-}$ can be distinguished by measuring $\hat{\sigma}_z$ for either spin 2 or spin 3 as discussed in subsection \ref{subsec:WBOptimalMeasurement}.

Another alternative is to probe the interface state through the spin current. The spin current is defined as the number of spin excitations decaying to the cold bath per unit of time
$\mathcal{J} = \gamma_1 \ev*{\hat{\sigma}^+_{1} \hat{\sigma}^-_{1} }_\mathrm{ss}$. This is equal to the population of the excited state of spin 1, $\mathrm{tr} \{ \hat{\sigma}_+^{(1)} \hat{\sigma}_-^{(1)} \hat{\rho}_{\text{ss}} \}$, times the rate of decay from the excited state, $\gamma_1$. This definition comes from looking at the rate of change of $\hat{\sigma}_n^z/2$ resulting in equations such as
\begin{align}
\frac{d\ev{\hat{\sigma}_1^z/2}_\mathrm{ss}}{dt} &= J_x \ev*{\hat{s}_{21}}_\mathrm{ss} + J_C \ev*{\hat{s}_{31}}_\mathrm{ss} - \gamma_1 \ev*{\hat{\sigma}_1^+ \hat{\sigma}_1^-}_\mathrm{ss} = 0,
\end{align}
where $\hat{s}_{ij} = \hat{\sigma}^{x}_{i}\hat{\sigma}^{y}_{j} - \hat{\sigma}^{y}_{i} \hat{\sigma}^{x}_{j}$. Similar to other chapters, this leads to the above definition of the current. Note the factor of 2 difference between the spin current here and in chapter \ref{chapter:Interference}. This a matter of preference. Since we want to emphasize the current as measurable, we chose the definition which results in a current with units of excitations, quanta of energy $\omega$, per unit of time. 

The heat, $\mathcal{K}$, exchanged between the system and the cold bath is equal to the spin current times the energy of a single spin excitation $\mathcal{K} \simeq \omega \mathcal{J}$ for $\omega \gg J_{23},J$. Therefore, $\mathcal{K}$ offers an alternative way of measuring $\mathcal{J}$. The entangled state results in a small current due to interference while the state $\ket{E_{\ds \ds}}$ allows for a current through $\ket{E_+}$. 

In Fig. \ref{figure:ChapterWBCurrents1}(a), the current is plotted as a function of $J_C$ for different values of $h_2$. First, we notice that the current peaks at the same value of the ratio $J_C$. 
Second, we notice that the current peaks at the expected $J_C = J_{C,0}$ where
\begin{equation}
J_{C,0} = J_x \left( 1+ \frac{h_2}{2J_{23}} \right). 
\end{equation}
As mentioned, this shows that the spin current can be used to probe $J_x$ without performing measurements on spins 2 and 3 directly. Therefore, the balance point for the quantum Wheatstone bridge is $J_C = J_{C,0}$.
In order to describe the width of the peak in current, we can again use the steady state, $\mathbf{W} \vec{P}_{\textbf{ss}}=0$, obtained in Eq.~\eqref{eq:WBSteadyStatePopulations}. 
Since spin 1 has been traced away, we instead look at the four transitions $|E_{\us \us} \rangle \rightarrow |E_{+} \rangle$, $|E_{\us \us} \rangle \rightarrow |E_{-} \rangle$, $|E_{+} \rangle \rightarrow |E_{\ds \ds} \rangle$, and $|E_{-} \rangle \rightarrow |E_{\ds \ds} \rangle$ driven by the cold bath. When any of the four transitions occur, a single spin excitation is absorbed by the cold bath. Therefore, the spin current is the sum of the rate for each transition multiplied by the probability of the interface being in the initial state 
\begin{align}
\mathcal{J} &= \Gamma_{|E_{\us \us} \rangle \rightarrow |E_{+} \rangle} P_{\text{ss}}(\ket{E_{\us \us}}) + \Gamma_{|E_{\us \us} \rangle \rightarrow |E_{-} \rangle} P_{\text{ss}}(\ket{E_{\us \us}}) \\ 
& \hspace{3cm} + \Gamma_{|E_{+} \rangle \rightarrow |E_{\ds \ds} \rangle} P_{\text{ss}}(\ket{E_{+}}) + \Gamma_{|E_{-} \rangle \rightarrow |E_{\ds \ds} \rangle} P_{\text{ss}}(\ket{E_{-}}). \nonumber
\end{align}
Both the populations $P_{\text{ss}}(\ket{E_+})$ and $P_{\text{ss}}(\ket{E_{\us \us}})$ are of order $J^2/J_{23}^2$. From the rates \eqref{eq:WBRatesFirst}-\eqref{eq:WBRatesLast}, it is evident that the spin current going into the cold bath is dominated by the decay of the state $\ket{E_+}$ through the channel $\Gamma_{\ket{E_+}\rightarrow \ket{E_{\ds \ds}}}$. The spin current can be written
\begin{align}
\mathcal{J} \simeq \gamma_1 P_{\text{ss}}(\ket{E_+}) \simeq \frac{\mathcal{J}_{\infty} (J_C-J_{C,0})^2 + \mathcal{J}_{0} \frac{\Lambda^2}{4}}{(J_C-J_{C,0})^2 + \frac{\Lambda^2}{4}}. \label{eq:WBApproximateCurrent}
\end{align}
The solution is found as a function of $J_C$ for $|J_{C,0} - J_C| \ll J$, reflecting the fact that the operation of the quantum Wheatstone bridge involves varying $J_C$ in order to determine $J_x$. This can be done since to linear order
\begin{align}
J_x-J_{x,0} = -(J_C - J_{C,0}).
\end{align}
The two constants in the expression, $\mathcal{J}_0$ and $\mathcal{J}_\infty$, are the currents for $J_C = J_{C,0}$, and for $|J_C| \rightarrow \infty$, respectively,
\begin{align}
\mathcal{J}_0 &= \frac{n (2n+1 )}{ 3n+1} \frac{8 \gamma_4 J^2}{ \eta_4^2 }, \\
\mathcal{J}_{\infty} &= \frac{(n+1) \gamma_1 h_2^2}{16 n J_{23}^2}.
\end{align}
Note that this expression is derived under the assumption of $|J_C - J_{C,0}| \ll J$ and hence not valid for $| J_C| \rightarrow \infty$, and therefore, $\mathcal{J}_\infty$ should not be taken literally. The full width at half maximum is the same as before, Eq.~\eqref{eq:WBWidth}.
This value of the current is plotted along side the exact value in Fig.~\ref{figure:ChapterWBCurrents1}(a). Furthermore, the exact current is plotted as a function of $J_C$ for different values of both $\gamma_1$ and $n$ in Fig. \ref{figure:ChapterWBCurrents1}(b). We observe that larger $\gamma_1$ results in a larger width, $\Lambda$, and a larger $\mathcal{J}_\infty$. Additionally, a larger $n$ results in a larger $\mathcal{J}_0$ and a slightly larger $\Lambda$. This is generally the behavior expected from the expressions of $\mathcal{J}_0$, $\mathcal{J}_\infty$, and $\Lambda$ from above. Contrary to what we would expect, a larger $\gamma_1$ changes $\mathcal{J}_0$. To capture this behavior, higher orders have to be included in Eq.~\eqref{eq:WBRatesFirst}-\eqref{eq:WBRatesLast}. 
To verify that the quantum Wheatstone bridge functions as intended for other values of the unknown parameter, the current is plotted as a function of $J_C$ for different values of $J_x$ in Fig.~\ref{figure:ChapterWBCurrents2}(a). The maximum does indeed occur at the same value of $J_C/J_x$.

Finally, we study the effect of calibration errors for all parameters. $h_2$ is already known to shift and widen the Lorenzian, so we will focus on the other parameters. In Fig.~\ref{figure:ChapterWBCurrents2}(b), $P_\text{ss}(\ket{\Psi_-})$ is plotted as a function of $J_C$ for random normal distributed errors on all parameters except $h_2$ and $J_x$. The standard deviations are $0.02J$ for couplings, $0.2J$ for magnetic fields, $0.1J$ for bath coupling rates, and $0.1$ for $n$. The errors on the coupling of spin 2 to spin 4 and the coupling of spin 3 to spin 4 where sampled separately. The error on the magnetic field for spins 2 and 3 is the same to keep $h_2=0$, and an error was also added to the magnetic field for spin 4. Decoherence was implemented through $\mathcal{L} \rightarrow \mathcal{L} + \mathcal{D}_2 + \mathcal{D}_3$, where 
\begin{align}
\mathcal{D}_i[\hat{\rho}] = T^{-1} \mathcal{M}[\hat{\sigma}^-_{i}, \hat{\rho}] + T^{-1} \mathcal{M}[\hat{\sigma}^z_{i}/2, \hat{\rho}],
\end{align}
for $i=2,3$, and $TJ = 4\cdot 10^4$. Remarkably, the minimum is at $J_C=J_x$ for all 10 samples, and the calibration errors do not result in systematic errors. Furthermore, $h_2 = 0$ should give an infinitely narrow drop, however, due to decoherence the drop will have finite width even for perfect parameters. Thus, $h_2 = 0$ can be found by minimizing the width with respect to $h_2$.

\cleardoublepage
\chapter{Giant magnetoresistance in boundary-driven spin chains}
\label{chapter:GMR}
\epigraph{\footnotesize This chapter is based on Ref.~\cite{PhysRevLett.126.077203}.
Sections \ref{sec:GMRHamiltonian}-\ref{sec:GMRZCoupling} and all figures herein have been reproduced with permission from Ref.~\cite{PhysRevLett.126.077203}.}{}

\noindent In chapters \ref{chapter:QutritDiode}-\ref{chapter:WB}, we used dissipation engineering to achieve certain mechanisms such as rectification. We used the unidirectionality of cold baths in combination with quantum transitions. The quantum transitions traversed by the systems were controlled by either energy conservation or the matrix elements. However, these techniques are more general and can be used in a wide range of situations beyond steady-state engineering. In fact, they are even useful in closed quantum systems, e.g., for transistors \cite{marchukov2016}. As a fundamental example, we can look at the familiar two spin 1/2 particles, labeled L and R for left and right, respectively
\begin{align}
\hat{H} = \ J \left( \hat{\sigma}_L^x \hat{\sigma}_R^x + \hat{\sigma}_L^y \hat{\sigma}_R^y \right) + h \hat{\sigma}_R^z.
\end{align}
Starting with a spin excitation on the left spin, $\ket{\us \ds}$, the population for the spin excitation to travel to the right spin is
\begin{align}
P(\ket{\ds \us}, t) = \frac{4J^2 \sin^2(\sqrt{4J^2 + h^2} t)}{4J^2 + h^2}.
\end{align}
This is a very basic example of a transition between two states at different energies. This solution is the same as the more famous situation of off-resonant Rabi oscillations \cite{bransden2003physics}. However, it illustrates the effect of an energy barrier in transport, and it is an important example to keep in mind as we look at larger systems. It might be illustrative to evaluate the population in the two regimes
\begin{align}
|h| \ll |J|,& \quad P(\ket{\ds \us}, t) = \sin^2(2J t), \\
|h| \gg |J|,& \quad P(\ket{\ds \us}, t) = \frac{4J^2}{h^2} \sin^2(h t).
\end{align}
Therefore, transport is proportional to $1/h^2$ for a large energy barrier or for a small transition matrix element $2J$. Importantly, no transport occurs for $J=0$. 

This chapter will study the effects of magnetic fields on spin transport in a general linear spin chain coupled to a spin reservoir at either end. The chain consists of weakly connected chains of strongly connected spins. First, one chain of strongly connected spins is studied, and the transport is understood using resonance mechanisms. Next, several chains are connected, and the same mechanism is found to explain the transport even for larger systems. Finally, anisotropy is included, making the current as a function of different magnetic field strengths much more complicated. This is explained by a combination of the resonance mechanism and transition matrix elements, as discussed above.

\begin{figure}[t]
\centering
\includegraphics[width=0.92\linewidth, angle=0]{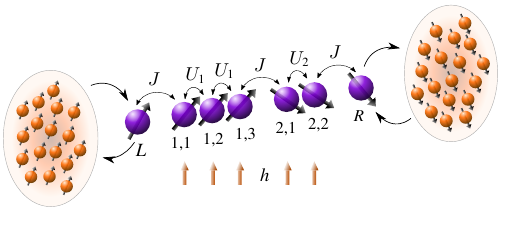}
\caption{ Illustration of the model with an example consisting of $N=2$ chains, the first containing $n_1 = 3$ spins and the second $n_2 = 2$ spins. The setup is coupled to spin reservoirs at each end, one with an abundance of spin excitations (left) and one with an abundance of spin excitation holes (right). The exchange coupling between the spins in the first chain is $U_1$, while the exchange between the spins in the second chain is $U_2$. The exchange between the two chains and outer spins is $J$. The numbering is shown below the spins, and the magnetic field is shown with red arrows. Reproduced with permission from Phys. Rev. Lett. \textbf{126}, 077203 (2021) \cite{PhysRevLett.126.077203}. Copyright 2021 American Physical Society.}\label{figure:ChapterGMRSetup}
\end{figure}

\section{Hamiltonian and master equation} 
\label{sec:GMRHamiltonian}
The general model studied is a set of $N$ linear spin-1/2 chains, where the $i$th chain is composed of $n_i$ spins coupled strongly to each other through the Hamiltonian
\begin{align}
\hat{H}_0 &= \sum_{i = 1}^{N} \left\{ \sum_{j = 1}^{n_i-1} U_i \left( \hat{\sigma}^x_{i,j} \hat{\sigma}^x_{i,j+1} + \hat{\sigma}_{i,j}^{y} \hat{\sigma}_{i, j+1}^{y} + \Delta_{U_i} \hat{\sigma}_{i,j}^{z} \hat{\sigma}_{i, j+1}^{z} \right) +\, h \sum_{j = 1}^{n_i} \hat{\sigma}^z_{i,j} \right\}.
\end{align}
The Pauli matrices for the $j$th spin within the $i$th chain is $\hat{\sigma}^\alpha_{i,j}$ for $\alpha = x,y,z$.
The exchange coupling between spins in the $i$th chain is $U_i$, the anisotropy is $\Delta_{U_i}$, and $h$ sets the spin excitation energy for the spins. 
We make these strongly coupled chain segments a part of a larger chain by adding two extra spins labeled $L$ and $R$. 
The extra two spins are described by the Pauli matrices $\hat{\sigma}_{L}^\alpha$ and $\hat{\sigma}_{R}^\alpha$ for $\alpha = x,y,z$. Finally, we couple these two spins and the strongly interacting chains weakly to each other through the Hamiltonian
\begin{align}
\hat{H}_{LR} &= J \left( \hat{\sigma}^x_{L} \hat{\sigma}^x_{1,1} + \hat{\sigma}_{L}^{y} \hat{\sigma}_{1,1}^{y} + \Delta_J \hat{\sigma}_{L}^{z} \hat{\sigma}_{1,1}^{z} \right) \\ & \hspace{3cm} + J \left( \hat{\sigma}^x_{N, n_N} \hat{\sigma}^x_{R} + \hat{\sigma}_{N, n_N}^{y} \hat{\sigma}_{R}^{y} + \Delta_J \hat{\sigma}_{N, n_N}^{z} \hat{\sigma}_{R}^{z} \right), \nonumber\\
\hat{H} =& \, \hat{H}_0 + \hat{H}_{LR} + J \sum_{i=1}^{N-1} \left( \hat{\sigma}^x_{i,n_i} \hat{\sigma}^x_{i+1,1} + \hat{\sigma}_{i,n_i}^{y} \hat{\sigma}_{i+1, 1}^{y} + \Delta_J \hat{\sigma}_{i,n_i}^{z} \hat{\sigma}_{i+1, 1}^{z} \right),
\end{align}
where the exchange coupling between chains $J$ must be smaller than the interchain exchanges $J \ll U_i$ and $|\Delta_J|, |\Delta_{U_i}| < 1$. An example of such a setup can be seen in Fig. \ref{figure:ChapterGMRSetup}.

To study spin transport in the system, we couple it to spin reservoirs through spin $L$ on the left and spin $R$ on the right; see Fig. \ref{figure:ChapterGMRSetup}. The evolution of the system is described by the local master equation on Lindblad form
\begin{align}
\frac{\partial \hat{\rho}}{\partial t} = \mathcal{L}[\hat{\rho}] = -i [\hat{H}, \hat{\rho}] + \mathcal{D}_L [\hat{\rho}] + \mathcal{D}_R [\hat{\rho}].
\label{Lindblad}
\end{align}
Unlike in previous chapters the baths are not thermal baths but rather spin reservoirs. As a consequence, the nature of the bath interaction is slightly different
\begin{align}
\mathcal{D}_{L,R} [\hat{\rho}] &= \gamma \left( \frac{1 + f}{2}  \mathcal{M}[\hat{\sigma}^+_{L,R} , \hat{\rho}] + \frac{1 - f}{2} \mathcal{M} [\hat{\sigma}^-_{L,R}, \hat{\rho}] \right),
\end{align}
where $\mathcal{M}[\hat{A}, \hat{\rho}] = \hat{A} \hat{\rho} \hat{A}^\dag - \frac{1}{2} \{ \hat{A}^\dag \hat{A}, \hat{\rho} \}$. $\gamma$ is the strength of the interaction with the baths, and $f$ determines the nature of the interaction. 
The characteristics of these reservoirs are determined by the parameter $f$. 
One reservoir has an abundance of spin excitations and forces the adjacent spin into a statistical mixture of predominantly up $\ev*{\hat{\sigma}^z_L} = f$, while the other has an abundance of excitation holes and forces the adjacent spin into a statistical mixture of predominantly down $\ev*{\hat{\sigma}^z_R} = -f$.
If $f>0$, on average, spin excitations are created on the left, transported through the chain, and decays on the right, resulting in a current flowing from left to right. However, if $f<0$, the current will tend to flow from right to left. 

Unless otherwise stated throughout this chapter, we will use the set of parameters 
\begin{align}
U_i = 10J, \quad h=0, \quad \Delta_J = \Delta_{U_i}, \quad \Delta_{U_i} = 0, \quad \gamma = J, \quad f = 0.5. \label{eq:GMRDefaultValues}
\end{align}
Once again, we are interested in the steady state (ss), $d \hat{\rho}_{\text{ss}} /d t = 0$, and its properties.
In particular, we will focus on the spin current defined as 
\begin{align}
\mathcal{J} = \tr (\hat{s}_{L}\hat{\rho}_{\text{ss}}) = \tr (\hat{s}_{R}\hat{\rho}_{\text{ss}}),
\end{align} 
where 
\begin{align}
\hat{s}_{L} &= 2J \left( \hat{\sigma}^{x}_{L}\hat{\sigma}^{y}_{1,1} - \hat{\sigma}^{y}_{L}\hat{\sigma}^{x}_{1,1}\right), \\
\hat{s}_{R} &= 2J \left( \hat{\sigma}^{x}_{N, n_N}\hat{\sigma}^{y}_{R} - \hat{\sigma}^{y}_{N, n_N}\hat{\sigma}^{x}_{R}\right).
\end{align}
This is the same spin current as defined and used in chapter \ref{chapter:Interference}, and therefore, a factor 2 larger than the one used in chapter \ref{chapter:WB}. Once again, the same excitation energy can be added to all spins, $\hat{H} \rightarrow \hat{H} + \frac{\omega}{2} \sum_{\alpha} \hat{\sigma}^z_\alpha$, without changing the spin current or the theory presented here. 
In this case, the spins $L$ and $R$ act as filters, allowing only excitations of frequency $\omega$ to pass. 
This $\omega$ could be an intrinsic excitation energy or a homogeneous magnetic field over the entire system.

\begin{figure}[t]
\centering
\includegraphics[width=1\linewidth, angle=0]{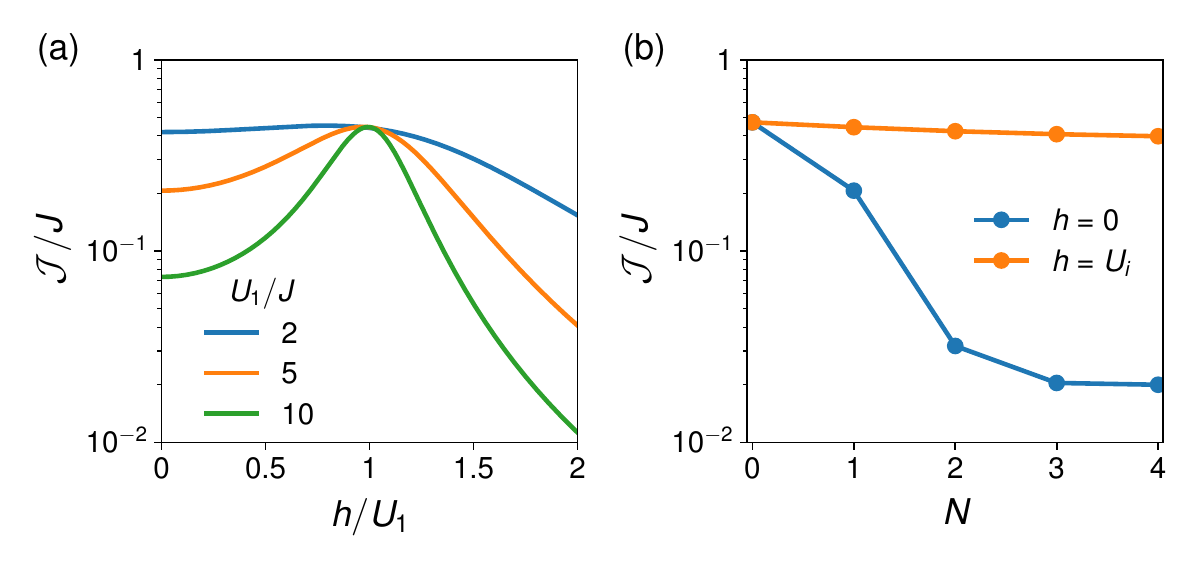}
\caption{ (a) $\mathcal{J}$ as a function of $h/U_1$ for a simple model of only $N=1$ chain consisting of $n_1 = 2$ strongly coupled spins. (b) $\mathcal{J}$ as a function of the number of chains $N$ each consisting of $n_i=2$ strongly coupled spins both on resonance $h=U$ and off resonance $h=0$. For this $U_i = U_1 = 5J$ was used. See Eq.~\eqref{eq:GMRDefaultValues} for the values of the remaining system parameters. Reproduced with permission from Phys. Rev. Lett. \textbf{126}, 077203 (2021) \cite{PhysRevLett.126.077203}. Copyright 2021 American Physical Society.}\label{figure:ChapterGMRn2}
\end{figure}

\section{A single chain of $n_1 = 2$ spins}
 
First, we study the simplest case with $N=1$ chain of $n_1 = 2$ spins coupled strongly to each other with coupling strength $U_1$ and with no anisotropy $\Delta_{U_1}= \Delta_J = 0$. 
This gives a total chain of four spins described by $\hat{\sigma}_L$, $\hat{\sigma}_{1,1}$, $\hat{\sigma}_{1,2}$, and $\hat{\sigma}_{R}$ similar to the example in Fig. \ref{figure:ChapterGMRSetup}. 
For this system, an analytical solution can be found for $f=0.5$. The steady-state current can be found to be 
\begin{align}
\mathcal{J}(h) = \frac{2^7 U_1^2( h^2 + 17 U_1^2) + 2^3 17 U_1^2 J^2}{2^8 \frac{U_1^2}{J^2} (h^2 - U_1^2)^2 + 2^5(33 h^2 + 129 U_1^2) U_1^2 + 513 U_1^2 J^2 + 2^4 J^4} J.
\end{align}
This current is plotted for different values of $U_1$ in Fig.~\ref{figure:ChapterGMRn2}(a). The largest current is obtained for $h = \pm U_1$, where the current is $\mathcal{J}(h=\pm U_1) = \frac{4}{9}J$ and, thus, independent of $U_1$. Furthermore, for no magnetic field $h=0$ the current is $\mathcal{J}(h=0) \simeq \frac{17 J^2}{2 U_1^2} J $ to lowest order in $J/U_1$ and, thus, heavily suppressed for large $U_1$. We, therefore, get giant magnetoresistance even for this minimal model. 

To explain this, we first diagonalize $\hat{H}_0$ to obtain the four states $\ket{\ds \ds}$, $\ket{\Psi_+}$, $\ket{\Psi_-}$, and $\ket{\us \us}$ for spins (1,1) and (1,2) with corresponding energies $E_{\ds\ds} = -2h$, $E_{\Psi_-} = -2U_1$, $E_{\Psi_+} = 2U_1$, and $E_{\us\us} = 2h$, where $\ket{\Psi_\pm} = (\ket{\us \ds} \pm \ket{\ds \us})/\sqrt{2}$. Next, we write the total Hamiltonian $\hat{H}$ in the single excitation basis $\ket{\us \ds \ds \ds}$, $\ket{\ds \!\Psi_+\! \ds}$, $\ket{\ds \!\Psi_-\! \ds}$, and $\ket{\ds \ds \ds \us}$:
\begin{align*}
H = 2 \begin{pmatrix}
-h & \frac{J}{\sqrt{2}} & \frac{J}{\sqrt{2}} & 0\\
\frac{J}{\sqrt{2}} & U_1 & 0 & \frac{J}{\sqrt{2}} \\
\frac{J}{\sqrt{2}} & 0 & -U_1 & -\frac{J}{\sqrt{2}} \\
0 & \frac{J}{\sqrt{2}} & -\frac{J}{\sqrt{2}} & -h
\end{pmatrix}.
\end{align*}
These four states are, therefore, eigenstates with the diagonal being the corresponding eigenenergies of the Hamiltonian to lowest order in $J/U_1$. For a spin excitation created at one end to propagate to the other end, it needs to pass the middle two spins. This is suppressed if the energies of an excitation at either end and an excitation at the middle chain are far from resonance with each other. 
This also corresponds to an excitation being localized, whereas on resonance, for $h = \pm U_1$, an excitation becomes delocalized over all four spins. 
Therefore, we would expect a maximum spin current for $h = \pm U_1$ as is also observed in Fig.~\ref{figure:ChapterGMRn2}(a). 

Remarkably, we see only peaks in the spin current at these two values. Because of the baths, we can expect multiple excitation states to be important. For the simple case of $n_1 = 2$, these can easily be included. An excitation at the left spin can also propagate to the middle two spins through the two transitions
\begin{align*}
\ket{\us \Psi_\pm \ds} \leftrightarrow \ket{\ds \us \us \ds}.
\end{align*}
These are likewise at resonance for $h = \pm U$, explaining why only two resonances are observed.

\begin{figure}[t]
\centering
\includegraphics[width=1\linewidth, angle=0]{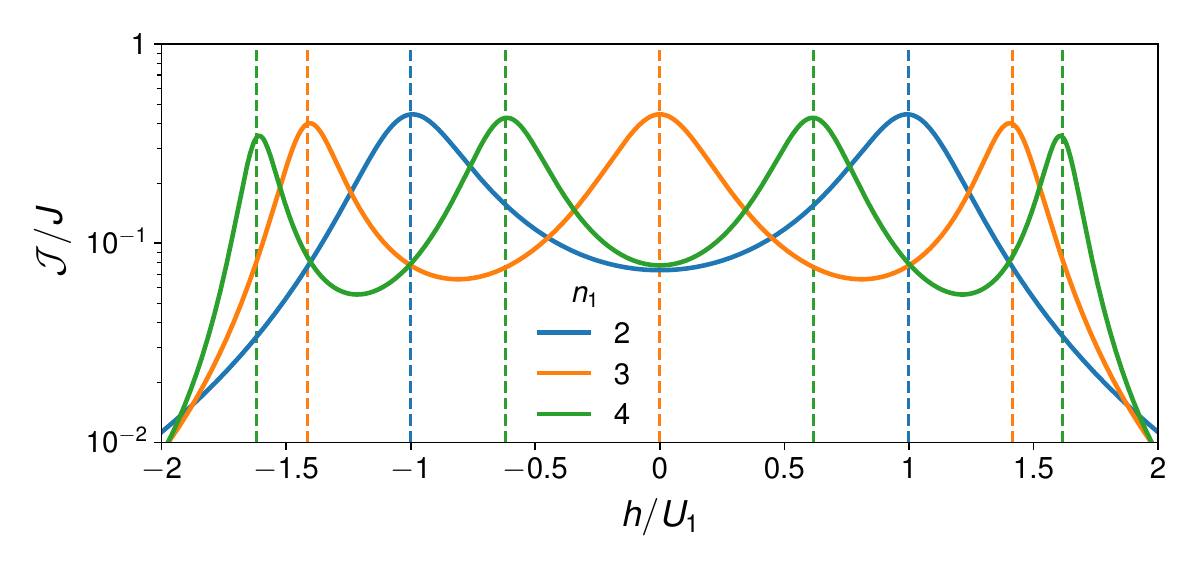}
\caption{ $\mathcal{J}$ as a function of $h/U_1$ for a single chain ($N=1$) consisting of a different number of strongly coupled spins $n_1$ interacting with an exchange of $U_1 = 10J$. The expected resonances are shown with vertical dashed lines (see the text). See Eq.~\eqref{eq:GMRDefaultValues} for the values of the remaining system parameters. Reproduced with permission from Phys. Rev. Lett. \textbf{126}, 077203 (2021) \cite{PhysRevLett.126.077203}. Copyright 2021 American Physical Society.}\label{figure:ChapterGMRnxN1}
\end{figure}

\section{Going beyond a single chain of $n_1 = 2$ spins} 
Keeping $\Delta_{U_i}, \Delta_J = 0$, there are two natural extensions of this, both of which are explored in Figs.~\ref{figure:ChapterGMRn2}(b) and \ref{figure:ChapterGMRnxN1}. First, the number of chains is varied keeping $n_i=2$ spins per chain in Fig.~\ref{figure:ChapterGMRn2}(b). Second, the number of spins is varied keeping only $N=1$ chain in Fig.~\ref{figure:ChapterGMRnxN1}. 

\subsection{Multiple chains of $n_i=2$ spins}
First, we look at a different number of chains $N$ while keeping $n_i = 2$. We also set all the strong exchange couplings equal $U_i = U_1$. The individual strongly coupled chains diagonalize just as before, and we still expect the strongest current for $h = \pm U_1$. 
The current both off ($h=0$) and at ($h = U_1$) resonance is plotted for a different number of pairs $N$ in Fig.~\ref{figure:ChapterGMRn2}(b). 
Off resonance, the spin current is heavily suppressed at first but then levels out for larger $N$, whereas on resonance the current is almost constant. In the limit $N \gg 1$, we likewise expect suppressed current for $h = 0$ and a larger current for $h=U_1$. If instead $n_i$ and $U_i$ are all picked at random, most segments will be off resonant for any $h$ and the current should be suppressed. Therefore, excitations cannot pass between segments similar to Anderson localization \cite{PhysRevLett.65.88}.

\subsection{A single chain of $n_1 > 2$ spins}
Next, we keep $N=1$ and instead vary $n_1$. Following the same process as before, we first diagonalize $\hat{H}_0$. Let $\ket{n}$ be the single excitation state with spin $(1,n)$ flipped. Keeping to this one excitation basis, the Hamiltonian $H_0$ can be written as
\begin{align*}
H_0 = \begin{pmatrix}
(2 - n_1)h & 2U_1 & 0 & \cdots & 0\\
2U_1 & (2 - n_1)h & 2U_1 & \cdots & 0\\
0 & 2U_1 & (2 - n_1)h & \cdots & 0\\
\vdots & \vdots & \vdots & \ddots & \vdots\\
0 & 0 & 0 & \cdots & (2 - n_1)h\\
\end{pmatrix}.
\end{align*} 
One can show that the eigenenergies become \cite{losonczi1992eigenvalues, yueh2005eigenvalues}
\begin{align*}
E_k = 4U_1 \cos \left( \frac{\pi k}{n_1 + 1} \right) + (2 - n_1)h, \quad \quad 1 \leq k \leq n_1.
\end{align*}
The corresponding states become eigenstates for $\hat{H}$ to lowest order in $J/U_1$. The states $\ket{\us \ds \ds \! . . .}$ and $\ket{. . . \! \ds \ds \us }$ have energy $-h n_1$ to lowest order. Therefore, an excitation at the ends is resonant with an excitation in the chain when 
\begin{align}
h = 2U_1 \cos \left( \frac{\pi k}{n_1 + 1} \right), \quad 1 \leq k \leq n_1. \label{eq:GMRResonanceCondition}
\end{align}
For $n_1 = 2$, this reduces to $h/U_1 = \pm 1$ as expected. As a few more examples, we get $h/U_1 = 0,\pm \sqrt{2}$ for $n_1 = 3$ and the four solutions $h/U_1 = \pm (\sqrt{5} \pm 1)/4$ for $n_1 = 4$. 
The spin current is plotted for these three examples in Fig.~\ref{figure:ChapterGMRnxN1}, and the resonances found above are plotted with vertical dashed lines. 
The maxima in the current occur at the resonances as expected, while the current is suppressed away from them. 

Once again, we observe maxima only at the values found above even though our analysis only includes single excitation states. Remarkably, all the allowed transitions overlap with Eq.~\eqref{eq:GMRResonanceCondition}. 
Like for the simple case, the entire Hilbert space for the strongly interacting spins could be included, but this becomes increasingly difficult as $n_1$ increases. 
It has been checked numerically that all resonances occur at the values shown in Eq.~\eqref{eq:GMRResonanceCondition} for all cases $n_1 \leq 15$; see Ref.~\cite{PhysRevLett.126.077203}. 

The current is heavily suppressed for $|h| > 2U_1 \cos \left( \frac{\pi}{n_1 + 1} \right)$. In the thermodynamic limit $n_1 \gg 1$, the single excitation spectrum for the strongly interacting chain approaches a continuum in the interval $-4U_1 < E_k < 4U_1$, and an appreciable current is expected for $-2U_1 < h < 2U_1$, while a hard dropoff should occur for $|h| > 2U_1$.

\begin{figure}[t]
\centering
\includegraphics[width=0.625\linewidth, angle=0]{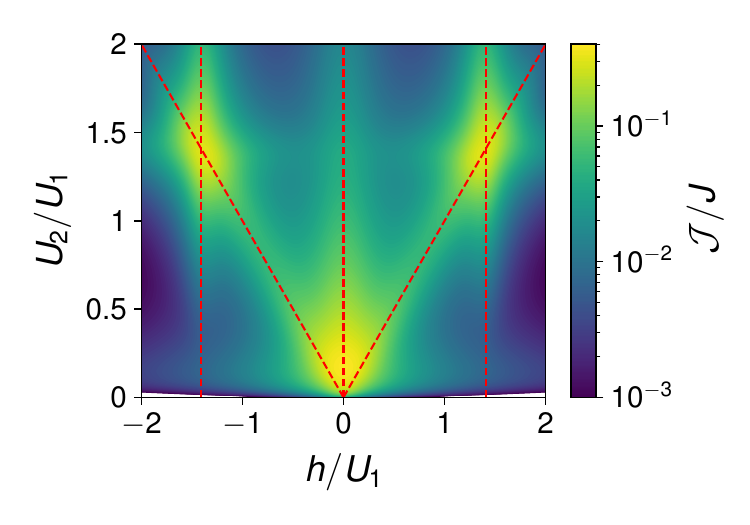}
\caption{ $\mathcal{J}$ as a function of $h/U_1$ and $U_2/U_1$ with $U_1 = 5J$ for the system illustrated in Fig.~\ref{figure:ChapterGMRSetup}. The dashed lines show the expected resonances for both the first chain $h/U_1 = 0,\pm \sqrt{2}$ and the second chain $h/U_2 = \pm 1$. See Eq.~\eqref{eq:GMRDefaultValues} for the values of the remaining system parameters. Reproduced with permission from Phys. Rev. Lett. \textbf{126}, 077203 (2021) \cite{PhysRevLett.126.077203}. Copyright 2021 American Physical Society.}\label{figure:ChapterGMRn32N2}
\end{figure}

\begin{figure}[t]
\centering
\includegraphics[width=0.75\linewidth, angle=0]{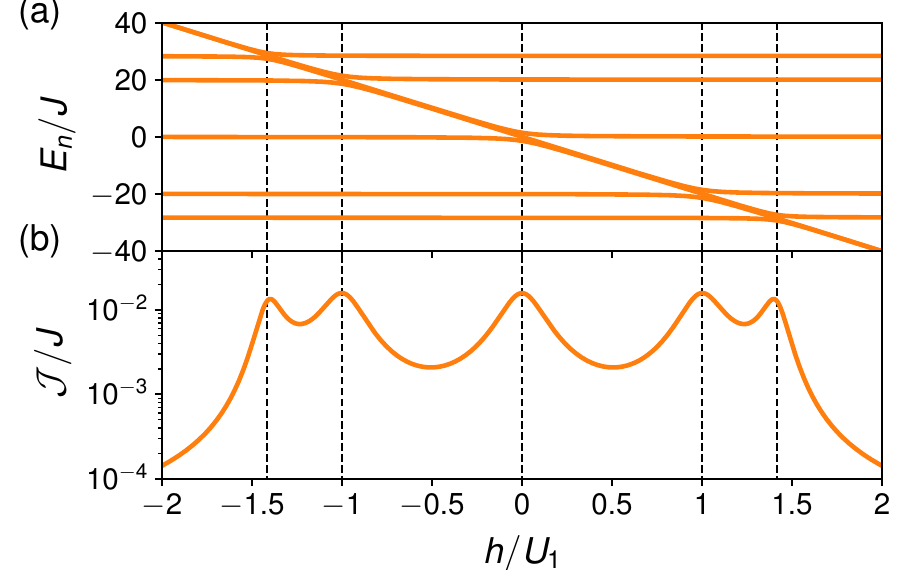}
\caption{ Single excitation spectrum, (a), and current $\mathcal{J}$, (b), plotted for the system illustrated in Fig. \ref{figure:ChapterGMRSetup}. The expected resonances are shown with dashed lines for $h/U_1 = 0, \pm 1, \pm \sqrt{2}$. See Eq.~\eqref{eq:GMRDefaultValues} for the values of the remaining system parameters. Reproduced with permission from Phys. Rev. Lett. \textbf{126}, 077203 (2021) \cite{PhysRevLett.126.077203}. Copyright 2021 American Physical Society.}\label{figure:ChapterGMRSpectrum}
\end{figure}

\section{Two chains of $n_1 = 3$ and $n_2 = 2$ spins}

In the more general case, we look at $N=2$ chains consisting of $n_1 = 3$ and $n_2 = 2$ strongly coupled spins, respectively, as seen in Fig. \ref{figure:ChapterGMRSetup}. 
At first, we keep $U_1 \neq U_2$. The first chain will then be at resonance with the ends for $h/U_1 = 0, \pm \sqrt{2}$, while the second chain will be at resonance with the ends for $h/U_2 = \pm 1$. 
However, only when both the chains individually are at resonance with each other so that a spin excitation can propagate between them do we expect the largest current. This is the case when both of the above conditions are upheld or rather when $U_2 = \pm \sqrt{2} U_1$ or $U_2 \sim 0$. 
To see that this is true, we plot the current as a function of both $U_2/U_1$ and $h/U_1$ in Fig.~\ref{figure:ChapterGMRn32N2} with the expected resonances plotted as dashed lines. Here we see that lines of high current run along the expected lines and that the current is extra large when the resonances meet. 

To illustrate the role of the single excitation spectrum explored above, we set $U_1 = U_2 = U$ and plot both the current and the single excitation spectrum as a function of $h/U$ in Figs. \ref{figure:ChapterGMRSpectrum}(b) and \ref{figure:ChapterGMRSpectrum}(a), respectively. 
Again, we plot the expected resonances with dashed lines. The two eigenenergies that are linearly dependent on $h/U$ corresponds to eigenstates that are close to $\ket{\us \ds ... \ds}$ and $\ket{\ds ... \ds \us}$, whereas the others are close to eigenstates that correspond to a spin excitation within the strongly coupled chains. 
Here it is clearly seen that, when the energies of the states describing excitations at the ends cross the energy of the states with excitations within the chains, a higher current is observed.
Hence, we see that the giant magnetoresistance is attributed to a set of resonance conditions that can be predicted for particular setups. 

This leads to several generalizations. First, for a large number of strongly interacting chains $N\gg 1$ with a random number of spins $n_i$, the excitation will be scattered at most boundaries, thus resulting in poor conductivity.
Second, if the spins $L$ and $R$ are substituted for general systems, a resonance will be observed when the frequencies of these systems are resonant with the neighboring strongly interacting chain.

\begin{figure}[t]
\centering
\includegraphics[width=1\linewidth, angle=0]{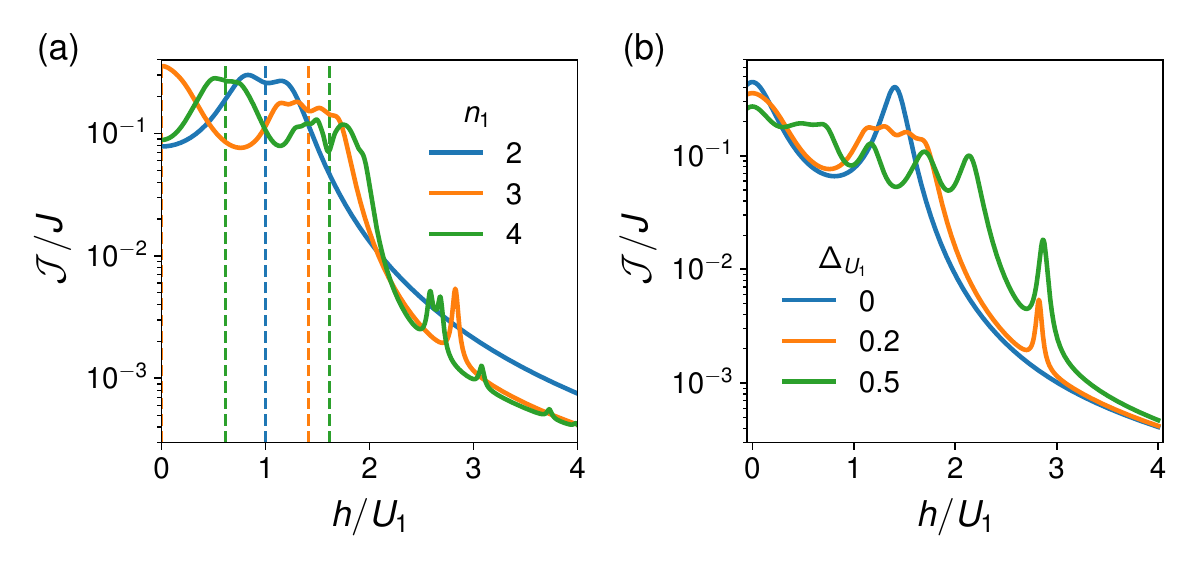}
\caption{ (a) $\mathcal{J}$ as a function of $h/U_1$ for $N=1$ chain of $n_1$ strongly coupled spins where $\Delta_{U_1} = \Delta_J = 0.5$. The resonances from Eq.~\ref{eq:GMRResonanceCondition} are plotted as dashed lines. (b) $\mathcal{J}$ as a function of $h/U_1$ for a model of only $N=1$ chain consisting of $n_1 = 3$ strongly interacting spins. See Eq.~\eqref{eq:GMRDefaultValues} for the values of the remaining system parameters. Reproduced with permission from Phys. Rev. Lett. \textbf{126}, 077203 (2021) \cite{PhysRevLett.126.077203}. Copyright 2021 American Physical Society.}\label{figure:ChapterGMRZCoupling}
\end{figure}

\section{Including Z-coupling} 
\label{sec:GMRZCoupling}
Finally, we include anisotropy for a model of only $N=1$ chain of a varying number of strongly interacting spins in Fig.~\ref{figure:ChapterGMRZCoupling}(a). Additionally, the old resonances from Eq.~\ref{eq:GMRResonanceCondition} are plotted as dashed lines. The addition of anisotropy has two main effects. First, the spectrum is perturbed, moving and splitting up the resonances. Second, new peaks appears. The second effect only happens for $n_1>2$, so we can start by explaining the results for $n_1 = 2$ and more forward from there. 

For the case of $n_1 = 2$ strongly interacting spins, there are four different ways for an excitation to travel from the left spin to the two middle spins. These are
\begin{align}
\ket{\us \ds \ds \ds} & \leftrightarrow \ket{\ds \Psi_- \ds}, \\
\ket{\us \ds \ds \ds} & \leftrightarrow \ket{\ds \Psi_+ \ds}, \\
\ket{\us \Psi_- \ds} & \leftrightarrow \ket{\ds \us \us \ds}, \\
\ket{\us \Psi_+ \ds} & \leftrightarrow \ket{\ds \us \us \ds}.
\end{align}
For $\Delta_{U_1} = 0$, these four transitions results in two resonances. The first and fourth transition obey energy conservation for $h=U$, while the second and third obeys energy conservation for $h=-U$. For $\Delta_{U_1} \neq 0$, the four transitions obey energy conservation for different $h$ values thus resulting in four resonances. 

For $n_1 = 3$, the current is plotted for different anisotropies in Fig.~\ref{figure:ChapterGMRZCoupling}(b) to better see the effects. The same splitting of resonances occur for $n_1 = 3$, however, an additional peak also appear for $h/U_1 \simeq 2 \sqrt{2}$ due to the transition 
\begin{align}
\ket{\us \Lambda_{\us \ds \ds} \ds} \rightarrow \ket{\ds \Lambda_{\us \us \ds} \ds} \rightarrow \ket{\ds \Lambda_{\us \ds \ds} \us},
\end{align}
where
\begin{align}
\ket{\Lambda_{\us \ds \ds}} &= \frac{1}{2} \left( \ket{\us \ds \ds} - \sqrt{2} \ket{\ds \us \ds} + \ket{\ds \ds \us} \right) + O(\Delta_{U_1}),\\
\ket{\Lambda_{\us \us \ds}} &= \frac{1}{2} \left( \ket{\us \us \ds} + \sqrt{2} \ket{\us \ds \us} + \ket{\ds \us \us} \right) + O(\Delta_{U_1}),
\end{align}
to lowest order in $\Delta_{U_1}$. The matrix element for the first transition is
\begin{align}
\mel{\ds \Lambda_{\us \us \ds} \ds}{\hat{H}}{\us \Lambda_{\us \ds \ds} \ds} = -\frac{\Delta_{U_1}}{4} + O(\Delta_{U_1}^3).
\end{align}
For this, higher order terms of $\ket{\Lambda_{\us \ds \ds}}$ and $\ket{\Lambda_{\us \us \ds}}$ were included. The matrix element is zero for $\Delta_{U_1}= 0$, explaining why the resonance is absent for this case. The transition is at resonance for 
\begin{align}
h/U_1 = 2\sqrt{2} \left[ 1+\frac{\Delta_{U_1}^2}{16} + O(\Delta_{U_1}^4) \right].
\end{align}
For $n_1=4$, as seen in Fig.~\ref{figure:ChapterGMRZCoupling}(a), the current as a function of the magnetic field becomes very chaotic, and it is difficult to identify the individual resonances. One could go through the same calculation as we did for $n_1=3$. However, the number of different possible transitions increase exponentially for larger $n_1$, and the information becomes less usefull as the resonances become indistinguishable.

\cleardoublepage
\chapter{Maxwell's demon assisted by non-Markovian effects}
\label{chapter:Maxwell}
\epigraph{\footnotesize This chapter is based on Ref.~\cite{PhysRevE.105.044141}.
Sections \ref{sec:MaxwellSetup}-\ref{sec:MaxwellMultiShot} and all figures herein have been reproduced with permission from Ref.~\cite{PhysRevE.105.044141}.
}{}

\noindent In the previous chapters, we engineered steady-state properties predominantly with the aim of controlling heat or spin transport. The thermodynamics of these systems is fairly simple as, generally, the only energy added is in the form of heat. If the aim is shifted into controlling the useful energy or work, the thermodynamics gets much more complicated. This becomes even more relevant with the possibility to control and measure individual quantum states where the control is conditioned on a previous measurement. Therefore, variations of Maxwell's demon are more relevant than ever. Fortunately, the omnipresence of Maxwell's demon in quantum thermodynamics also makes the interplay between information and work far more intuitive in the context of quantum mechanics than in the original formulation. 

For the simplest version of a quantum demon, consider a two-level system with energy $\omega_M$ acting as a medium for the demon \cite{PhysRevA.56.3374}. The medium is connected to a thermal bath at temperature $T_H \gg \omega_M$, resulting in the density matrix
\begin{align}
\hat{\rho}_M &= \frac{1}{1+e^{-\omega_M/T_H}} \op{0_M} + \frac{e^{-\omega_M/T_H}}{1+e^{-\omega_M/T_H}} \op{1_M} \\
& \simeq \frac{1}{2} \op{0_M} + \frac{1}{2} \op{1_M}. \nonumber
\end{align}
It is easy to see that a measurement of the two-level system followed by a $\pi$-pulse if the measurement yields $\ket{1_M}$, would result in an average extracted work of
\begin{align}
\mathcal{W} \simeq \frac{\omega_M}{2}.
\end{align}
This lowers the entropy of the two-level system by $\ln 2$ in apparent contradiction with the second law of thermodynamics. 

However, this analysis ignores a vital part of the setup, namely the demon's memory. Since the measurement yields a binary result, the demon's memory is included as another two-level system with energy $\omega_D$. The three steps of the demon process are
\begin{center}
\begin{tabular}{ l l }
Step 1. & Information on the medium is stored in the demon's memory. \\ 
Step 2. & The information is used to extract work. \\  
Step 3. & The demon memory is either reset or replaced.
\end{tabular}
\end{center}
Using the density matrices of the medium and memory, the three steps are
\begin{align}
& \left(\frac{1}{2} \op{0_M} + \frac{1}{2} \op{1_M} \right) \op{0_D} \nonumber \\
& \xrightarrow{\mathrm{step}\,1} \frac{1}{2} \op{0_M 0_D} + \frac{1}{2} \op{1_M 1_D} \\
& \xrightarrow{\mathrm{step}\,2}  \op{0_M} \left(\frac{1}{2} \op{0_D} + \frac{1}{2} \op{1_D}\right) \nonumber \\
& \xrightarrow{\mathrm{step}\,3}  \op{0_M 0_D}. \nonumber
\end{align}
The third step is performed by letting the demon memory interact with a thermal bath at temperature $T_D \ll \omega_D$. During the first step, the energy $\omega_D/2$ is added to the system through work, while the energy is decreased by $\omega_M/2$ through work during the second step. The final step dumps the energy $\omega_D/2$ into the bath. The total heat extracted is, therefore,
\begin{align}
\mathcal{W} \simeq \frac{\omega_M-\omega_D}{2}.
\end{align}
Simultaneously, the entropy is $\ln 2$ throughout steps 1-2, while it is decreased to zero during the third step. The total entropy change is 
\begin{align}
\Delta S \simeq -\ln 2.
\end{align}
However, the entropy decrease is now during the interaction with a thermal bath. The entropy of the thermal bath will increase to compensate, and the apparent violation of the second law of thermodynamics is resolved. In the above example, information is used to extract work. It is, therefore, a variation of the Szilard engine rather than a Maxwell's demon \cite{bennett1987demons}. 

This chapter will study a variation of Maxwell's demon where information is used to transfer the heat from a cold bath to a warm bath. The two baths are non-Markovian, allowing for information to flow back into the medium from the baths. First, a single operation of the demon acquiring and using information is studied, and the thermodynamics of the operation is analyzed. Second, the demon is allowed to operate twice, allowing for the exploitation of the non-Markovian effects. Finally, the demon operates repeatedly with a constant period. After sufficient time, the number of transferred excitations by the demon converges, and the importance of the timing of the demon can be examined.

\begin{figure}[t]
\centering
\includegraphics[width=0.92\linewidth, angle=0]{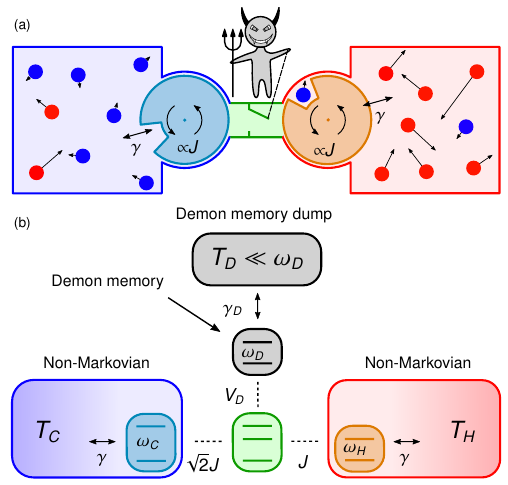}
\caption{ (a) Classical analog of the demon setup where rotating wheels illustrate the added predictability of a non-Markovian bath.
(b) Illustration of the demon setup where two non-Markovian baths are connected by a qutrit. The cold non-Markovian bath consists of a qubit whose correlation functions decay due to the Markovian bath of temperature $T_C$ and likewise for the non-Markovian hot bath. A third qubit is the demon's memory, which can decay through interaction with the memory dump. Reproduced with permission from Phys. Rev. E \textbf{105}, 044141 (2022) \cite{PhysRevE.105.044141}. Copyright 2022 American Physical Society.}\label{figure:ChapterMaxwellSetup}
\end{figure}

\section{Setup for demon and non-markovian baths}
\label{sec:MaxwellSetup}

The setup consists of two non-Markovian baths connected by a qutrit acting as the medium, as seen in Fig.~\ref{figure:ChapterMaxwellSetup}. The non-Markovian baths are comprised of two parts: first, a Markovian bath of temperature $T_{C/H}$ and, second, a qubit with frequency $\omega_{C/H}$. A third qubit is used for the demon's memory. The Hamiltonian of the qutrit and the three qubits is given by
\begin{align}
\hat{H}_0 &= \omega_C \Big[ \op{1_C}  +  \op{2_M}\Big] + \omega_H \Big[ \op{1_M} + \op{1_H}\Big]  \\
& \hspace{8cm} + \omega_D \op{1_D}. \nonumber
\end{align}
The qutrit states are denoted $\ket{0_M}$, $\ket{1_M}$, and $\ket{2_M}$; the cold qubit states are denoted $\ket{0_{C}}$ and $\ket{1_{C}}$; the hot qubit states are denoted $\ket{0_{H}}$ and $\ket{1_{H}}$; and the demon-memory states are denoted $\ket{0_D}$ and $\ket{1_D}$. The two qubits are coupled to the qutrit with strength $J$. If the qubit frequencies are picked such that $|\omega_C - \omega_H|, |\omega_C|, |\omega_H| \gg |J|$, the cold qubit can only couple the qutrit states $\ket{0_M}$ and $\ket{2_M}$, and the hot qubit can only couple the qutrit states $\ket{0_M}$ and $\ket{1_M}$. The full Hamiltonian becomes
\begin{align}
\hat{H} &= \hat{H}_0  + \sqrt{2} J\left( \hat{\sigma}_C^- |2_M\rangle \langle 0_M| + \hat{\sigma}_C^+ |0_M\rangle \langle 2_M|  \right) \label{hamilton} \\
&\hspace{4.5cm} + J\left( |0_M \rangle \langle 1_M| \hat{\sigma}_H^+ +  |1_M \rangle \langle 0_M| \hat{\sigma}_H^- \right), \nonumber
\end{align}
where $\hat{\sigma}_{C(H)}^- = |0_{C(H)}\rangle \langle 1_{C(H)}|$ and $\hat{\sigma}_{C(H)}^+ = |1_{C(H)}\rangle \langle 0_{C(H)}|$. The factor of $\sqrt{2}$ is due to the cold qubit interacting with the second excited state of the qutrit. 

The evolution of the system is described by the local master equation on Lindblad form
\begin{equation}
\frac{d \hat{\rho}}{d t} = -i [\hat{H}+\hat{V}_D (t), \hat{\rho}] + \mathcal{D}_C[\hat{\rho}] + \mathcal{D}_H[\hat{\rho}] + \mathcal{D}_D [\hat{\rho}](t). \label{me:1}
\end{equation}
The Markovian baths are modeled using the non-unitary parts
\begin{align}
\mathcal{D}_{C}[\hat{\rho}] &= \gamma (n_{C} + 1)  \mathcal{M}[ \hat{\sigma}^-_{C}, \hat{\rho}] + \gamma n_{C} \mathcal{M}[ \hat{\sigma}^+_{C}, \hat{\rho}], \\
\mathcal{D}_{H}[\hat{\rho}] &= \gamma (n_{H} + 1)  \mathcal{M}[ \hat{\sigma}^-_{H}, \hat{\rho}] + \gamma n_{H} \mathcal{M}[ \hat{\sigma}^+_{H}, \hat{\rho}],\\ 
\mathcal{D}_D[\hat{\rho}](t) &= \gamma_D(t) \mathcal{M}[ \hat{\sigma}^-_D, \hat{\rho}],
\end{align}
where $\mathcal{M}[\hat{A}, \hat{\rho}] = \hat{A} \hat{\rho} \hat{A}^\dag - \{\hat{A}^\dag \hat{A}, \hat{\rho}\}/2$. The coupling strength between the Markovian baths and the cold and hot qubit is $\gamma$, the coupling of the demon memory to the memory dump is $\gamma_D(t)$, and the mean number of excitations in the bath mode of energy $\omega_C$ and $\omega_H$, respectively,~are
\begin{equation}
n_{C} = \left( e^{\omega_C/T_{C}} -1 \right)^{-1} \quad \mathrm{and} \quad n_{H} = \left( e^{\omega_H/T_{H}} -1 \right)^{-1}.
\end{equation}
To study the effects of non-Markovianity, we can keep the qutrit-bath coupling, $J$, constant while varying the rate of decay of the bath correlation functions through $\gamma$. The Markovian limit for the cold (hot) bath is $\gamma (n_{C(H)} + 1/2) \gg J$ as we will see in subsection \ref{subsec:MaxwellMarkovianLimit}.
If the system is left alone, i.e., $\hat{V}_D (t)= \gamma_D(t) = 0$, and the demon memory is reset to $\ket{0_D}$, the density matrix will eventually reach a unique steady state $\hat{\rho}_{\mathrm{ss}}$.

Unless otherwise stated, the parameters are suitably picked for superconducting circuits \cite{doi:10.1063/1.5089550, doi:10.1146/annurev-conmatphys-031119-050605, PhysRevX.11.021010} to be 
\begin{align}
& \omega_C = 3500J, \quad \omega_H = 2000J, \quad T_C = 4/7\omega_C, \quad T_H = 1.5\omega_H,  \label{eq:MaxwellDefaultValues} \\ 
& \hspace{1.5cm} \gamma_D \in \{0, 8J\}, \quad \tau_Y = 0.02J^{-1}, \quad \tau_{CZ} = 0.1 J^{-1} . \nonumber
\end{align}
We set $\gamma_D = 8J$ when the demon's memory is interacting with the memory dump and $\gamma_D = 0$ otherwise. The demon controlls the memory and medium through Y-rotations and controlled phase gates with gate times $\tau_Y$ and $\tau_{CZ}$, respectively.

In summary, we study a cold non-Markovian bath interacting with the second excited state of the qutrit and a hot non-Markovian bath interacting with the first excited state of the qutrit. Excitations can thus be sorted from the cold to the hot bath by forcing the transition $\ket{2_M} \rightarrow \ket{1_M}$. This could be done by coupling the transition to an even colder bath. However, in this case, the entropy of the even colder bath would increase obeying the second law of thermodynamics.

\section{Single operation of the demon} 
Instead, we wish to elucidate the interplay between entropy and information using a Maxwell's demon.
The demon's memory is modeled by the qubit with frequency $\omega_D$. The demon operates in three steps:

\bigskip

\noindent \begin{minipage}{0.15\columnwidth}
Step 1.
\smallskip

Step 2.\\
\smallskip

Step 3.\\

\end{minipage}
\begin{minipage}{0.85\columnwidth}
Information on the qutrit is stored in the demon's memory.
\smallskip

The information is used to transfer one excitation from the cold bath to the hot bath.
\smallskip

The demon's memory is either reset or a clean memory slot is accessed.
\end{minipage}
\smallskip

\noindent For the qutrit in a general statistical mixture, the steps are
\begin{align}
&\big(p_0 \op{0_M} + p_1 \op{1_M} + p_2 \op{2_M} \big) \op{0_D} \nonumber \\
&\xrightarrow{\mathrm{step}\,1} p_0 \op{0_M 0_D} + p_1 \op{1_M 0_D} + p_2 \op{2_M 1_D} \\ &\xrightarrow{\mathrm{step}\,2} p_0 \op{0_M 0_D} + p_1 \op{1_M 0_D} + p_2 \op{1_M 1_D} \nonumber \\ 
& \xrightarrow{\mathrm{step}\,3} \big(p_0 \op{0_M} + (p_1 + p_2) \op{1_M} \big) \op{0_D} \nonumber
\end{align}
The first two steps constitute controlled NOT gates. These three steps will add energy to the system through work. Step 1 does average work $p_2 \omega_D$, and step 2 does average work $-p_2 (\omega_C - \omega_H)$. The work done through step 3 depends on how it is carried out. If the demon memory is reset through coupling to a cold bath, energy is subtracted through heat, and if a new demon memory is accessed, no heat or work is done. In either case, the total average work performed during the three steps is $p_2 (\omega_D + \omega_H-\omega_C)$. Thus work is performed to transfer heat similar to a refrigerator. However, looking at the special case $\omega_D = \omega_C-\omega_H$, we see that no work is done and the system does indeed implement a Maxwell's demon. 

For concreteness, we use a superconducting qubit platform to model an experimental implementation. 
The CNOT gate can be implemented by supplementing the native controlled-phase gate \cite{DiCarlo2009} with single-qubit Y-gates. The two CNOT gates in the first two steps can, therefore, be summarized in a circuit for the medium density matrix $\hat{\rho}_M$ and the demon memory density matrix $\hat{\rho}_D$.

\begin{figure}[t]
\centering
\includegraphics[width=0.875\linewidth, angle=0]{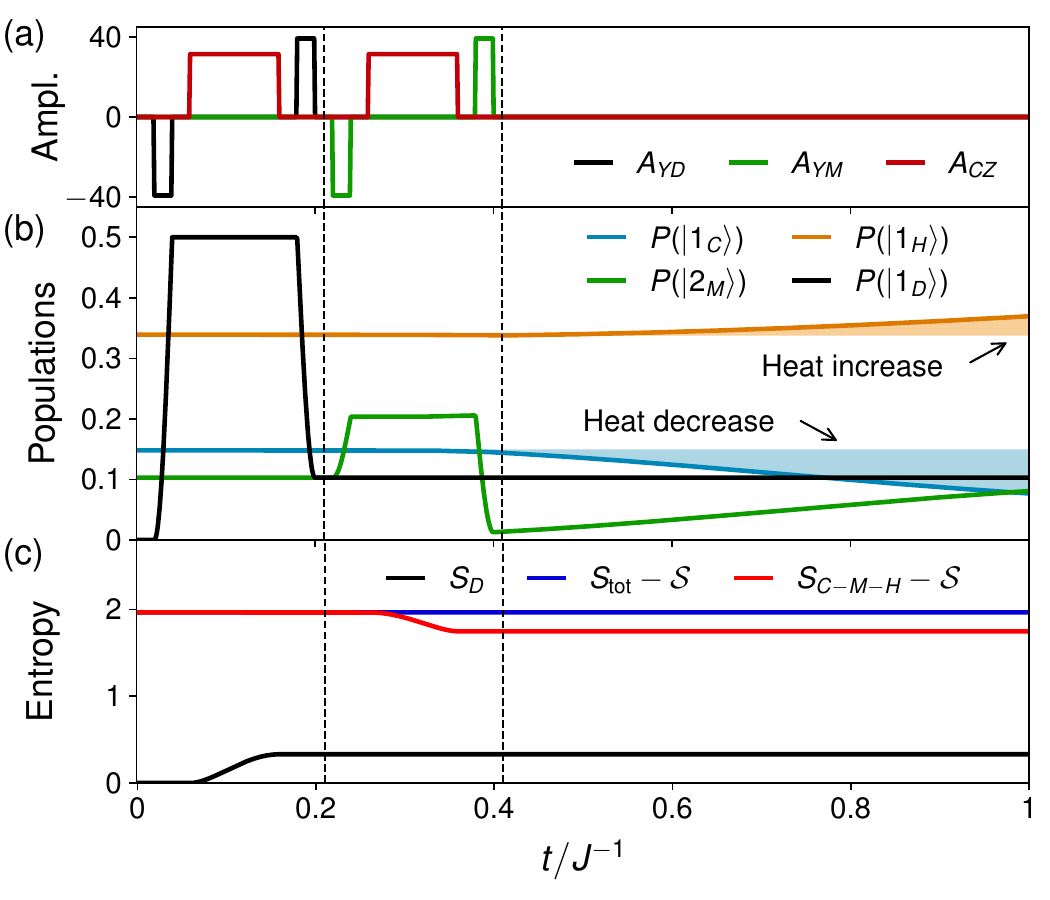}
\caption{(a) Amplitudes, as seen in the control Hamiltonian \eqref{eq:MaxwellControl}, for implementing steps 1 and 2 of the demon protocol.
(b) Populations for the excited states as a function of time starting from the system steady state at $t=0$. The orange and blue shadows show the difference between the population and the steady-state population for the hot and cold qubits. (c) Entropy of the baths and qutrit system $S_{C-M-H}$, the demon $S_D$, and the entire system $S_{tot}$ as a function of time. $\mathcal{S}$ is the constant entropy of the Markovian baths. For this simulation, $\gamma = 10^{-3}J$. The dashed lines separate step 1, step 2, and the subsequent free evolution. See Eq.~\eqref{eq:MaxwellDefaultValues} for the values of the remaining system parameters. Reproduced with permission from Phys. Rev. E \textbf{105}, 044141 (2022) \cite{PhysRevE.105.044141}. Copyright 2022 American Physical Society.}\label{figure:ChapterMaxwellSingleShot}
\end{figure}

\begin{picture}(100.,65.)(-250, -2)
\put(-166,0){\includegraphics[scale=0.37]{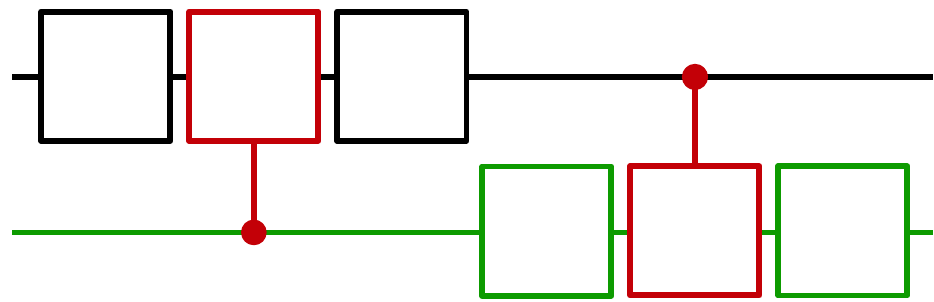}}
\put(-185,38){$\hat{\rho}_M$}
\put(-185,12.0){$\hat{\rho}_D$}
\put(-156, 39){$Y_{\text{-}\frac{\pi}{2}}^{01}$}
\put(-129.5,37){$Z^{01}$}
\put(-102.5,39){$Y_{\frac{\pi}{2}}^{01}$}
\put(-78,11.){$Y_{\text{-}\frac{\pi}{2}}^{\tiny{12}}$}
\put(-51,9.){$Z^{12}$}
\put(-25,11.){$Y_{\frac{\pi}{2}}^{12}$}
\end{picture}

\noindent The superconducting circuit control Hamiltonian relevant for this proposal can be written as
\begin{equation}
\begin{aligned}
\hat{V}_D(t) &= A_{YM}(t) \left(i |2_M \rangle \langle 1_M| e^{-i (\omega_C -\omega_H)t} - i |1_M \rangle \langle 2_M| e^{i (\omega_C -\omega_H) t} \right) \\ &\hspace{1cm}+ A_{YD}(t) \left( i |1_D \rangle \langle 0_D| e^{-i \omega_D t} - i |0_D \rangle \langle 1_D| e^{i \omega_D t} \right)\label{eq:MaxwellControl} \\ &\hspace{1cm}+ A_{CZ}(t) \op{2_M 1_D}.
\end{aligned}
\end{equation}
\noindent The three amplitudes $A_{YM}$, $A_{YD}$, and $A_{CZ}$ define the demon protocol. These are picked such that the single-qubit Y-rotation gate time is $\tau_Y$ and the controlled-phase gate time is $\tau_{CZ}$. 

To show this process in action, the system is left alone for times $t<0$ such that the system reaches steady state $\hat{\rho}_{\mathrm{ss}}$, at $t=0$. Afterwards, step 1 and step 2 are implemented using the protocol shown in Fig.~\ref{figure:ChapterMaxwellSingleShot}(a).
The populations, $P(\ket{\alpha}) = \mathrm{tr}\Big\{ \op{\alpha} \hat{\rho} \Big\}$ for $\alpha~\in~\{ 1_C, 2_M, 1_H, 1_D \}$, are plotted for this process in Fig.~\ref{figure:ChapterMaxwellSingleShot}(b) with $\hat{\rho}(t)=\hat{\rho}_{\mathrm{ss}}$ at $t=0$. From Fig. \ref{figure:ChapterMaxwellSingleShot}(b), we notice several things. 
After step 1, the demon-memory population reaches the value of the qutrit population, $P(\ket{1_D}) \sim P(\ket{2_M})$. 
After step 2, $P(\ket{2_M}) \sim 0$ and an excitation has been transferred from the cold to the hot bath, thus lowering the entropy of the baths and qutrit system. The transferred heat is also visible in the increase of $P(\ket{1_H})$ and the decrease of $P(\ket{1_C})$.
Without step 3, the demon memory is left in a statistical mixture giving $\ket{1_D}$ if an excitation was transferred and $\ket{0_D}$ otherwise. 

The entropy of the baths and qutrit system, $S_{C-M-H}$, the entropy of the memory, $S_D$, and the total entropy, $S_{\mathrm{tot}}$, during the operation of the demon is plotted in Fig.~\ref{figure:ChapterMaxwellSingleShot}(c). The entropy of a system described by a density matrix $\hat{\rho}$ is defined by
\begin{equation}
S = - \tr \{ \hat{\rho} \ln \hat{\rho} \}.
\end{equation} 
The entropy $S_{C-M-H}$ does indeed decrease during the operation of the demon, the entropy of the demon increases, and the total entropy remains constant. Furthermore, the difference $S_{C-M-H} + S_{D} - S_{\mathrm{tot}}$ quantifies the mutual information between the qutrit-baths system and the demon memory. This mutual information is largest between steps 1 and 2, but it remains non-zero even after step 2. Since the structure of the Markovian baths is unknown, their entropy is denoted $\mathcal{S}$, and the rate $\gamma \ll J$ is kept small enough that $\mathcal{S}$ can be assumed constant during the simulation. 

This implies that if we run the demon protocol once, as in Fig.~\ref{figure:ChapterMaxwellSingleShot}(b), all populations will eventually return to the steady state $\hat{\rho}_{\mathrm{ss}}$, 
as $T_C$ and $T_H$ are fixed. To calculate the change in temperature due to the exchange of energy quanta would require knowledge
of the heat capacity of the baths and depends on the concrete physical realizations.
Without step 3, the demon protocol can only be run once, and the average number of excitations transferred will be less than 
\begin{equation}
\mathrm{tr}\Big\{ \op{2_M} \hat{\rho}_{\mathrm{ss}} \Big\} = \frac{e^{-\omega_C/T_C}}{1+e^{-\omega_H/T_H}+e^{-\omega_C/T_C}}.
\end{equation}
This does not exhibit any non-Markovian behavior since bath memory can not be seen through a single interaction.

\begin{figure}[t]
\centering
\includegraphics[width=0.875\linewidth, angle=0]{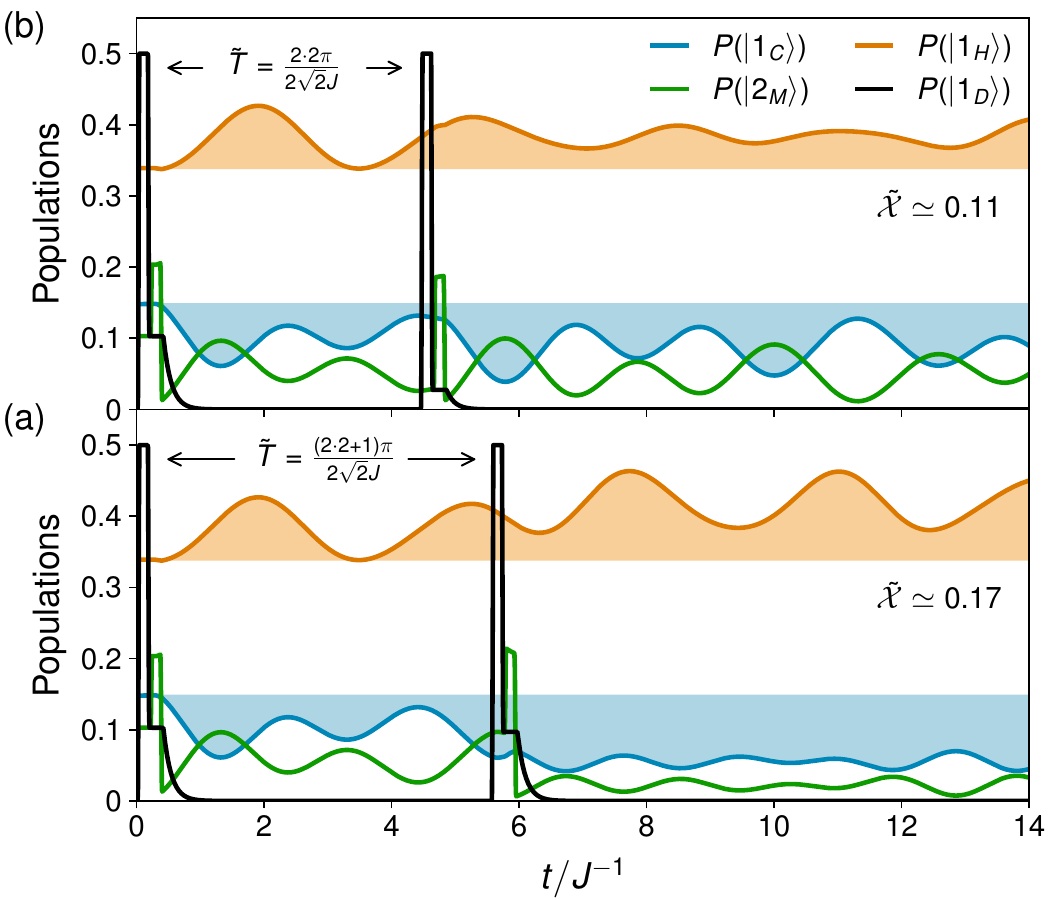}
\caption{Populations for the excited states as a function of time, starting from the system steady state at $t=0$. The figure is similar to Fig.~\ref{figure:ChapterMaxwellSingleShot}(b), however, the demon works twice here. The time between demon cycles is $\tilde{T}$ and the total number of transferred excitations is $\tilde{\mathcal{X}}$. For this simulation, $\gamma =10^{-3}J$. See Eq.~\eqref{eq:MaxwellDefaultValues} for the values of the remaining system parameters. Reproduced with permission from Phys. Rev. E \textbf{105}, 044141 (2022) \cite{PhysRevE.105.044141}. Copyright 2022 American Physical Society.}\label{figure:ChapterMaxwellDoubleShot}
\end{figure}

\section{Double operation of the demon} 
\label{sec:MaxwellDouble}

The simplest operation where non-Markovian effects become important is a double operation of the demon. The first operation puts the system out of steady state making energy oscillate between the baths and the medium. Depending on the timing of the second demon operation, it will transfer more or less energy. The populations for this simulation can be seen in Fig.~\ref{figure:ChapterMaxwellDoubleShot}. The protocol for the demon is the same as in Fig.~\ref{figure:ChapterMaxwellSingleShot}. However, in order for the demon to operate twice, the demon memory is allowed to decay between operations giving the new circuit

\begin{picture}(100.,65.)(-10, -2)
\put(0,0){\includegraphics[scale=0.37]{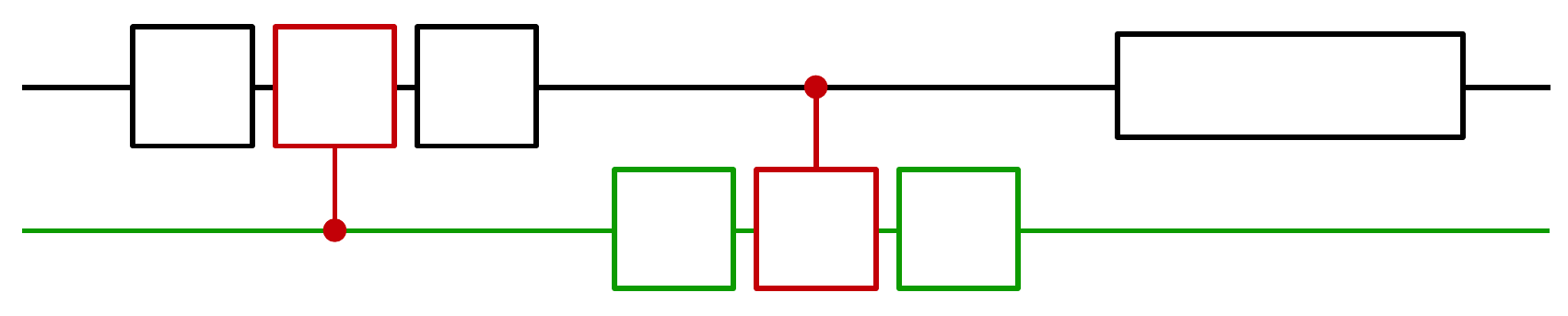}}
\put(-15,44){$\hat{\rho}_M$}
\put(-15,15.0){$\hat{\rho}_D$}
\put(28.5, 43.5){$Y_{\text{-}\frac{\pi}{2}}^{01}$}
\put(56.5,42){$Z^{01}$}
\put(84,43.5){$Y_{\frac{\pi}{2}}^{01}$}
\put(121.5,15.){$Y_{\text{-}\frac{\pi}{2}}^{12}$}
\put(149,13.5){$Z^{12}$}
\put(177.5,15.){$Y_{\frac{\pi}{2}}^{12}$}
\put(225,42){$\gamma_D(t) \neq 0$}
\end{picture}

\noindent The time between the two operations is denoted $\tilde{T}$. 

From the first demon operation and until the second demon operation, the populations are the same as in Fig.~\ref{figure:ChapterMaxwellSingleShot}(b). 
Here the oscillations between the qubits and the qutrit are clearly visible. If the cold qubit is excited at $t = 0$, it will oscillate back and forth between the cold qubit and the qutrit. 
The excitation will be at the qutrit at times $t = \frac{\pi}{2 \sqrt{2}J} (1+2k)$, where $k \geq 0$ is a whole number.

In Fig.~\ref{figure:ChapterMaxwellDoubleShot}(a), the second demon operation is at $\tilde{T} = \frac{2\cdot 2 \pi}{2 \sqrt{2}J}$, which is the time it takes one excitation at the cold qubit to oscillate to the qutrit and back twice. The number of excitations can then be calculated as
\begin{align}
\tilde{X} = P_\mathrm{ss}(\ket{1_C}) + P_\mathrm{ss}(\ket{2_M}) - \left[P(\ket{1_C})(t=\tilde{t}) + P(\ket{2_M})(t=\tilde{t})\right],
\end{align}
where $\tilde{t}$ is any time after step 2 is completed for the second operation and $\tilde{t} \ll \gamma^{-1}$. This results in a total of $\tilde{\mathcal{X}}\simeq 0.11$ transferred excitations. 

In Fig.~\ref{figure:ChapterMaxwellDoubleShot}(b), the second demon operation is at $\tilde{T} = \frac{(2\cdot 2+1)\pi}{2 \sqrt{2}J}$, which is the time it takes one excitation to perform 2.5 oscillations. This results in a total of $\tilde{\mathcal{X}}\simeq 0.17$ transferred excitations. Thus, by exploiting the non-markovian memory effects of oscillating excitations in and out of the bath, an additional $\sim 50\%$ excitations can be transferred.

\begin{figure}[t]
\centering
\includegraphics[width=0.625\linewidth, angle=0]{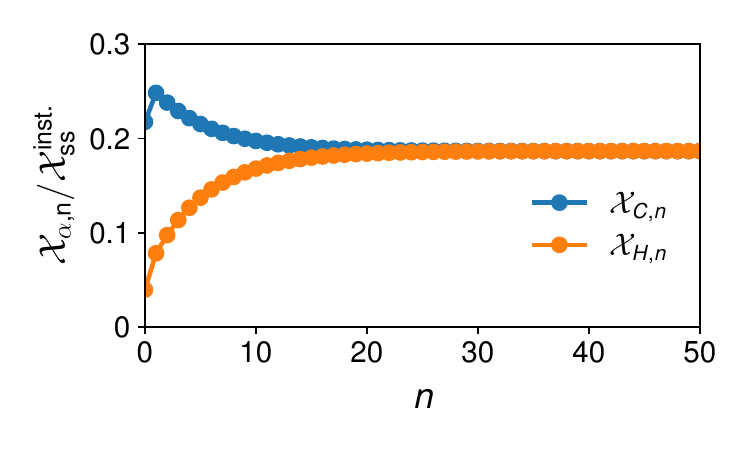}
\caption{ Transferred number of excitations between the cold bath and qutrit, $\mathcal{X}_{C,n}$, and between the qutrit and hot bath, $\mathcal{X}_{H,n}$, during the $n$th cycle as a function of $n$. Here, $\gamma = 10J$ and $T = J^{-1}$. See Eq.~\eqref{eq:MaxwellDefaultValues} for the values of the remaining system parameters. Reproduced with permission from Phys. Rev. E \textbf{105}, 044141 (2022) \cite{PhysRevE.105.044141}. Copyright 2022 American Physical Society.}\label{figure:ChapterMaxwellConvergence}
\end{figure}

\section{Repeated operation of the demon} 
\label{sec:MaxwellMultiShot}

We are now in a possition to study repeated operations of the demon at constant intervals $T$. There are two ways to repeat the operation of the demon. First, the demon memory can be expanded. If the demon memory consists of $N$ qubits, the protocol can be repeated $N$ times. Second, information stored in the demon memory can be erased allowing it to be reused. The demon memory is erased by letting it interact with the memory dump, i.e., $\gamma_D \neq 0$ like in section \ref{sec:MaxwellDouble}. 

We wish to study how the timing of the demon and the non-Markovian nature of the baths affect the transferred heat. Therefore, all three steps of the demon are repeated without allowing the qubits to thermalize between cycles.
We let $T$ be the total time to perform all three steps. The three steps are repeated $n$ times such that when step 3 is finished, step 1 is performed once again. Since steps 1 and 2 take constant time, $T$ is varied through step 3.

To quantify the transport between the cold and hot baths, we define the  excitation current from the cold qubit to the qutrit as $\mathcal{J}_C = \mathrm{tr} \left\{ \hat{s}_C \hat{\rho} \right\}$, and the excitation current from the qutrit to the hot qubit as $\mathcal{J}_H = \mathrm{tr} \left\{ \hat{s}_H \hat{\rho} \right\}$, where
\begin{align}
\hat{s}_C &= -\sqrt{2}iJ\Big( \hat{\sigma}_C^- |2_M\rangle \langle 0_M |  - \hat{\sigma}_C^+ |0_M \rangle \langle 2_M| \Big),\\
\hat{s}_H &= -iJ\Big( |0_M\rangle \langle 1_M| \hat{\sigma}_H^+ - |1_M\rangle \langle 0_M| \hat{\sigma}_H^- \Big).
\end{align}
These are determined in the same way as in previous chapters. Since the Hamiltonian is time-dependent, this will vary in time. To get a good measure of the number of transferred excitations, this is integrated over a single demon cycle,
\begin{equation}
\mathcal{X} = \lim_{n \rightarrow \infty} \int_{nT}^{(n+1)T} \mathcal{J}_C (t)\, dt = \lim_{n \rightarrow \infty} \int_{nT}^{(n+1)T} \mathcal{J}_H (t)\, dt.
\end{equation}
The integral above is the transferred excitations during the $n$th cycle of the demon. To check that this limit does indeed exist, we look instead at
\begin{align}
\mathcal{X}_{C,n} &= \int_{nT}^{(n+1)T} \mathcal{J}_C (t)\, dt,\\
\mathcal{X}_{H,n} &= \int_{nT}^{(n+1)T} \mathcal{J}_H (t)\, dt.
\end{align}
$\mathcal{X}_{C,n}$ and $\mathcal{X}_{H,n}$ are plotted as a function of $n$ in Fig.~\ref{figure:ChapterMaxwellConvergence} for $\gamma = 10J$ and $T = J^{-1}$. It is seen that they both converge to the same value as expected. We let $n \geq 100$ throughout. 

From this, we also define the average excitation current, $\mathcal{J}_{\mathrm{av}} = \mathcal{X}/T$, driven by the demon. For large $T$, the system reaches the steady state $\hat{\rho}_\mathrm{ss}$ between each cycle and the transferred number of excitations is
\begin{equation}
\lim_{T \rightarrow \infty} \mathcal{X} \leq \mathcal{X}_{ \mathrm{ss}}^{\mathrm{inst}} = \mathrm{tr}\Big\{ \op{2_M} \hat{\rho}_{\mathrm{ss}} \Big\} = \frac{e^{-\omega_C/T_C}}{1+e^{-\omega_H/T_H}+e^{-\omega_C/T_C}}.
\end{equation}
$\chi_{\text{ss}}^{\text{inst}}$ is the transferred number of excitations only for instantaneous gates. In a realistic setting, the system is allowed to evolve during steps 1 and 2 resulting in less excitations transferred. Therefore, the actual number of transferred excitations, even in steady state, will be less than $\mathcal{X}_{ \mathrm{ss}}^{\mathrm{inst}}$.

\subsection{Markovian limit}
\label{subsec:MaxwellMarkovianLimit}

For the purpose of comparison, we wish to calculate the Markovian limit of the two baths. This follows the same approach as chapters \ref{chapter:QutritDiode} and \ref{chapter:WB}. Unlike in chapter \ref{chapter:WB}, we will do the full derivation here. To do this, the cold and hot qubits are assumed to have quickly decaying correlation functions such that they can be traced away.
First, we study the Markovian limit for just the cold qubit. The Hamiltonian of just the cold qubit is
\begin{align}
\hat{H}_C = \omega_C \op{1_C} .
\end{align}
In the case where this qubit is only weakly coupled to the rest of the system but strongly coupled to the heat bath, $J \ll \gamma$, the evolution of the density matrix of just the cold qubit $\hat{\rho}_C$ will predominantly be determined by the heat bath,
\begin{align}
\frac{d\hat{\rho}_C}{dt} &= -i [\hat{H}_C, \hat{\rho}_C]  + \gamma n_C \left( \hat{\sigma}_C^+ \hat{\rho}_C \hat{\sigma}_C^- - \frac{1}{2} \{\hat{\sigma}_C^- \hat{\sigma}_C^+, \hat{\rho}_C\} \right) \\& \hspace{2.3cm} + \gamma (n_C + 1) \left( \hat{\sigma}_C^- \hat{\rho}_C \hat{\sigma}_C^+ - \frac{1}{2} \{\hat{\sigma}_C^+ \hat{\sigma}_C^-, \hat{\rho}_C\} \right). \nonumber
\end{align}
The state of the cold qubit will after sufficient time approach the thermal state,
\begin{align}
\hat{\rho}_C(t \rightarrow \infty) &= \frac{e^{-\beta \hat{H}_C}}{\mathrm{tr} \{ e^{-\beta \hat{H}_C} \}} = (1-\lambda_C) \op{0}{0} + \lambda_C \op{1}, \\
\lambda_C &= \left(1 + e^{\omega_C/T_C}\right)^{-1} = \frac{n_C}{2n_C+1}.
\end{align}
In the Markovian limit, the cold qubit is assumed to remain in this state even for $J\neq 0$. The coherences between the qubit and the qutrit will decay exponentially in $\gamma$ and can, therefore, be neglected. In the Heisenberg picture, an operator acting on the cold qubit $\hat{B}$ will evolve as
\begin{align}
\frac{d}{dt}\hat{B}(t) &= i [\hat{H}_C, \hat{B}(t)]  + \gamma n_C \left( \hat{\sigma}_C^- \hat{B}(t) \hat{\sigma}_C^+ - \frac{1}{2} \{\hat{\sigma}_C^- \hat{\sigma}_C^+, \hat{B}(t)\} \right) \\
&\hspace{2cm}+ \gamma (n_C + 1) \left( \hat{\sigma}_C^+ \hat{B}(t) \hat{\sigma}_C^- - \frac{1}{2} \{\hat{\sigma}_C^+ \hat{\sigma}_C^-, \hat{B}(t)\} \right). \nonumber
\end{align}
The Heisenberg picture is shown through the explicit time dependence. This can be solved for the ladder operators giving
\begin{align}
\hat{\sigma}_C^-(t) &= \hat{\sigma}_C^- e^{-i\omega_C t - \gamma (n_C + 1/2) t},\\
\hat{\sigma}_C^+(t) &= \hat{\sigma}_C^+ e^{i\omega_C t - \gamma(n_C + 1/2) t}.
\end{align}
With this the time correlation function, $\langle \hat{B}^\dag (t) \hat{B} \rangle$, for these two operators can be found to be
\begin{align}
\langle \hat{\sigma}_C^+(t) \hat{\sigma}_C^- \rangle &= \mathrm{tr}\{ \hat{\sigma}_C^+(t) \hat{\sigma}_C^- \hat{\rho}_C \} \label{eq:MaxwellCorrelationFunctions}\\
&= \frac{n_C}{2n_C+1} e^{i\omega_C t - \gamma(n_C + 1/2) t}, \nonumber \\ 
\langle \hat{\sigma}_C^-(t) \hat{\sigma}_C^+ \rangle &= \frac{n_C+1}{2n_C+1} e^{-i\omega_C t - \gamma(n_C + 1/2) t}. 
\end{align}
Note that the initial condition, for $t=0$, is just either the excited state population or the ground state population in the thermal state. The one-sided Fourier transforms are thus
\begin{align}
\Gamma^+_C(\omega) &= \int_0^\infty dt\, e^{-i\omega t} \langle \hat{\sigma}_C^+(t) \hat{\sigma}_C^- \rangle \\
&= \frac{n_C}{2n_C+1} \int_0^\infty dt\, e^{i(\omega_C-\omega)t - \gamma(n_C+1/2) t} \nonumber \\
&= \frac{n_C}{2n_C+1} \frac{i}{ \omega_C- \omega +i \gamma(n_C + 1/2)} \nonumber \\
&= \frac{n_C}{2n_C+1} \frac{\gamma(n_C+1/2) + i(\omega_C- \omega)}{ (\omega_C- \omega)^2  + \gamma^2 (2n_C + 1)^2/4}, \nonumber \\
\Gamma^-_C(\omega) &= \int_0^\infty dt\, e^{-i\omega t} \langle \hat{\sigma}_C^-(t) \hat{\sigma}_C^+ \rangle\\
&= \frac{n_C+1}{2n_C+1} \frac{\gamma(n_C + 1/2) - i(\omega_C+ \omega)}{ (\omega_C + \omega)^2  + \gamma^2(2n_C + 1)^2/4}. \nonumber
\end{align}
And thus
\begin{align}
\gamma^+_C (\omega) &= \Gamma^+_C + \Gamma^{+*}_C
=\frac{\gamma n_C }{ (\omega_C- \omega)^2  + \gamma^2 (2n_C + 1)^2/4},\\
\gamma^-_C (\omega) &= \Gamma^-_C + \Gamma^{-*}_C
= \frac{\gamma (n_C+1)}{ (\omega_C+ \omega)^2  + \gamma^2 (2n_C+1)^2/4}.
\end{align}
The same calculation can be carried out for the hot qubit,
\begin{align}
\gamma^+_H (\omega) &=\frac{\gamma n_H }{ (\omega_H - \omega)^2  + \gamma^2 (2n_H + 1)^2/4},\\
\gamma^-_H (\omega) &= \frac{\gamma (n_H+1)}{ (\omega_H + \omega)^2  + \gamma^2 (2n_H+1)^2/4}.
\end{align}
The interactions between the qutrit and two qubits are given by the terms
\begin{align}
\hat{H}_{C-M} &= \sqrt{2} J\Big(\hat{\sigma}_C^+ |0_M \rangle \langle 2_M| + \hat{\sigma}_C^- |2_M \rangle \langle 0_M|\Big),\\
\hat{H}_{M-H} &= J\Big( |1_M\rangle \langle 0_M| \hat{\sigma}_H^-  + |0_M \rangle \langle 1_M| \hat{\sigma}_H^+ \Big).
\end{align}
Treating the two qubits as environments and using the Redfield equation, after the Born-Markov and secular approximations; see Eq.~\eqref{eq:MasterEquationsGlobalSecular}, the master equation becomes
\begin{align}
\frac{d \hat{\rho}}{d t} &= -i [\hat{H}_{0,m} + \hat{V}_D(t), \hat{\rho}] + \mathcal{D}_C[\hat{\rho}] +\mathcal{D}_H[\hat{\rho}] + \mathcal{D}_D (t)[\hat{\rho}],\\
\mathcal{D}_C[\hat{\rho}] &=  \frac{8 (n_C+1) J^2 }{ \gamma (2n_C +1)^2} \left( |0_M \rangle \langle 2_M| \hat{\rho} |2_M \rangle \langle 0_M| - \frac{1}{2} \{ \op{2_M}, \rho \} \right)\\ 
&\hspace{0.8cm} + \frac{8 n_C J^2}{ \gamma (2n_C +1)^2} \left( |2_M \rangle \langle 0_M| \hat{\rho} |0_M \rangle \langle 2_M| - \frac{1}{2} \{ \op{0_M}, \rho \} \right), \nonumber \\ 
\mathcal{D}_H[\hat{\rho}] &= \frac{ 4 (n_H+1) J^2}{ \gamma (2n_H +1)^2} \left( |0_M \rangle \langle 1_M| \hat{\rho} |1_M \rangle \langle 0_M| - \frac{1}{2} \{ \op{1_M}, \rho \} \right) \\
& \hspace{0.8cm} + \frac{4 n_H J^2}{ \gamma (2n_H +1)^2} \left( |1_M\rangle \langle 0_M| \hat{\rho} |0_M \rangle \langle 1_M| - \frac{1}{2} \{ \op{0_M}, \rho \} \right), \nonumber \\
\mathcal{D}_D[\hat{\rho}] &= \gamma_D(t) \left( \hat{\sigma}^-_D \hat{\rho} \hat{\sigma}^+_D - \frac{1}{2} \{ \hat{\sigma}^+_D \hat{\sigma}^-_D, \rho \} \right),\\
\hat{H}_{0,m} &= \omega_C \op{2_M} + \omega_H \op{1_M} + \omega_D \op{1_D}.
\end{align}
This approximation is valid when the correlation functions of the bath from Eq.~\eqref{eq:MaxwellCorrelationFunctions} decay much faster than the dynamics of the system. Therefore, the inequalities that need to be fulfilled are
\begin{align}
\gamma (n_C + 1/2) \gg \sqrt{2} J, \quad \mathrm{and} \quad \gamma (n_H + 1/2) \gg J,
\end{align}
for the cold and hot qubit, respectively. So the Markov approximation is not only valid for large $\gamma$, but also for large temperatures $T_C$ and $T_H$.

\begin{figure}[t]
\centering
\includegraphics[width=0.875\linewidth, angle=0]{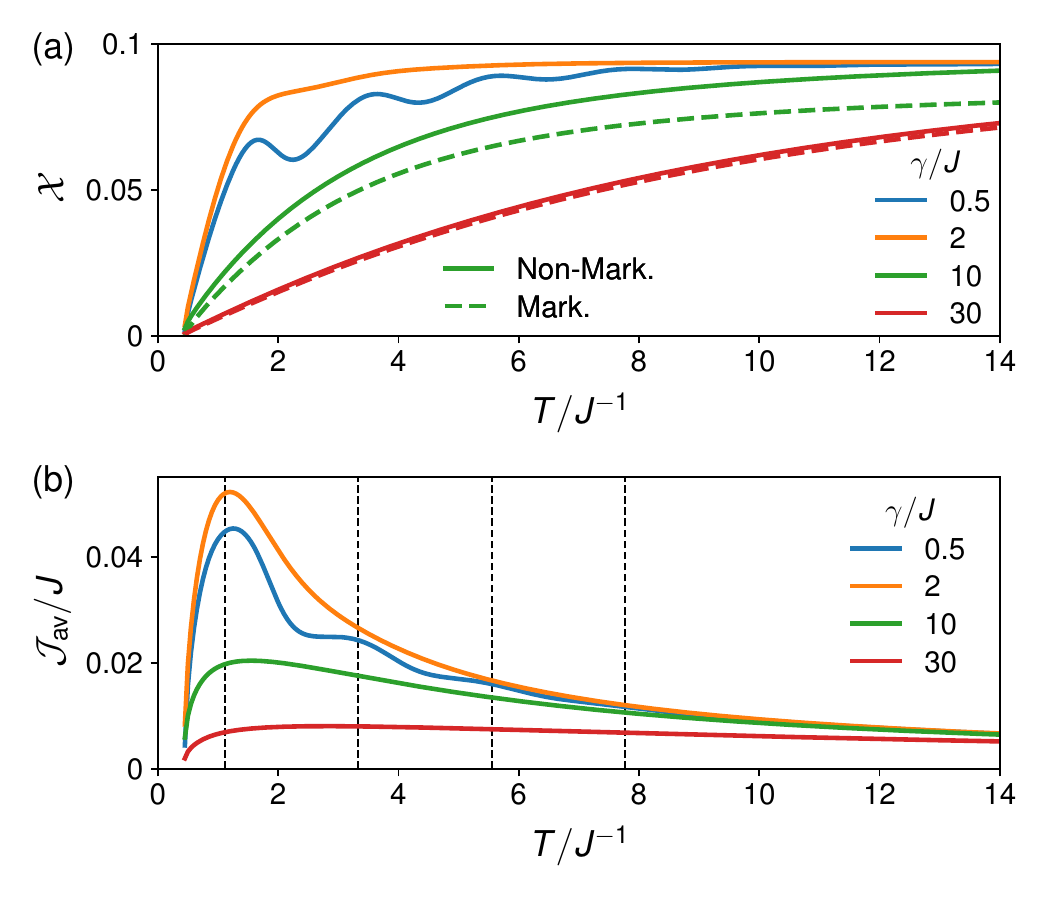}
\caption{ (a) Transferred excitations, $\mathcal{X}$, as a function of $T$ for different rates $\gamma$. This is plotted for both the full treatment (solid lines) and using a Markovian approximation on the cold and hot qubit (dashed lines). (b) Average excitation current, $\mathcal{J}_{\mathrm{av}}$, as a function of $T$ for different rates $\gamma$. See Eq.~\eqref{eq:MaxwellDefaultValues} for the values of the remaining system parameters. Reproduced with permission from Phys. Rev. E \textbf{105}, 044141 (2022) \cite{PhysRevE.105.044141}. Copyright 2022 American Physical Society.}\label{figure:ChapterMaxwellMultiShot}
\end{figure}

\begin{figure}[t]
\centering
\includegraphics[width=1\linewidth, angle=0]{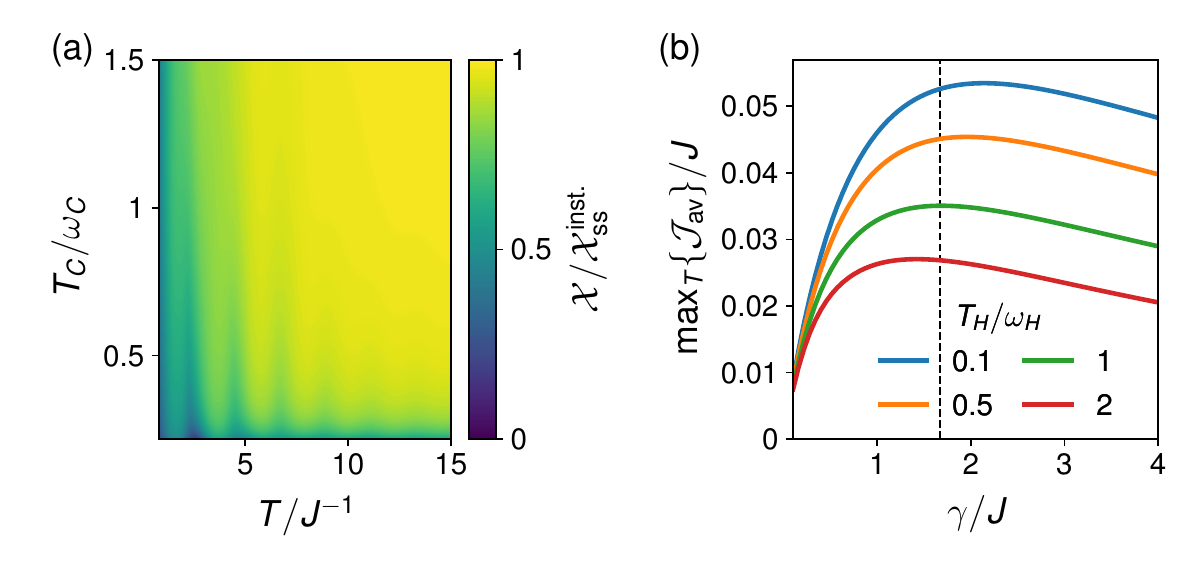}
\caption{ (a) Transferred excitations $\mathcal{X}$ as a function of both cold bath temperature, $T_C$, and $T$ for $\gamma = 0.5J$ and $T_H = 1000J$. (b) The average current maximized over $T$, $\text{max}_T \{\mathcal{J}_{\text{av}}\}$, as a function of $\gamma$ for different hot bath temperatures and $n_C = 0.1$. The dashed line corresponds to $\gamma (n_C+1/2) = J$. See Eq.~\eqref{eq:MaxwellDefaultValues} for the values of the remaining system parameters. Reproduced with permission from Phys. Rev. E \textbf{105}, 044141 (2022) \cite{PhysRevE.105.044141}. Copyright 2022 American Physical Society.}\label{figure:ChapterMaxwellTC}
\end{figure}

\subsection{Influence of demon timing on transport}

With the Markovian limit calculated, we plot $\mathcal{X}$ as a function of $T$ in Fig.~\ref{figure:ChapterMaxwellMultiShot}(a) for different values of $\gamma$. The dashed lines denote the Markovian limit which was calculated in subsection \ref{subsec:MaxwellMarkovianLimit}. 
Remarkably, the largest $\mathcal{X}$, and thus the largest entropy decrease, is achieved for non-Markovian baths, $\gamma = 2J$. The explanation for this is as follows: For small $\gamma$, the demon is limited by the small rate of excitation of the cold qubit. For large $\gamma$, the correlations between the cold qubit and the qutrit are suppressed resulting in a suppressed effective coupling between them that is similar to the quantum Zeno effect \cite{breuer2002theory}.

When the qubits start turning Markovian, $\gamma \geq 10J$, the full description predicts a larger $\mathcal{X}$ than the Markovian theory.
For $\gamma =30J$, the Markov approximation is valid and the results overlap. 
As $\gamma$ becomes small, $\mathcal{X}$ oscillates with $T$ due to non-Markovian effects or memory in the qubits. The Markovian solution in subsection \ref{subsec:MaxwellMarkovianLimit} contains no mechanism for producing these oscillations, and therefore, it is clearly a non-Markovian effect. For $T$ sufficiently large, the system reaches the steady state between updates and the transferred excitations are the same for all $\gamma$. 

Another interesting quantity is the average current, $\mathcal{J}_{\mathrm{av}}$, which is plotted in Fig.~\ref{figure:ChapterMaxwellMultiShot}(b). 
Here the oscillations in $\mathcal{X}$ for smaller $\gamma$ are again clearly seen.
As a reminder, if the cold qubit is excited at $t = 0$, it will oscillate back and forth between the cold qubit and the qutrit. 
The excitation will be at the qutrit at times $t = \frac{\pi}{2 \sqrt{2}J} (1+2k)$, where $k \geq 0$ is a whole number. The first four of these times are drawn as dashed lines in Fig.~\ref{figure:ChapterMaxwellMultiShot}(b), which are close to the maxima in the oscillations. 
These oscillations are thus due to the non-Markovian nature of the cold bath.
The period of oscillation between the qutrit and hot bath is $\pi/J$. However, this period is not present in Figs.~\ref{figure:ChapterMaxwellMultiShot}(a) and \ref{figure:ChapterMaxwellMultiShot}(b), suggesting that the non-Markovian effects are predominantly due to the cold bath. 

To further back this up, $\mathcal{X}$ is plotted as a function of both $T$ and the cold bath temperature $T_C$ in Fig.~\ref{figure:ChapterMaxwellTC}(a). 
For $T_C \ll \omega_C$, the cold bath is non-Markovian and the oscillations are observed. 
For $T_C > \omega_C$, the cold bath starts turning Markovian and the oscillations disappear as expected.
As an aside, note that $T_H = 1000J \simeq 0.29\omega_C$ in Fig.~\ref{figure:ChapterMaxwellTC}(a) such that $T_C > T_H$ in some cases. 
Since $\mathcal{X} > 0$ for both forward bias, $T_C > T_H$, and reverse bias, $T_C < T_H$, the system also implements a device of negative rectification, $\mathcal{R} = -\mathcal{J}_{\mathrm{av,f}}/\mathcal{J}_{\mathrm{av,r}} < 0$. Here, $\mathcal{J}_{\mathrm{av,f}}$ is the average current in forward bias, and $\mathcal{J}_{\mathrm{av,r}}$ is the average current in reverse bias.

\begin{figure}[t]
\centering
\includegraphics[width=0.625\linewidth, angle=0]{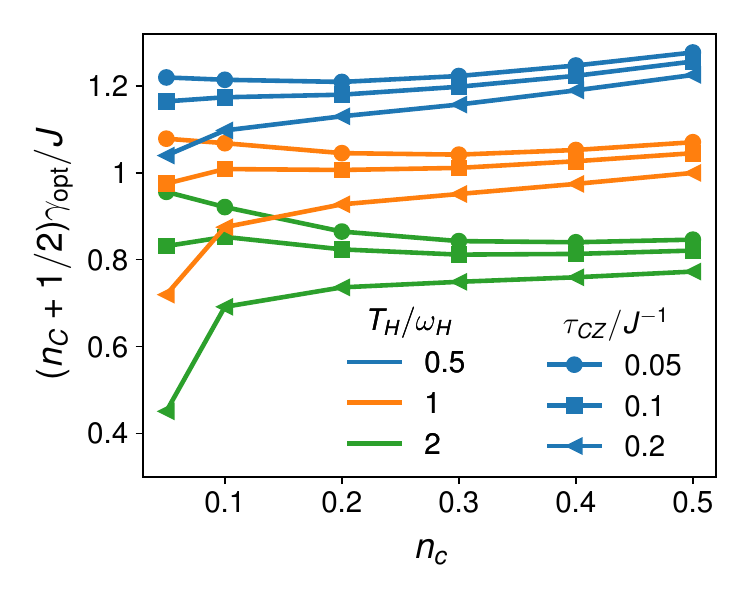}
\caption{ (a) The product $(n_C +1/2)\gamma_{\text{opt}}$ as a function of $n_C$ for different values of $T_H$ and $\tau_{CZ}$. See Eq.~\eqref{eq:MaxwellDefaultValues} for the values of the remaining system parameters. Reproduced with permission from Phys. Rev. E \textbf{105}, 044141 (2022) \cite{PhysRevE.105.044141}. Copyright 2022 American Physical Society.}\label{figure:ChapterMaxwellOptimalGamma}
\end{figure}

\subsection{Optimal bath-qubit coupling rate}

The optimal coupling rate is the value of $\gamma$ that allows for the largest average current induced by the demon, assuming that $T$ can be chosen freely. We have already seen that the optimal $\gamma$ is a balance, but we have yet to see how this balance depends on other system parameters. 

In Fig.~\ref{figure:ChapterMaxwellTC}(b), the average current maximized over the timing $T$, i.e., $\max_T
 \{\mathcal{J}_\mathrm{av}\}$ is plotted as a function of $\gamma$. 
From Eq.~\eqref{eq:MaxwellCorrelationFunctions}, it is seen that the Markovianity of the cold bath is determined by the product $\gamma (n_C + 1/2)$. Therefore, the dashed line corresponds to $\gamma (n_C + 1/2)=J$ for comparison. The optimal $\gamma$ seems to be around $\gamma (n_C + 1/2)=J$. However, it is also seen that the precise value of $\gamma$ is not important. If $\gamma (n_C+1/2) = J$ is picked, the average current is within a few percent of the maximum, that can be achieved, in all four cases seen in Fig.~\ref{figure:ChapterMaxwellTC}(b).

To examine the dependence of the optimal $\gamma$ on other parameters, we define the optimal coupling as
\begin{align}
\gamma_{\text{opt}} = \text{argmax}_\gamma \Big\{ \text{max}_T \{\mathcal{J}_\text{av} \}\Big\},
\end{align}
where $\mathrm{argmax}_x \{f(x)\}$ is the value of $x$ that maximizes $f(x)$. The product $\gamma_{\text{opt}} (n_C + 1/2)$ is plotted as a function of $n_C$ for different values of $T_H$ and $\tau_{CZ}$ in Fig.~\ref{figure:ChapterMaxwellOptimalGamma}.
Again, the optimal coupling is seen to be around $\gamma_{\text{opt}} (n_C+1/2)= J$. However, the precise value depends on both the hot qubit temperature and the controlled-phase gate time. 
This is to be expected since the quality of the gates is influenced by both.

\cleardoublepage
\chapter{Mott-insulator to superfluid phase transition in a lattice of transmons}
\label{chapter:PhaseTransition}

\noindent Classical phase transitions, such as the evaporation of water into vapor, happen as the temperature is increased. The temperature at which the transition occurs is called the critical temperature. On the other hand, quantum phase transitions occur at zero temperature, i.e., for the ground state, and the phase transition occurs as some relevant parameter of the Hamiltonian is increased. The Mott insulator to superfluid phase transition is a quantum phase transition between two phases. The Mott insulating phase is insulating due to a strong inter-particle interaction, while the superfluid phase is famous for its ability to flow without frictions.

These phase transitions are a popular area of study for quantum simulators. Particularly, neutral atoms trapped in an optical lattice have proven to be very powerful quantum simulators. The first quantum simulation of the Mott-insulator to superfluid phase transition was performed with neutral atoms in Ref.~\cite{Greiner2002}, after the theoretical proposal in Ref.~\cite{PhysRevLett.81.3108}. In this experiment, the canonical ensemble, i.e., with constant particle number, was explored. Since then, it has become possible to measure individual atoms in the lattice \cite{doi:10.1126/science.1192368, 10.1093/nsr/nww023} and produce higher quality lattices with a well-defined chemical potential \cite{Schafer2020, Mazurenko2017}. Likewise, disorder has been studied by adding fermionic particles \cite{PhysRevLett.96.180402, PhysRevLett.96.180403} or through the lattice itself \cite{PhysRevLett.98.130404, Billy2008, Roati2008}.

Similarly, superconducting circuits have been used for quantum simulation, e.g., for the dissipative preparation of the Mott-insulating state in Ref.~\cite{Ma2019}. However, so far, the phase transition has not been achieved. Unlike for neutral atoms in optical lattices, superconducting circuits have precise single-site control over the chemical potential, the on-site particle interaction, and the inter-site particle interaction, as well as the ability to measure in any basis for each site. Most importantly, superconducting circuits have better integration with particle reservoirs, allowing for the simulation of the phase transition in the grand canonical ensemble. In other words, an additional dimension in the phase diagram can be reached.

This chapter will study the Mott insulator to superfluid phase transition in the canonical and grand canonical ensemble for a lattice of transmons. First, the Hamiltonian for the Bose-Hubbard model is introduced, and the phases that it supports are described. Next, the phase transition is studied in the canonical ensemble through the energy gap and variance in particle number. Then, the full phase diagram is studied in the grand canonical ensemble using several different order parameters. Finally, the connection to transmons is made, fully simulating the phase diagram for adiabatic state preparation using the transmon Hamiltonian and including decoherence.

\section{Hamiltonian and phase diagram} 

The system of interest, here, is a 2D Bose-Hubbard model with the Hamiltonian \cite{sachdev2011quantum}
\begin{align}
\hat{H}_{BH} = -\mu \sum_i^{M} \hat{n}_i + \frac{U}{2} \sum_i^M \hat{n}_i \left(\hat{n}_i -1 \right) -J \sum_{ \langle i, j \rangle} \left( \hat{a}_i \hat{a}_j^\dag + \hat{a}_i^\dag \hat{a}_j \right). \label{eq:PhaseTransitionHamiltonianIntroduction}
\end{align}
Here, $M$ is the number of lattice sites, $ \hat{a}_i$ is the boson annihilation operator for the $i$th lattice site, $\hat{n}_i$ is the number operator for the $i$th site, and the last sum is over nearest neighbors. $\mu$ is the chemical potential resulting in a decrease of $\mu$ in energy if one particle is added to the lattice. $U>0$ is a repulsion between particles resulting in an energy increase of $U$ if two particles occupy the same site. $J$ is the hopping between nearest neighbor sites. 

This Hamiltonian supports two phases: the Mott insulating and superfluid phases. However, with a minor tweak, the Bose glass phase is also supported. Phase transitions only occur in the thermodynamic limit $M\rightarrow \infty$; however, this is computationally difficult. Some Mean-Field \cite{sachdev2011quantum} and Monte-Carlo \cite{RevModPhys.83.1405} techniques are applicable, but they have their limitations. Moreover, gaining information about the phase transition is still possible, even for a smaller finite lattice.

For $J=0$, the ground state is easily found to be a state with a whole number of particles, $n$, per site. For $M=2 \times 2$, these states are
\begin{align}
\left|\, \begin{matrix}
0 & 0\\ 0 & 0
\end{matrix} \right\rangle, \, 
\left|\, \begin{matrix}
1 & 1\\ 1 & 1
\end{matrix} \right\rangle, \,  
\left|\, \begin{matrix}
2 & 2\\ 2 & 2
\end{matrix} \right\rangle, \,... \,.
\end{align}
The ground state has $n$ particles per site where
\begin{align}
n &= 0 \quad \mathrm{for} \quad \mu \leq 0,\\
n &= k \quad \mathrm{for} \quad  k-1 \leq \frac{\mu}{U} \leq k. \nonumber
\end{align}
The state with $n=0$ is the vacuum state, which is a special case. In the situation with $n > 0$, an energy of $U$ is required to move any particle to another site. The movement of the particles is trapped by this energy barrier induced by the repulsion between particles. Such a state is called a Mott insulator. 
Additionally, even in the thermodynamic limit, adding or subtracting a particle requires work of between $0$ and $U$. For the special case where $\mu/U$ is a whole number, the ground state is degenerate, and one particle can be either added or subtracted but never both. Therefore, the Mott-insulating state is called incompressible. In summary, some of the defining features of the Mott-insulating phase is that it has an integer number of particles, is insulating, and is incompressible.

\begin{figure}[t]
\centering
\includegraphics[width=1\linewidth, angle=0]{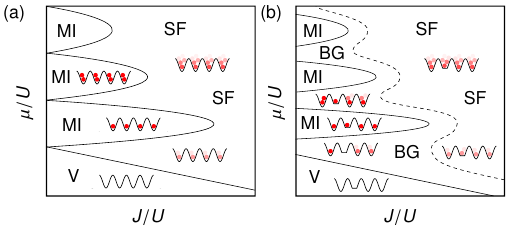}
\caption{ Sketch of the phase diagram for the Bose-Hubbard model without disorder, (a), and with disorder, (b). The abbreviations denoting the different phases are V for the vacuum state, MI for the Mott insulating phase, BG for the Bose glass phase, and SF for the superfluid phase. The pictograms show examples of the phases at different points in the phase diagram.}\label{figure:ChapterPhaseTransitionPhaseDiagram}
\end{figure}

For $\mu = U = 0$, the ground state is generally more complicated to derive. For $M \rightarrow \infty$, the single particle case is equivalent to the tight-binding model in a 2D square lattice \cite{grosso2013solid}, and the energies become
\begin{align}
E = -2t( \cos(k_1) + \cos(k_2)), \quad -\pi < k_1,k_2 \leq \pi. \label{eq:PhaseTransitionLimitEnergy}
\end{align}
However, for $\mu=U=0$, there is no penalty to adding more particles, and the full single particle ground state can be written
\begin{align}
\left( \sum_i \hat{a}_i^\dag\right)^{N} \ket{0},
\end{align}
where $N$ is the total number of particles and $\ket{0}$ is the vacuum state. Since the particles are completely evenly dispersed on the lattice $n=N/M$. So unlike the Mott-insulating phase, there is no restriction to the average number of particles per site, and $n$ can be any positive irrational number. The overall ground state contains infinite particles as a consequence of $U~=~0$. However, for a constant number of particles, an infinitesimal energy is required to reach the first excited state. In other words, any infinitesimally small nudge to the perfect ground state will set the particles into motion forever. Additionally, particles can be added or subtracted without doing any work, and as a consequence, the superfluid phase is compressible. In summary, some of the defining features of the superfluid phase is that it can have any positive $n$, is not insulating, and is compressible.

A sketch of the phase diagram for $M \rightarrow \infty$ is shown in Fig.~\ref{figure:ChapterPhaseTransitionPhaseDiagram}(a). In most experimental settings, disorder occurs naturally through $\mu$ and $U$, resulting in slight deviations of these parameters per site. With disorder, a third phase occurs called the Bose glass phase. Due to the disorder, the phase transition from Mott insulator to superfluid occurs gradually at different points in the parameter space at different locations in the lattice. Therefore, isolated puddles of superfluidity occur, resulting in an insulating and compressible phase. Similar to the superfluid phase, there is no restriction to the number of particles, and $n$ can be any positive irrational number. A sketch of the phase diagram with disorder can be seen in Fig.~\ref{figure:ChapterPhaseTransitionPhaseDiagram}(b).

The ground state of the Hamiltonian in Eq.~\eqref{eq:PhaseTransitionHamiltonianIntroduction} is most easily prepared using adiabatic state preparation. For example, the Mott insulating state can be prepared for $J=0$, and $J$ can subsequently be turned on slowly to prepare the superfluid state. For a closed system, energy can only be added through work, i.e., the time-dependence of the Hamiltonian. For a sinusoidal driving, $\cos(\omega t)$, transitions of frequency $\omega$ are driven as was seen in the introduction to chapter \ref{chapter:GMR}. We let $\Delta E$ be the minimum energy splitting between the ground state and first excited state throughout the traversal of $J$. The system will stay in the ground state if the time dependence is slow enough that this transition can not be driven, i.e., the Fourier transform of $J(t)$ should only have nonzero values for $\omega \ll \Delta E$. 

In most experimental setups, the number of particles remains constant during this adiabatic ramp. This corresponds to the canonical ensemble, where the system can exchange energy but not particles with a reservoir. In the canonical ensemble, $\mu$ becomes irrelevant, and the phase transition depends only on $U$ and $J$. For the full phase diagram, the adiabatic state preparation has to allow the system to exchange particles with a reservoir. This would correspond to the grand canonical ensemble, where the system can exchange both energy and particles with a reservoir.

\begin{figure}[t]
\centering
\includegraphics[width=1\linewidth, angle=0]{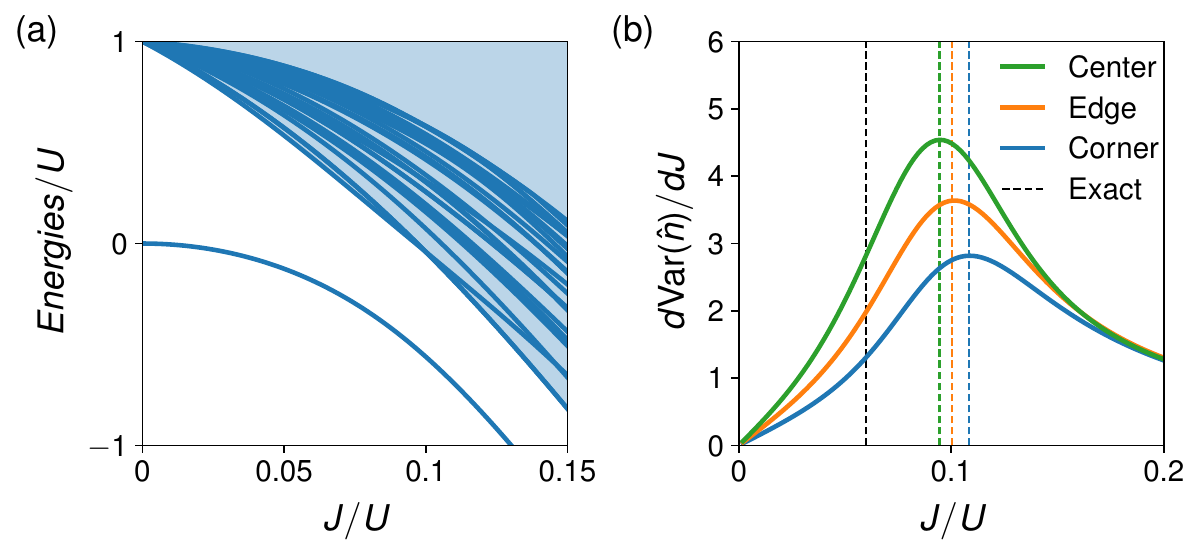}
\caption{ (a) Lowest 30 Eigenenergies of the Hamiltonian in Eq.~\eqref{Eq:PhaseTransitionCanonicalHamiltonian} with $M =3\times 3$ sites and $N=9$ particles. (b) The derivative of the variance of the single site particle number as a functions of the hopping $J$ for a center, edge, or corner site. The dashed vertical lines correspond to the maximum values, and the black dashed line corresponds to the exact critical hopping. Four levels were included per site.}\label{figure:ChapterPhaseTransitionCanonical}
\end{figure}

\section{The canonical ensemble}

In the canonical ensemble, the Bose-Hubbard lattice can exchange energy but not particles with some bath at zero temperature. This will drive the state into the lowest energy state at some fixed particle number determined by the initial state. Since the particle number is constant, $\mu$ is irrelevant, and the Hamiltonian becomes
\begin{align}
\hat{H}_{BH}(\mu=0) = \frac{U}{2} \sum_i^{M} \hat{n}_i \left(\hat{n}_i-1 \right) -J \sum_{ \langle i, j \rangle} \left( \hat{a}_i \hat{a}_j^\dag + \hat{a}_i^\dag \hat{a}_j \right). \label{Eq:PhaseTransitionCanonicalHamiltonian}
\end{align}
For unity filling the critical value is $(J/U)_{c}\simeq 0.06$ \cite{Lacki2016, Krauth_1991}. 

Since only particle subspaces with $N=kM$ for any whole number $k>0$ support the Mott insulator state, we will study the particle subspace with $N=9$ particles in a $M=3 \times 3$ lattice. In Fig.~\ref{figure:ChapterPhaseTransitionCanonical}(a), the 30 lowest eigenenergies are plotted. For $J=0$, there is a clear band gap between the energy of the lowest state and the other energies. As $J$ increases, the excited states split into a distinct band. Since the system is finite, it can be difficult to identify bands. However, as the ground state in the superfluid phase is part of an energy band, the band gap is expected to close at the critical hopping $J$. Looking at Fig.~\ref{figure:ChapterPhaseTransitionCanonical}(a), the band gap is getting smaller. However, estimating the critical value from this spectrum is impossible. 

In Ref.~\cite{PhysRevA.77.015602}, it was proven that the critical value can be estimated through the value of $J$ maximizing 
\begin{equation}
\frac{d}{dJ} \mathrm{Var} (\hat{n}_i), \quad \mathrm{where} \quad \mathrm{Var} (\hat{n}_i) = \langle n_i^2 \rangle - \langle n_i \rangle^2.
\end{equation}
This quantity is plotted in Fig.~\ref{figure:ChapterPhaseTransitionCanonical}(b) as a function of $J$, and the maximum is shown with dashed lines. The black dashed line is the exact value of $(J/U)_{c}\simeq 0.06$ \cite{Lacki2016, Krauth_1991}. A three-by-three lattice is too small to give a precise estimate of the critical coupling. However, it does yield a decent estimate. Additionally, the center site yields the best estimate while also being the only site with four nearest neighbors. In an infinite 2D lattice, every site has four nearest neighbors, and we would, therefore, expect a better estimate from the center.

\begin{table}[t]
\centering
\begin{tabular}{|c|c|c|c|}
\hline
\quad & Mott insulator & Superfluid & Bose-Glass \\ \hline
$n$ & $\in \mathbb{N}$ & $\in \mathbb{R}$ & $\in \mathbb{R}$\\ \hline
$a$ & 0 & $> 0$ & $>0$\\ \hline
$\kappa$ & $0$ & $\simeq 0$ & $> 0$\\ \hline
\end{tabular}
\caption{Order parameter values for the three phases supported by the Bose-Hubbard model.}
\label{table:ChapterPhaseTransitionOrderParameters}
\end{table}

\section{The grand canonical ensemble}
\label{sec:PhaseTransitionGrand}

In the grand canonical ensemble, the Bose-Hubbard lattice can exchange energy and particles with some bath at zero temperature. This will drive the state into the overall ground state, allowing us to explore the entire phase diagram sketched in Fig.~\ref{figure:ChapterPhaseTransitionPhaseDiagram}(a). Therefore, $\mu$ is relevant again, and we need the full Hamiltonian
\begin{align}
\hat{H}_{BH} = -\mu \sum_i^M \hat{n}_i + \frac{U}{2} \sum_i^M \hat{n}_i \left(\hat{n}_i -1 \right)   -J \sum_{ \langle i, j \rangle} \left( \hat{a}_i \hat{a}_j^\dag + \hat{a}_i^\dag \hat{a}_j \right). \label{eq:eq:PhaseTransitionGrandHamiltonian}
\end{align}
To distinguish the three different phase, there are three common order parameters
\begin{align}
n = \frac{1}{M} \sum_i \langle \hat{n}_i \rangle, \quad
a = \Big| \frac{1}{M} \sum_i \langle \hat{a}_i \rangle \Big|, \quad
\kappa = \frac{1}{M} \sum_i \langle \hat{n}_i \rangle^2 - \Big( \frac{1}{M} \sum_i \langle \hat{n}_i \rangle \Big)^2. \label{Eq:1}
\end{align}
$n$ is the average number of particles per site, $a$ is the norm of the average value of $\langle \hat{a}_i \rangle$ for the entire lattice, and $\kappa$ is called the Edwards-Anderson order parameter. The Edwards-Anderson order parameter is the classical variance of $\langle \hat{n}_i \rangle$ taken over all sites, and therefore, it measures how uneven the particles are distributed. 

\begin{figure}[t]
\centering
\includegraphics[width=1\linewidth, angle=0]{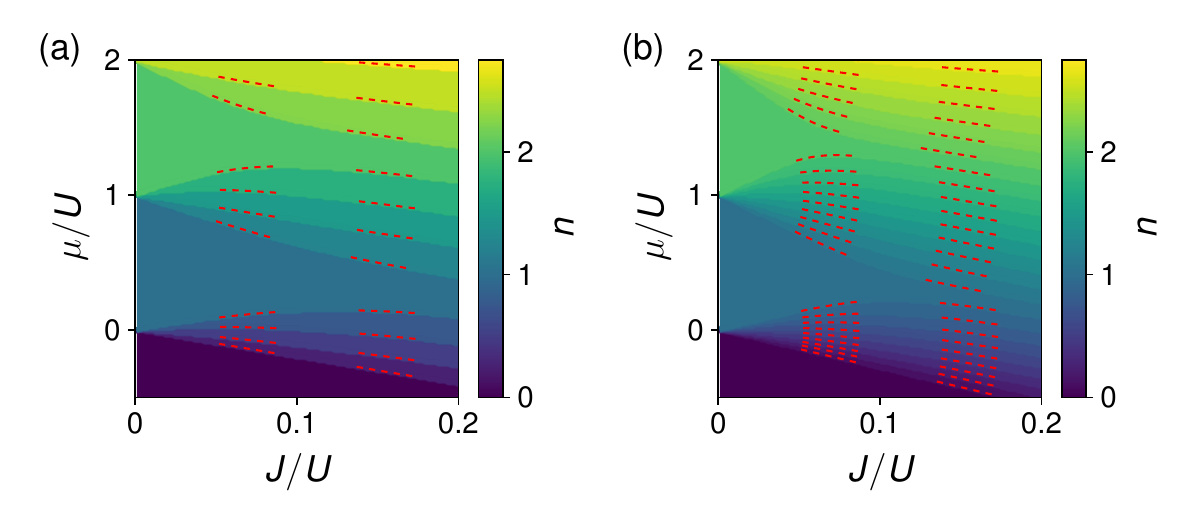}
\caption{ The order parameter $n$ as a function of both $J$ and $\mu$ for a two by two lattice, (a), and a three-by-three lattice, (b). Five levels were included per site.}\label{figure:ChapterPhaseTransitionGrand}
\end{figure}

A subtle but important point is connected to $a$. The Bose-Hubbard Hamiltonian is invariant under an overall phase, $\theta$, on the ladder operators.
\begin{align}
\hat{a}_i \rightarrow e^{i\theta} \hat{a}_i.
\end{align}
This is called a global $U(1)$ symmetry. As a consequence, the Hamiltonian is particle preserving, and a set of eigenstates with a well-defined particle number can be constructed, resulting in $a=0$. A nonzero $a$ is only possible for a degenerate ground state that mixes two or more different total particle numbers. 
In other words, the ground state has to be a superposition of two or more states with particle numbers that differ by 1. If one state has a total particle number $k$, the other has to have a total particle number $k-1$ or $k+1$. 
This is equivalent to the state being compressible, and a superfluid, therefore, always has $a \neq 0$. 

A full list of order parameter values in the three phases can be seen in table \ref{table:ChapterPhaseTransitionOrderParameters}. 
The first order parameter $n$ is plotted for $M=2\times 2$ and $M=3\times 3$ in Fig.~\ref{figure:ChapterPhaseTransitionGrand}. 
The two first Mott insulator regions are identifiable as the two regions with $n=1$ and $n=2$. The dashed red lines marks when the degeneracy requirement discussed above is fulfilled, i.e., $a\neq 0$. 
Looking at the difference between panels (a) and (b) in Fig.~\ref{figure:ChapterPhaseTransitionGrand}, we see the Mott insulating regions become smaller and more pronounced. The order parameter $n$ can take more values, and the number of lines with $a\neq 0$ increases. Furthermore, the second Mott insulator region with $n=2$ is smaller than the first Mott insulating region with $n=1$. 
This is as expected from the sketch in Fig.~\ref{figure:ChapterPhaseTransitionPhaseDiagram}. 

As $M \rightarrow \infty \times \infty$, the Mott insulating regions should narrow even further. Additionally, there will be infinitely many lines of $a\neq 0$ eventually forming a region of $a\neq 0$ and the steps in $n$ will become smaller, and eventually, $n$ becomes continuous. 
This region of $a\neq 0$ and continuous change in $n$ is the region of superfluidity. Therefore, Fig.~\ref{figure:ChapterPhaseTransitionPhaseDiagram}(b) is a precursor to the phase diagram, and the superfluid part of the phase diagram will only have $a\neq 0$, be truly compressible, and have no band gap in the thermodynamic limit, i.e., $M \rightarrow \infty \times \infty$. 
However, Fig.~\ref{figure:ChapterPhaseTransitionPhaseDiagram}(b) is still very instructive and shows a good approximation of the phase diagram at a fairly small number of sites.

\begin{figure}[t]
\centering
\includegraphics[width=1\linewidth, angle=0]{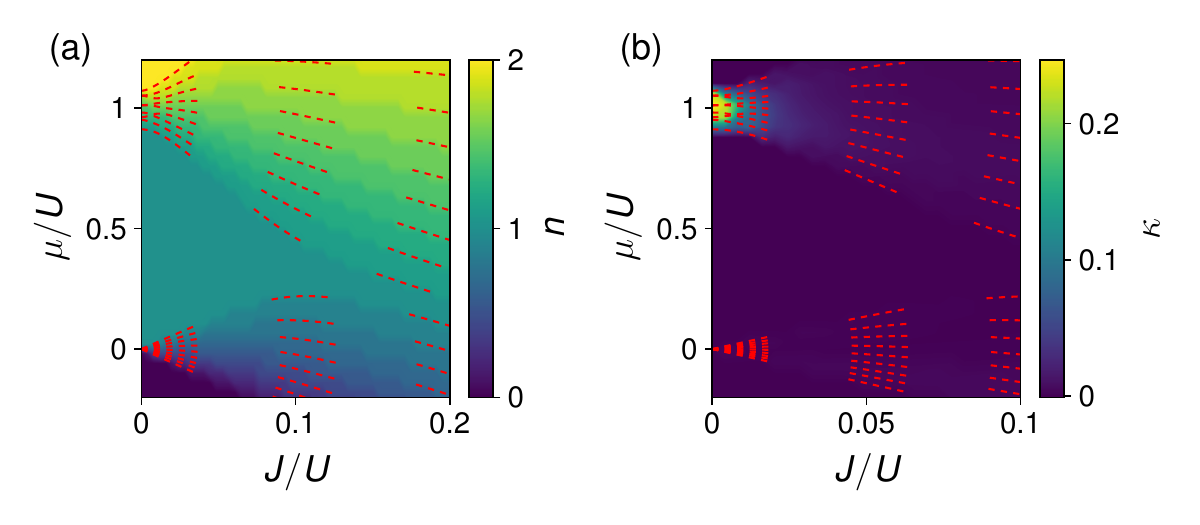}
\caption{ Average number of particles, (a), and the Edwards-Anderson order parameter, (b), as a function of both $J$ and $\mu$. A lattice of $M=3\times 3$ site was used, and four levels were included per site. }\label{figure:ChapterPhaseTransitionDissorder}
\end{figure}

\subsection{Disorder and the Bose-glass phase}
\label{subsec:ChapterPhaseTransitionDissorder}

So far, we have kept all parameters constant across the lattice, however, this is never the case in a real setup. Instead, there will be small variations, allowing for the third phase discussed in section \ref{sec:PhaseTransitionGrand}. In a lattice of transmons, the onsite chemical potential and intersite couplings can be controlled; see section \ref{sec:PhaseTransitionLatticeOfTransmons} for more information. Therefore, the disorder in $\mu$ and $J$ can be calibrated away. On the other hand, $U$ is not controllable, and we will consider disorder in $U$ through the Hamiltonian
\begin{align}
\hat{H}_{BH,\mathrm{dis}} = -\mu \sum_i \hat{n}_i + \sum_i \frac{U + \delta U_i}{2} \hat{n}_i \left(\hat{n}_i -1 \right)   -J \sum_{ \langle i, j \rangle} \left( \hat{a}_i \hat{a}_j^\dag + \hat{a}_i^\dag \hat{a}_j \right).
\end{align}
We let the lattice have $M=3\times 3$ sites with disorder
\begin{align}
&\delta U_1 \simeq 0.05U, \, \delta U_2 \simeq -0.02U, \, \delta U_3 \simeq 0.01U, \nonumber \\ 
& \delta U_4 \simeq 0.07U, \, \delta U_5 \simeq 0.05U,\, \delta U_6 \simeq -0.09U, \\
& \delta U_7 \simeq 0.01U, \, \delta U_8 \simeq -0.05U, \, \delta U_9 \simeq -0.04U. \nonumber
\end{align}
These were sampled randomly such that 
\begin{align}
\mathrm{Mean} (\delta U_i) =\frac{1}{M} \sum_{i}^M \delta U_i = 0, \quad \mathrm{and} \quad \mathrm{Var} (\delta U_i) = \frac{1}{M} \sum_{i}^M \delta U_i^2 \simeq 0.05U.
\end{align}
The two order parameters $n$ and $\kappa$ are plotted in Fig.~\ref{figure:ChapterPhaseTransitionDissorder}(a) and Fig.~\ref{figure:ChapterPhaseTransitionDissorder}(b), respectively. Comparing Fig.~\ref{figure:ChapterPhaseTransitionDissorder}(a) with Fig.~\ref{figure:ChapterPhaseTransitionGrand}(b), it is clear that the Mott insulator region gets more narrow in $\mu$. From Fig.~\ref{figure:ChapterPhaseTransitionDissorder}(b), it is clear that this new region, which used to be a Mott insulator, has $\kappa > 0$. This is the hallmark of the Bose-glass phase, which has been induced by the disorder in $U$. 

Note that the Bose-glass phase only appears around $\mu=U$ and not $\mu=0$ as shown in Fig.~\ref{figure:ChapterPhaseTransitionPhaseDiagram}(b). Since $U$ is unimportant when there is less than one particle per site, we would not expect a region of Bose glass to appear around $\mu=0$. The phase diagram seen in Fig.~\ref{figure:ChapterPhaseTransitionPhaseDiagram}(b) was drawn for the case where disorder is included for $\mu$ instead of $U$.

\section{Lattice of transmons}
\label{sec:PhaseTransitionLatticeOfTransmons}

So far, we have identified the regions of Mott insulator, Bose glass, and superfluid within the phase diagram for a three-by-three lattice. Such a lattice can be built from the transmon superconducting qubit as seen in Ref.~\cite{Karamlou2022}. The Hamiltonian for a lattice of transmons is \cite{Yanay2020, PhysRevLett.126.180503, PRXQuantum.2.040204}
\begin{align}
\hat{H}_{transmon} &= \omega \sum_i \hat{n}_i - \frac{U}{2} \sum_i \hat{n}_i (\hat{n}_i - 1) + J \sum_{\langle i,j \rangle} \left( \hat{a}_i \hat{a}_j^\dag + \hat{a}_i^\dag \hat{a}_j \right) \\ 
&\hspace{4.5cm} + \sum_i \chi \left( \hat{a}_i e^{i(\omega - \mu)t} + \hat{a}_i^\dag e^{-i(\omega - \mu)t} \right). \nonumber
\end{align}
$\omega$ is the transmon frequency, $U$ is called the anharmonisity, $J$ is the nearest-neighbor coupling, and $\chi$ is a tunable single qubit coupling. These parameters typically take values close to $\omega/2\pi \sim 10\mathrm{GHz}$, $U/2\pi \sim 100\mathrm{MHz}$, $J/2\pi \sim 10\mathrm{MHz}$, and $\chi/2\pi \sim 10\mathrm{MHz}$ \cite{10.1063/1.5089550}. As mentioned in subsection \ref{subsec:ChapterPhaseTransitionDissorder}, $\omega$, $J$, and $\chi$ are all tunable, and disorder in these parameters can be calibrated away. If the transmons are coupled through tuneable couplers, the hopping can traverse the interval $0 \leq J <0.3U$ \cite{PhysRevApplied.10.054062, PhysRevX.11.021058}. The term containing $\chi$ is achieved by coupling each transmon to a microwave field and is usually used for single qubit gates. Here all transmons couples to the microwave field with the same strength. 

This Hamiltonian has obvious similarities with the Bose-Hubbard model; However, there are a few problems to solve, e.g., the transmon frequency is much larger than the anharmonicity $\omega \sim 100U$. To solve these problems, we first go into the interaction picture with respect to \(\hat{H}_0 = (\omega - \mu)\sum_i \hat{n}_i \) and make the transformation \(\hat{H}_I \rightarrow -\hat{H}_I\)
\begin{equation}
\hat{H}_{I} = -\mu \sum_i \hat{n}_i + \frac{U}{2} \sum_i \hat{n}_i (\hat{n}_i - 1) - J \sum_{\langle i,j \rangle} \left( \hat{a}_i \hat{a}_j^\dag + \hat{a}_i^\dag \hat{a}_j \right) - \sum_i \chi_i(t) \left( \hat{a}_i + \hat{a}_i^\dag \right).
\end{equation}
For $\chi_i=0$, this is exactly the Bose-Hubbard model. Because of the transformation above, the ground state of the Bose-Hubbard model is the highest excited state of the transmon Hamiltonian in the interaction picture. However, since we wish to use adiabatic state preparation to prepare the ground state of the Bose-Hubbard model, this is not a problem.

\subsection{Adiabatic state preparation}
\label{subsec:PhaseTransitionAdiabatic}

The ground state of the Bose Hubbard model is prepared for some set of parameters where the ground state is known. If the parameters are then changed slowly enough to the desired values, the system will stay in the ground state unless some symmetry prevents it. Since the Hamiltonian for the Bose-Hubbard model is particle preserving, the lattice is coupled to a particle reservoir through $\chi_i \neq 0$. This reservoir is in a coherent state and not a thermal state, so it will not induce any non-unitary dynamics. Instead, it can provide particles whenever the ground state changes from one particle subspace to another. This is what allows us to traverse the entire phase diagram. More specifically, the interaction with the particle reservoir is the first thing to be turned on and the last thing to be turned off. 

In order to not presume any prior information on the phase diagram, the adiabatic state preparation is started in the vacuum state. For $\mu \ll -|J|$, the cost of adding a particle is large enough that no particles are present in the ground state. In fact, since the Bose-Hubbard model can be easily solved for one particle, the boundary between the vacuum state and the superfluid region can be calculated. The boundaries for a few different lattice sizes are
\begin{align}
&M = 2\times 2 : \quad \quad \mu = -2J, \\
&M = 3\times 2 : \quad \quad \mu = -(1+\sqrt{2})J \simeq 2.41J, \\
&M = 3\times 3 : \quad \quad \mu = -2\sqrt{2}J \simeq -2.83J, \\
&M = 5\times 5 : \quad \quad \mu = -2\sqrt{3}J \simeq -3.46J, \\
&M = \infty \times \infty : \quad \mu = -4J.
\end{align}
For the thermodynamic limit, the spectrum given by Eq.~\ref{eq:PhaseTransitionLimitEnergy} was used. 

\begin{figure}[t]
\centering
\includegraphics[width=1\linewidth, angle=0]{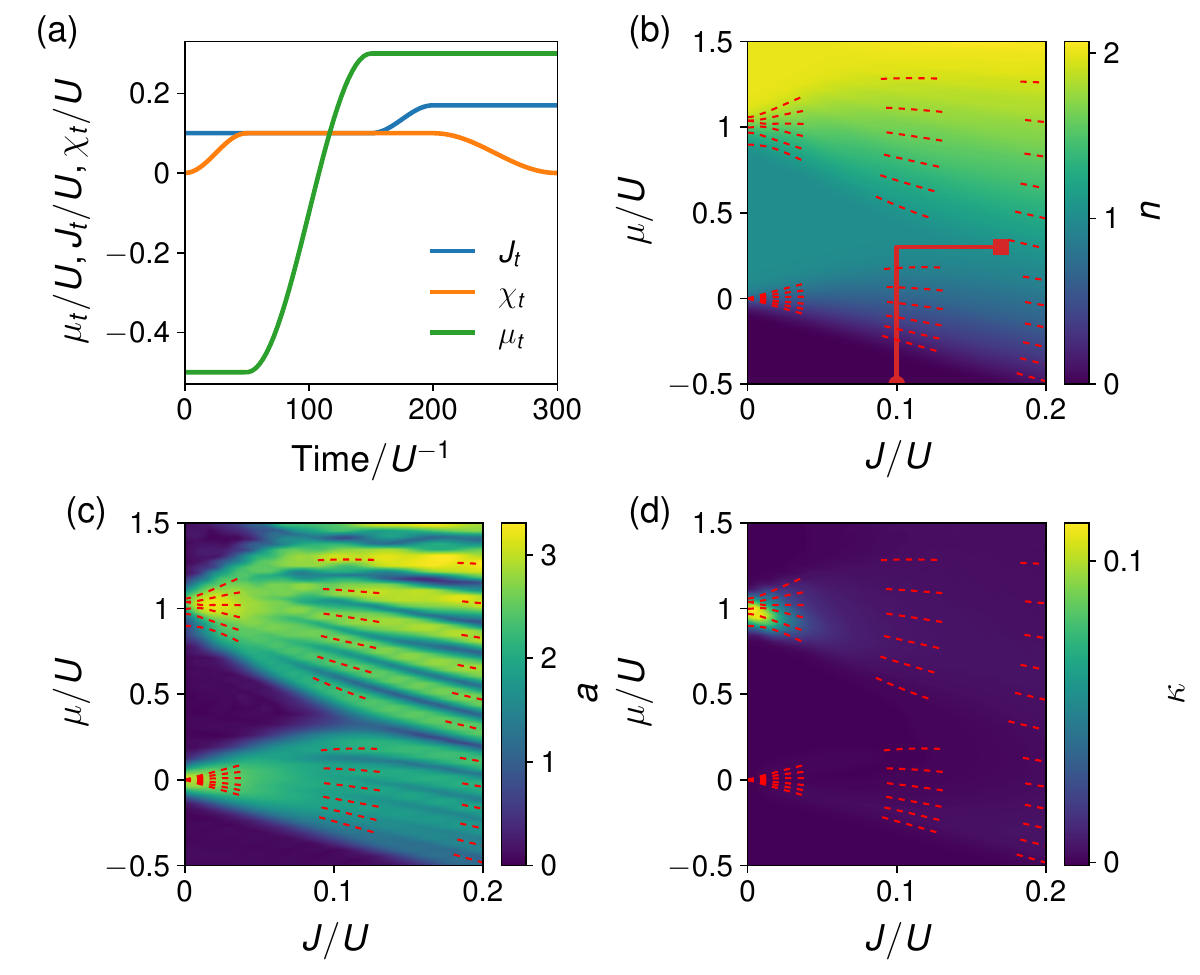}
\caption{ (a) The time-evolution of the three system parameters during adiabatic state preparation for the endpoint $\mu_{t_4}=0.3$ and $J_{t_4}=0.17$. (b)-(d) The three order parameters $n$, $a$, and $\kappa$, respectively, as a function of both $\mu$ and $J$. The circle, solid line, and square in panel (b) show the start, path, and end of the adiabatic state preparation shown in panel (a). A lattice of $M=3\times 2$ site was used, and four levels were included per site.}\label{figure:ChapterPhaseTransitionAdiabatic6}
\end{figure}

Denoting the time-dependent system parameters $\mu_t$, $J_t$, and $\chi_t$, the adiabatic state preparation is started at $\mu_0 = -0.5U$ and $J_0 = 0.1U$, which is in the vacuum region for any size lattice. The adiabatic state preparation is split into four steps, each ending at times $t_1$, $t_2$, $t_3$, and $t_4$, respectively. 
\begin{itemize}
\item[] Step 1: $\chi_t$ is ramped from $\chi_{0} = 0$ to $\chi_{t_1} = 0.1U$.
\item[] Step 2: $\mu_t$ is ramped from $\mu_{t_1} = -0.5U$ to $\mu_{t_2}=\mu$.
\item[] Step 3: $J_t$ is ramped from $J_{t_2} = 0.1U$ to $J_{t_3}=J$.
\item[] Step 4: $\chi_t$ is ramped from $\chi_{t_3} = 0.1U$ to $\chi_{t_4} = 0$.
\end{itemize}

\noindent Additionally, we use a cosine ramp, and the duration of the four steps are $t_1 = 50U^{-1}$, $t_2 - t_1 = 100U^{-1}$, $t_3 - t_2 = 50U^{-1}$, and $t_4 - t_3 = 100U^{-1}$, respectively. The total state preparation time is $t_4 = 300U^{-1}$. An example of the system parameters during the adiabatic state preparation can be seen in Fig.~\ref{figure:ChapterPhaseTransitionAdiabatic6}(a). 

In Figs.~\ref{figure:ChapterPhaseTransitionAdiabatic6}(b)-(d), the three main order parameters are plotted after the adiabatic state preparation. The plots for $n$ and $\kappa$ are very similar to Fig.~\ref{figure:ChapterPhaseTransitionGrand}; however, both are more smooth. The effectiveness of adiabatic state preparation depends on the duration versus the minimum energy splitting to the first excited state. Close to where the ground state changes the total number of particles, shown by dashed lines in Fig.~\ref{figure:ChapterPhaseTransitionAdiabatic6}, the final energy splitting between the ground state and the first excited state is very small. Therefore, the final state is a superposition of the ground state and the first excited state, resulting in a less abrupt change in $n$. Additionally, a non-zero $a$ is possible around a point of degeneracy, and $a$ can now be plotted; This is done in Fig.~\ref{figure:ChapterPhaseTransitionAdiabatic6}(c). The symmetry breaking of the superfluid phase ($a\neq0$) is now clearly visible as a consequence of the imperfect ground state prepared by the adiabatic state preparation.

\begin{figure}[t]
\centering
\includegraphics[width=1\linewidth, angle=0]{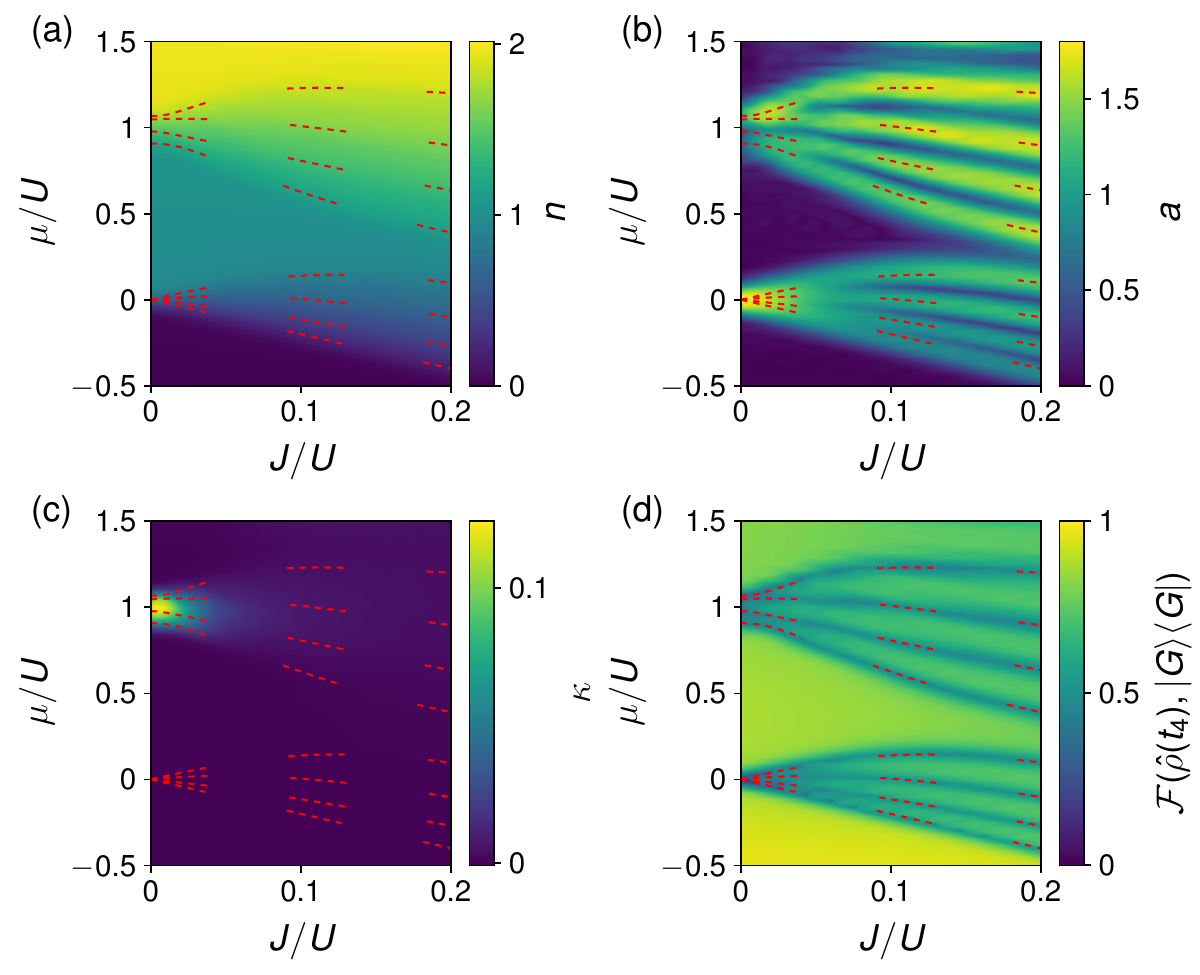}
\caption{ (a)-(c) The three order parameters $n$, $a$, and $\kappa$, respectively, as a function of both $\mu$ and $J$. (d) The fidelity between the overall ground state of the Bose-Hubbard model and the state prepared through adiabatic state preparation. A lattice of $M=2\times 2$ site was used, and four levels were included per site. }\label{figure:ChapterPhaseTransitionAdiabatic4}
\end{figure}

\subsection{Adiabatic state preparation with decoherence}

Finally, we will test the feasibility of the phase diagram simulation by including decoherence. Similar to other chapters, the state of the system is described by the density matrix $\hat{\rho}$, and the evolution of the density matrix is governed by the local master equation on Lindblad form
\begin{equation}
\frac{d \hat{\rho}}{d t} = -i [\hat{H}, \hat{\rho}] + \sum_i^M \mathcal{D}_i[\hat{\rho}].
\end{equation}
The decoherence it described through
\begin{align}
\mathcal{D}_i[\hat{\rho}] = \gamma_1 \mathcal{M}[\hat{a}_i, \hat{\rho}] + 2\gamma_2 \mathcal{M}[\hat{a}_i^\dag \hat{a}_i, \hat{\rho}],
\end{align}
where $\gamma_1^{-1} = 15000U^{-1}$ and $\gamma_2^{-1} = 3000U^{-1}$. Since the dephasing rate is much larger than the decay rate, the decay and dephasing coherence rates are $T_1 = \gamma_1^{-1}$ and $T_2 \simeq \gamma_2^{-1}$. For $U/2\pi \sim 100\mathrm{MHz}$, these coherence times are $T_1 \sim 24\mathrm{\mu s}$ and $T_2 \sim 5\mathrm{\mu s}$ which is much lower than state-of-the-art. However, with a large lattice of transmons coupled to individual control and measurement, this is to be expected.

Simulating open systems is much more computationally difficult, so we will only study a lattice of $M=2\times 2$ sites. The three order parameters are plotted in Figs.~\ref{figure:ChapterPhaseTransitionAdiabatic4}(a)-(c). The order parameters are very similar to the results found in subsection~\ref{subsec:PhaseTransitionAdiabatic}. Since $n$ can only take three values between each Mott insulating region, the plateaus can be seen again. Likewise, the individual lines of $a\neq 0$ are clearly identifiable. 

Finally, we let $\ket{G}$ be the overall ground state of $\hat{H}_{BH}$ for a given set of parameters. To properly identify the effect of decoherence, the fidelity between the state prepared using adiabatic state preparation and $\ket{G}$ is plotted in Fig.~\ref{figure:ChapterPhaseTransitionAdiabatic4}(d). The fidelity used is defined as
\begin{align}
\mathcal{F}(\hat{\rho} (t_4), \op{G}) = \langle G| \hat{\rho} (t_4) |G \rangle.
\end{align}
The fidelity drops around the lines of degeneracy, as expected. Near these lines, the remaining fidelity lies in the first excited state, which is very close to the ground state in energy. Outside these lines, the fidelity is large for smaller $\mu$ but decreases for larger $\mu$. This is because a larger $\mu$ results in states with more particles. Both decay and dephasing occur faster for states with more particles. In a superconducting circuit, dephasing does not necessarily behave this way \cite{10.1063/1.5089550}. Dephasing is not fully understood, and the aim is not to simulate decoherence in the most realistic way. Instead, the aim is to test the robustness of the results towards random decay and dephasing. 

Finally, measurement was not taken into account here. The measurement itself can be imperfect, and since performing a measurement is a slow process, the state can decay during this time. Measurements are always performed in the Z-basis; Therefore, dephasing is unimportant. For measurement of, e.g., $a$, the measurement consists of a basis change through single-site gates followed by a measurement in the Z-basis. The basis change is fast, and dephasing is again of little concern.

\cleardoublepage
\chapter{Conclusion and final remarks}
\label{chap:conclusion}

This thesis touches on many different aspects of transport in quantum systems, measurement and information theory, and statistical physics. As a consequence, it is difficult to make a single coherent conclusion for the entire thesis. Instead, this final chapter will present a brief summary of the main results from each chapter. These conclusions are not self-contained, and the reader is referred to the individual chapters for a full discussion on the subject. Additionally, I will give some final remarks on the PhD project and future directions of the field.

\section{Chapter \ref{chapter:Introduction}: \nameref{chapter:Introduction}}

A diverse range of subjects are discussed in this thesis, each with a long history of prior results. The theoretical context is long and goes more than a century back with thought experiments such as Maxwell's demon and Schrödinger's cat. After the invention of the computer, numerical experiments of larger systems have become possible, and transport properties such as rectification have been studied in a diverse set of systems. Rectification is a reasonably common property in systems as small as one two-level system. However, for large rectification values, more complicated systems are required. With recent advances in quantum computing and superconducting circuits, experimental results on boundary-driven quantum systems and realizations of Maxwell's demon have begun appearing, making theoretical studies even more relevant.

\section{Chapter \ref{chapter:MasterEquations}: \nameref{chapter:MasterEquations}}

A quantum master equation is an equation that describes the time evolution of a quantum system in the presence of one or more baths. Several different master equations exist for different purposes and in different regimes. The global master equation can be derived from a general model of the system and bath under certain assumptions, e.g., the Born-Markov approximation. For a sufficiently flat one-sided Fourier transform of the bath correlation functions, the global master equation becomes local in the interaction with the system. Alternatively, the global master equation can be simplified using the secular approximation. Numerical results show that these three master equations are valid in the expected regimes.

\section{Chapter \ref{chapter:QutritDiode}: \nameref{chapter:QutritDiode}}

Rectification is present in systems as small as a single two-level system, although with fairly small rectification values. A perfect rectifier is found with only three levels through careful engineering of the bath interactions. The system is driven into a dark state in reverse bias, which completely blocks transport. An implementation of this three-level rectifier can be built using a qutrit and two harmonic oscillators. This implementation approaches the perfect diode in the limit of large anharmonicity for the qutrit. The rectification results are supported by an analytic solution obtained in the limit of strong bath system coupling. Additionally, the rectification is protected from decoherence. The rectifier is then put as a part of a larger circuit, namely a full wave bridge rectifier. The output bias is independent of the input bias as intended.

\section{Chapter \ref{chapter:Interference}: \nameref{chapter:Interference}}

The dark-state-induced perfect rectifier can also be implemented using a dark state that relies on interference instead of energy conservation. This is done through two linear chain segments of spin connected via a two-way interface. For an appropriate set of parameters, the interface is driven into an entangled state in reverse bias. The explanation for this is twofold. First, the entangled state blocks transport through interference, which both prevents the state from decaying into the cold bath and blocks transport from the hot bath. Second, the entangled state develops in reverse bias because of a recovery mechanism that relies on the cold bath. The system is found to exhibit rectification even using the global master equation without any global excitation energy for all spins. For the local master equation, rectification values of $>10^6$ are found, while for the global master equation, rectification values of $>10^8$ are found.

\section{Chapter \ref{chapter:WB}: \nameref{chapter:WB}}

A similar two-way spin chain as for the interference rectifier is found to implement a quantum version of the Wheatstone bridge. An entangled state is again developed in steady state for a particular set of parameters. This mechanism is broken when a controllable coupling is tuned to the balance point where the entangled state is destroyed. The criteria for the balance point is calculated, and remarkably, it only depends on three system parameters: the unknown coupling, the controllable coupling, and an external magnetic field. The sensitivity can be measured using the quantum Fisher information, the maximum of which can be controlled via the external magnetic field. The entangled state can be measured using the spin current. An approximate solution so the state, Fisher information, and spin current can be found, and both the state and current have a Lorentzian shape. Finally, the mechanism is shown to persist even with decoherence and uncertainties on all parameters.

\section{Chapter \ref{chapter:GMR}: \nameref{chapter:GMR}}

Giant magnetoresistance is found in a chain of weakly interacting chains of strongly interacting spins. First, the transport dependence on an external magnetic field is studied in a small chain of two strongly interacting spins. The peaks of large current are explained through a resonance condition, and an analytic solution for this simple case is found. The same resonance condition applies to multiple weakly interacting chains of two spins. For larger spin chains, more spin current resonances are found. These are explained by finding the single excitation spectrum for the strongly interacting chain and finding the appropriate magnetic field that allows an excitation to travel all the way through the chain. 
The derived resonances even work for a mix of strongly interacting chains with a different number of spins. Finally, it is shown that the addition of a Z-coupling has two main effects. The resonances from before split up into several new resonances, and entirely new ones appear that were forbidden before. The spin current profile generally becomes chaotic for chains of more than four strongly interacting spins.

\section{Chapter \ref{chapter:Maxwell}: \nameref{chapter:Maxwell}}

A Maxwell's demon setup can be constructed using a qutrit and two baths. The cold bath is allowed to interact with the second excited state, and the hot bath is allowed to interact with the first excited state. The demon can then sort excitations from the cold to the hot bath by transferring the qutrit from the second excited state to the first excited state. This is done by letting the demon acquire and use information on the state of the qutrit. The demon's memory is modeled using a separate two-level system. Additionally, the two baths are non-Markovian, allowing for information to flow back from the bath to the qutrit. If the demon operates more than once, the effectiveness of the demon depends greatly on the timing of the second operation. This effect is less pronounced but still present for infinite operations of the demon in the long time limit. This non-Markovian enhancement of the transferred number of excitations is primarily due to non-Markovian effects in the cold bath.

\section{Chapter \ref{chapter:PhaseTransition}: \nameref{chapter:PhaseTransition}}

The phase transition from a Mott insulator to a superfluid occurs in the Bose-Hubbard model at zero temperature. The Bose-Hubbard model is almost identical to the Hamiltonian for a lattice of transmons, which is one of the leading architectures for superconducting qubits. For a constant number of particles, the critical value for the system parameters can be estimated. For a variable number of particles, the phase diagram for the Bose-Hubbard model can be probed using order parameters such as the average number of particles per site. If disorder is included in the lattice, the third state of Bose glass is also present in the phase diagram. For a lattice of three-by-three sites, the phase diagram resembles the actual phase diagram. In the context of superconducting circuits, the ground state of the Bose-Hubbard model can be prepared using adiabatic state preparation. If the circuit is allowed to interact with a coherent field, particles can be supplied, and the overall ground state can be prepared. The adiabatic state preparation is started for a large negative chemical potential for which the vacuum is the ground state. The ground state could be prepared numerically with large fidelity values using adiabatic state preparation both without and with decoherence.

\section{Final remarks}

The intersection between statistical physics and quantum mechanics is a fascinating and puzzling area of physics. With the approaching second quantum revolution, the subject is slowly turning from fundamental physics into a vital subject in the engineering of quantum technologies. Additionally, statistical quantum systems are excellent systems to study on the first generation of quantum computers due to their similarity with the quantum hardware. This was exploited in chapter \ref{chapter:PhaseTransition} or in the simulation of the transverse field Ising model from Ref.~\cite{Kim2023}.

The research process is never linear, even though it is often presented that way, and the same is true for my work during this PhD. In fact, the system described in chapter \ref{chapter:Interference} was the very first theme of my study. The almost three years of detective work to finally figure out the mechanism for rectification in this system was one of the most rewarding puzzles I have solved.

It was very important to me that everything should be realizable. This goal gets easier to achieve as my knowledge of experiments increases. Knowing the progression of the projects, starting with the system in chapter \ref{chapter:Interference} and finishing in chapter \ref{chapter:PhaseTransition}, this progression in my experimental knowledge becomes apparent. However, I do still believe that everything in this thesis is realizable in experiments using current technologies with some margin to spare. In fact, to this day, it still blows my mind how Maxwell's demon has become completely standard in modern experimental physics in the form of active resetting or some forms of error correction. I am sure that this technological development will continue and that I will be just as amazed by the state of quantum computers in 20 years.

%

\backmatter


\footnotesize
\setlength{\bibsep}{1pt plus 0.3ex}
\bibliography{bibliography}


\end{document}